\documentclass{aa}
\usepackage[varg]{txfonts}
\usepackage{graphicx}
\usepackage{lscape}

\begin{document} 

\def\kms{km\,s$^{-1}$}
\def\teff{$T_{\rm eff}$}
\def\logg{$\log{g}$}
\def\vmic{$\upsilon_t$}

\titlerunning{On a new and homogeneous metallicity scale for Galactic classical Cepheids}
\authorrunning{B. Proxauf et al.}

\title{On a new and homogeneous metallicity scale \\ for Galactic classical Cepheids}

\subtitle{I. Physical parameters
\thanks{Partly based on observations made with ESO Telescopes at the La Silla/Paranal 
Observatories under program IDs: 072.D-0419, 073.D-0136 and 190.D-0237 for HARPS spectra; 
084.B-0029, 087.A-9013, 074.D-0008, 075.D-0676 and 60.A-9120 for FEROS spectra; 
081.D-0928, 082.D-0901, 089.D-0767 and 093.D-0816 for UVES spectra.}$^,$
\thanks{Partly based on data obtained with the STELLA robotic observatory in Tenerife, 
an AIP facility jointly operated by AIP and IAC.}
}

\author{B. Proxauf\inst{1} \and
        R. da Silva\inst{2,3} \and
        V.V. Kovtyukh\inst{4,5} \and
        G. Bono\inst{1,2} \and
        L. Inno\inst{6} \and
        B. Lemasle\inst{7} \and
        J. Pritchard\inst{23} \and
        N. Przybilla\inst{8} \and \\
        J. Storm\inst{9} \and
        M.A. Urbaneja\inst{8} \and
        E. Valenti\inst{23} \and
        M. Bergemann\inst{6} \and
        R. Buonanno\inst{1,10} \and
        V. D'Orazi\inst{11,12} \and
        M. Fabrizio\inst{2,3} \and
        I. Ferraro\inst{2} \and
        G. Fiorentino\inst{13} \and
        P. Fran\c cois\inst{14,15} \and
        G. Iannicola\inst{2} \and
        C.D. Laney\inst{16,17} \and
        R.-P. Kudritzki\inst{18,19,20} \and
        N. Matsunaga\inst{21} \and
        M. Nonino\inst{22} \and
        F. Primas\inst{23} \and
        M. Romaniello\inst{23} \and
        F. Th\'evenin\inst{24}
        }

\institute{Dipartimento di Fisica, Università degli Studi di Roma Tor Vergata, via della Ricerca Scientifica 1, 00133 Rome, Italy \\ \email{giuseppe.bono@roma2.infn.it}
           \and INAF-Osservatorio Astronomico di Roma, via Frascati 33, 00078 Monte Porzio Catone, Rome, Italy
           \and Agenzia Spaziale Italiana, via del Politecnico snc, 00133 Rome, Italy
           \and Astronomical Observatory, Odessa National University, Shevchenko Park, 65014 Odessa, Ukraine
           \and Isaac Newton Institute of Chile, Odessa Branch, Shevchenko Park, 65014 Odessa, Ukraine
           \and Max-Planck Institute for Astronomy, D-69117, Heidelberg, Germany
           \and Astronomisches Rechen-Institut, Zentrum f\"ur Astronomie der Universit\"at Heidelberg, M\"onchhofstr. 12-14, 69120 Heidelberg, \\ Germany
           \and Institute for Astro- and Particle Physics, University of Innsbruck, Technikerstr. 25/8, 6020 Innsbruck, Austria
           \and Leibniz-Institut f\"ur Astrophysik Potsdam, An der Sternwarte 16, 14482 Potsdam, Germany
           \and INAF-Osservatorio Astronomico di Teramo, via Mentore Maggini snc, 64100 Teramo, Italy
           \and INAF-Osservatorio Astronomico di Padova, vicolo dell'Osservatorio 5, 35122, Padova, Italy
           \and Monash Centre for Astrophysics, School of Physics and Astronomy, Monash University, Melbourne, VIC 3800, Australia
           \and INAF-Osservatorio Astronomico di Bologna, via Ranzani 1, 40127 Bologna, Italy
           \and GEPI, Observatoire de Paris, CNRS, Universit\'e Paris Diderot, Place Jules Janssen, 92190 Meudon, France
           \and UPJV, Universit\'e de Picardie Jules Verne, 33 rue St. Leu, 80080 Amiens, France
           \and Department of Physics and Astronomy, N283 ESC, Brigham Young University, Provo, UT 84601, USA
           \and South African Astronomical Observatory, PO Box 9, Observatory 7935, South Africa
           \and Institute for Astronomy, University of Hawaii, 2680 Woodlawn Drive, Honolulu, HI 96822, USA
           \and Max-Planck-Institute for Astrophysics, Karl-Schwarzschild-Str.1, 85741 Garching, Germany
           \and University Observatory Munich, Scheinerstr. 1, 81679 Munich, Germany
           \and Department of Astronomy, School of Science, The University of Tokyo, 7-3-1 Hongo, Bunkyo-ku, Tokyo 113-0033, Japan
           \and INAF-Osservatorio Astronomico di Trieste, via G. B. Tiepolo 11, 34143 Trieste, Italy
           \and European Southern Observatory, Karl-Schwarzschild-Str. 2, 85748 Garching bei M\"unchen, Germany
           \and Laboratoire Lagrange, CNRS/UMR 7293, Observatoire de la C\^ote d'Azur, Bd de l'Observatoire, CS 34229, 06304 Nice, France
           }

\abstract{
We gathered more than 1130 high-resolution optical spectra for more than 250 Galactic classical Cepheids. The spectra were collected with different optical spectrographs: UVES at VLT, HARPS at 3.6m, FEROS at 2.2m MPG/ESO, and STELLA. To improve the effective temperature estimates, we present more than 150 new line depth ratio (LDR) calibrations that together with similar calibrations already available in the literature allowed us to cover a broad range in wavelength (5348 $\le$ $\lambda$ $\le$ 8427~\AA) and in effective temperatures (3500 $\le$ \teff\ $\le$ 7700~K). This means the unique opportunity to cover both the hottest and coolest phases along the Cepheid pulsation cycle and to limit the intrinsic error on individual measurements at the level of $\sim$100~K. Thanks to the high signal-to-noise ratio of individual spectra we identified and measured hundreds of neutral and ionized lines of heavy elements, and in turn, have the opportunity to trace the variation of both surface gravity and microturbulent velocity along the pulsation cycle. The accuracy of the physical parameters and the number of \ion{Fe}{i} (more than one hundred) and \ion{Fe}{ii} (more than ten) lines measured allowed us to estimate mean iron abundances with a precision better than 0.1~dex. Here we focus on 14 calibrating Cepheids for which the current spectra cover either the entire or a significant portion of the pulsation cycle. The current estimates of the variation of the physical parameters along the pulsation cycle and of the iron abundances agree quite well with similar estimates available in the literature. Independent homogeneous estimates of both physical parameters and metal abundances based on different approaches that can constrain possible systematics are highly encouraged.}


\maketitle

\section{Introduction}

Radially variable stars played a crucial role in the transition from qualitative 
to quantitative astrophysics. The reasons are manifold. They are simultaneously 
excellent primary distance indicators and very robust stellar tracers. The 
most popular ones are:
$i)$ RR~Lyrae: old ($t > 10$~Gyr), low-mass stars;
$ii)$ Mira: intermediate-age (from a few hundred Myr to several Gyr) stars; and
$iii)$ classical Cepheids: young (from tens of Myr to a few hundred Myr) stars.
Pulsation and evolutionary observables have been adopted for more than one century
to constrain Galactic stellar populations \citep{Baade1958}, and in particular to
improve our knowledge of the physical mechanisms driving their pulsation properties 
and evolution 
\citep{Kraft1957,Preston1964,Prestonetal1965,Wallerstein1972,Wallerstein1979}.

\begin{table*}
\caption{Calibrating Cepheids, for which high-resolution spectra cover either a substantial fraction or the entire pulsation cycle.}
\label{calibceph}
\centering
{\small 
\begin{tabular}{l r@{ }l cc r@{}l c r@{ }l ccccc}
\noalign{\smallskip}\hline\hline\noalign{\smallskip}
Name &
\multicolumn{2}{c}{\parbox[c]{1.1cm}{\centering $R_{\rm G}^a$ $\pm$ $\sigma$ [pc]}} &
$\alpha_{\rm ICRS}$ & $\delta_{\rm ICRS}$ &
\multicolumn{2}{c}{\parbox[c]{1.0cm}{\centering Period$^b$ [days]}} &
\parbox[c]{1.9cm}{\centering $T_0^b - 2\,400\,000$ [days]} &
\multicolumn{2}{c}{[Fe/H]$_{\rm lit}^a$ $\pm$ $\sigma$} &
$N_\mathrm{F}$ & $N_\mathrm{H}$ & $N_\mathrm{U}$ & $N_\mathrm{S}$ & $N_\mathrm{tot}$ \\
\noalign{\smallskip}\hline\noalign{\smallskip}
\object{V340\,Ara}    & 4657 & $\pm$ 427 & 16:45:19.112 & $-$51:20:33.393 & 20&.80876 & 44881.2740 &    0.33 & $\pm$ 0.09 &  26 & ... &   6 & ... &  32 \\
\object{$\eta$\,Aql}  & 7750 & $\pm$ 452 & 19:52:28.368 & $+$01:00:20.370 &  7&.17679 & 43368.8611 &    0.14 & $\pm$ 0.02 & ... & ... & ... &  11 &  11 \\
\object{S\,Cru}       & 7593 & $\pm$ 451 & 12:54:21.998 & $-$58:25:50.214 &  4&.68973 & 44301.5560 &    0.08 & $\pm$ 0.10 &   1 &  12 & ... & ... &  13 \\
\object{$\beta$\,Dor} & 7936 & $\pm$ 451 & 05:33:37.517 & $-$62:29:23.369 &  9&.84308 & 47913.0970 & $-$0.06 & $\pm$ 0.10 & ... &  46 & ... & ... &  46 \\
\object{$\zeta$\,Gem} & 8273 & $\pm$ 452 & 07:04:06.531 & $+$20:34:13.074 & 10&.15072 & 50139.4010 & $-$0.11 & $\pm$ 0.10 & ... &  47 & ... &  81 & 128 \\
\object{Y\,Oph}       & 7141 & $\pm$ 452 & 17:52:38.702 & $-$06:08:36.870 & 17&.12415 & 44083.4490 &    0.12 & $\pm$ 0.04 & ... &   8 & ... & ... &   8 \\
\object{RS\,Pup}      & 8585 & $\pm$ 444 & 08:13:04.216 & $-$34:34:42.696 & 41&.44002 & 53014.2808 &    0.21 & $\pm$ 0.10 & ... &  15 & ... & ... &  15 \\
\object{UZ\,Sct}      & 5309 & $\pm$ 448 & 18:31:22.368 & $-$12:55:43.350 & 14&.7482  & 45496.3631 &    0.33 & $\pm$ 0.08 &  28 & ... &   6 & ... &  34 \\
\object{AV\,Sgr}      & 5980 & $\pm$ 454 & 18:04:48.780 & $-$22:43:56.600 & 15&.41153 & 53109.1989 &    0.35 & $\pm$ 0.17 &  28 & ... &   5 & ... &  33 \\
\object{VY\,Sgr}      & 5862 & $\pm$ 453 & 18:12:04.568 & $-$20:42:14.580 & 13&.55845 & 50891.6007 &    0.33 & $\pm$ 0.12 &  30 & ... &   4 & ... &  34 \\
\object{XX\,Sgr}      & 6706 & $\pm$ 453 & 18:24:44.501 & $-$16:47:49.816 &  6&.42414 & 44822.6740 & $-$0.01 & $\pm$ 0.06 & ... & ... &   5 & ... &   5 \\
\object{Y\,Sgr}       & 7483 & $\pm$ 452 & 18:21:22.986 & $-$18:51:36.002 &  5&.77335 & 40762.4310 &    0.11 & $\pm$ 0.03 & ... &  20 & ... &   3 &  23 \\
\object{R\,TrA}       & 7519 & $\pm$ 451 & 15:19:45.713 & $-$66:29:45.742 &  3&.38924 & 52365.1127 &    0.16 & $\pm$ 0.11 &   1 &  14 & ... & ... &  15 \\
\object{RZ\,Vel}      & 8249 & $\pm$ 445 & 08:37:01.303 & $-$44:06:52.848 & 20&.39689 & 45003.4620 &    0.19 & $\pm$ 0.10 &   1 &  11 & ... & ... &  12 \\
\hline
\end{tabular}
}
\tablefoot{From left to right the columns give name, galactocentric distance, right ascension, declination, pulsation period, and zero-phase reference epoch of maximum light in the V-band. Column seven lists the iron abundance available in the literature. The columns from eight to eleven show the number of optical spectra used for each spectrograph: $N_\mathrm{F}$, FEROS; $N_\mathrm{H}$, HARPS; $N_\mathrm{U}$, UVES; $N_\mathrm{S}$, STELLA. The last column lists the total number of spectra per target.}
\tablebib{
\tablefoottext{a}{\citet{Genovalietal2014}};
\tablefoottext{b}{This investigation.}
}
\end{table*}

In this context, classical Cepheids (CCs) have been the cross-road of a paramount 
theoretical \citep{Bonoetal1999a,Bonoetal1999b,Fiorentinoetal2007,Marconietal2005,
Andersonetal2016} and observational \citep{Riessetal2016,FreedmanMadore2010,
Gierenetal2013,Pietrzynskietal2013,Soszynskietal2017} effort.
They are the most popular primary distance indicators used to 
calibrate secondary indicators, and to estimate the Hubble constant.  
They are bright ($-2 \le M_{\rm V} \le -7$~mag), and recent photometric investigations 
based on ground-based and/or space facilities provide accurate mean magnitudes 
for Cepheids located in external galaxies in the Local Group and in the Local Volume 
\citep[][and references therein]{Bonoetal2010,Macrietal2015,Hoffmannetal2016}.

However, the spectroscopic investigations are lagging, and indeed they have 
been mainly focussed on Galactic Cepheids
\citep[][and references therein]{Lucketal2011,LuckLambert2011,Wallersteinetal2015,
Genovalietal2014,Genovalietal2015,Lemasleetal2013,daSilvaetal2016}
and on a few nearby stellar systems like the Magellanic Clouds \citep{Lucketal1998,
Romanielloetal2008,Lemasleetal2017}. CCs are also excellent physics laboratories, and
indeed they have been used to investigate their dynamical properties along the pulsation cycle. 
They have been investigated both in the optical \citep{Struve1944,Kraft1956} and in the 
near-infrared (NIR) regime \citep{Sasselovetal1989,SasselovLester1990b}. More recently,
they have also been studied by \citet{Nardettoetal2009} to constrain the variation of the
projection factor, in a very exhaustive paper by \citet{Wallersteinetal2015}, and in
the validation of the quasi-static approximation by \citet{Vasilyevetal2017a,Vasilyevetal2017b}.

The elemental abundances are in a positive status since we are approaching 
an almost complete spectroscopic census of the currently known Galactic Cepheids
($\sim$450) based on high-resolution and high signal-to-noise (S/N) optical 
spectra (Bono et al., in prep.). Our group has been involved in a 
long-term project (DYONISOS) aimed at providing a homogeneous metallicity 
scale for field and cluster Galactic and Magellanic Cepheids. The current 
analyses mainly rely on the classical quasi-static approximation, in which the
spectra of a CC, randomly collected along the pulsation cycle, are approximated 
with the physical properties of a static star with similar effective temperature, 
surface gravity and microturbulent velocity.   

One of the key problems in dealing with spectroscopy of variable
stars in the Cepheid instability strip is that the effective temperature, 
when moving from minimum to maximum light, changes by roughly 1\,000~K. 
At the same time, the surface gravity also changes by up to 0.8-0.9~dex.
These variations are correlated with the luminosity amplitude. The 
quasi-static approximation becomes more severe in dealing with spectra 
collected across pulsation phases affected by non-linear phenomena (formation
and propagation of shocks), i.e., the phases along the rising branch or just 
before maximum compression. The reader is referred to \citet{Bonoetal2000b}
for a detailed discussion concerning these phenomena and their interplay with 
the Hertzsprung progression.  

The effective temperature of CCs can be estimated using 
color-temperature relations, but this approach requires very accurate 
optical and NIR photometry. Moreover, this approach is prone to possible 
systematics introduced by reddening uncertainties and/or metallicity 
dependence. A very promising independent approach has been recently 
provided by \citet{Kervellaetal2004} and \citet{Merandetal2015} using optical 
and NIR interferometric measurements of the diameter of nearby CCs. 
The same applies to the infrared surface brightness (IRSB) method by
\citet{Stormetal2011a,Stormetal2011b} and by \citet{Groenewegen2008} 
using optical/NIR photometry and radial velocities to constrain the 
angular diameter variations.

In this context, a temperature diagnostic that appears quite robust is 
the Line Depth Ratio (LDR). It relies on plain physical assumptions: 
the depth ratio of several pairs of absorption lines is strongly correlated 
with the effective temperature. To minimize the dependence of the abundance 
on the surface gravity and possible uncertainties in the continuum location, 
the lines forming these pairs should come from the same (or a similar) element,
have similar wavelengths, be weak, non saturated, and come from neutral 
species \citep{Gray2005}.

\begin{figure*}
\centering
\includegraphics[angle=-90,width=\hsize]{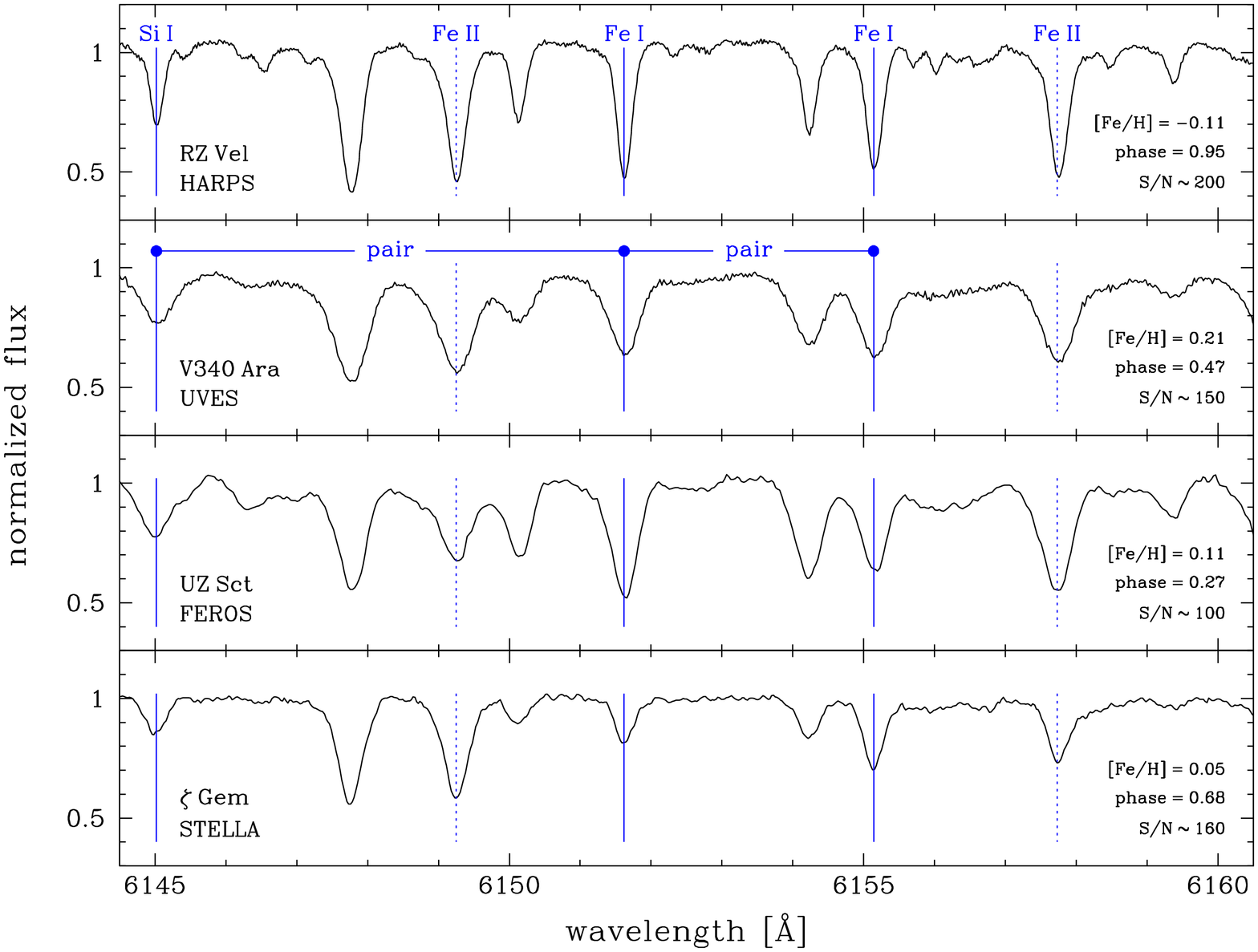}
\caption{Examples of high-resolution spectra for different calibrating classical Cepheids collected with different spectrographs: HARPS, UVES, FEROS, and STELLA. The vertical lines display the position either of iron lines only used for the metallicity determination (dashed) or of lines adopted also in the LDR method (solid). The name of the Cepheid, its mean metallicity, the phase, and the signal-to-noise ratio of these example spectra around 6000~\AA\ are also labelled.}
\label{signoisereg}
\end{figure*}

The use of the LDR method to estimate the effective temperature of CCs was pioneered by several 
authors: \citet{SasselovLester1990a}, \citet{Krockenbergeretal1998}, 
\citet[][hereinafter KG00]{KovtyukhGorlova2000}, and by
\citet[][hereinafter K03]{Kovtyukhetal2003b}.
In particular, KG00 provided a set of calibrations based on 32 pairs of lines. K03 provided 105 new LDR calibrations using more than 180 FGK main sequence stars covering $\sim$1~dex in iron abundance ($-$0.5 $\le$ [Fe/H] $\le$ 0.5) and for which were available high-resolution ($R \sim$42\,000), high S/N spectra, together with accurate trigonometric parallaxes from HIPPARCOS and effective temperature estimates with an accuracy of the order of 1\%. Subsequently, \citet[][hereinafter K07]{Kovtyukh2007} obtained a set of 131 LDR calibrations using 161 FGK supergiants, increasing the valid range in temperature to about 3600-7800~K (F0\,I-K5\,I). The equivalent width (EW) measurements of each pair provide an independent estimate of the effective temperature. These LDRs have been quite successful, being used in many recent spectroscopic investigations of CCs \citep[e.g.][]{Andrievskyetal2002a,Andrievskyetal2002b,Lemasleetal2007,Lemasleetal2008,Genovalietal2013,Genovalietal2014}, with a typical precision of the order of 150~K.

The main aim of this investigation is to provide new and homogeneous estimates of the intrinsic 
parameters (effective temperature, surface gravity, microturbulent velocity) for CCs and to 
constrain their impact on the iron abundance. The key advantage of the current investigation 
when compared with similar analyses available in the literature is that we are dealing with 
14 calibrating CCs covering a broad range in pulsation period (0.53 $\le$ $\log{P}$ $\le$ 1.62) 
and in metal abundance ($-$0.11 $\le$ [Fe/H] $\le$ 0.35~dex). The current sample was defined as 
"calibrating" CCs, since the optical high-resolution spectra cover the pulsation cycle either
fully or for the most part. Indeed, the number of spectra per object range from five
(\object{XX\,Sgr}) to more than one hundred (\object{$\zeta$\,Gem}).

The structure of the paper is the following. In \S~\ref{spec} we present the entire spectroscopic dataset and we discuss the S/N and the wavelength range covered by the spectra. In \S~\ref{datareduc} we discuss the different strategies adopted to pre-reduce and to calibrate the spectra. Section~\ref{equiwidths} deals with the adopted linelists and with the approach adopted to measure the equivalent widths. In \S~\ref{params_abund} we discuss the determination of the atmospheric parameters and the radial velocities, including the LDR calibrations used to derive the effective temperature and their validity range. The determination of the iron abundances is presented in \S~\ref{metal}, together with the variations along the pulsation cycle. The summary of the results and the future perspective of this project are given in \S~\ref{concl}.  

\section{Spectroscopic datasets}
\label{spec}

The spectroscopic datasets partly analyzed in the current paper, and that will be used for a new series of papers, are spectra collected at three different telescopes of the \textit{European Southern Observatory} (ESO): the \textit{Ultraviolet and Visual Echelle Spectrograph} \citep[UVES,][]{Dekkeretal2000} at the VLT, the \textit{High Accuracy Radial velocity Planet Searcher} \citep[HARPS,][]{Mayoretal2003} at the 3.6m, and the \textit{Fiber-fed Extended Range Optical Spectrograph} \citep[FEROS,][]{Kauferetal1999} at the 2.2m MPG/ESO. A list of CCs was defined for the three spectrographs and the related spectra were downloaded from the ESO archive, forming the datasets UVES~TS (for {\it This Study}, 32 spectra, 3 targets), HARPS (199 spectra, 9 targets), and FEROS~TS (486 spectra, 169 targets).

The quoted spectra were complemented with spectra collected by 
Inno et al. (ID: 093.D-0816, dataset UVES~IN, 154 spectra, 46 targets), 
\citet[dataset UVES~KO, 9 spectra, 1 target, and dataset FEROS~KO, 2 spectra, 2 targets]{Kovtyukhetal2016}, 
\citet[dataset UVES~GE, 120 spectra, 73 targets]{Genovalietal2015}. We also included 134 high-resolution spectra for 5 targets collected with the STELLA Echelle Spectrograph 
\citep[SES,][]{Strassmeieretal2004,Strassmeieretal2010}.
As a whole, we ended up with 1136 high-resolution spectra 
for 251 Cepheids, explicitly accounting for the multiplicity of objects among different spectrographs.

The spectral resolution of the quoted spectrographs for the instrument settings used are: 
$R\sim40\,000$ (UVES), $R\sim115\,000$ (HARPS), $R\sim48\,000$ (FEROS), 
and $R\sim55\,000$ (STELLA). The corresponding wavelength ranges for our sample are:
$i)$ UVES:
$\sim$3050--3870~\AA;
$\sim$3760--4985~\AA, $\sim$5684--7520~\AA, $\sim$7663--9458~\AA;
$\sim$4786--5750~\AA, $\sim$5833--6806~\AA;
$\sim$4980--5952~\AA, $\sim$6035--7002~\AA;
$\sim$6700--8523~\AA, $\sim$8659--10\,422~\AA;
$ii)$ HARPS:
$\sim$3781--5304~\AA, $\sim$5337--6912~\AA;
$iii)$ FEROS:
$\sim$4000--9216~\AA; and
$iv)$ STELLA:
$\sim$3872--8813~\AA.

The current spectroscopic dataset can be divided into three different subgroups:

{\em Calibrating Cepheids:}

$a)$ Phase dependence -- For 14 targets in our sample the spectra cover either a significant part
or the entire pulsation cycle. This is the subsample of CCs that we analyze in the present 
paper and that we will
adopt to constrain the accuracy of the intrinsic parameters and, in particular, their impact on
the iron abundance. Note that the bulk of CCs are strictly periodic on long evolutionary time
scales. This means that we take advantage of the strict periodicity in cyclic variations.

$b)$ Cluster Cepheids -- Our sample includes 14 CCs that are candidate cluster variables. 
The number of spectra per target ranges from one to more than one hundred. 
These targets
will be adopted to link field and cluster CCs on the same metallicity scale.

{\em Cepheids with new iron abundances:}
This subgroup includes roughly 50 Cepheids for which we secured high-resolution UVES 
spectra and for which no metallicity estimate is available in the literature. The number 
of spectra per target ranges from 1 to more than fifty. 

{\em Cepheids with homogeneous iron abundances:}
We derive homogeneous iron abundances based on high-resolution spectra for 216 CCs. 
The number of spectra per target ranges from 1 to almost fifty.

In the current investigation we focus on the 14 calibrating CCs with multi-epoch spectroscopic measurements. Details on the number of spectra are given in Table~\ref{calibceph}. The spectra of \object{EV\,Sct} and \object{X\,Sgr} were initially included in our analysis, but afterwards excluded. In a detailed spectroscopic investigation based on high-resolution spectra, \citet{KovtyukhAndrievsky1999} found that \object{EV\,Sct} shows strong line asymmetries and even split lines. They suggested that this object might be a binary system with two short period Cepheids, in spite of the observed phase coherence. Concerning \object{X\,Sgr}, \citet{Mathiasetal2006}, on the basis of both optical and NIR high-resolution spectroscopy, found strong dynamical variations in the outermost layers. Moreover, optical and NIR interferometric data \citep{Licausietal2013,Gallenneetal2014} showed evidence of a possible companion, which is also suggested by the orbital velocity curve determined by \citet{Feastetal2008}. The interested reader is also referred to \citet{Szabados1990,Szabados2003,Evans1992,Groenewegen2008} for a more detailed discussion. These two CCs were also included in the list of candidate non-radial pulsators by \citet{Kovtyukhetal2003a}.

\section{Data reduction and analysis}
\label{datareduc}

The spectra have to be prepared for the later analysis by doing an initial pre-reduction (up to the wavelength calibration step). The spectra from UVES and HARPS (phase-3) were already pre-reduced by their own pipeline. FEROS spectra were reduced with a modified version of the FEROS-DRS pipeline developed by one of us (J. Pritchard). Several FEROS spectra taken before 2004 could not be reduced due to a change in the CCD software architecture from the BIAS controller to FIERA and therefore a different file structure. The current FEROS sample includes 355 spectra, with the remaining 133 not yet included in the analysis.

\begin{table*}
\caption{Mean parameters derived for the calibrating Cepheids.}
\label{meanparams}
\centering
\begin{tabular}{l r@{ }l r@{ }l r@{ }l r@{ }l r@{ }l r@{ }l c}
\noalign{\smallskip}\hline\hline\noalign{\smallskip}
Name &
\multicolumn{2}{c}{\parbox[c]{1.4cm}{\centering $\langle$\teff$\rangle$ $\pm$ $\sigma$ [K]}} &
\multicolumn{2}{c}{\parbox[c]{1.6cm}{\centering $\langle$\logg$\rangle$ $\pm$ $\sigma$}} &
\multicolumn{2}{c}{\parbox[c]{1.2cm}{\centering $\langle$\vmic$\rangle$ $\pm$ $\sigma$ [\kms]}} &
\multicolumn{2}{c}{\parbox[c]{1.8cm}{\centering [\ion{Fe}{i}/H] $\pm$ $\sigma$}} &
\multicolumn{2}{c}{\parbox[c]{1.8cm}{\centering [\ion{Fe}{ii}/H] $\pm$ $\sigma$}} &
\multicolumn{2}{c}{\parbox[c]{1.6cm}{\centering [Fe/H] $\pm$ $\sigma$} (std)} &
$N_{\rm spec}$ \\
\noalign{\smallskip}\hline\noalign{\smallskip}
\object{V340\,Ara} 	& 5293 & $\pm$ 42 & 1.14 & $\pm$ 0.11 & 4.77 & $\pm$ 0.19 &    0.24 & $\pm$ 0.07 &    0.22 & $\pm$ 0.05 &    0.23 & $\pm$ 0.04 (0.07) &   7 \\
\object{$\eta$\,Aql} 	& 5479 & $\pm$ 39 & 1.11 & $\pm$ 0.09 & 3.43 & $\pm$ 0.15 &    0.26 & $\pm$ 0.05 &    0.23 & $\pm$ 0.03 &    0.24 & $\pm$ 0.03 (0.09) &  11 \\
\object{S\,Cru} 	& 6014 & $\pm$ 21 & 1.64 & $\pm$ 0.08 & 3.08 & $\pm$ 0.14 &    0.09 & $\pm$ 0.03 &    0.08 & $\pm$ 0.04 &    0.09 & $\pm$ 0.03 (0.04) &  13 \\
\object{$\beta$\,Dor} 	& 5557 & $\pm$ 13 & 1.35 & $\pm$ 0.04 & 3.78 & $\pm$ 0.07 & $-$0.04 & $\pm$ 0.02 & $-$0.02 & $\pm$ 0.02 & $-$0.03 & $\pm$ 0.01 (0.05) &  46 \\
\object{$\zeta$\,Gem} 	& 5494 & $\pm$  7 & 1.12 & $\pm$ 0.03 & 3.22 & $\pm$ 0.04 &    0.15 & $\pm$ 0.01 &    0.17 & $\pm$ 0.01 &    0.16 & $\pm$ 0.01 (0.05) & 128 \\
\object{Y\,Oph} 	& 5612 & $\pm$ 33 & 0.99 & $\pm$ 0.11 & 3.22 & $\pm$ 0.18 &    0.06 & $\pm$ 0.05 &    0.09 & $\pm$ 0.06 &    0.08 & $\pm$ 0.04 (0.05) &   8 \\
\object{RS\,Pup} 	& 5381 & $\pm$ 27 & 0.84 & $\pm$ 0.08 & 4.66 & $\pm$ 0.13 &    0.13 & $\pm$ 0.04 &    0.15 & $\pm$ 0.03 &    0.14 & $\pm$ 0.02 (0.07) &  14 \\
\object{UZ\,Sct} 	& 5038 & $\pm$ 36 & 1.24 & $\pm$ 0.11 & 4.86 & $\pm$ 0.18 &    0.13 & $\pm$ 0.07 &    0.07 & $\pm$ 0.09 &    0.11 & $\pm$ 0.05 (0.09) &   8 \\
\object{AV\,Sgr} 	& 5228 & $\pm$ 38 & 1.18 & $\pm$ 0.11 & 4.86 & $\pm$ 0.18 &    0.25 & $\pm$ 0.07 &    0.32 & $\pm$ 0.03 &    0.31 & $\pm$ 0.02 (0.08) &   8 \\
\object{VY\,Sgr} 	& 5340 & $\pm$ 35 & 0.98 & $\pm$ 0.09 & 4.59 & $\pm$ 0.16 &    0.21 & $\pm$ 0.06 &    0.27 & $\pm$ 0.04 &    0.25 & $\pm$ 0.03 (0.08) &  10 \\
\object{XX\,Sgr} 	& 5843 & $\pm$ 41 & 1.30 & $\pm$ 0.13 & 2.98 & $\pm$ 0.22 &    0.09 & $\pm$ 0.06 &    0.03 & $\pm$ 0.06 &    0.06 & $\pm$ 0.04 (0.02) &   5 \\
\object{Y\,Sgr} 	& 5924 & $\pm$ 26 & 1.75 & $\pm$ 0.06 & 4.11 & $\pm$ 0.11 &    0.07 & $\pm$ 0.03 & $-$0.02 & $\pm$ 0.02 &    0.00 & $\pm$ 0.01 (0.06) &  22 \\
\object{R\,TrA} 	& 6039 & $\pm$ 25 & 1.97 & $\pm$ 0.08 & 4.01 & $\pm$ 0.13 &    0.02 & $\pm$ 0.03 & $-$0.04 & $\pm$ 0.03 & $-$0.01 & $\pm$ 0.02 (0.03) &  15 \\
\object{RZ\,Vel} 	& 5479 & $\pm$ 29 & 1.23 & $\pm$ 0.09 & 4.62 & $\pm$ 0.14 &    0.09 & $\pm$ 0.04 &    0.08 & $\pm$ 0.04 &    0.08 & $\pm$ 0.03 (0.06) &  12 \\
\hline
\end{tabular}
\tablefoot{From left to right the columns give name, effective temperature, surface gravity, microturbulent velocity, iron abundances, and number of spectra used to compute the mean values. These are the weighted mean and its uncertainty computed from the values in Table~\ref{paramscontrolsample}. The standard deviation of the mean computed using individual abundances of both \ion{Fe}{i} and \ion{Fe}{ii} is also shown.}
\end{table*}

The next step, the continuum normalization, was required for UVES (except UVES~GE, already continuum-normalized), HARPS and FEROS. Before the normalization, UVES and HARPS spectra were split into blue and red spectral parts due to the big central gap present therein. The continuum normalization was performed using the {\it Image Reduction and Analysis Facility} (IRAF\,\footnote{Distributed by the National Optical Astronomy Observatories (NOAO), USA.}) by fitting cubic spline functions to a set of continuum windows visually selected in the spectra. For UVES and HARPS spectra we normally used 1st order functions, but for FEROS spectra high-order (20-50) cubic spline functions were required given their large wavelength range.

The radial velocity of the objects was determined using IRAF by cross-correlating the target spectrum with an observed solar template spectrum in the rest frame \citep[Solar Flux Atlas,][]{Kuruczetal1984} degraded to the UVES resolution. The UVES~GE spectra were already in the rest frame and served as templates. For FEROS spectra, because of possible contamination due to bleeding during the wavelength calibration \citep{ferosuserman}, we preferred to adopt the radial velocities derived by the routine used to measure the EWs (see Sect.~\ref{equiwidths}). For STELLA spectra, the radial velocities come directly from the STELLA reduction pipeline \citep{Weberetal2012}, which is based on IRAF. It performs the standard data reduction steps including scattered light removal and continuum normalization. The radial velocities are based on cross-correlation with a synthetic template spectrum. In the case of Cepheids a G-type star template is used. The resulting radial velocities have estimated uncertainties of about 0.2~\kms.

The spectra were examined for EW measurements by checking the S/N in different continuum regions in the spectra. From the preliminary S/N estimates derived from these blocks in combination with a visual check, the different spectra were classified as low, intermediate, or high-quality exposures, and the usability for further analysis, especially the metallicity determination, was evaluated. The UVES~GE sample has already been marked with sufficient S/N by the original authors. The UVES~KO and UVES~IN S/N ratios could be taken directly from the ESO archive. For FEROS the S/N ratios range from 25 to 475, for HARPS they are between 145 and 400, and for UVES~TS they cover a range between 235 and 480. We noticed that many of the FEROS spectra have been classified as low-quality exposures, and have then not been included in the metallicity and effective temperature determination. Examples of HARPS, UVES, FEROS, and STELLA spectra of different metallicities and with different S/N estimates (measured around 6000~\AA) are depicted in Fig.~\ref{signoisereg}.

\subsection{Equivalent width measurements}
\label{equiwidths}

The equivalent widths were measured using the {\it Automatic Routine for line Equivalent widths in stellar Spectra} \citep[ARES,][]{Sousaetal2007,Sousaetal2015}. First, a global set of common input parameters was used, and then the parameters were individually adjusted, giving better fits of the spectral line profiles. As mentioned in Sect.~\ref{datareduc}, ARES also performs an automatic estimate of the radial velocity, whose values were used in the case of FEROS spectra.

Three linelists were created:

$a)$ one built by combining four individual linelists received from Kovtyukh and used to derive the effective temperature (\teff) of the objects (153 lines);

$b)$ one from \citet{Genovalietal2013}, complemented with the {\it Gaia-ESO Survey} \citep[GES,][]{Gilmoreetal2012,Randichetal2013}, and cross-checked with the {\it Vienna Atomic Line Database} \citep[VALD3,][]{Ryabchikovaetal2015}, containing iron features (615 lines); 

$c)$ one from \citet{Genovalietal2015} and \citet{daSilvaetal2016} including lines 
belonging to other elements (113 lines, $\alpha$-, s- and r-elements).

The atomic lines used for effective temperature determination are listed in Table~\ref{teffcaliblist}. The linelist of the other elements will be discussed in a forthcoming paper.

\section{Atmospheric parameters}
\label{params_abund}

\subsection{New and old LDR calibrations}
\label{ldrcalib}

Although being widely used, the LDRs by KG00 are based on polynomial relations hampered by a 
limited range in effective temperature (4700-6700~K) and in wavelength (5670-6850~\AA). 
The former limitation affects the accuracy when dealing with spectra collected across the 
hottest pulsation phases, the latter one when dealing with spectra having higher S/N in 
redder wavelengths ($\lambda > 6500$~\AA). For this reason, K07 extended the number
of calibrations to a total of 131 pairs of lines located in the wavelength range
between 5348 and 6768~\AA. The key advantage of the new LDRs is that they allow effective 
temperature estimates up to 7800~K.

\begin{figure}
\centering
\includegraphics[width=\hsize]{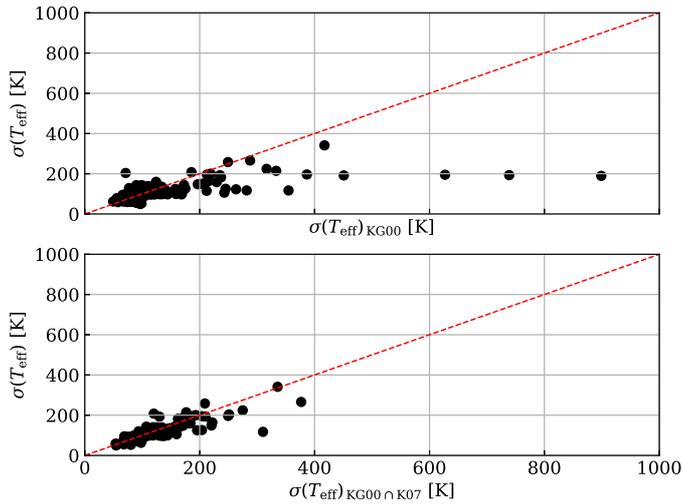}
\caption{Standard deviations of the effective temperature estimates for the 199 HARPS spectra.
{\it Top panel:} the entire set of 257 LDR calibrations used in the current investigation (KG00 $\cap$ K07 $\cap$ K17) is compared with the 32 calibrations provided by KG00.
{\it Bottom panel:} the same as in the top panel, but the comparison is between this study and the calibrations provided by KG00 $\cap$ K07.}
\label{stddevotherharps}
\end{figure}

To further improve the range in wavelength covered by the LDR calibrations, one of us (V. Kovtyukh) identified 151 new pairs of lines. Here we publish these new calibrations, which are based on effective temperatures estimated by K07. They fixed the temperature scale using non-variable supergiants for which the effective temperature was already available in the literature using independent approaches. Subsequently, they applied the new temperature scale to classical Cepheids and they only retained the LDRs for which the standard deviation was smaller than 110~K.

In case the same pair appears in different calibrations, we always selected the most recent one. All in all we ended up with 257 pairs of lines ranging from 5348 to 8427~\AA, and covering the temperature range from 3500 to 7700~K. To our knowledge this is the most complete list of LDRs ever compiled for F-K spectral type stars. Moreover, as shown in Fig.~\ref{stddevotherharps}, the standard deviations of the effective temperature determinations are clearly reduced with the increasing number of calibrations. The figure shows the results for the HARPS sample, but it is also valid for the other samples.

We are dealing therefore with three different sets of LDR calibrations:

$a)$ KG00: 32 calibrations;

$b)$ K07: discussed 131 calibrations, but the analytical functions are presented here for the first time;

$c)$ K17 (this investigation): 151 new calibrations.

From the 257 calibrations (multiplicity removed), based on 153 lines, 257 independent estimates of the effective temperature could be obtained. The list of the analytical relations for the LDRs used in this investigation is given in Table~\ref{teffcaliblist}.

\subsection{Effective temperature estimation}
\label{teffestimate}

Several calibrations were removed, since they provide effective temperatures 
significantly different from the bulk of the LDRs. Such outliers in the 
temperature distribution may be caused either by blends in the specific pair 
or by limited S/N. Each calibration also has a certain range of validity where 
the LDR can be used. For many of our sample stars we did not have a previous 
estimate of their effective temperature, therefore, we performed an iterative 
process. In each iteration, the calibrations had to pass a sigma clipping and 
were then accepted only if a defined interval around the median temperature 
was overlapping with the temperature range of the calibration. The clippings 
were stopped when no more calibrations were cut.

Once the clippings were finished, the mean and median temperature was computed from the remaining values. The number of calibrations that survived the sigma clipping was typically quite high ($\sim$100) so that a solid statistical basis was present and the standard deviations were relatively low ($<$ 100~K) for most of the spectra.

We performed a number of numerical simulations assuming no sigma clipping and 
we found that the mean/median values are minimally affected, while the standard 
deviation increases to $\sim$150~K. The approach we adopted to estimate the 
effective temperature for the individual spectra can be summarized as follows:

\begin{itemize}

\item[a)] Take all available LDR calibrations and calculate individual effective temperatures.

\item[b)] Calculate the effective temperature mean/median/$\sigma$.

\item[c)] If the effective temperature of a given LDR calibration is out of 2$\sigma$ from the median and $\sigma > 100$~K, discard the LDR; if no LDR calibration is discarded go to (d), otherwise go back to (b).

\item[d)] Check if the median value is within the validity range of the calibrations; discard all the LDR calibrations that do not cover the derived median; if no LDR calibration is discarded go to (e), otherwise go back to (b).

\item[e)] Calculate the final mean/median/$\sigma$.

\end{itemize}

The median values of effective temperatures and the standard deviations, together with other parameters, derived for individual spectra of the calibrating Cepheids are listed in Table~\ref{paramscontrolsample}. Spectra for which the effective temperature could not be estimated, as explained in Sect.~\ref{datareduc}, are listed in Table~\ref{noparamscontrolsample}.

\begin{figure}
\centering
\includegraphics[width=\hsize]{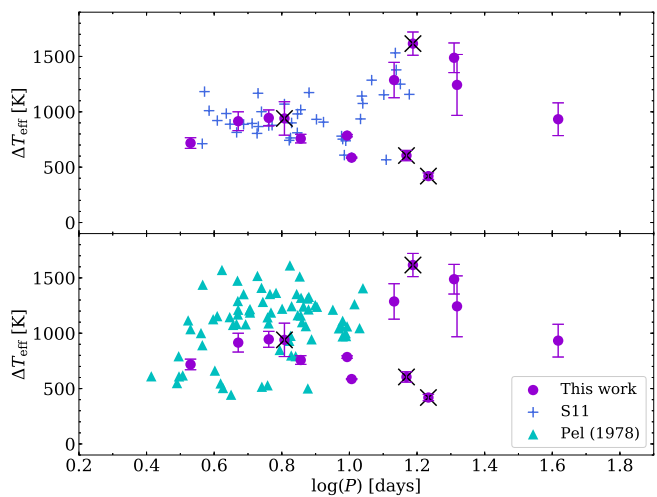}
\caption{Effective temperature amplitude ($\Delta${\teff} = \teff$^{\rm max}$ $-$ \teff$^{\rm min}$) as a function of the logarithmic period. The black crosses indicate stars for which the available spectra do not properly cover the maximum and minimum of the pulsation cycle.}
\label{delta_teff_logp}
\end{figure}

In order to validate the effective temperature estimates based on the new sets of LDRs, we performed a detailed comparison with similar estimates available in the literature. The top panel of Fig.~\ref{delta_teff_logp} shows the comparison between the current effective temperature amplitude 
($\Delta${\teff} = \teff$^{\rm max}$ $-$ \teff$^{\rm min}$) 
and similar estimates for 60 Galactic Cepheids for which \citet[][hereinafter S11]{Stormetal2011a} applied the IRSB method to estimate individual distances and intrinsic parameters. We estimated the $\Delta{T_{\rm eff}}$ by fitting the effective temperature curves with sinusoidal functions. Cepheids for which the phase coverage is not optimal (\object{Y\,Oph}, \object{UZ\,Sct}, \object{AV\,Sgr}, and \object{XX\,Sgr}) were marked with black crosses. The vertical error bars display the standard deviations on the fitted functions. Typical error bars on the $\Delta{T_{\rm eff}}$ values from S11 are smaller than the symbol size. Data plotted in this panel display the typical V-shape distribution \citep{Bonoetal2000a}, i.e., pulsation amplitudes display a well defined minimum across the so-called Hertzsprung progression \citep[$\log{ P }\sim1.0$,][]{Bonoetal2000b}. The agreement between the two data sets is quite good in the period range in common.

We also compared the current $\Delta{T_{\rm eff}}$ estimates with similar evaluations provided by 
\citet{Pel1978} using multi-band Walraven photometry (bottom panel of
Fig.~\ref{delta_teff_logp}). Note that CCs identified by the author as having known or suspected 
companions, or having other peculiarities, were not included in the figure. The agreement is 
once again good over the entire period range in common. Note that this approach is only based 
on photometric measurements.

\subsection{Surface gravity and microturbulent velocity}
\label{logg_vmic}

The surface gravity (\logg) was derived through the ionization equilibrium between
\ion{Fe}{i} and \ion{Fe}{ii} lines, and the microturbulent velocity (\vmic) was derived by 
minimizing the slope in the [\ion{Fe}{i}] vs. EW plot. This means that the \logg\ value is 
changed until the \ion{Fe}{i} and \ion{Fe}{ii} lines provide the same abundance, within the
errors, while the \vmic\ value is changed until the dependence of the derived abundances on
the EWs is removed. Indeed, weak and strong lines are supposed to provide the same
elemental abundances. In the iterative procedure, the \teff\ values (derived as described in Sect.~\ref{teffestimate}) are kept fixed, and the \logg\ and \vmic\ values are changed until the aforementioned conditions are satisfied.

For the determination of these parameters, we used the MOOG LTE radiative code \citep{Sneden2002} applied to model atmospheres derived by interpolation in the grid of \citet{CastelliKurucz2004}. We did not perform a specific test to constrain the difference when using different grids of atmosphere models. However, recent detailed results available in the literature \citep{HeiterEriksson2006,Gustafssonetal2008} support a very good agreement in the spectral range (F-K) typical of classical Cepheids. The standard solar abundances adopted by the MOOG code (version of July 2014) come from \citet{Asplundetal2009}. Though they have been recently revised by \citet{Grevesseetal2015} and by \citet{Scottetal2015a,Scottetal2015b}, we decided to use the abundances from \citet{Asplundetal2009} for consistency with our previous spectroscopic analyses.

Table~\ref{meanparams} lists the weighted mean of the surface gravity and microturbulent velocity computed for the 14 calibrating Cepheids using the multiple values listed in Table~\ref{paramscontrolsample}. In that table, the uncertainties on the individual estimates of \logg\ and \vmic\ are not listed, but they are expected to be of the order of $\sim$0.3~dex and $\sim$0.5~\kms, respectively \citep[see][]{Genovalietal2014}. In Table~\ref{meanparams}, the uncertainty on the weighted mean is shown.

\begin{figure}
\centering
\includegraphics[width=\hsize]{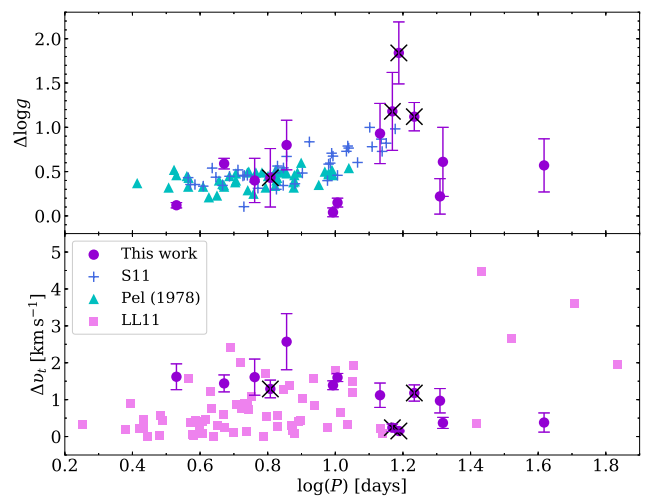}
\caption{The same as in Fig.~\ref{delta_teff_logp}, but showing the surface gravity amplitude
(top panel) and the microturbulent velocity amplitude (bottom panel).}
\label{delta_param_logp}
\end{figure}

In the top panel of Fig.~\ref{delta_param_logp} we show the comparison among the current
surface gravity amplitudes ($\Delta{\log{g}}$) and similar estimates by \citet{Pel1978} and by
S11. As in Fig.~\ref{delta_teff_logp}, stars from \citet{Pel1978} with known or suspected 
companions, or having other peculiarities, were not included. The three different datasets
agree quite well within the errors. Note that the two Cepheids with larger surface
gravity amplitudes (\object{UZ\,Sct} and \object{AV\,Sgr}) are among those for which the
phase coverage is not optimal. The bottom panel of the same figure shows the comparison
between the current microturbulent velocity amplitudes and similar estimates provided
by \citet[][hereinafter LL11]{LuckLambert2011} for Cepheids with multiple measurements.
The two datasets agree quite well once we take account of the fact that several targets 
in the LL11 sample only have a few spectra. The same applies for the
targets for which we do not have an optimal phase coverage. The surface gravity and 
microturbulent velocity amplitudes will be discussed in more detail in a forthcoming paper
(Urbaneja et al., in prep.) using an independent spectroscopic approach.

\begin{figure*}
\centering
\begin{minipage}[t]{0.33\textwidth}
\centering
\resizebox{\hsize}{!}{\includegraphics{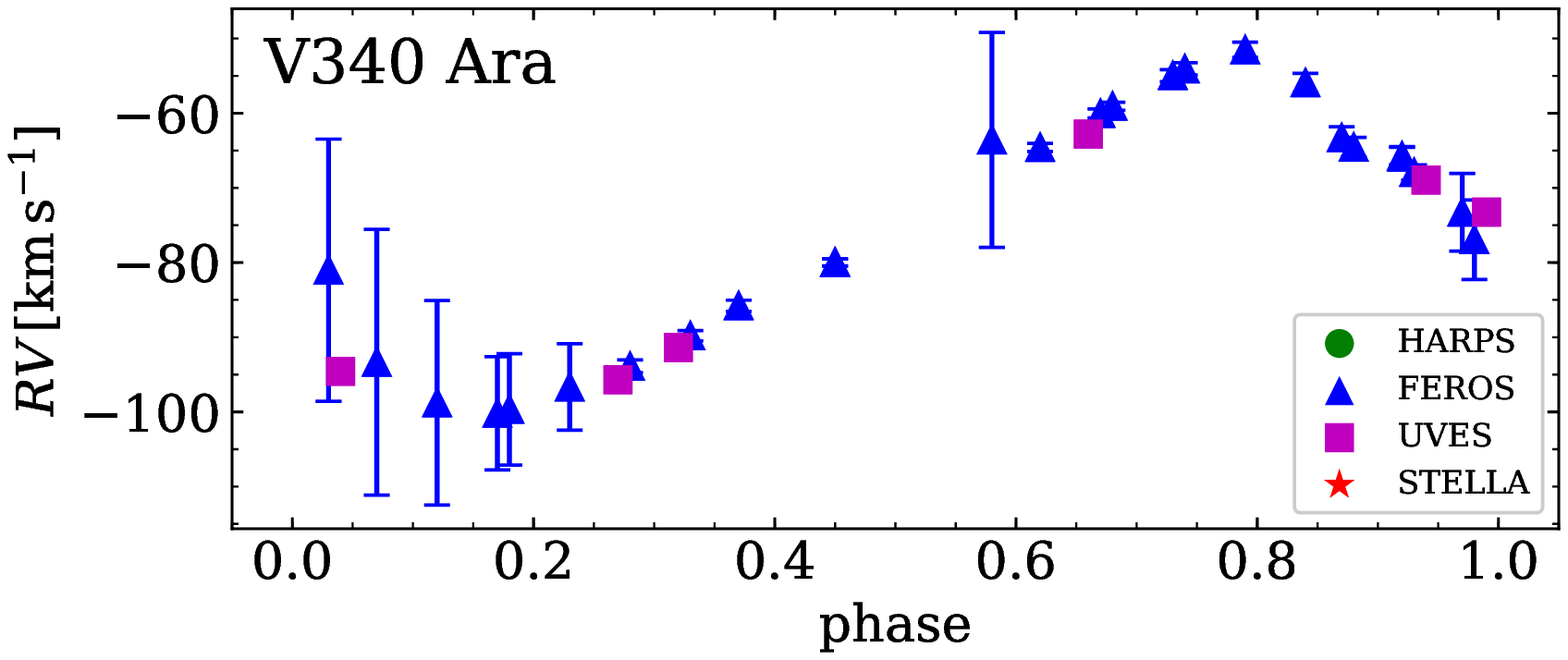}}
\end{minipage}
\begin{minipage}[t]{0.33\textwidth}
\centering
\resizebox{\hsize}{!}{\includegraphics{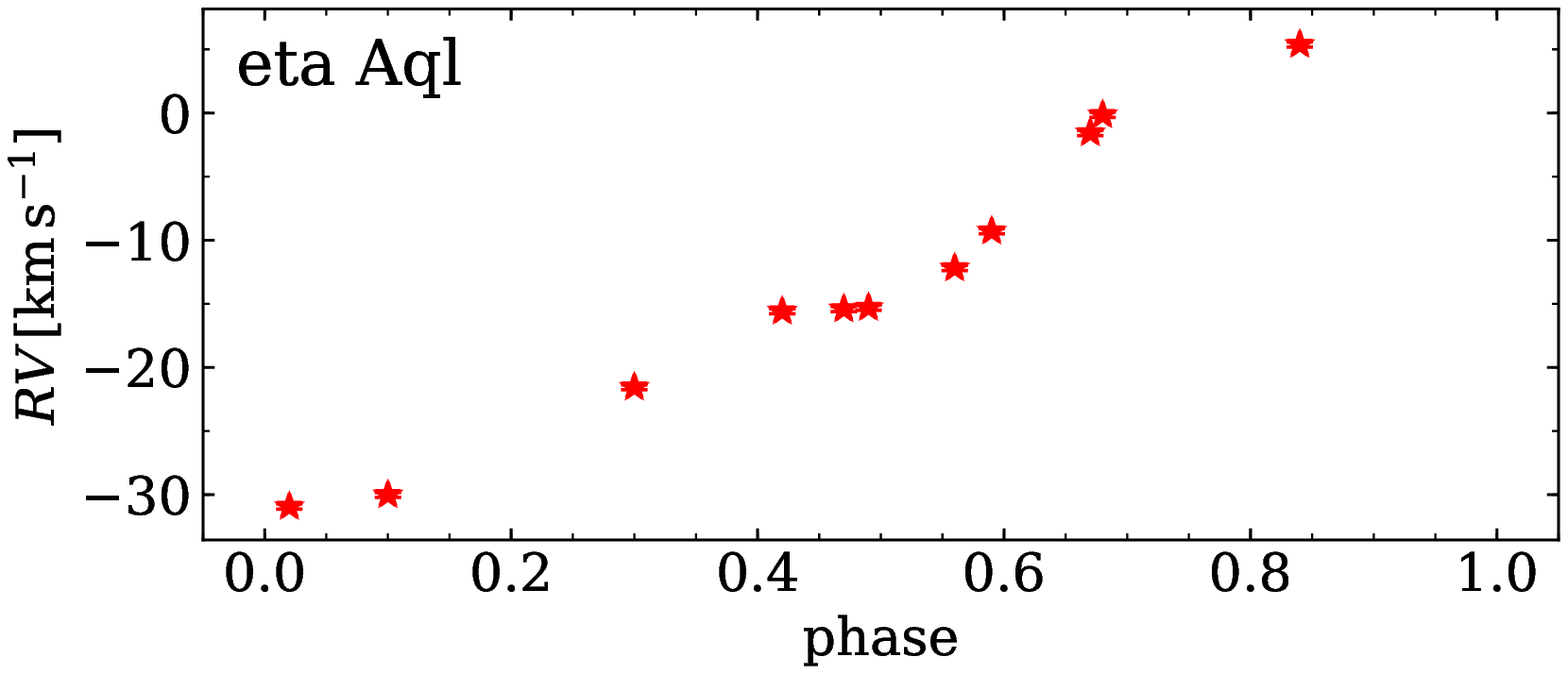}}
\end{minipage}
\begin{minipage}[t]{0.33\textwidth}
\centering
\resizebox{\hsize}{!}{\includegraphics{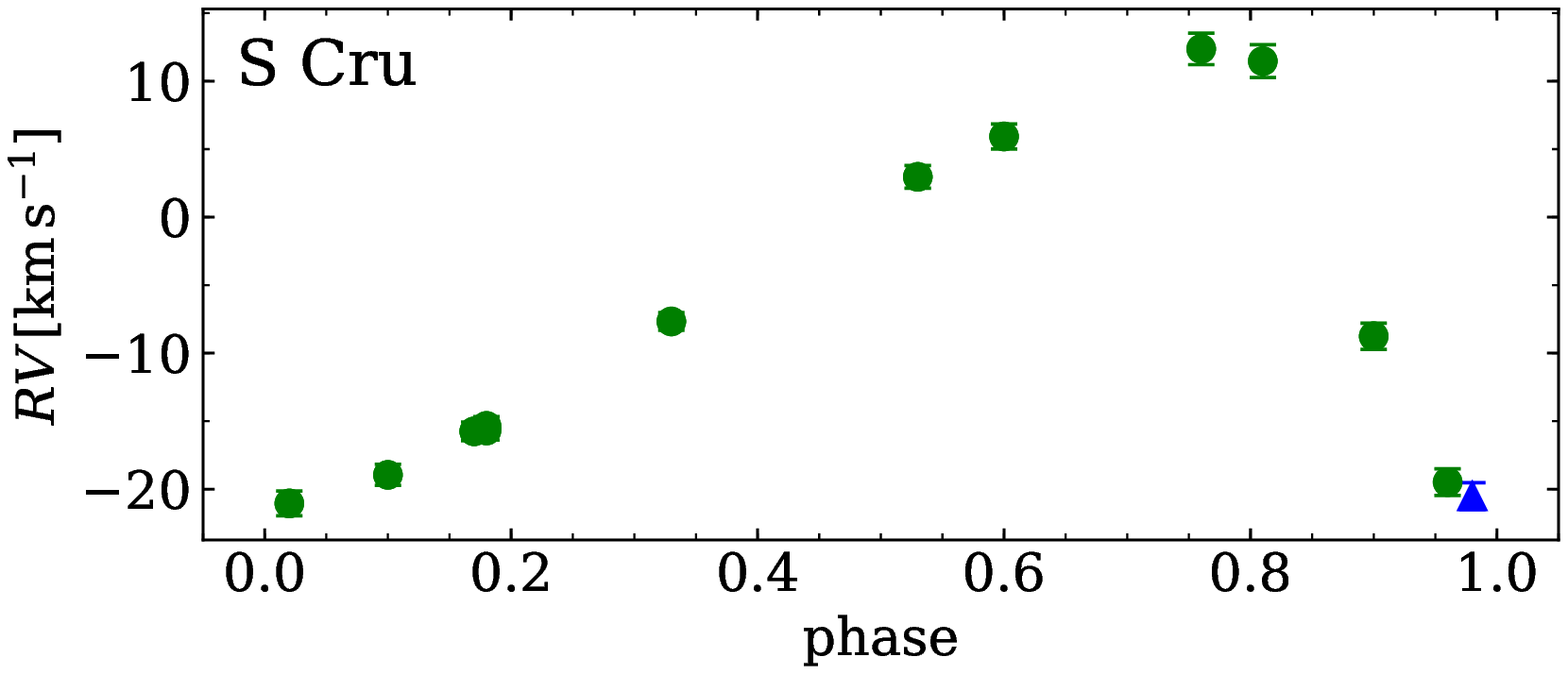}}
\end{minipage} \\
\begin{minipage}[t]{0.33\textwidth}
\centering
\resizebox{\hsize}{!}{\includegraphics{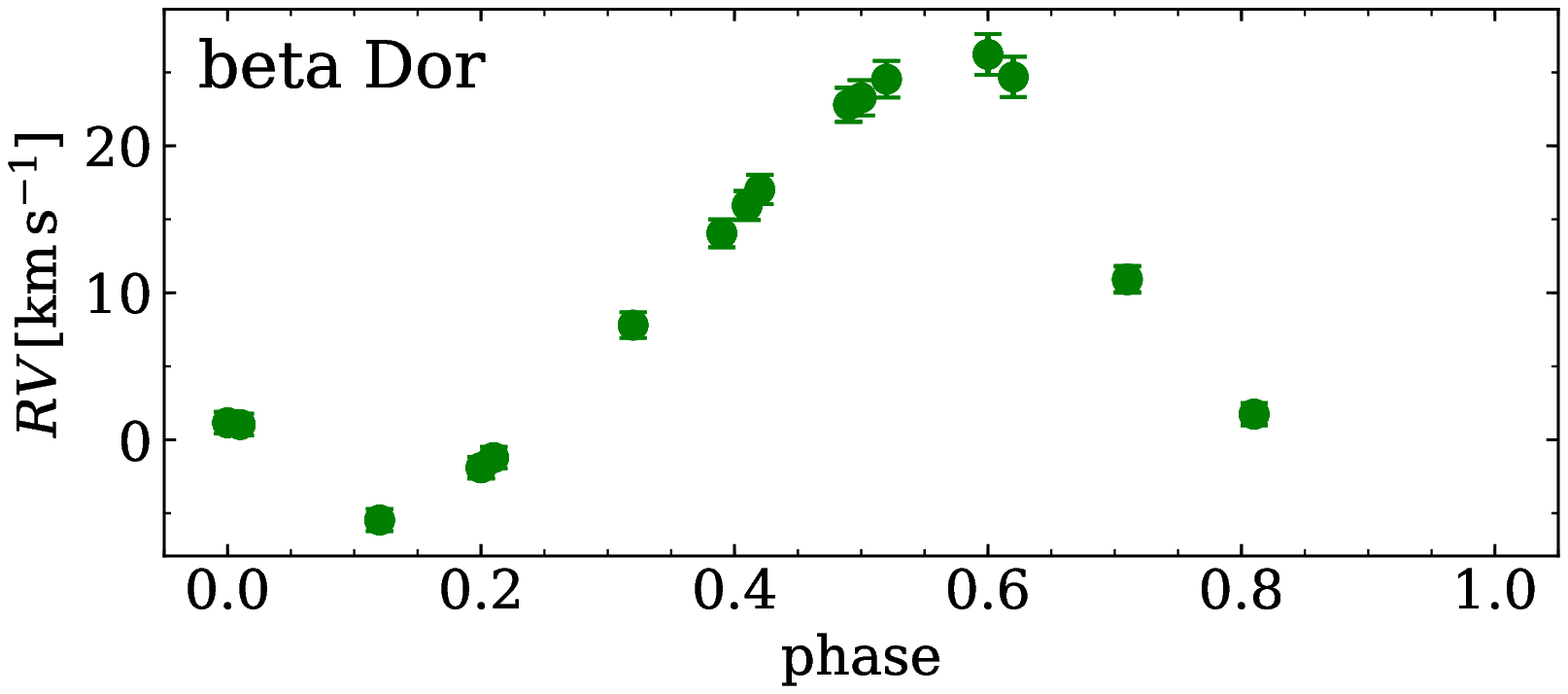}}
\end{minipage}
\begin{minipage}[t]{0.33\textwidth}
\centering
\resizebox{\hsize}{!}{\includegraphics{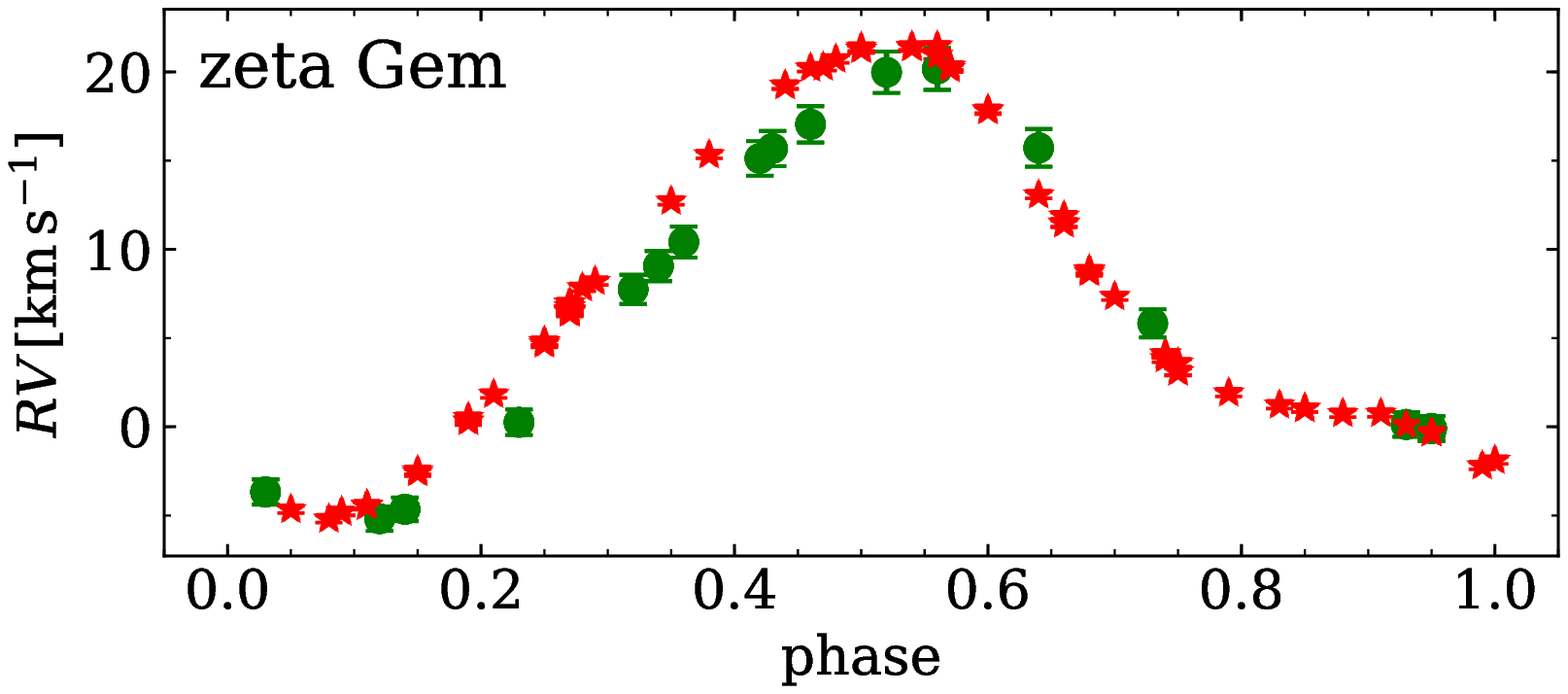}}
\end{minipage}
\begin{minipage}[t]{0.33\textwidth}
\centering
\resizebox{\hsize}{!}{\includegraphics{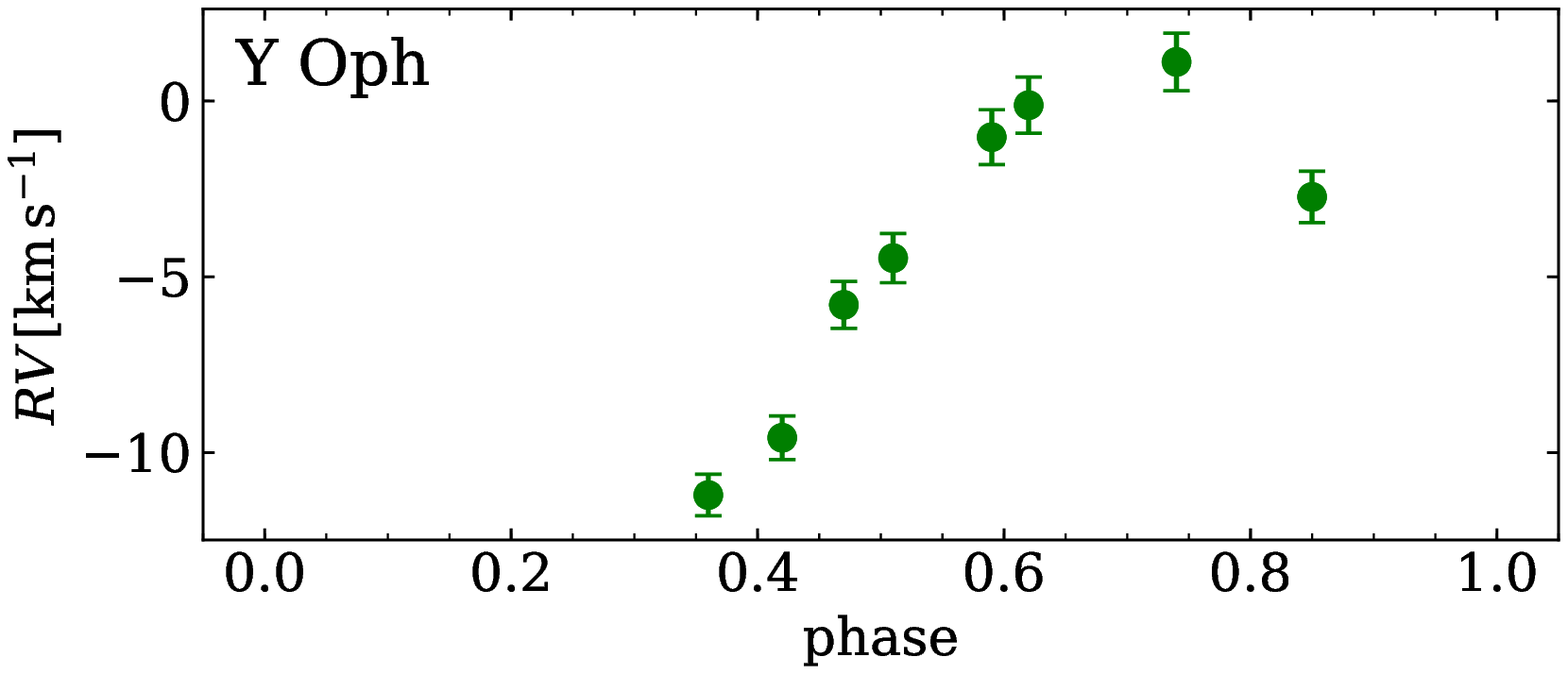}}
\end{minipage} \\
\begin{minipage}[t]{0.33\textwidth}
\centering
\resizebox{\hsize}{!}{\includegraphics{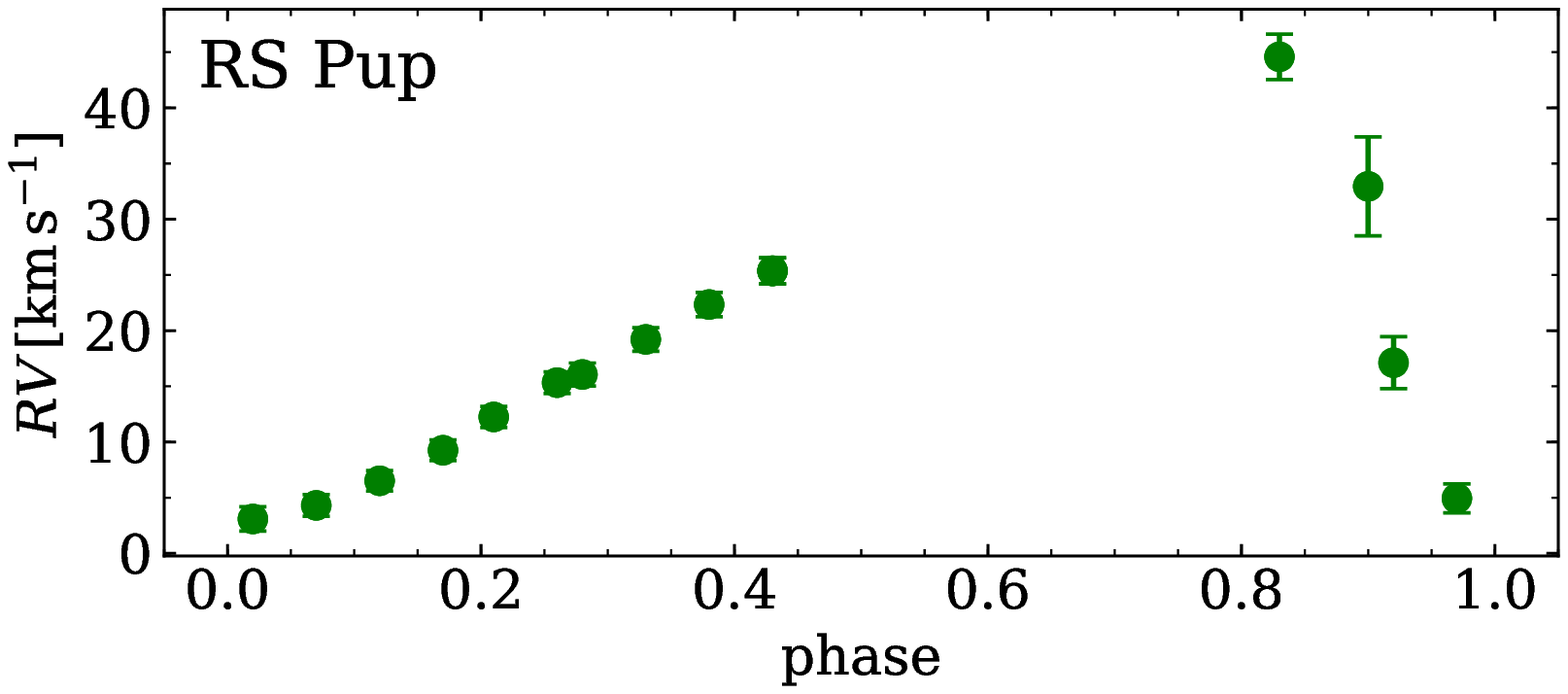}}
\end{minipage}
\begin{minipage}[t]{0.33\textwidth}
\centering
\resizebox{\hsize}{!}{\includegraphics{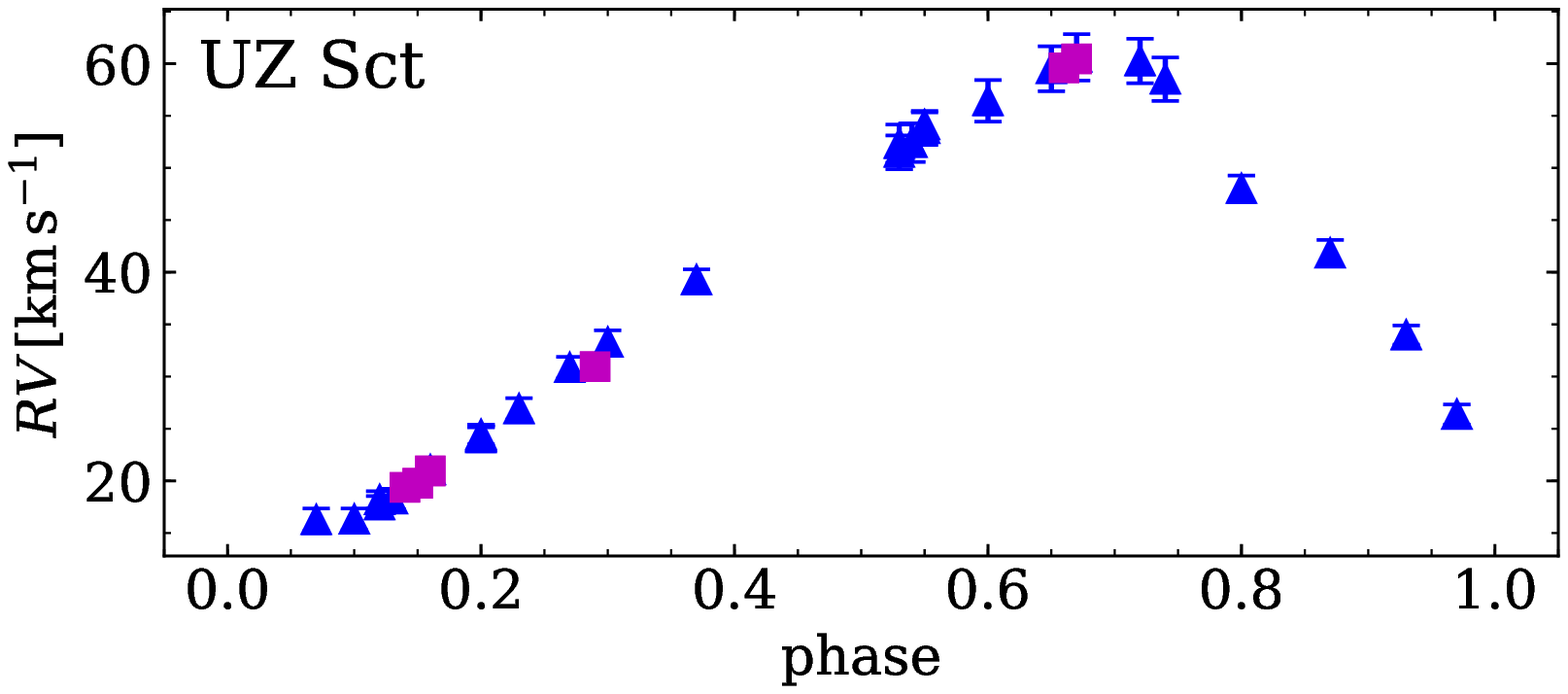}}
\end{minipage}
\begin{minipage}[t]{0.33\textwidth}
\centering
\resizebox{\hsize}{!}{\includegraphics{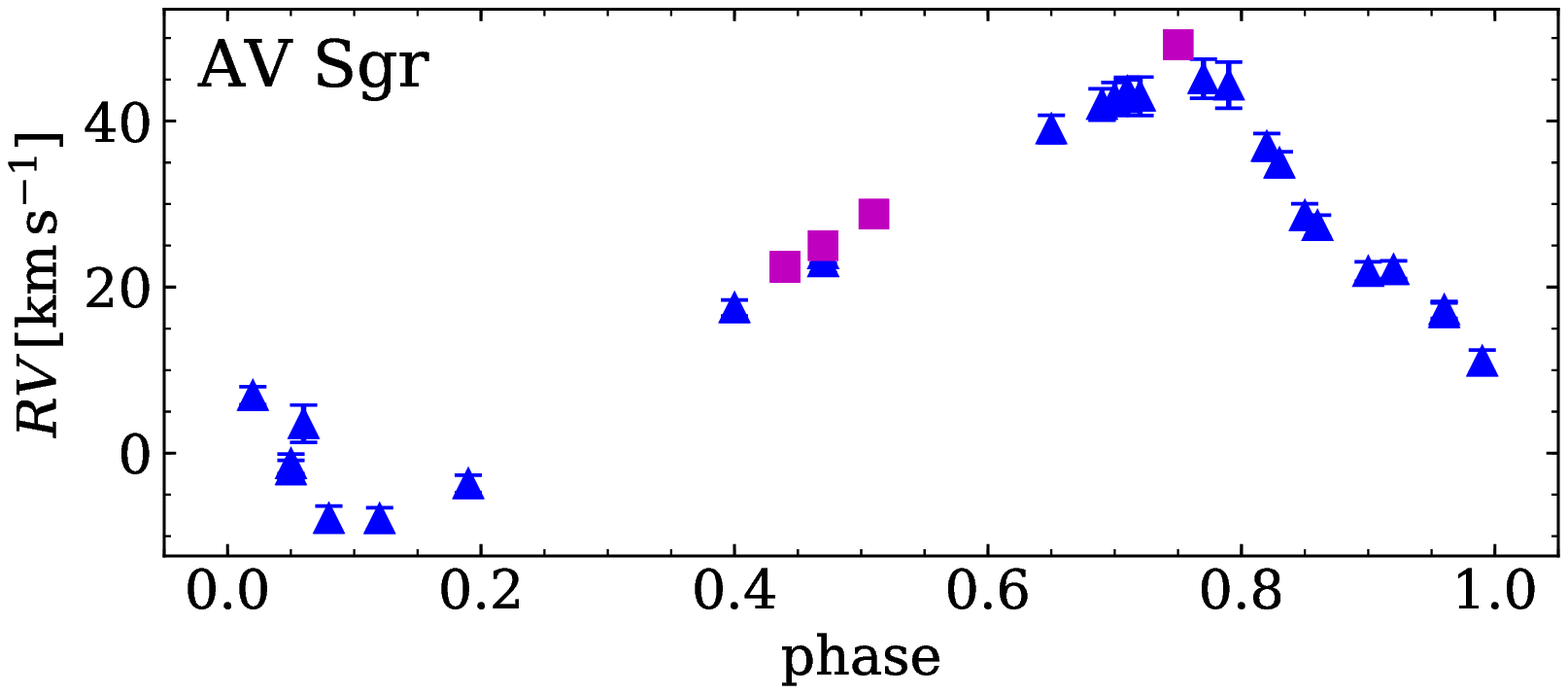}}
\end{minipage} \\
\begin{minipage}[t]{0.33\textwidth}
\centering
\resizebox{\hsize}{!}{\includegraphics{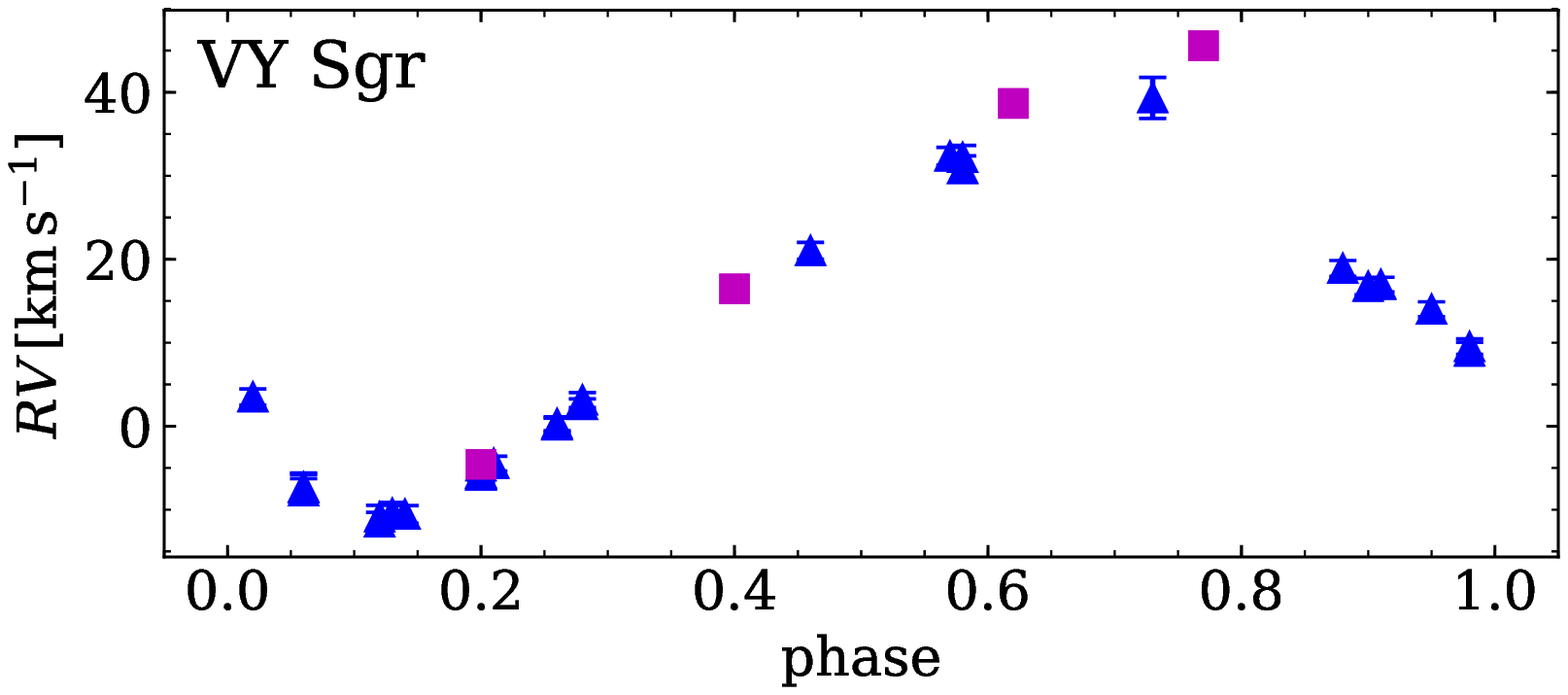}}
\end{minipage}
\begin{minipage}[t]{0.33\textwidth}
\centering
\resizebox{\hsize}{!}{\includegraphics{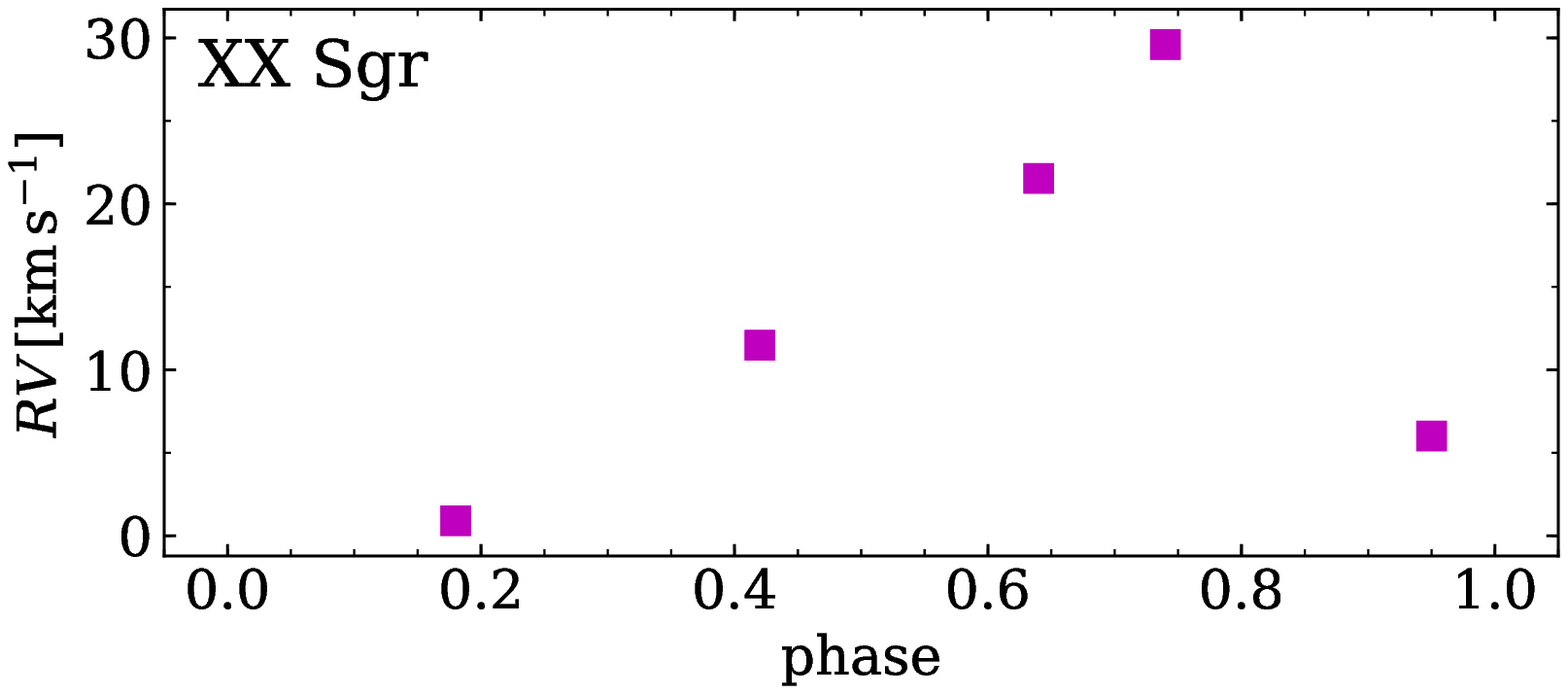}}
\end{minipage}
\begin{minipage}[t]{0.33\textwidth}
\centering
\resizebox{\hsize}{!}{\includegraphics{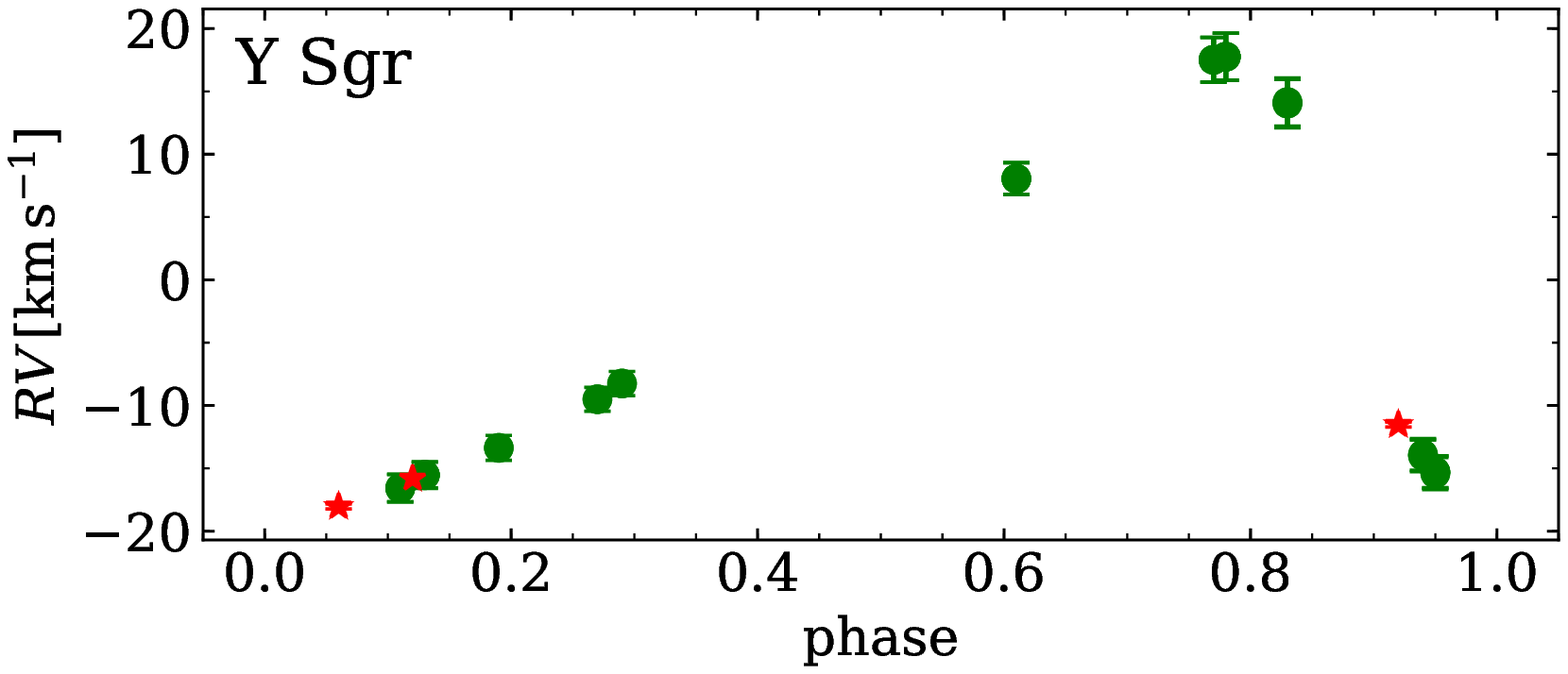}}
\end{minipage} \\
\begin{minipage}[t]{0.33\textwidth}
\centering
\resizebox{\hsize}{!}{\includegraphics{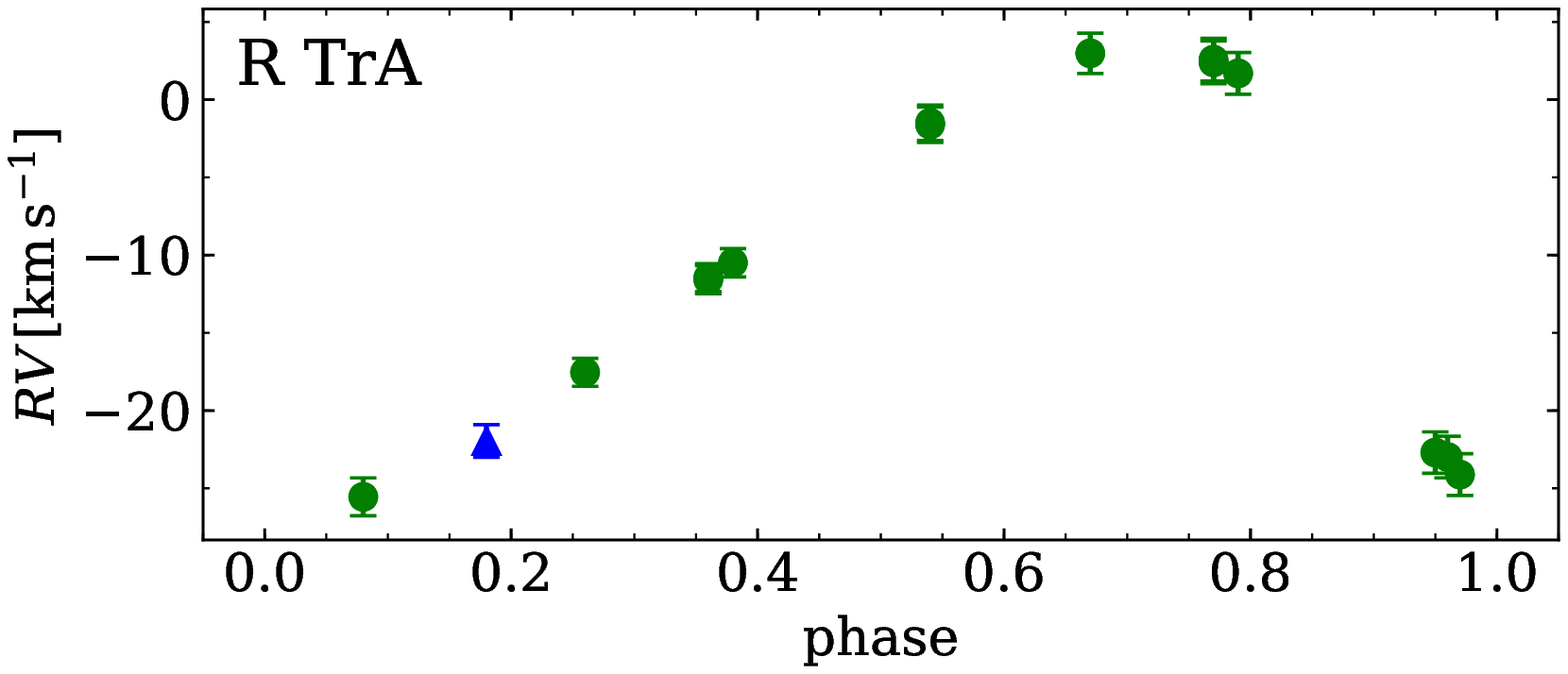}}
\end{minipage}
\begin{minipage}[t]{0.33\textwidth}
\centering
\resizebox{\hsize}{!}{\includegraphics{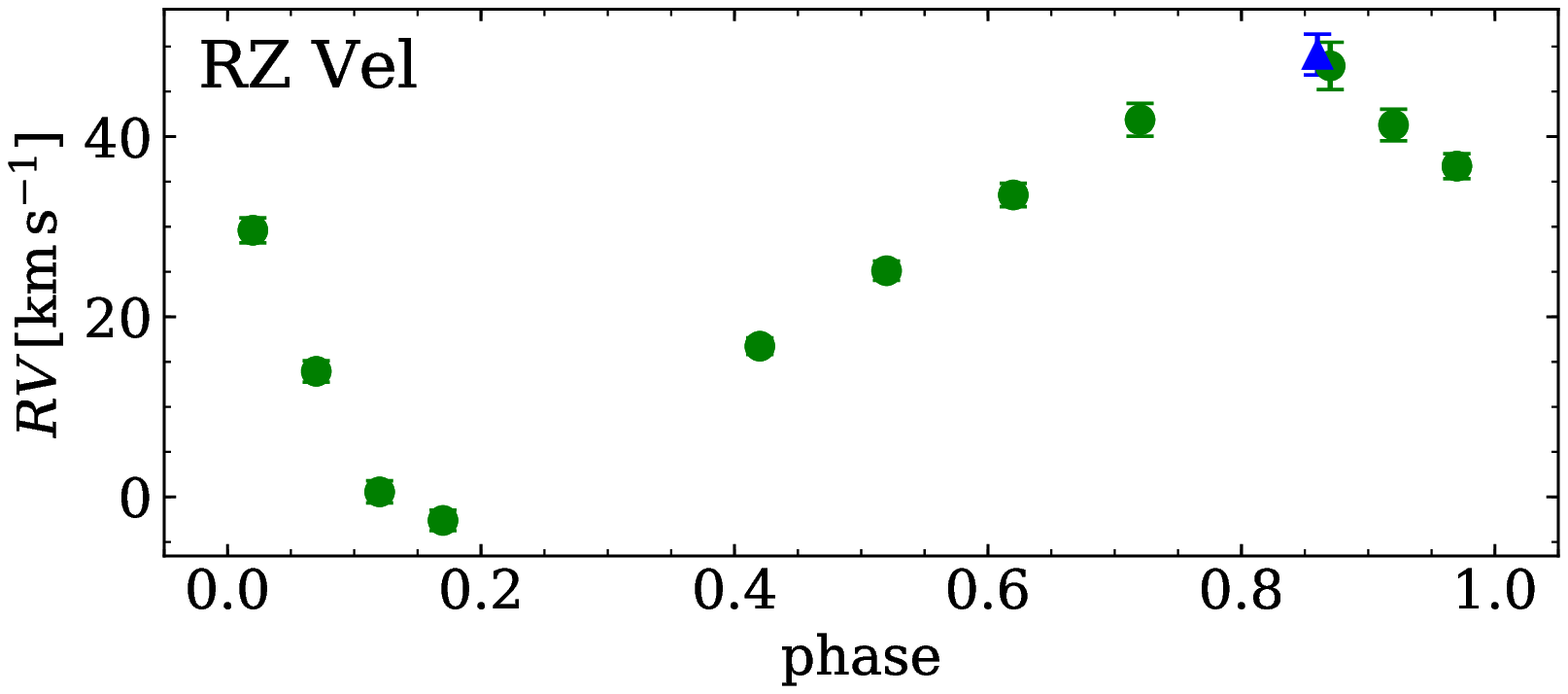}}
\end{minipage}
\begin{minipage}[t]{0.33\textwidth}
\centering
\strut
\end{minipage}
\caption{Radial velocities as a function of the pulsation phase. Measurements based on different spectrographs have been marked with different colors and symbols. The error bars in some cases are smaller than the symbol size.}
\label{vrad_phase}
\end{figure*}

\subsection{Radial velocity estimates}

Radial-velocity measurements were performed for the entire spectroscopic data set, i.e., we also included the spectra for which the S/N was not good enough for the spectroscopic analysis. The radial-velocity curves as a function of the pulsation phase are shown in Fig.~\ref{vrad_phase}. Typical radial-velocity errors are around 0.1~\kms, and normally smaller than 0.5~\kms, which is often smaller than the symbol size in the figure. The pulsation phase corresponding to each observed spectrum has been computed on the basis of the pulsation period and the photometric data (V-band) available in the literature \citep{Groenewegen2008,Stormetal2011a}. However, in most cases the time span between the photometric and the spectroscopic observation is larger than 30 years. On such long time-scales, Cepheids change their period due to evolutionary effects, and by using an outdated period we would introduce a scatter in the folded curves. In order to avoid this effect, we combined the photometric and spectroscopic data to compute a more accurate period, by using a generalized Lomb-Scargle algorithm. The new periods are listed in Table~\ref{calibceph}, together with the zero-phase reference epoch ($T_0$) corresponding to maximum light in the V-band.

\begin{figure}
\centering
\includegraphics[width=\hsize]{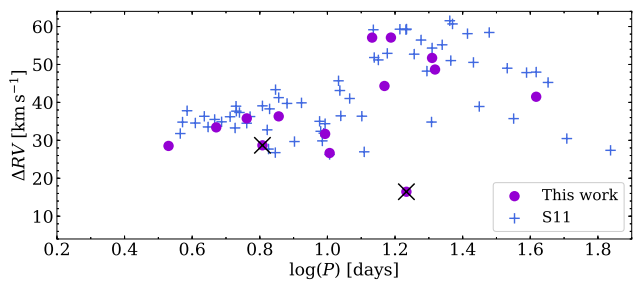}
\caption{The same as in Fig.~\ref{delta_teff_logp}, but showing the radial velocity amplitude.}
\label{delta_rv_logp}
\end{figure}

To validate the radial-velocity amplitudes estimated for the calibrating Cepheids,
Fig.~\ref{delta_rv_logp} shows the comparison between the current values and those provided by
S11 for IRSB Galactic Cepheids. Note that only two stars (\object{Y\,Oph} and \object{XX\,Sgr})
are marked with black crosses because the phase coverage is better in the RV plots. The two
datasets agree quite well over the entire period range and display the expected V-shape across
the Hertzsprung progression ($\log{P}\sim1.0$). In this context, we would like to draw the
attention to a small sample of Cepheids with periods ranging from $\log{P}\sim1.1$ to
$\log{P}\sim1.6$ that, at fixed period, display radial velocity amplitudes that are on 
average a factor of two smaller than the bulk of Cepheids.
One possible culprit could be the metallicity, since there is preliminary evidence that 
the amplitudes might decrease when moving into the more metal-poor regime
\citep[see Fig.~11 in][]{Genovalietal2014}. The quality and the homogeneity of
the spectra we are collecting will allow us, on a time scale of a few years, to provide more
quantitative constraints on this working hypothesis.

\section{Iron abundance determinations}
\label{metal}

The output file of the MOOG code provides the iron abundance for each one of the \ion{Fe}{i} and \ion{Fe}{ii} lines passed as input. The current estimates, when compared with similar estimates available in the literature, present several advantages:

\begin{itemize}

\item[a)] Our sample has between five and more than one hundred spectra per star. This is the reason why the current mean iron abundances have intrinsic errors smaller than 0.1~dex. Before the present work, the number of classical Cepheids for which multiple measurements were available were only a few \citep[see e.g.][]{LuckAndrievsky2004,Lucketal2008,Romanielloetal2008,Genovalietal2014}.

\item[b)] The current high-resolution spectra cover both the rising and decreasing branch. This means that they cover the pulsation phases during which Cepheids experience the largest variations in effective temperature, surface gravity and microturbulent velocity.

\item[c)] The current calibrating Cepheids roughly cover the period range typical of Galactic classical Cepheids, i.e., from $\sim$3 to more than 40~days. This means that the current sample in the Bailey Diagram (luminosity amplitude versus logarithmic period) covers both the low and the large amplitude regime. Moreover, we are also sampling the region of Bump Cepheids. Classical Cepheids with periods ranging from $\sim$7 to $\sim$10~days display a well defined bump either along the rising (shorter periods) or along the decreasing (longer periods) branch. This is the so-called Hertzsprung progression. The physical mechanisms driving the occurrence of this phenomenon are not fully understood yet, but there is mounting evidence that it is driven by nonlinear phenomena (shocks) across the entire envelope \citep{Bonoetal2000b}.

\end{itemize}

The individual \ion{Fe}{i} and \ion{Fe}{ii} abundances and their uncertainties, together with the number of lines used are listed in Table~\ref{paramscontrolsample}. Note that the number of lines measured for \ion{Fe}{i} lines ranges from a few tens to more than one hundred, while for \ion{Fe}{ii} it ranges from a few to almost two dozen. These are smaller than the total number of lines available (570 for \ion{Fe}{i} and 45 for \ion{Fe}{ii}) because the number of lines measured in each spectrum is limited, e.g., by the wavelength range covered by the instruments, by the intensity of the lines for a given spectral type, by their quality and by the level of blending. Moreover, for each spectrum, we performed a cleaning in order to remove lines that systematically provided too high/low abundances (outside 2$\sigma$) when compared with the average abundance.

The mean \ion{Fe}{i} and \ion{Fe}{ii} abundances together with the mean intrinsic parameters are listed in Table~\ref{meanparams}. A glance at the data given in this table indicates that mean abundances based either on neutral or on ionized iron have similar errors.

The comparison between the current mean iron abundances and similar estimates 
available in the literature (see Table~\ref{calibceph}) indicates that they agree quite well 
within the errors. The iron abundances and their errors listed in Table~\ref{calibceph} come 
from \citet{Genovalietal2014}, in which they derived spectroscopic abundances for the entire
sample of CCs based on high-resolution optical spectra.
Note that for the measurements for which the original authors did not 
provide an estimate of the error, they assumed a typical error of 0.1~dex. 
This means that the difference between the current mean iron abundances and 
similar estimates available in the literature is, on average, smaller than 
1$\sigma$. There is only one exception, \object{$\zeta$\,Gem}, for which the difference is 
of the order of 3$\sigma$. The reason for this difference is not clear. 
\object{$\zeta$\,Gem} is the object with the highest precision, since we analyzed 128
spectra and they cover the entire pulsation cycle. Moreover,   
the variation of the physical parameters (see Fig.~\ref{atmpar_phase}) is smooth during
both rising and decreasing branch, and both \ion{Fe}{i} and \ion{Fe}{ii} estimates display 
minimal variations along the entire pulsation cycle of this object (see Fig.~\ref{feh_phase}).
Nevertheless, we should notice that \object{$\zeta$\,Gem} is a peculiar Cepheid that has been investigated by \citet{Szabados1983} and classified as a variable star having secular period changes. Indeed, its radial velocity clearly changes with the epoch, as can be seen in Fig.~\ref{vrad_phase}, not just due to a velocity offset but real phase shifts seem to be observed, possibly caused by an unseen companion.

\begin{figure}
\centering
\includegraphics[width=\hsize]{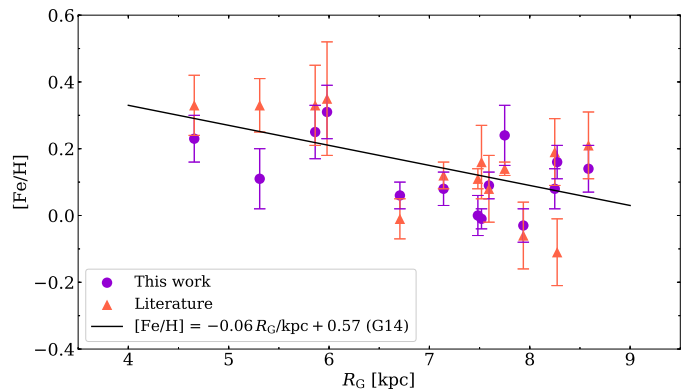}
\caption{Iron abundances as a function of Galactocentric distances for the calibrating Cepheids. Values derived in the present work (Table~\ref{meanparams}) are compared with those from the literature (Table~\ref{calibceph}). The error bars on our metallicity estimates are the largest value between the uncertainty on the weighted mean and the standard deviation. The metallicity gradient derived by \citet{Genovalietal2014} is also shown.}
\label{feh_RGal}
\end{figure}

To further quantify the difference between the current iron abundances and 
similar abundances available in the literature, Fig.~\ref{feh_RGal} displays 
the comparison with the metallicity gradient of Galactic Cepheids provided 
by \citet{Genovalietal2014}. Data plotted in this figure show some interesting results:
$a)$ The current sample follows quite well the global metallicity gradient and 
the new homogeneous mean abundances display a smaller dispersion compared 
with the literature ones (0.10 vs. 0.12~dex).
$b)$ The new mean abundances of the four innermost disk objects (\object{V340\,Ara},
\object{UZ\,Sct}, \object{VY\,Sgr}, \object{AV\,Sgr}) are systematically more metal-poor
than literature estimates. This means that the determination of the metallicity gradient
in the transition zone between the inner disk and the Bulge \citep{Bonoetal2013,
Genovalietal2014} will strongly benefit of more homogeneous and accurate mean iron
abundances (Inno et al., in prep.).

\subsection{Phase dependence}

In the determination of the effective temperature, high standard deviations are not automatically linked to problems in the line-depth measurements. Spectra of stars in the rising phase of their pulsation cycle (i.e., when their effective temperature is increasing) present themselves with intrinsically higher dispersions due to the star's variable nature (they are thus fixed by the physical structure). This could be checked only for the spectra with multiple measurements.

\begin{figure*}
\centering
\begin{minipage}[t]{0.33\textwidth}
\centering
\resizebox{\hsize}{!}{\includegraphics{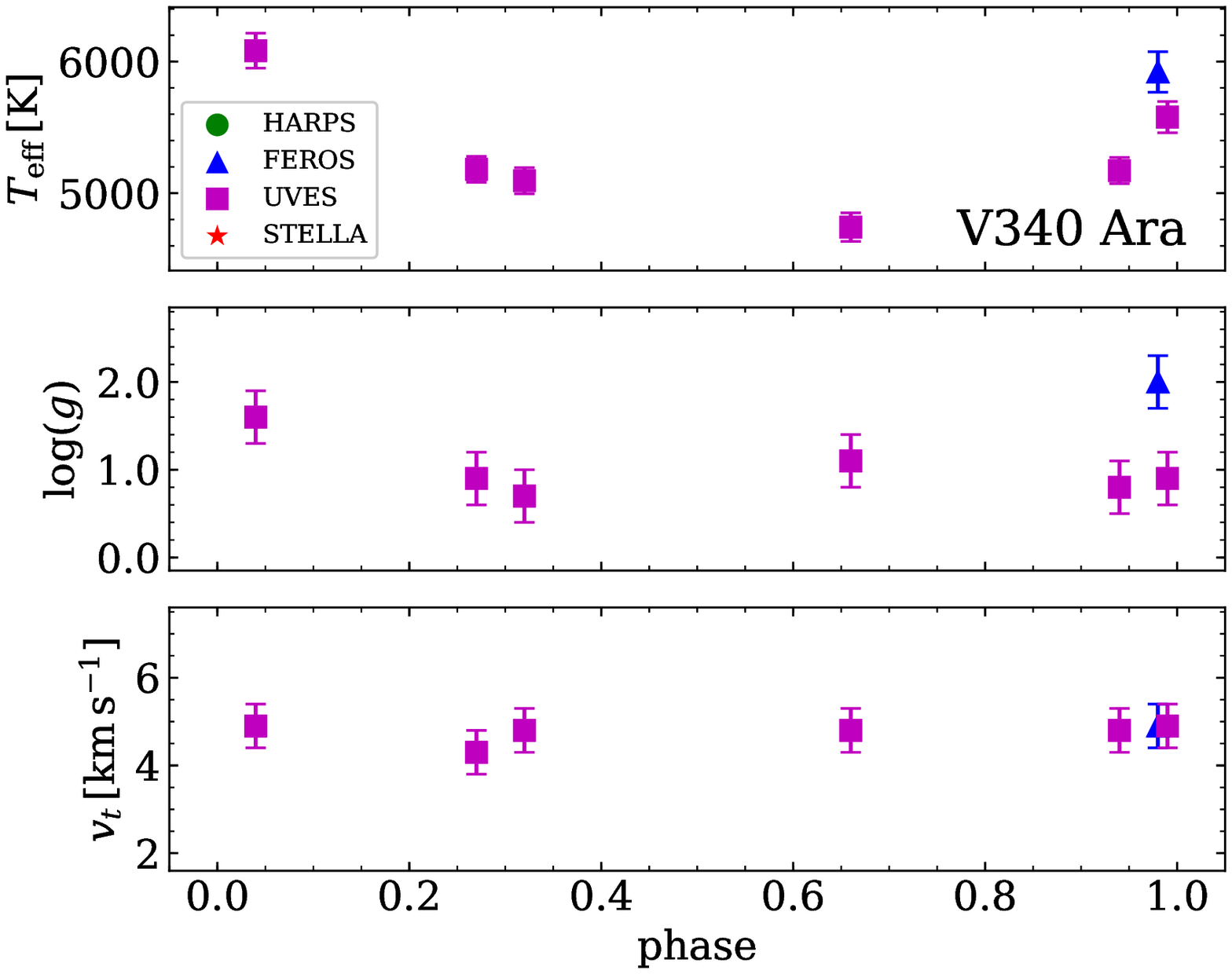}}
\end{minipage}
\begin{minipage}[t]{0.33\textwidth}
\centering
\resizebox{\hsize}{!}{\includegraphics{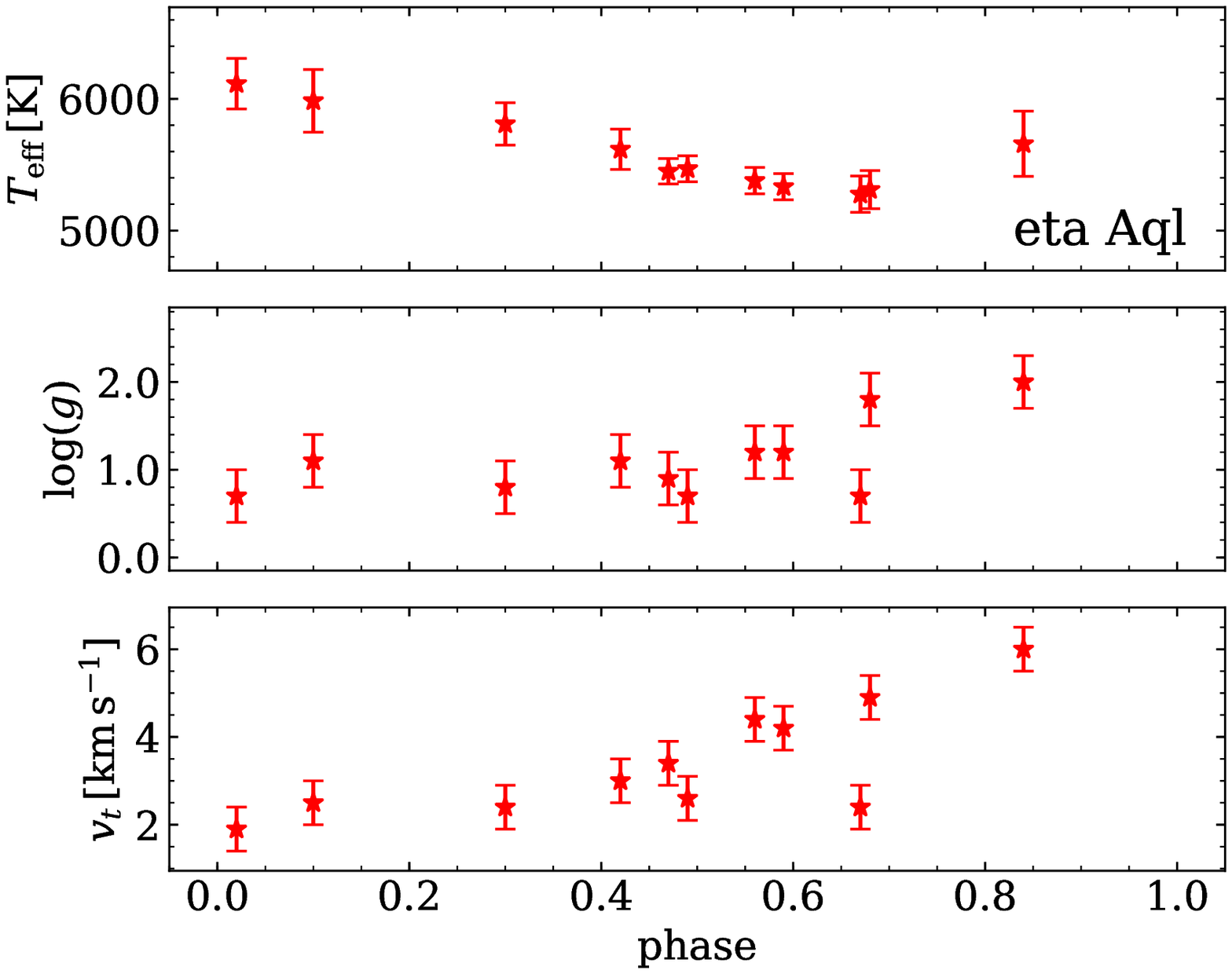}}
\end{minipage}
\begin{minipage}[t]{0.33\textwidth}
\centering
\resizebox{\hsize}{!}{\includegraphics{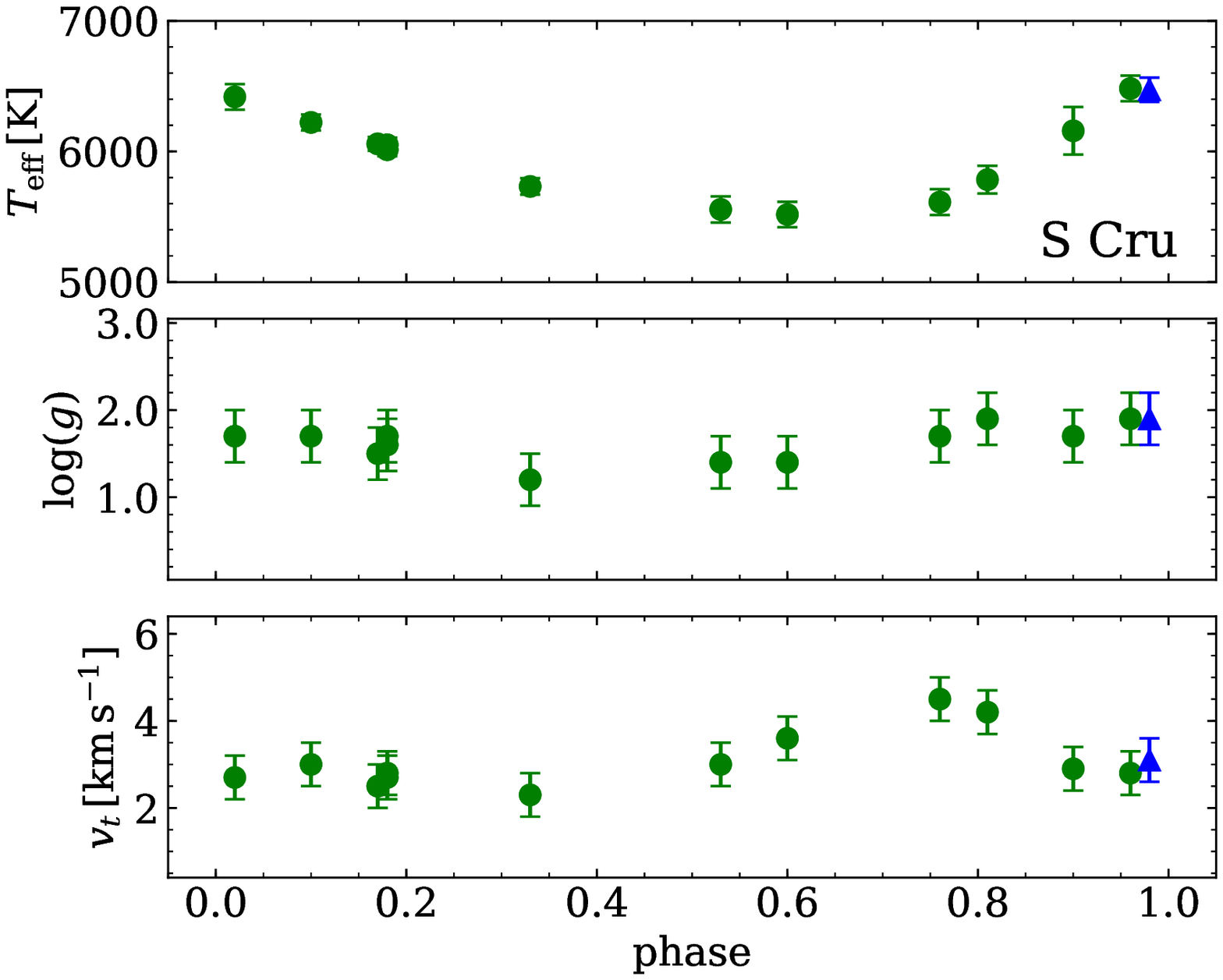}}
\end{minipage} \\
\begin{minipage}[t]{0.33\textwidth}
\centering
\resizebox{\hsize}{!}{\includegraphics{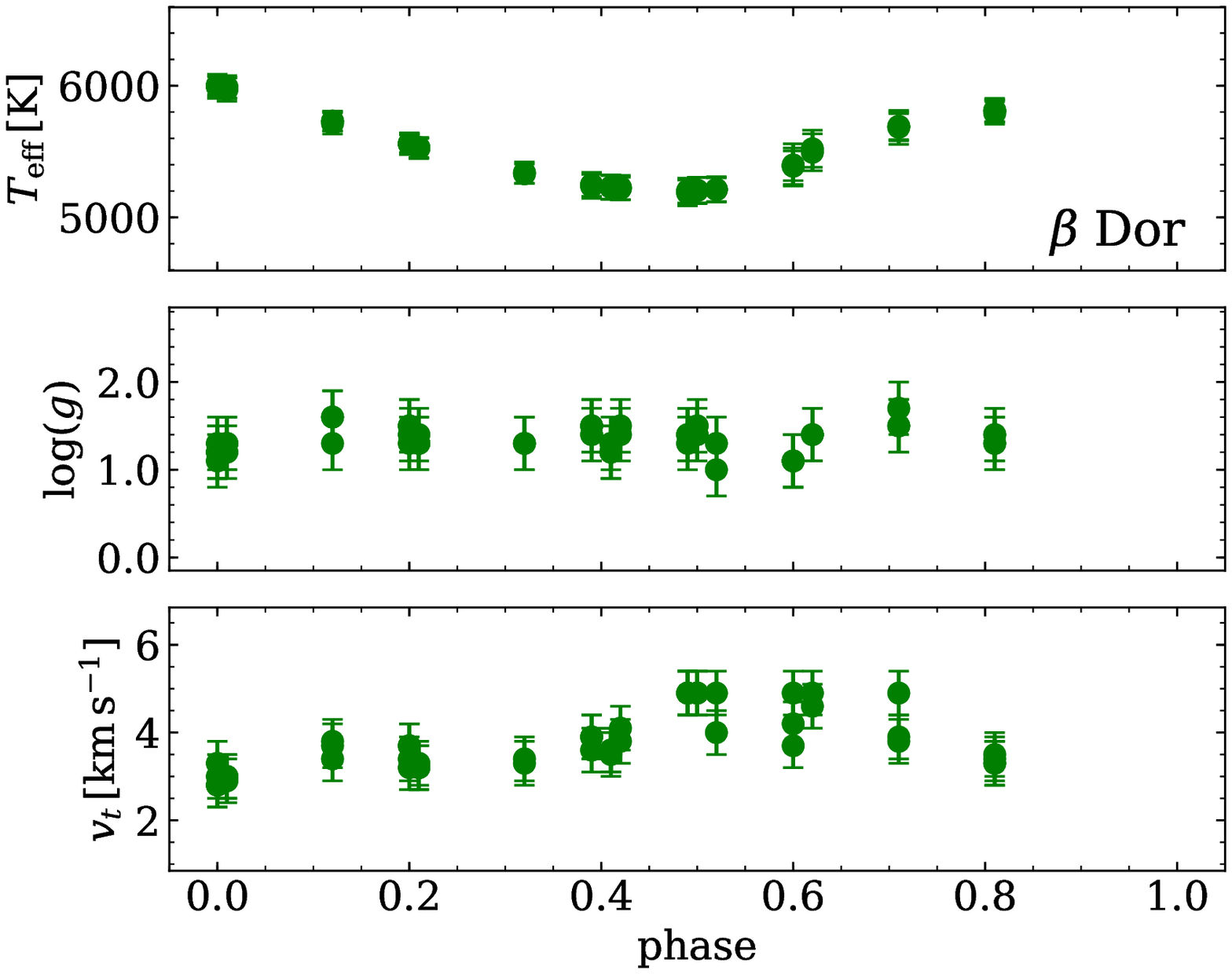}}
\end{minipage}
\begin{minipage}[t]{0.33\textwidth}
\centering
\resizebox{\hsize}{!}{\includegraphics{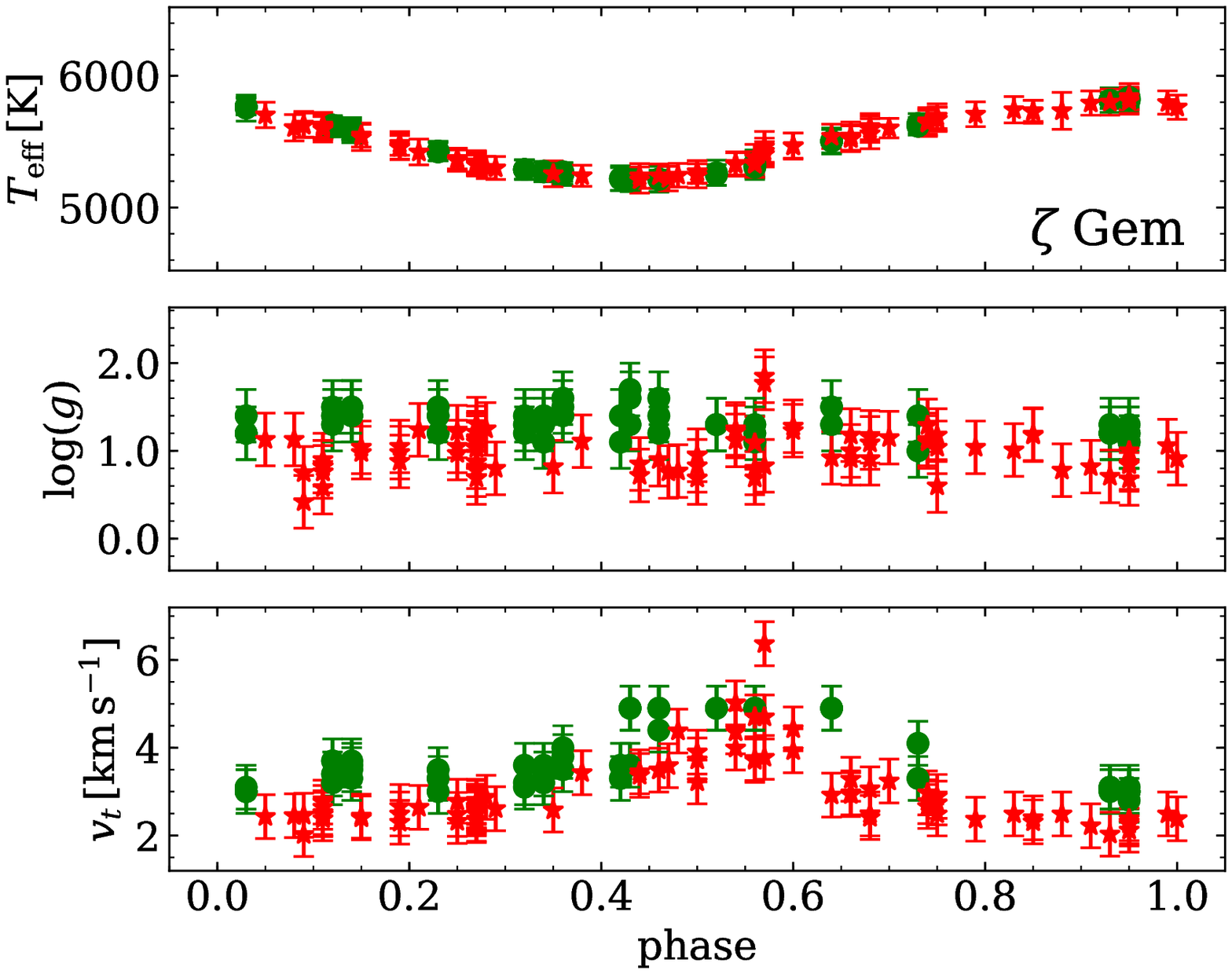}}
\end{minipage}
\begin{minipage}[t]{0.33\textwidth}
\centering
\resizebox{\hsize}{!}{\includegraphics{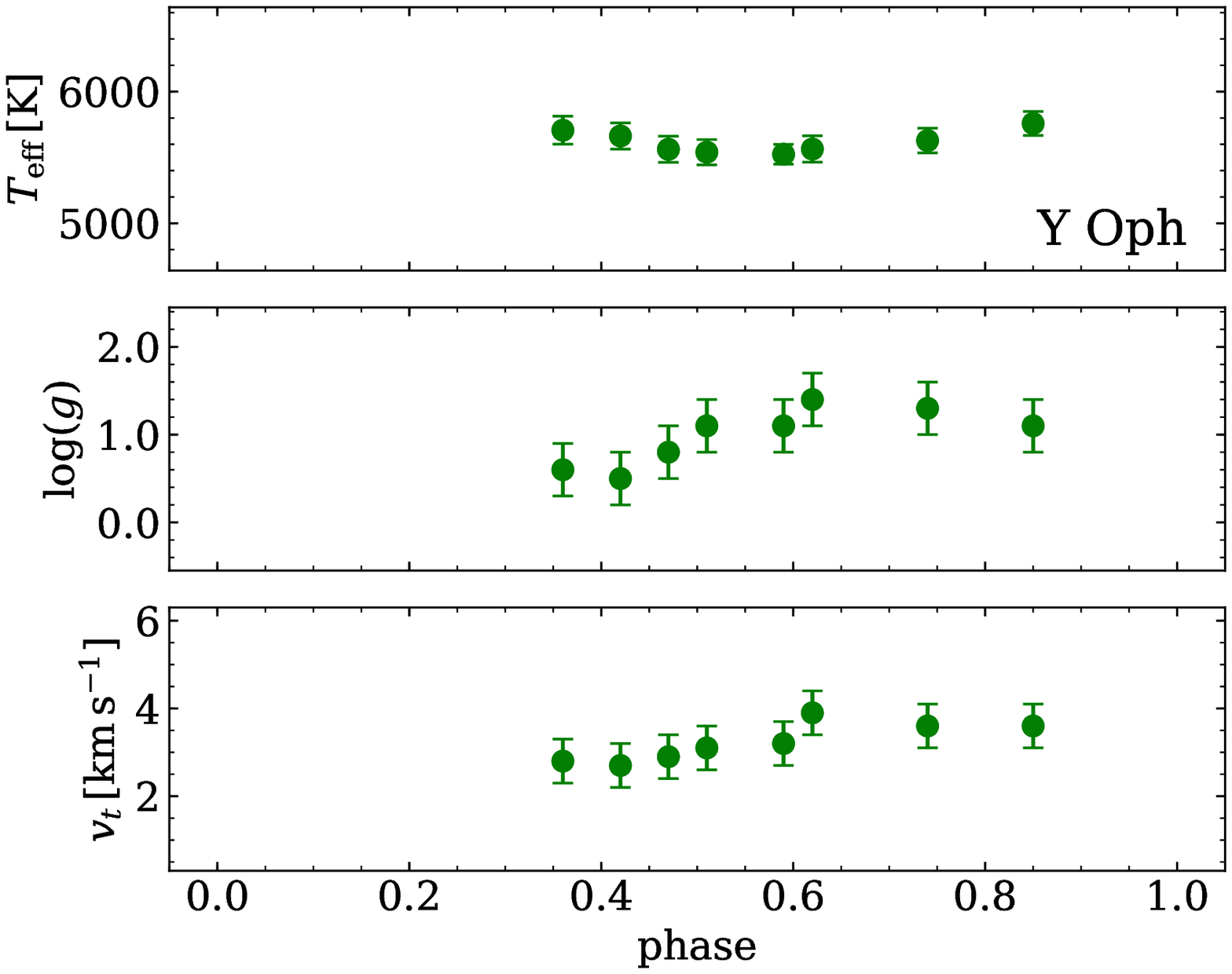}}
\end{minipage} \\
\begin{minipage}[t]{0.33\textwidth}
\centering
\resizebox{\hsize}{!}{\includegraphics{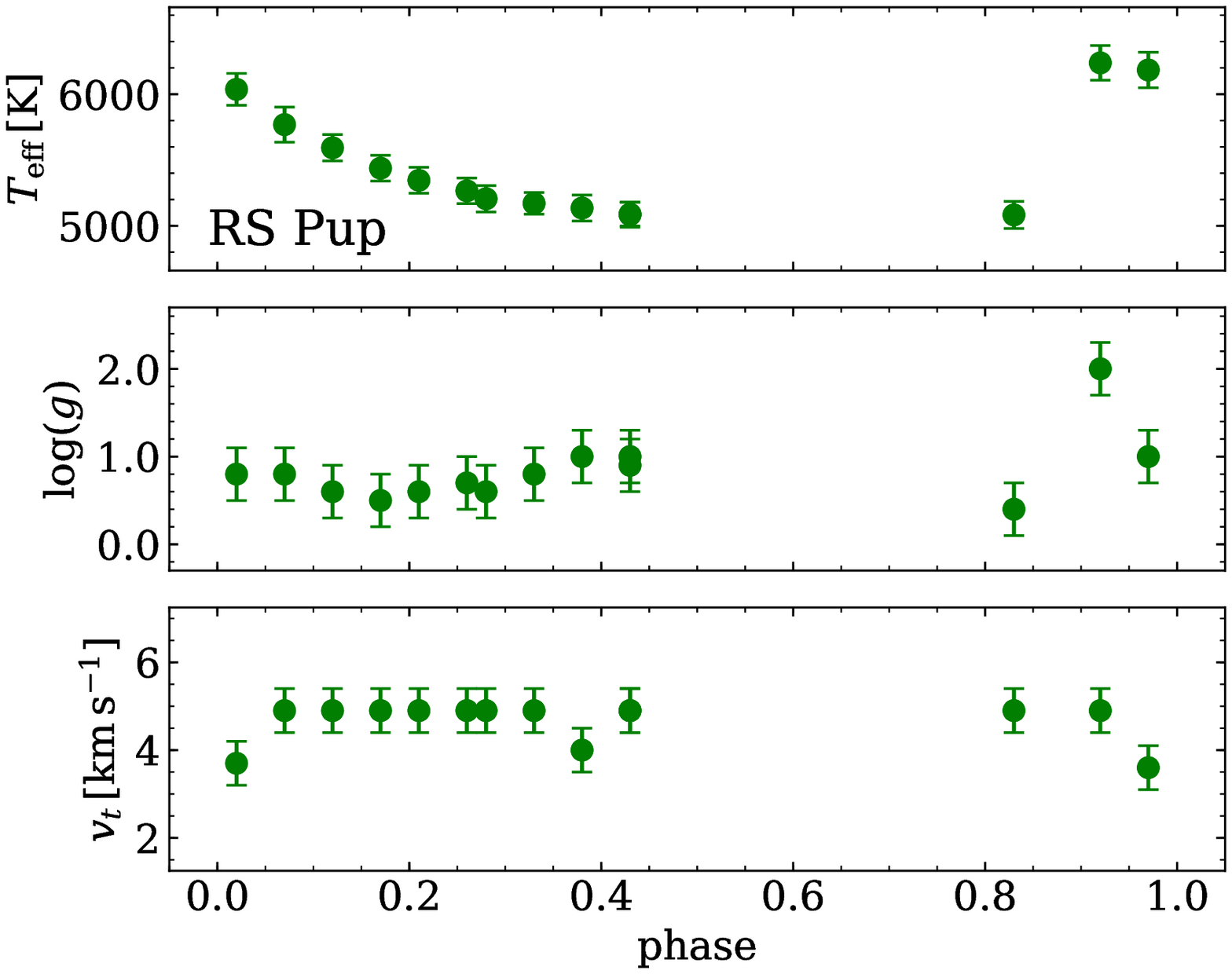}}
\end{minipage}
\begin{minipage}[t]{0.33\textwidth}
\centering
\resizebox{\hsize}{!}{\includegraphics{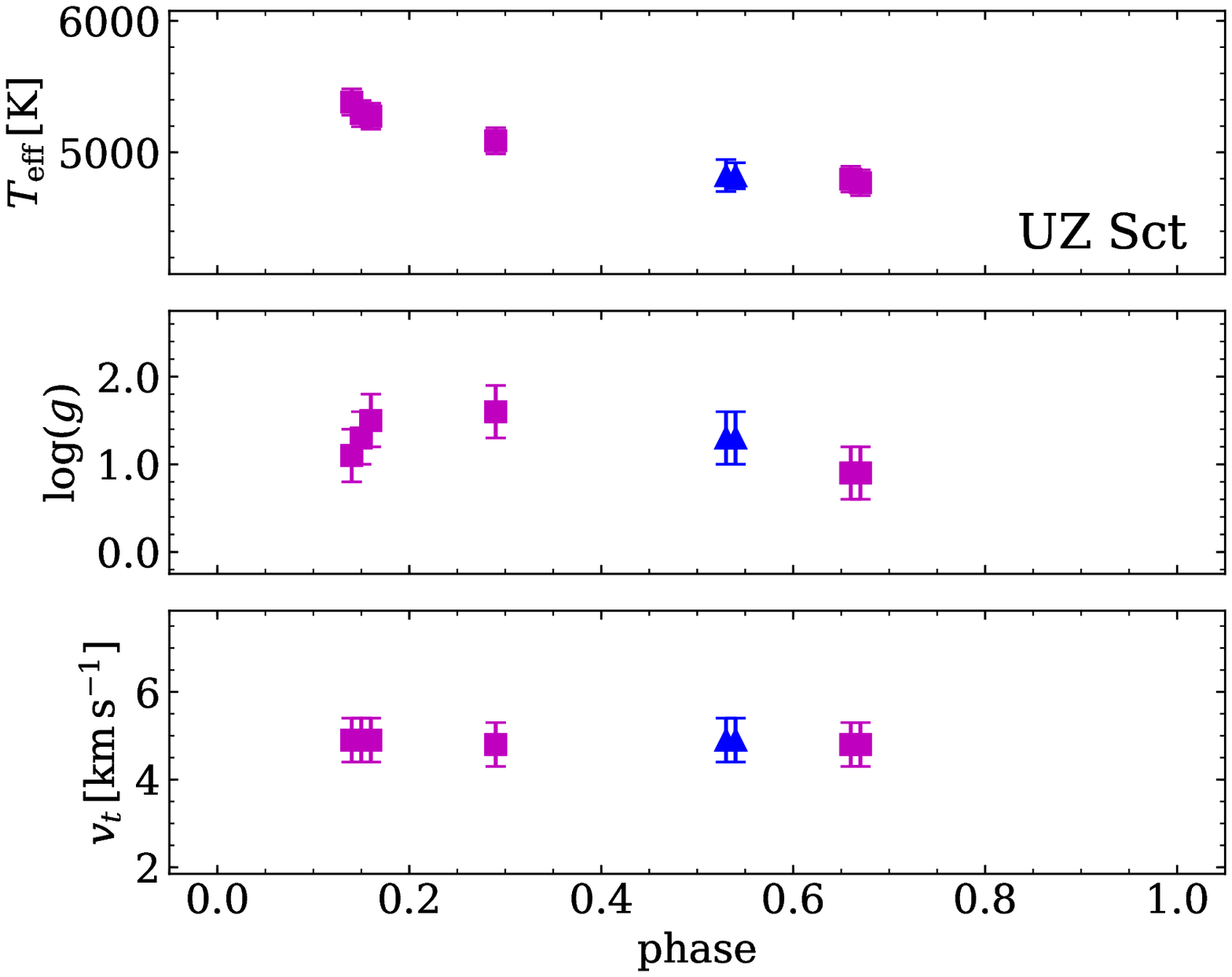}}
\end{minipage}
\begin{minipage}[t]{0.33\textwidth}
\centering
\resizebox{\hsize}{!}{\includegraphics{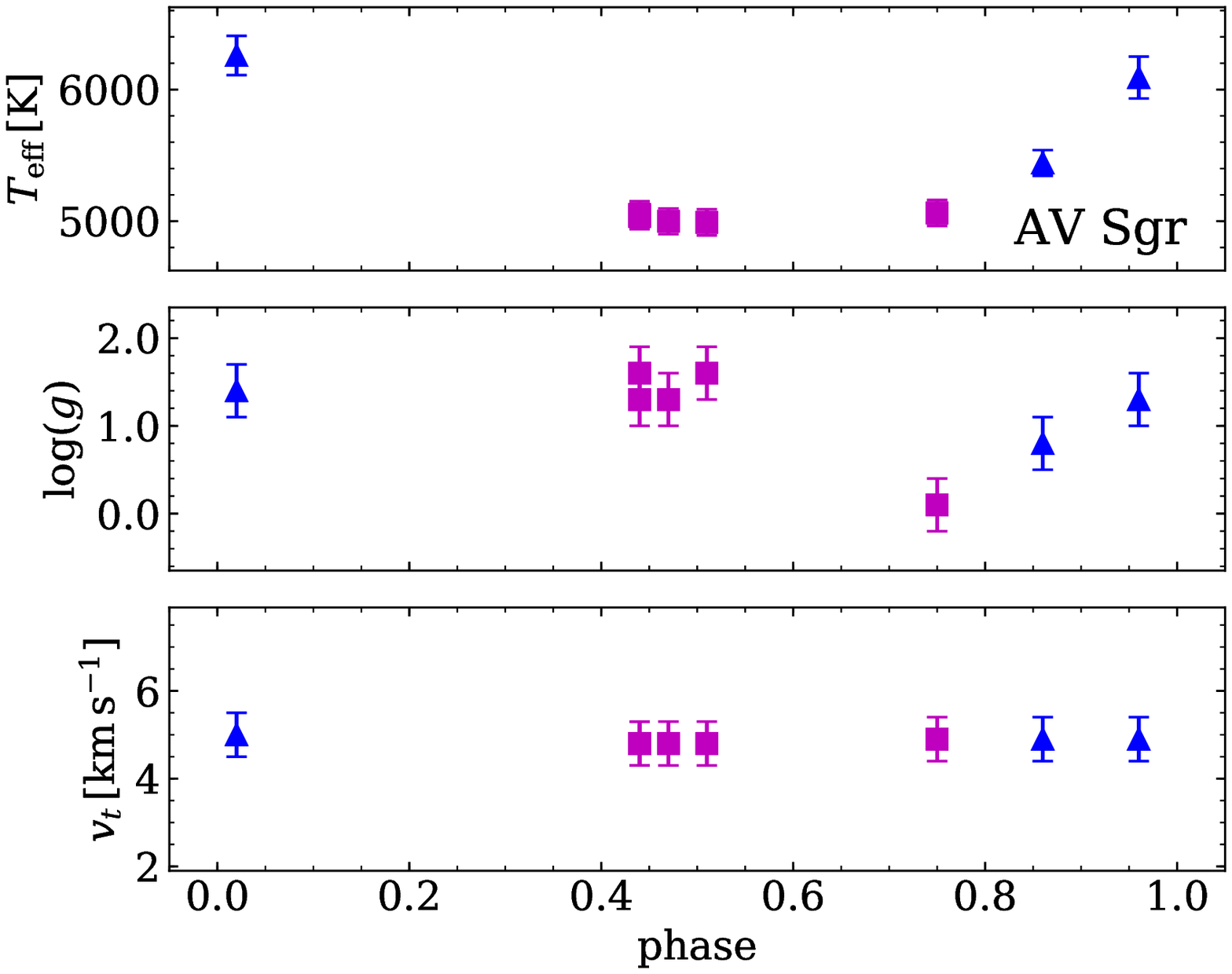}}
\end{minipage} \\
\begin{minipage}[t]{0.33\textwidth}
\centering
\resizebox{\hsize}{!}{\includegraphics{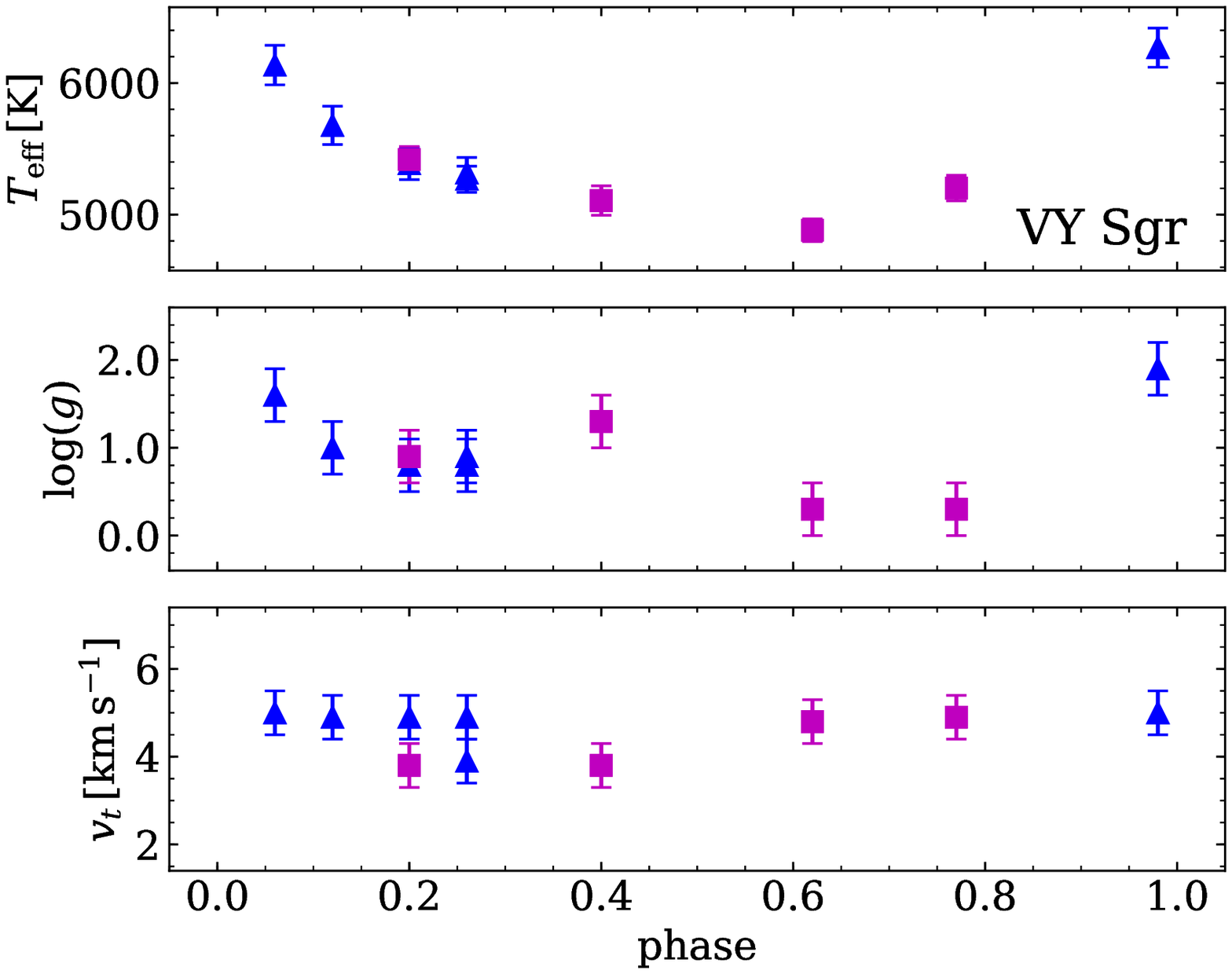}}
\end{minipage}
\begin{minipage}[t]{0.33\textwidth}
\centering
\resizebox{\hsize}{!}{\includegraphics{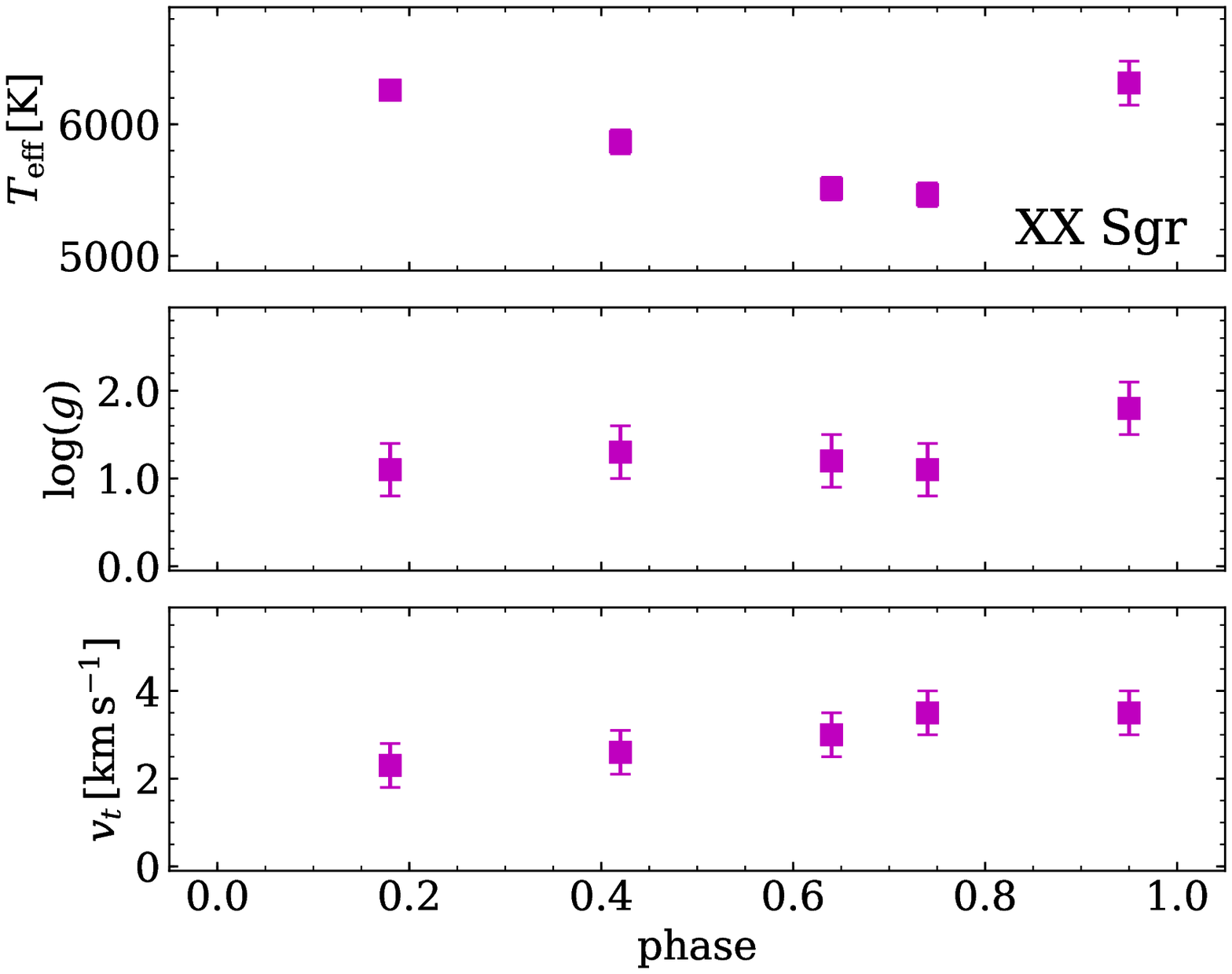}}
\end{minipage}
\begin{minipage}[t]{0.33\textwidth}
\centering
\resizebox{\hsize}{!}{\includegraphics{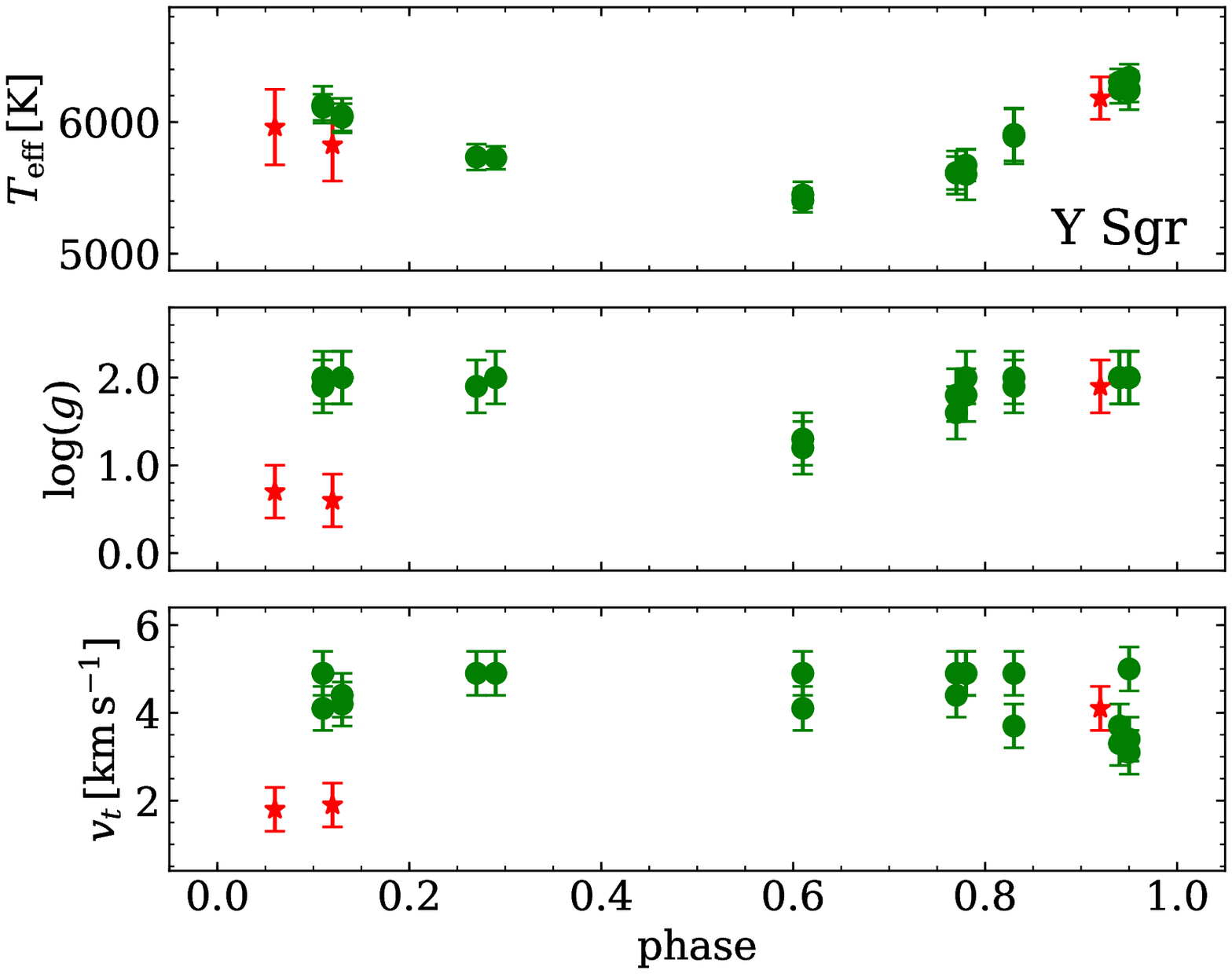}}
\end{minipage} \\
\begin{minipage}[t]{0.33\textwidth}
\centering
\resizebox{\hsize}{!}{\includegraphics{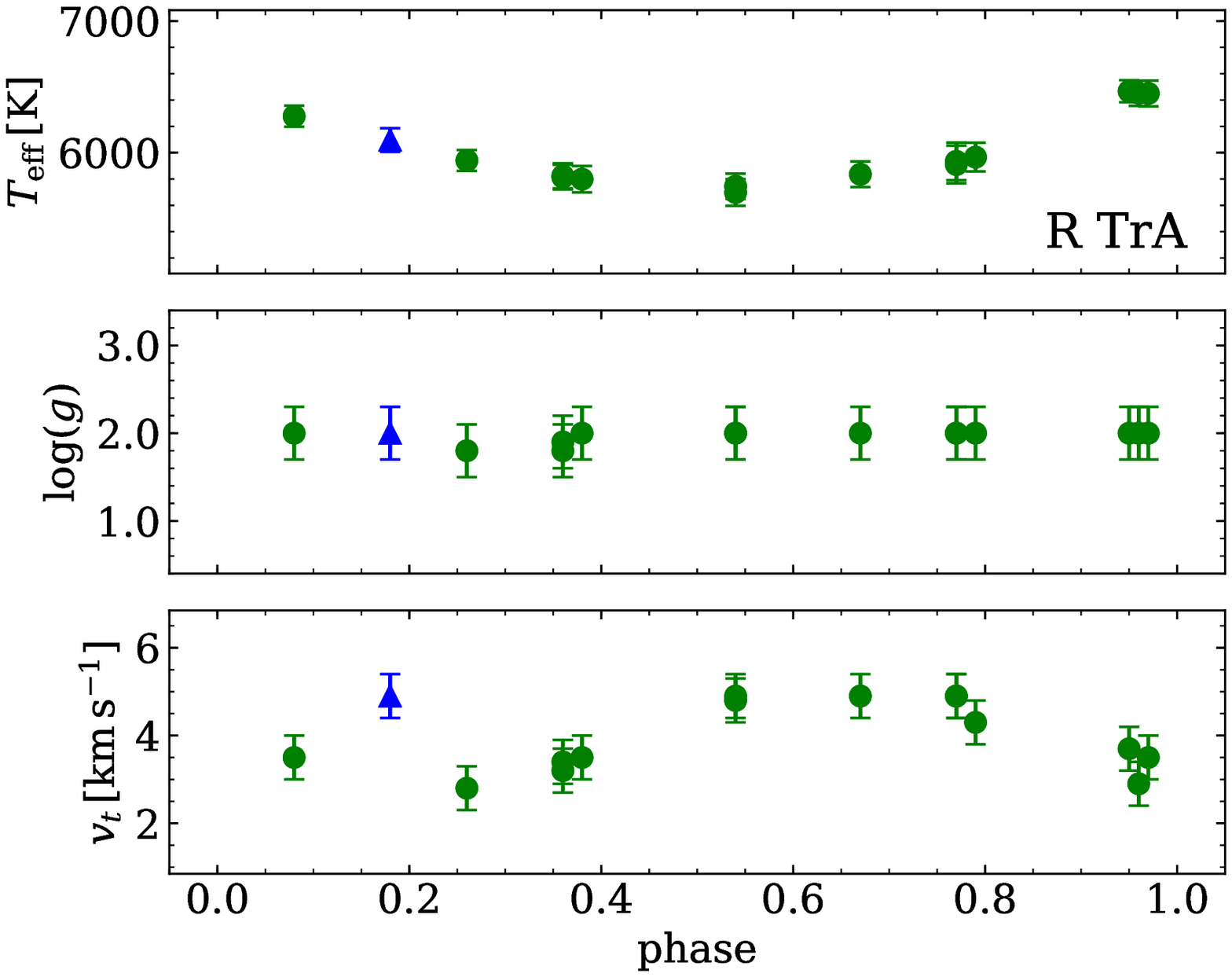}}
\end{minipage}
\begin{minipage}[t]{0.33\textwidth}
\centering
\resizebox{\hsize}{!}{\includegraphics{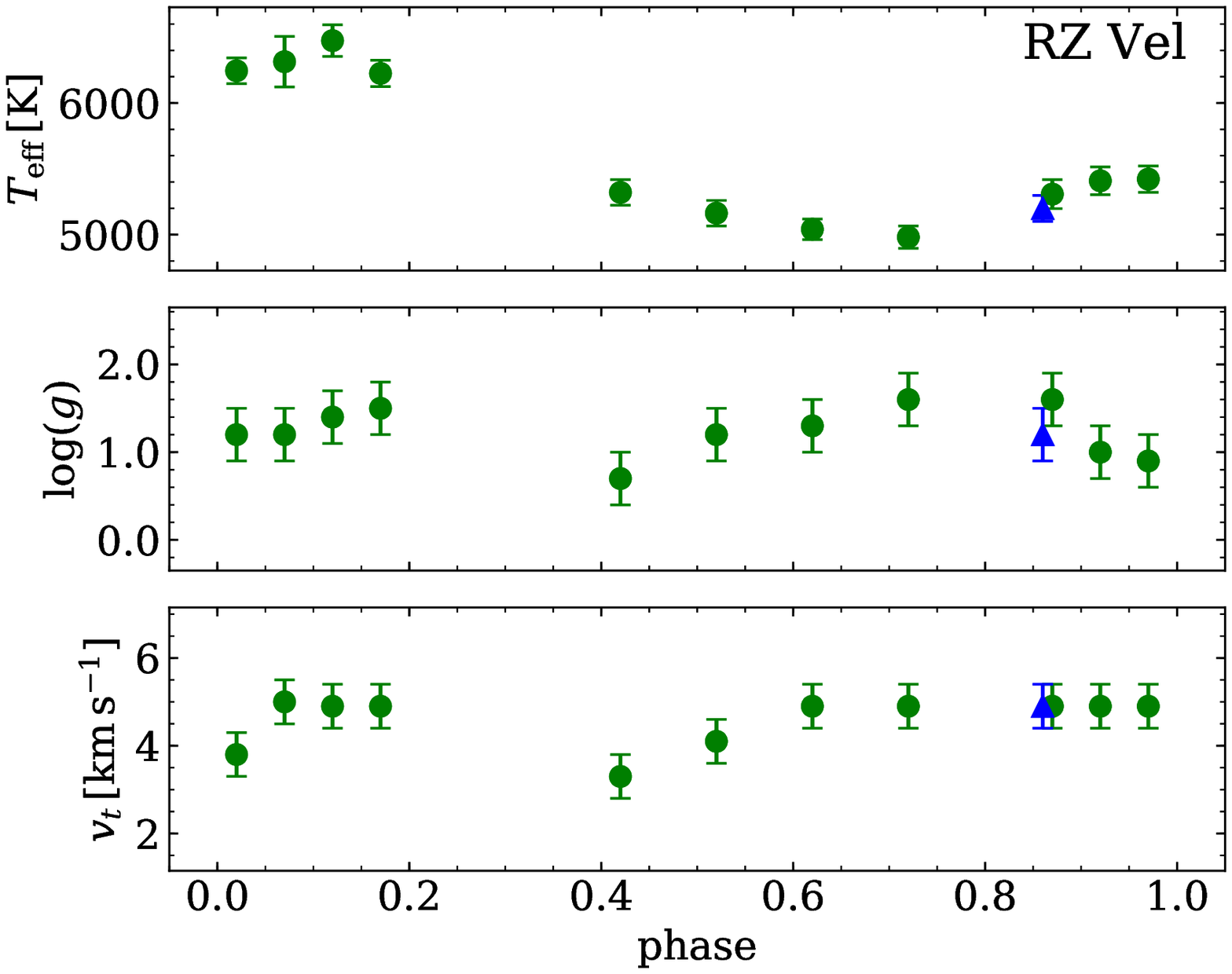}}
\end{minipage}
\begin{minipage}[t]{0.33\textwidth}
\centering
\strut
\end{minipage}
\caption{Atmospheric parameters as a function of the pulsation phase. Measurements from different spectrographs are marked with different colors and symbols. To help with the comparison, the panels are plotted with same y-axis range: 2000~K for \teff, 3~dex for \logg, and 6~\kms\ for \vmic.}
\label{atmpar_phase}
\end{figure*}

The dependence of the effective temperature on the pulsation phase 
is shown in Fig.~\ref{atmpar_phase}. The same figure also shows the 
variation of the surface gravity and of the microturbulent velocity with the 
pulsation phase, but the dependence is very weak given the uncertainties. 
It is worth mentioning that the microturbulent velocity peaks around the 
pulsation phases in which the Cepheid attains its lowest effective 
temperatures and soon after, i.e., the phases between $\sim$0.5 and 
$\sim$0.7/0.8 (see in Fig.~\ref{atmpar_phase}: \object{S\,Cru}, 
\object{$\beta$\,Dor}, \object{$\zeta$\,Gem}, \object{R\,TrA}). 
This evidence supports earlier findings by \citet{LuckAndrievsky2004}, 
\citet{Kovtyukhetal2005}, \citet{Andrievskyetal2005}, and \citet{Lucketal2008}, 
suggesting that the microturbulent velocity peaks around phases 0.6-0.8. 
A more quantitative comparison is hampered by the difference in the targets 
and in the phase coverage. We still lack detailed empirical constraints on 
the variation of the microturbulent velocity as a function of the pulsation 
period, and in particular, across the Hertzsprung progression. Homogeneous 
spectra covering the entire pulsation cycle and a broad period range are 
highly desired. The same outcome applies to the different approaches suggested 
to trace the variation of convective motions \citep{Gilletetal1999}.    

\begin{figure*}
\centering
\begin{minipage}[t]{0.33\textwidth}
\centering
\resizebox{\hsize}{!}{\includegraphics{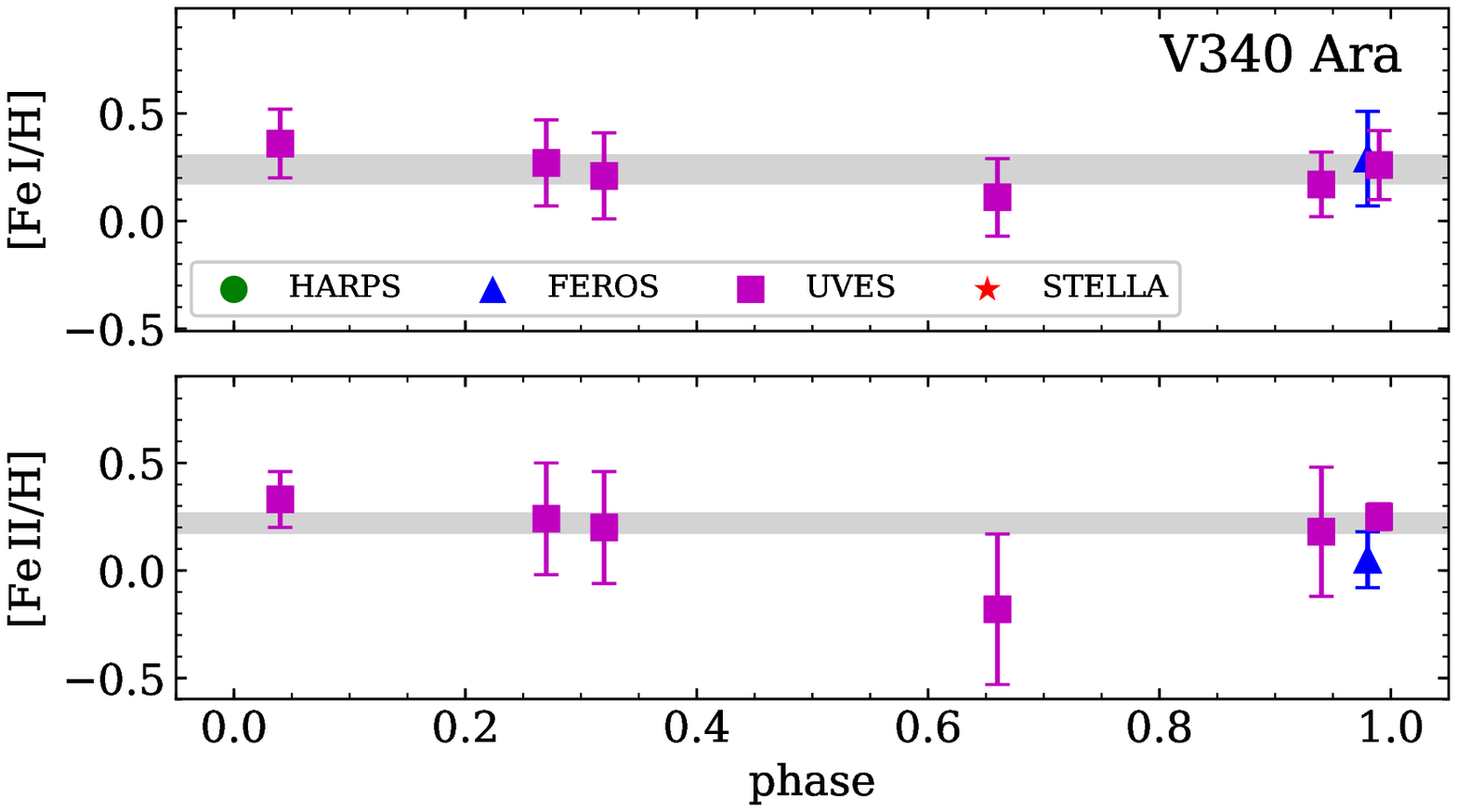}}
\end{minipage}
\begin{minipage}[t]{0.33\textwidth}
\centering
\resizebox{\hsize}{!}{\includegraphics{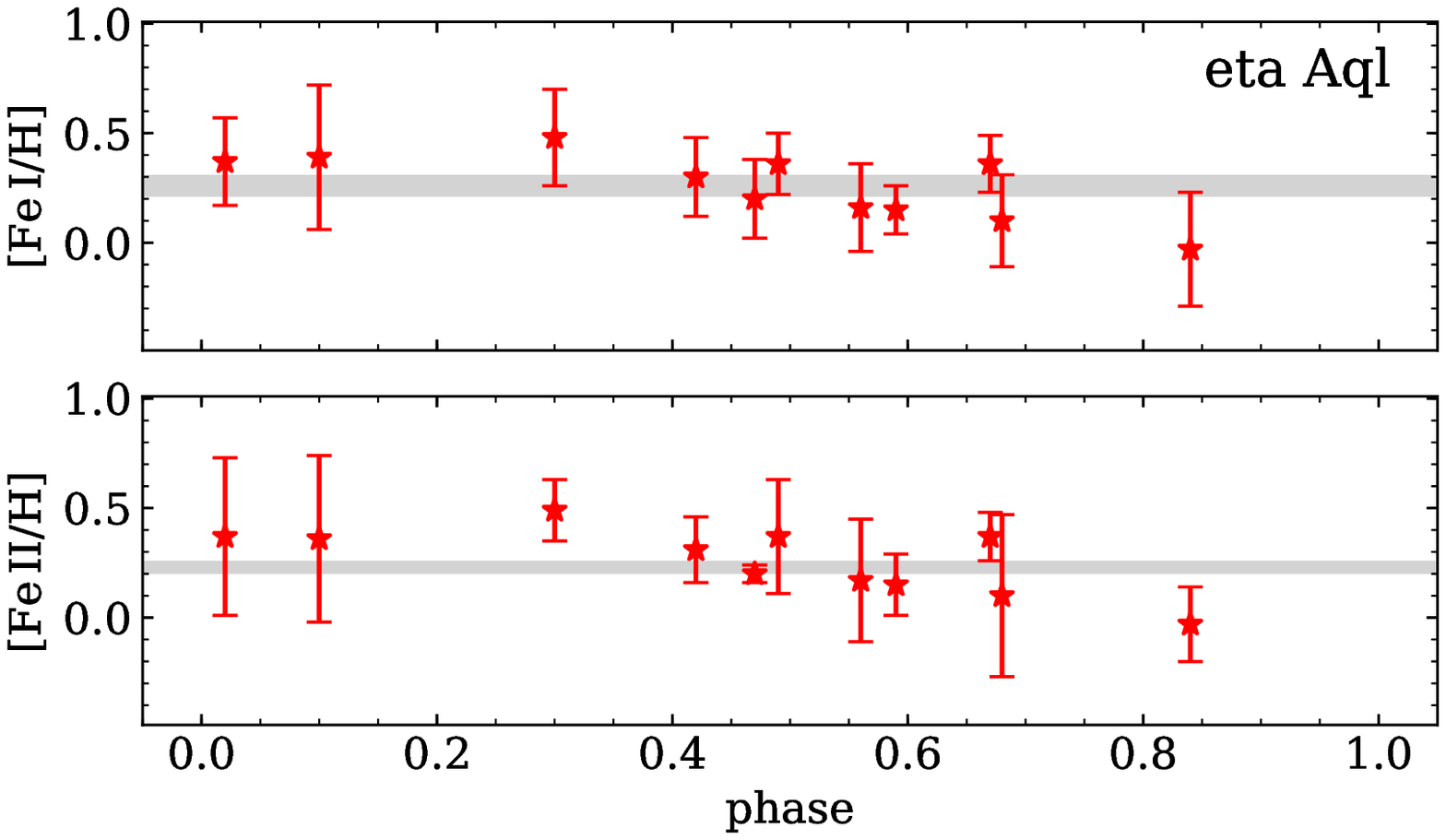}}
\end{minipage}
\begin{minipage}[t]{0.33\textwidth}
\centering
\resizebox{\hsize}{!}{\includegraphics{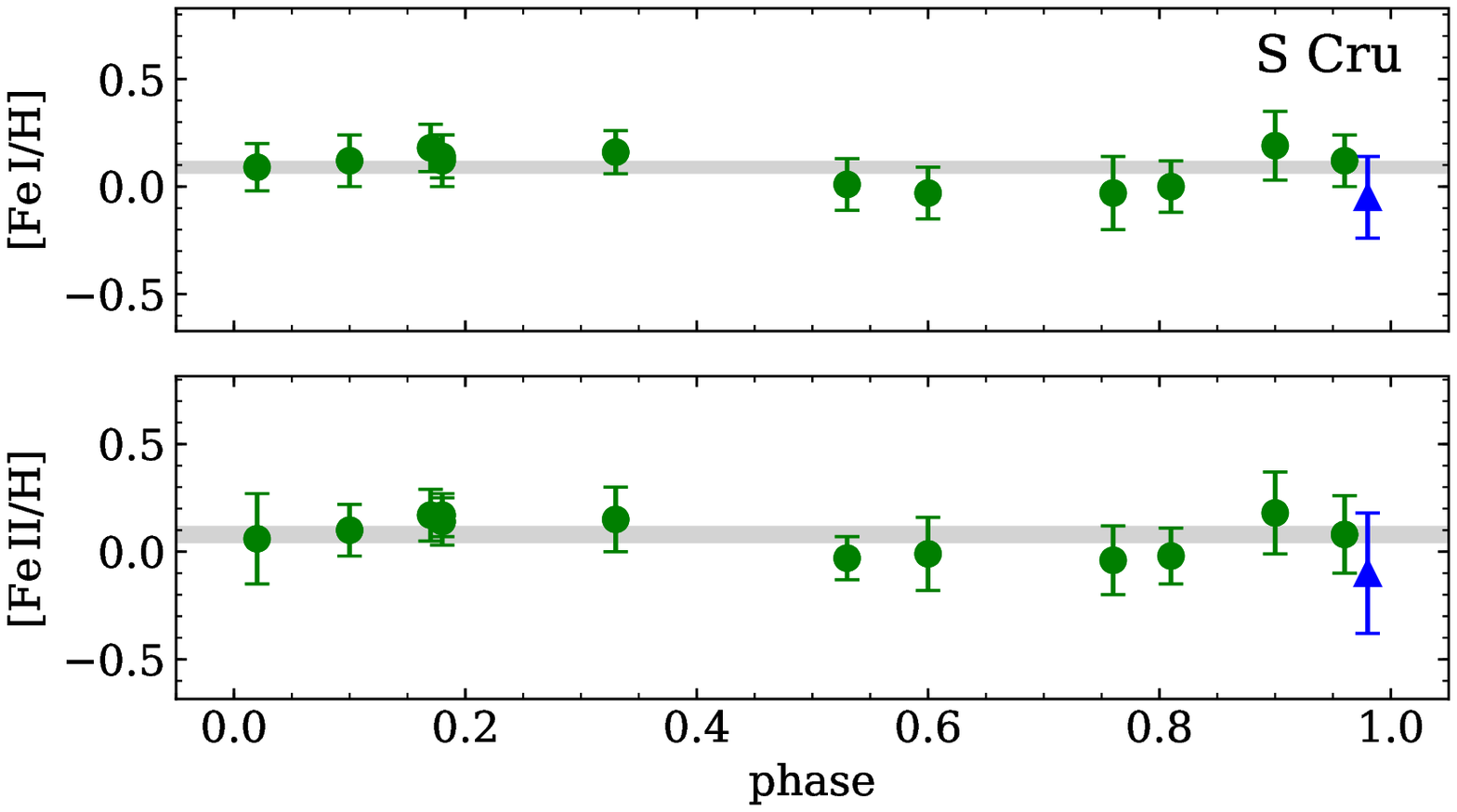}}
\end{minipage} \\
\begin{minipage}[t]{0.33\textwidth}
\centering
\resizebox{\hsize}{!}{\includegraphics{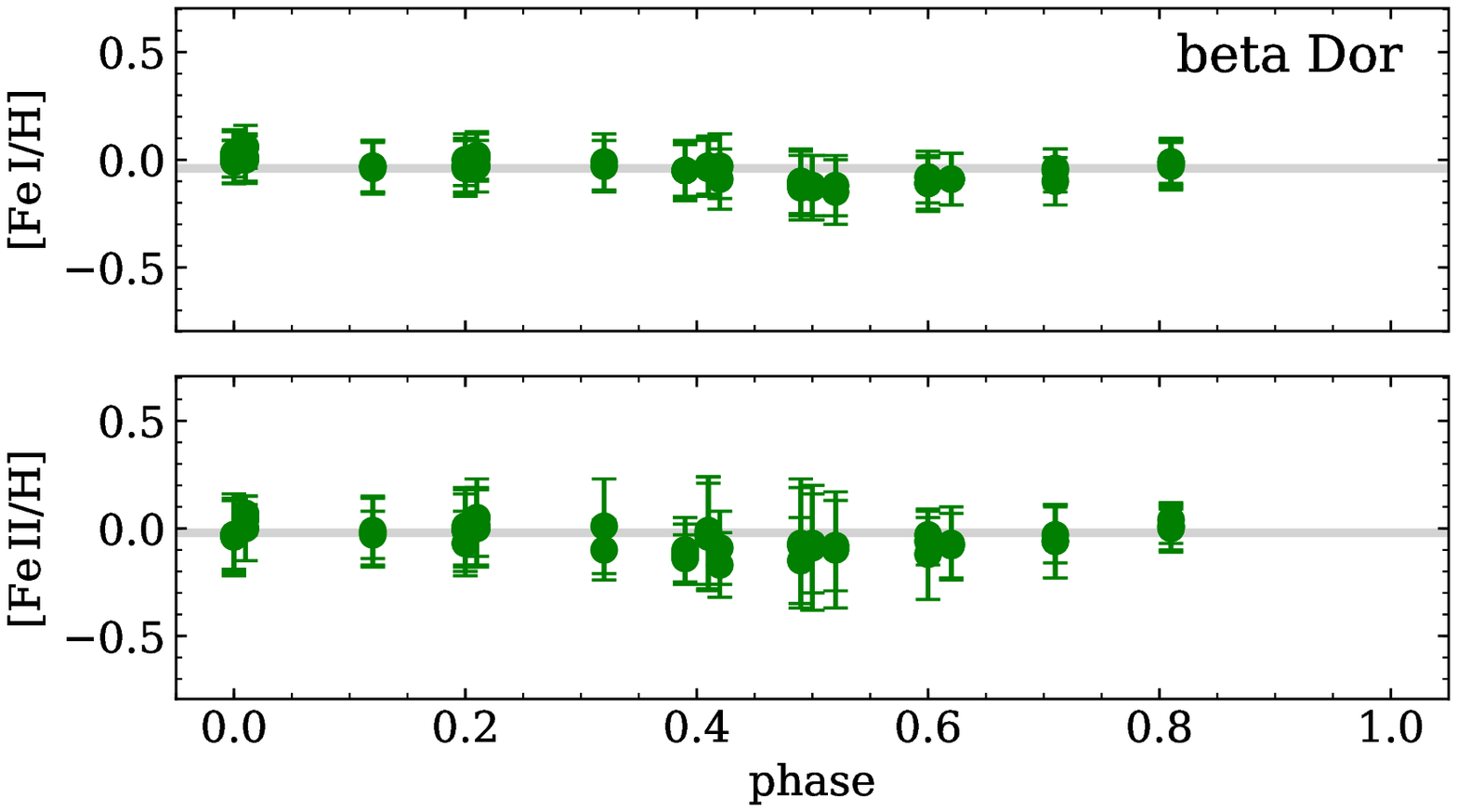}}
\end{minipage}
\begin{minipage}[t]{0.33\textwidth}
\centering
\resizebox{\hsize}{!}{\includegraphics{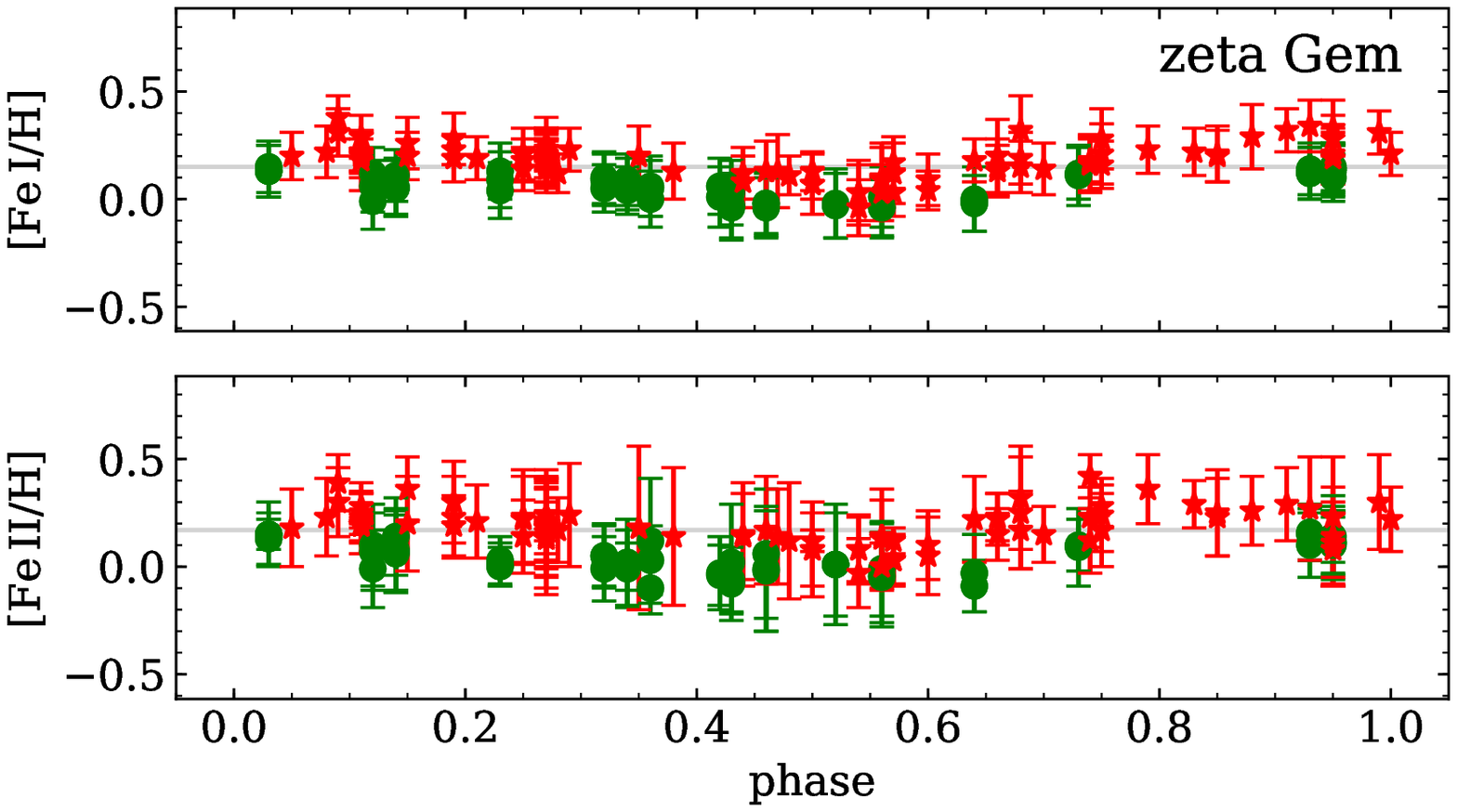}}
\end{minipage}
\begin{minipage}[t]{0.33\textwidth}
\centering
\resizebox{\hsize}{!}{\includegraphics{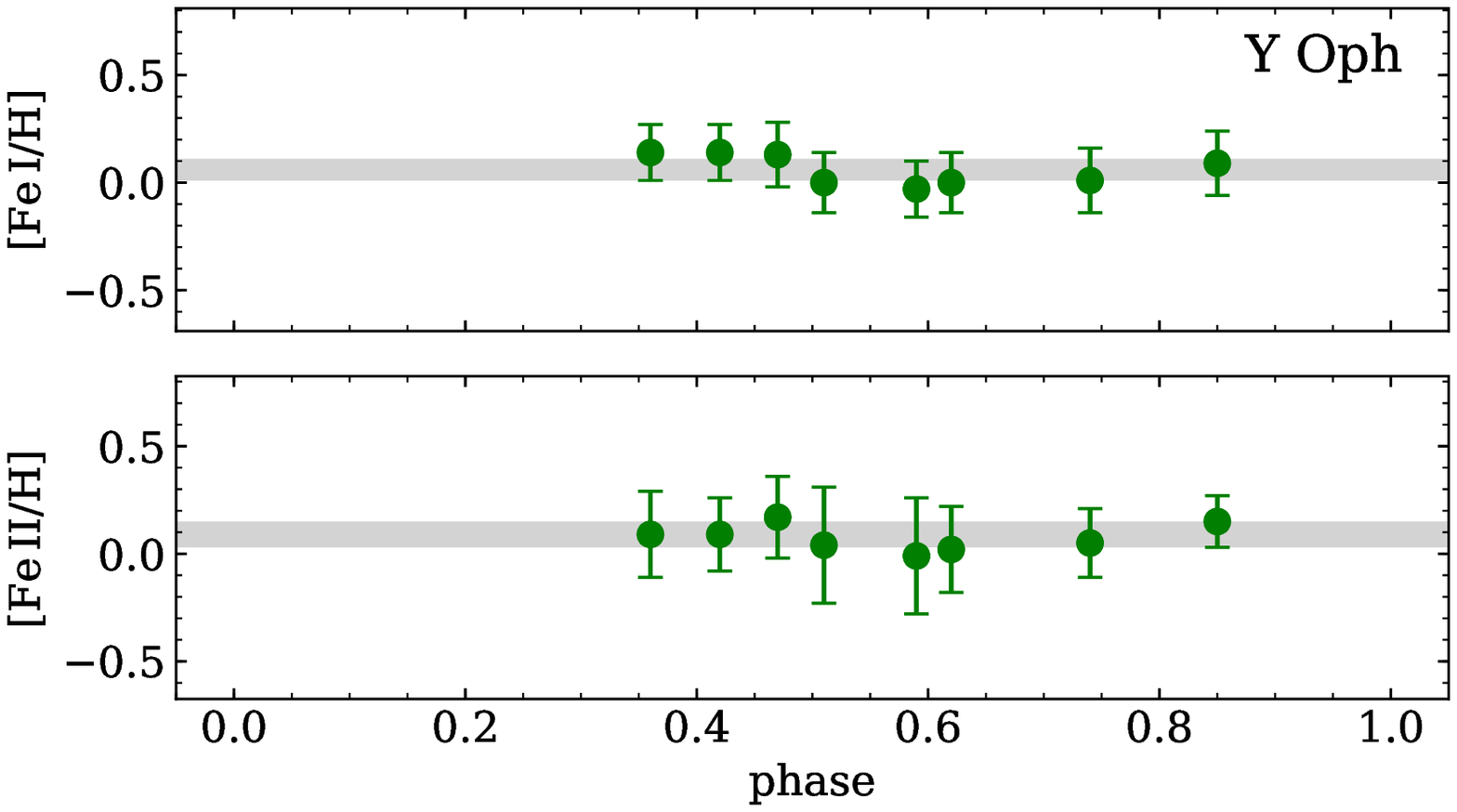}}
\end{minipage} \\
\begin{minipage}[t]{0.33\textwidth}
\centering
\resizebox{\hsize}{!}{\includegraphics{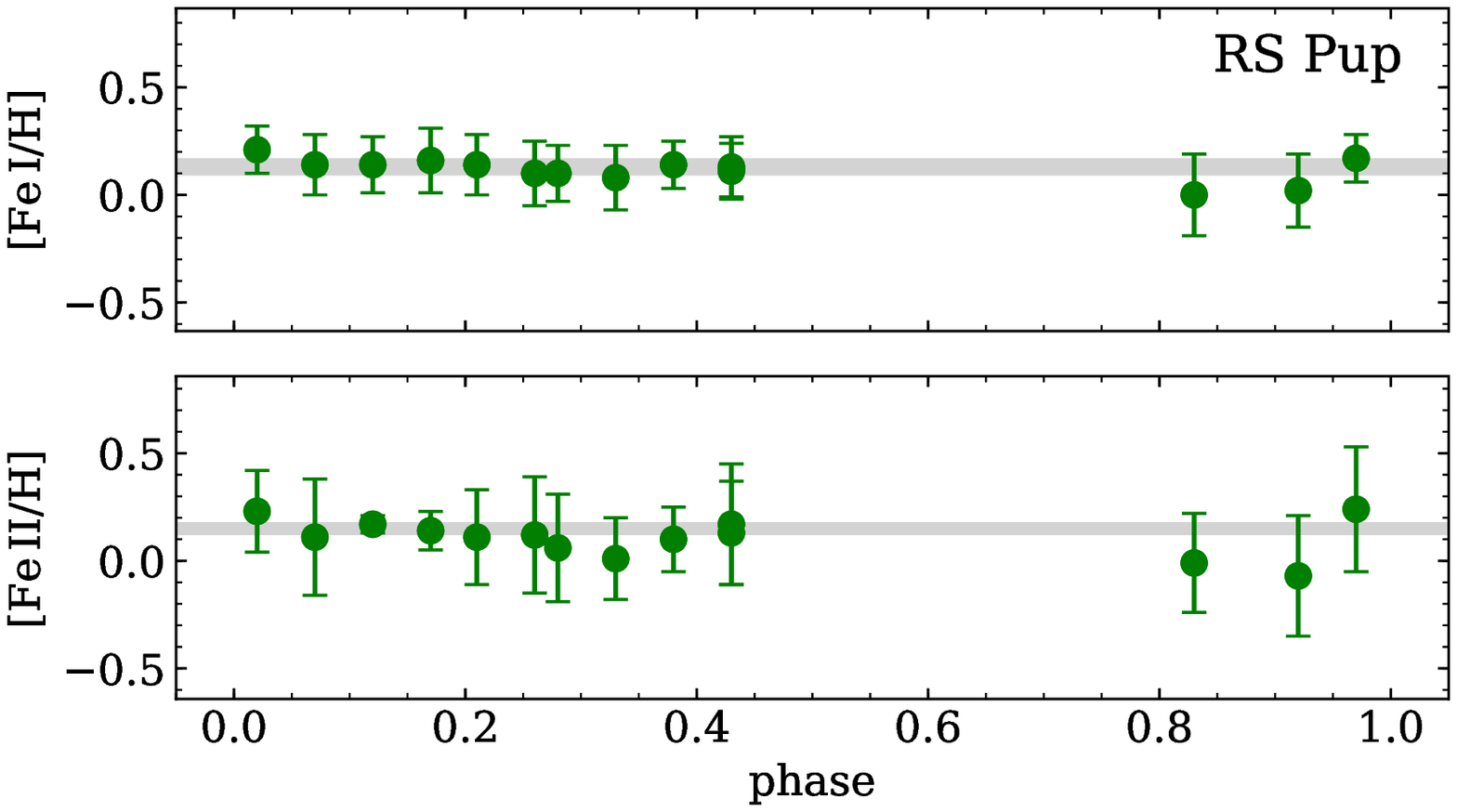}}
\end{minipage}
\begin{minipage}[t]{0.33\textwidth}
\centering
\resizebox{\hsize}{!}{\includegraphics{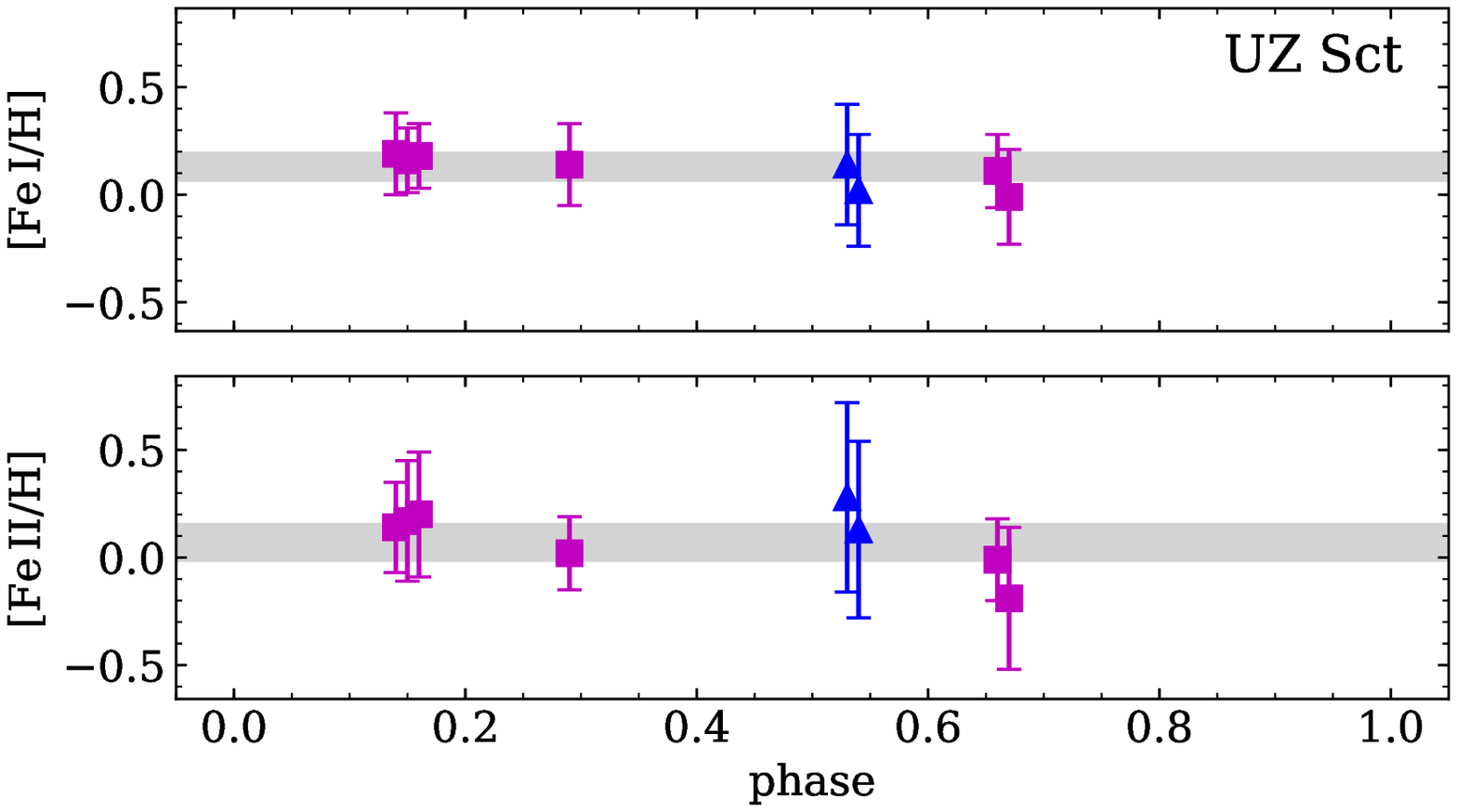}}
\end{minipage}
\begin{minipage}[t]{0.33\textwidth}
\centering
\resizebox{\hsize}{!}{\includegraphics{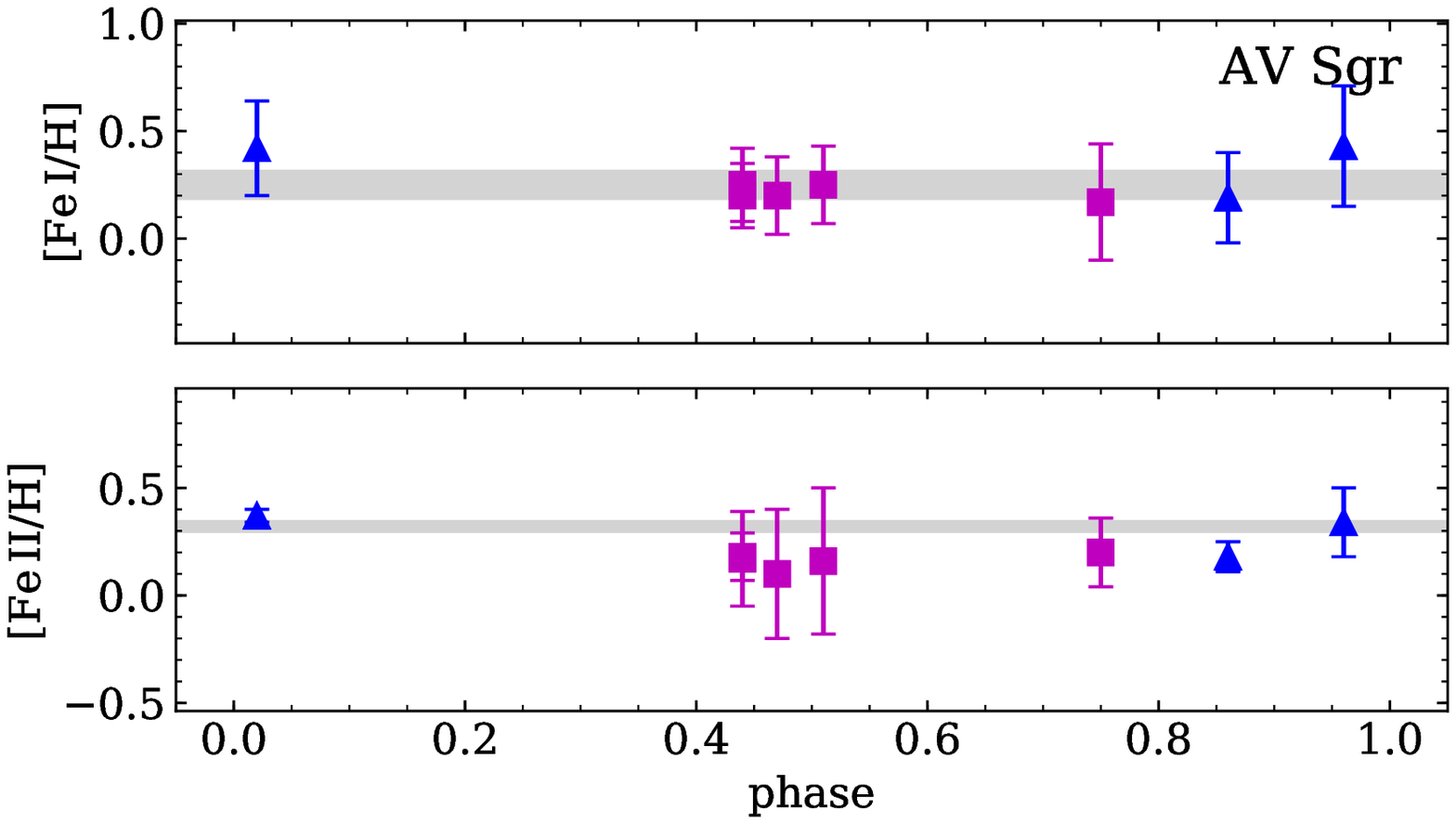}}
\end{minipage} \\
\begin{minipage}[t]{0.33\textwidth}
\centering
\resizebox{\hsize}{!}{\includegraphics{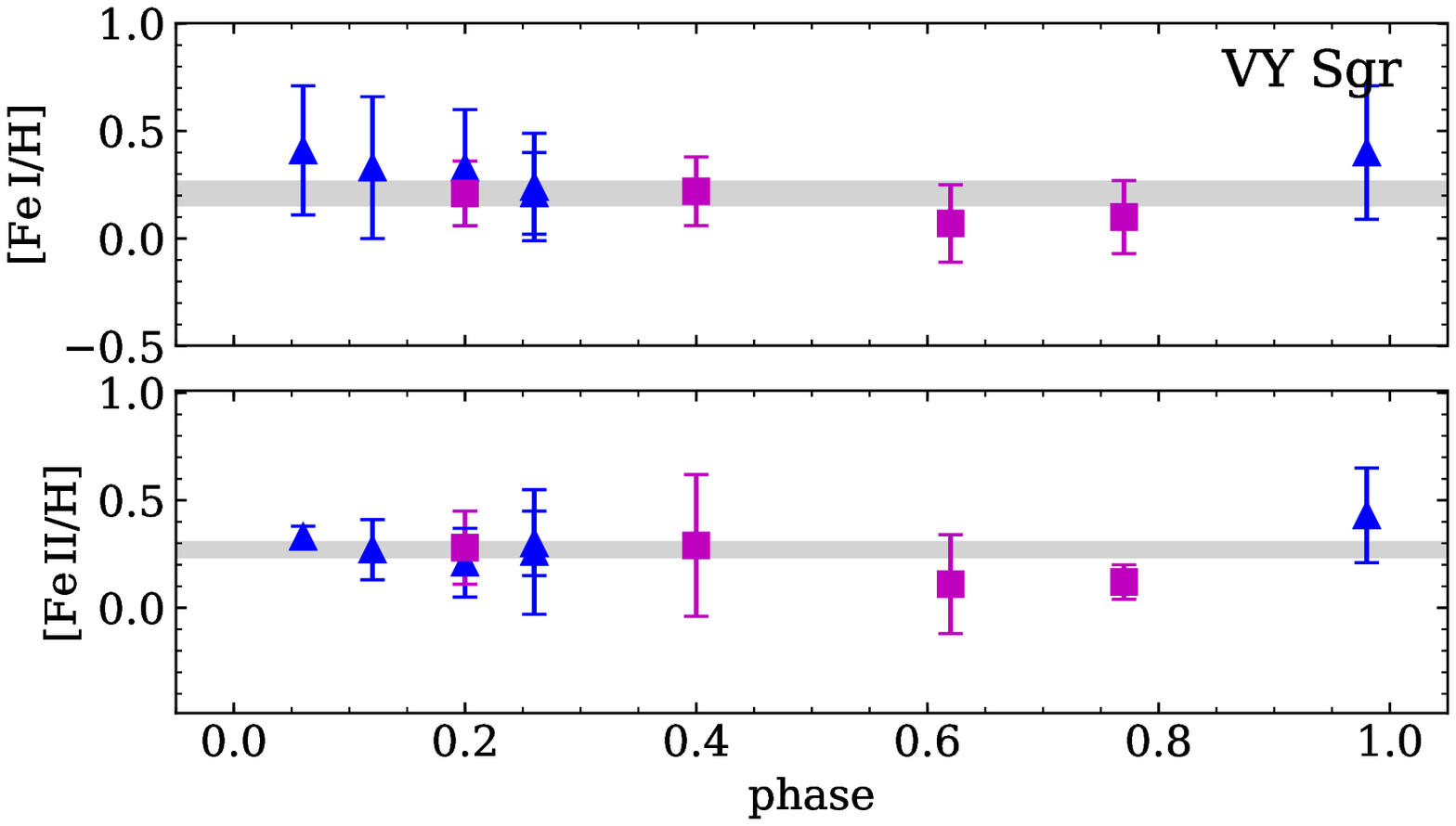}}
\end{minipage}
\begin{minipage}[t]{0.33\textwidth}
\centering
\resizebox{\hsize}{!}{\includegraphics{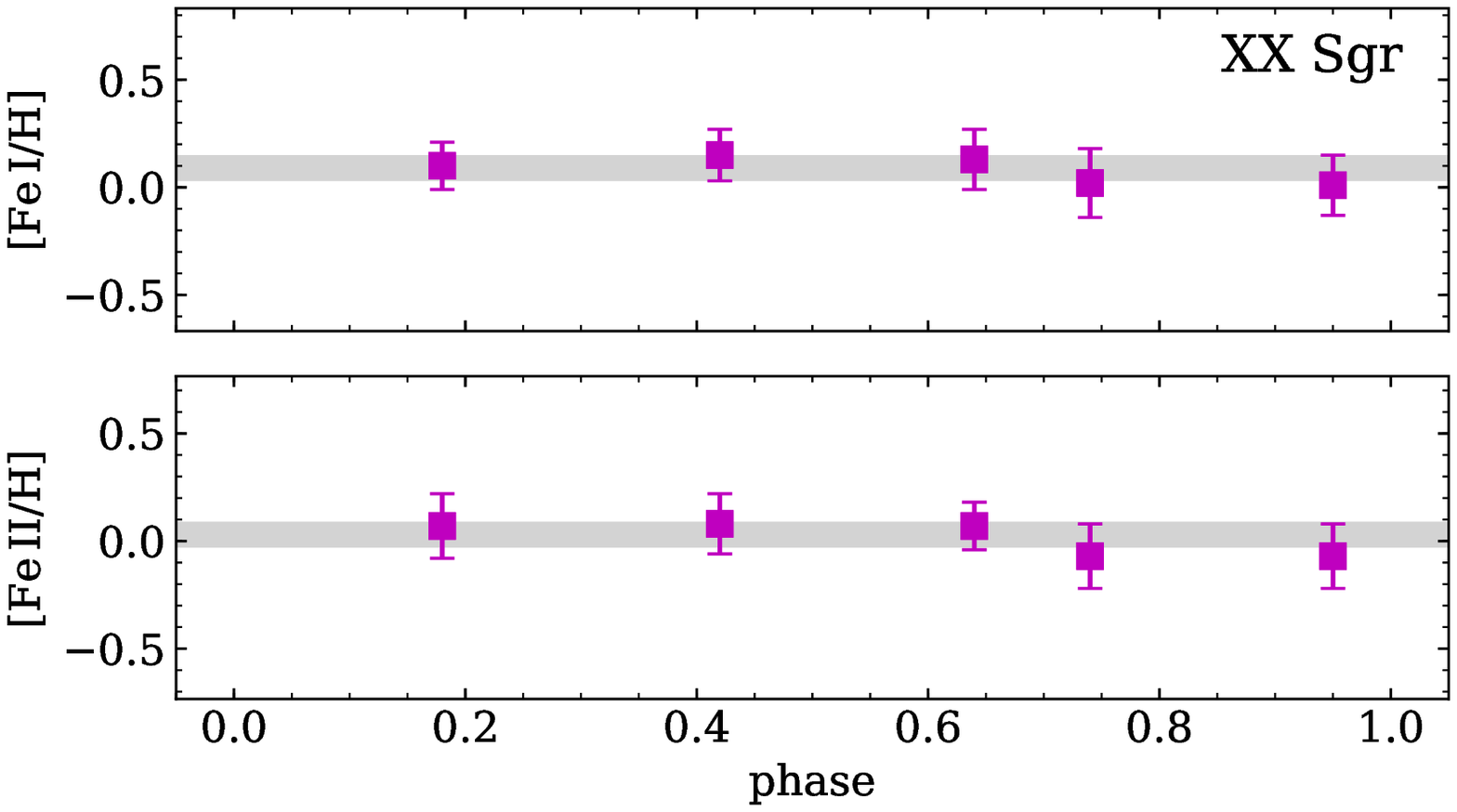}}
\end{minipage}
\begin{minipage}[t]{0.33\textwidth}
\centering
\resizebox{\hsize}{!}{\includegraphics{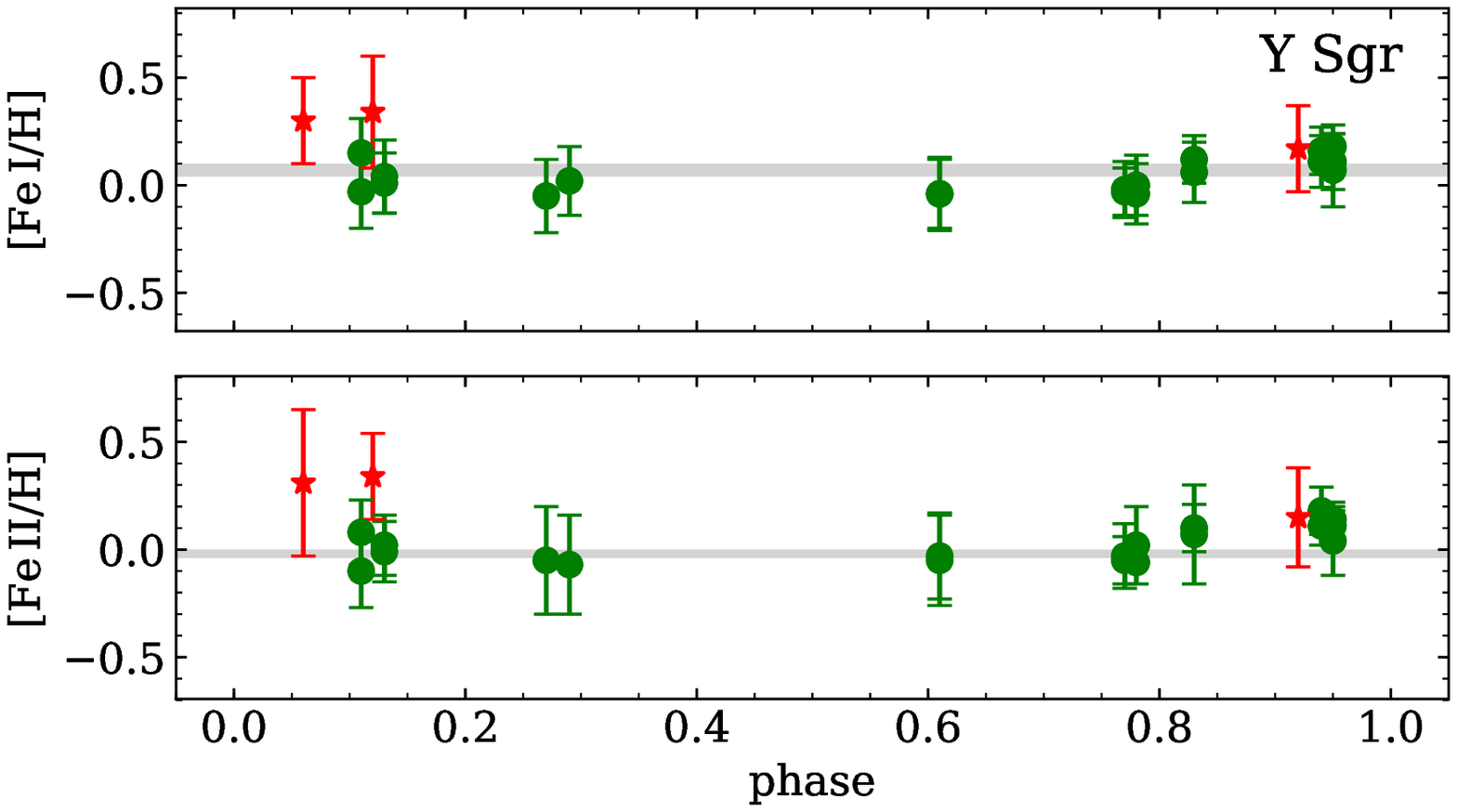}}
\end{minipage} \\
\begin{minipage}[t]{0.33\textwidth}
\centering
\resizebox{\hsize}{!}{\includegraphics{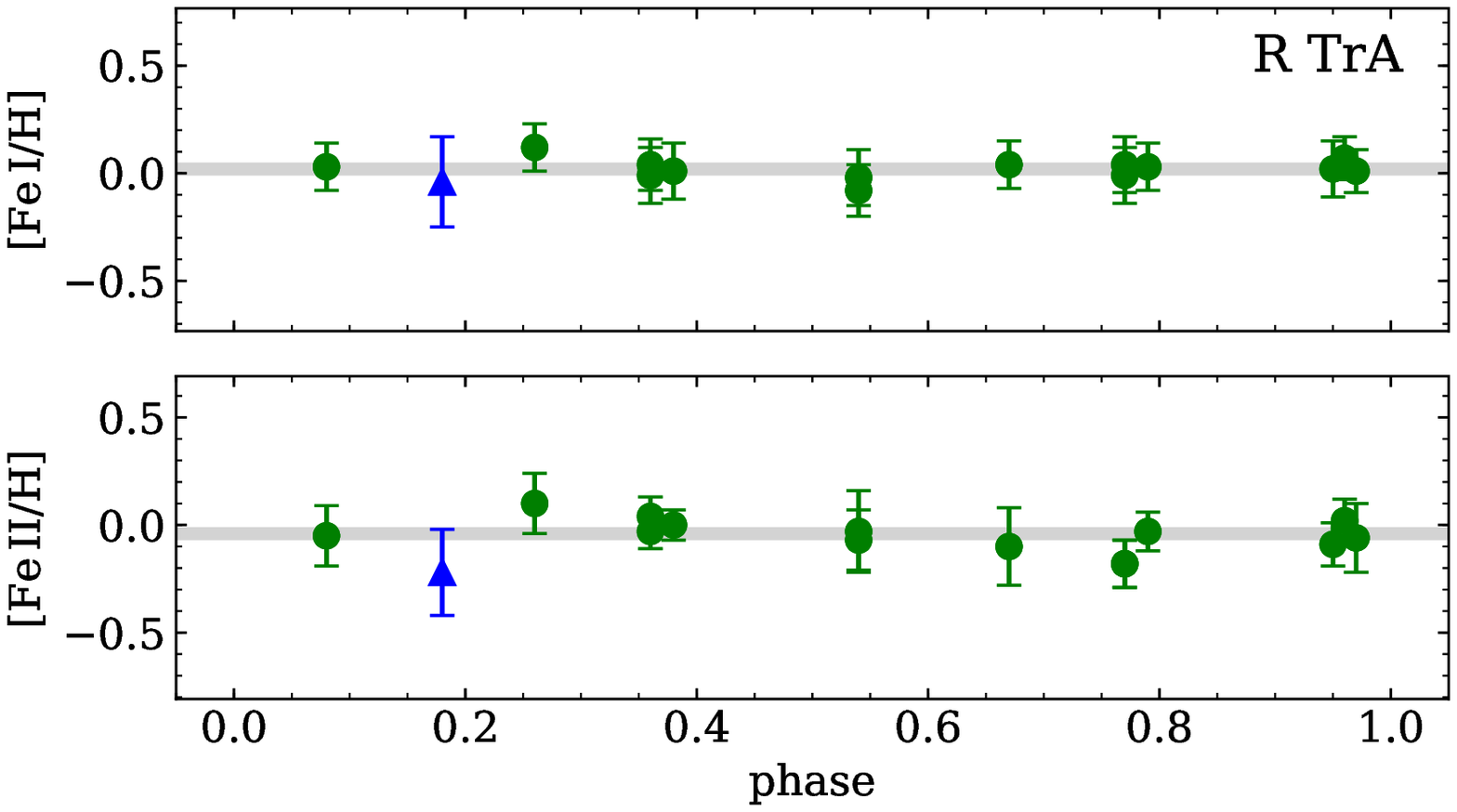}}
\end{minipage}
\begin{minipage}[t]{0.33\textwidth}
\centering
\resizebox{\hsize}{!}{\includegraphics{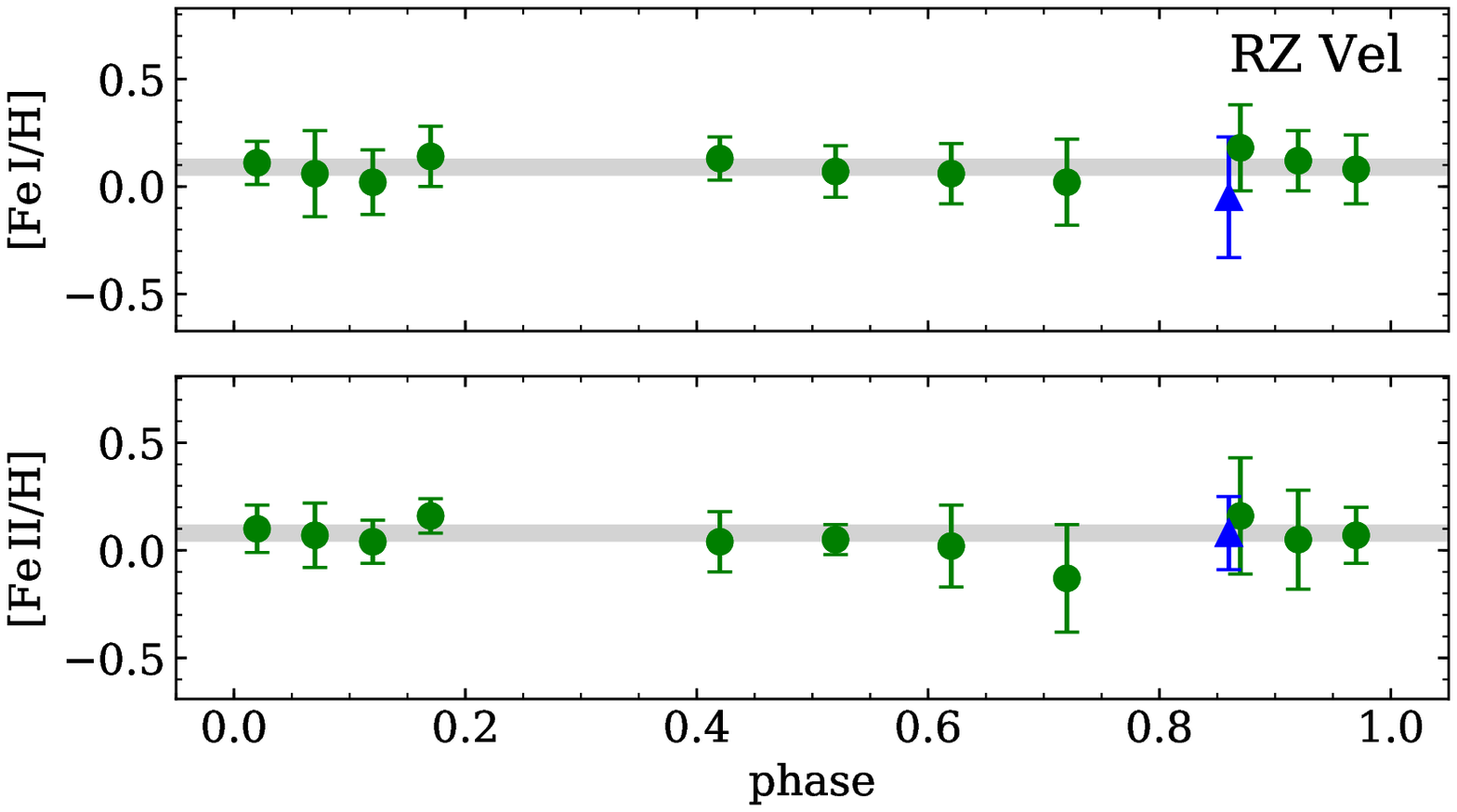}}
\end{minipage}
\begin{minipage}[t]{0.33\textwidth}
\centering
\strut
\end{minipage}
\caption{Abundances from \ion{Fe}{i} and \ion{Fe}{ii} lines as a function of the pulsation phase. 
The color coding of the different points is the same as in Fig.~\ref{atmpar_phase}. To help with the comparison, the panels are plotted with same y-axis range: 1.5~dex for both \ion{Fe}{i} and \ion{Fe}{ii} panels. The light grey shaded regions indicate the $\pm$1$\sigma$ uncertainty around the weighted mean (from columns 5 and 6 of Table~\ref{meanparams}).}
\label{feh_phase}
\end{figure*}

Finally, let us note that data plotted in Fig.~\ref{feh_phase} clearly show 
that \ion{Fe}{i} and \ion{Fe}{ii} abundances agree quite well within the 
errors. Moreover and even more importantly, they are independent of the 
pulsation phase.

\section{Summary and final remarks}
\label{concl}

The quoted results bring forward a few relevant issues worth being 
discussed in more detail. 

\begin{itemize}

\item[a)] High-resolution, high S/N optical spectra of variable stars 
allow us to provide precise estimates of both physical parameters 
and abundances along the pulsation cycle. This means that spectra 
collected at random phases can provide solid estimates of Cepheid 
elemental abundances. This argument applies for relative measurements. 
Solid constraints on the possible occurrence of systematics in the zero-point
of physical parameters and in elemental abundances do require independent 
spectroscopic approach based either on spectral synthesis and/or on an 
NLTE analysis.

\item[b)] The observational scenario concerning the LDRs in the NIR regime is lagging compared
with the optical one, after the seminal investigation by \citet{Sasselovetal1989} and
\citet{SasselovLester1990a,SasselovLester1990b} has been hampered by the lack of
efficient echelle NIR spectrographs. Fortunately, recent investigations are paving
the way for an extension of the LDR into the NIR regime.
\citet{Fukueetal2015} collected H-band spectra with a Subaru high-resolution camera and spectrograph \citep{Kobayashietal2000} for several G- and K-type giants and supergiants. Interestingly enough, they found that they can provide effective temperatures with an accuracy of the order of 60~K, in spite of the limited range in wavelengths (1.4-1.8~$\mu$m) covered by their spectra and the limited number of pairs (nine) they used.
\citet{Taniguchietal2018}, using high-resolution spectra collected with WINERED \citep{Ikedaetal2016} in the Y and J bands (0.9-1.35~$\mu$m) for ten early G- and M-type giants, found 81 LDR-\teff\ relations, achieving a precision of 10~K in the best cases.
These findings appear as a very promising opportunity for future developments 
of NIR spectrographs, such as WINERED \citep[see also][]{DOrazietal2018}, 
CRIRES+ \citep{Follertetal2014}, and GIANO \citep{Origliaetal2016}.

\item[c)] The new calibrations of the LDRs presented in this paper together with 
similar calibrations available in the literature span, for the first time, 
the range in effective temperature covered by CCs along their pulsation cycles.   
However, the range in metallicity covered by the current Cepheids is roughly 
half a dex around solar metallicity. New extensions into the more metal-poor/metal-rich 
regime are highly encouraged.

\item[d)] The estimate of the surface gravity using the ionization equilibrium between
\ion{Fe}{i} and \ion{Fe}{ii} lines is quite robust, but new approaches are required for
metal-poor objects and/or for NIR spectra in which the number of metallic lines is limited. 
The next Gaia release (DR2), by including accurate estimates of geometrical distances, 
photometry, and spectroscopy, will constrain the variation of surface gravity 
for static and variable stars. This is a unique opportunity to constrain possible systematics.  

\item[e)] The anti-correlation between microturbulent velocity and effective temperature is 
quite interesting. Further investigations to derive analytical relations can pave the 
way to a better understanding of the physical mechanisms (convective transport, 
non-linear phenomena) driving the efficiency of microturbulent velocity along the 
pulsation cycle. 1D non-LTE static atmosphere models and 3D dynamical atmosphere 
models \citep{Chiavassaetal2018} would be highly desirable to investigate the physical 
phenomena affecting line formation and abundances in variable stars. 

\item[f)] The current long term variability surveys are discovering hundreds/thousands 
of classical Cepheids along the obscured regions of the Galactic plane (Udalski et al. 2018, 
private communication). The new identifications together with fiber multi-object (4MOST, 
MOONS, APOGEE-South, WEAVE) and slit NIR spectrographs (CRIRES+, WINERED, GIANO, CARMENES) 
will provide a unique opportunity to investigate the chemical enrichment of young stellar 
populations across the Galactic thin disk.

\end{itemize}

\begin{acknowledgements}
L. Inno acknowledges the Sonderforschungsbereich SFB 881 "The Milky Way System" (subproject A3) of the German Research Foundation (DFG).
\end{acknowledgements}
\bibliographystyle{aa}
\bibliography{Proxaufetal2018.bib}

\begin{thebibliography}{101}
\expandafter\ifx\csname natexlab\endcsname\relax\def\natexlab#1{#1}\fi

\bibitem[{{Anderson} {et~al.}(2016){Anderson}, {Saio}, {Ekstr{\"o}m}, {Georgy},
  \& {Meynet}}]{Andersonetal2016}
{Anderson}, R.~I., {Saio}, H., {Ekstr{\"o}m}, S., {Georgy}, C., \& {Meynet}, G.
  2016, \aap, 591, A8

\bibitem[{{Andrievsky} {et~al.}(2002{\natexlab{a}}){Andrievsky}, {Bersier},
  {Kovtyukh}, {Luck}, {Maciel}, {L{\'e}pine}, \&
  {Beletsky}}]{Andrievskyetal2002a}
{Andrievsky}, S.~M., {Bersier}, D., {Kovtyukh}, V.~V., {et~al.}
  2002{\natexlab{a}}, \aap, 384, 140

\bibitem[{{Andrievsky} {et~al.}(2002{\natexlab{b}}){Andrievsky}, {Kovtyukh},
  {Luck}, {L{\'e}pine}, {Bersier}, {Maciel}, {Barbuy}, {Klochkova}, {Panchuk},
  \& {Karpischek}}]{Andrievskyetal2002b}
{Andrievsky}, S.~M., {Kovtyukh}, V.~V., {Luck}, R.~E., {et~al.}
  2002{\natexlab{b}}, \aap, 381, 32

\bibitem[{{Andrievsky} {et~al.}(2005){Andrievsky}, {Luck}, \&
  {Kovtyukh}}]{Andrievskyetal2005}
{Andrievsky}, S.~M., {Luck}, R.~E., \& {Kovtyukh}, V.~V. 2005, \aj, 130, 1880

\bibitem[{{Asplund} {et~al.}(2009){Asplund}, {Grevesse}, {Sauval}, \&
  {Scott}}]{Asplundetal2009}
{Asplund}, M., {Grevesse}, N., {Sauval}, A.~J., \& {Scott}, P. 2009, \araa, 47,
  481

\bibitem[{{Baade}(1958)}]{Baade1958}
{Baade}, W. 1958, Ricerche Astronomiche, 5, 165

\bibitem[{{Bono} {et~al.}(1999{\natexlab{a}}){Bono}, {Caputo}, {Castellani}, \&
  {Marconi}}]{Bonoetal1999a}
{Bono}, G., {Caputo}, F., {Castellani}, V., \& {Marconi}, M.
  1999{\natexlab{a}}, \apj, 512, 711

\bibitem[{{Bono} {et~al.}(2010){Bono}, {Caputo}, {Marconi}, \&
  {Musella}}]{Bonoetal2010}
{Bono}, G., {Caputo}, F., {Marconi}, M., \& {Musella}, I. 2010, \apj, 715, 277

\bibitem[{{Bono} {et~al.}(2000{\natexlab{a}}){Bono}, {Castellani}, \&
  {Marconi}}]{Bonoetal2000a}
{Bono}, G., {Castellani}, V., \& {Marconi}, M. 2000{\natexlab{a}}, \apj, 529,
  293

\bibitem[{{Bono} {et~al.}(1999{\natexlab{b}}){Bono}, {Marconi}, \&
  {Stellingwerf}}]{Bonoetal1999b}
{Bono}, G., {Marconi}, M., \& {Stellingwerf}, R.~F. 1999{\natexlab{b}}, \apjs,
  122, 167

\bibitem[{{Bono} {et~al.}(2000{\natexlab{b}}){Bono}, {Marconi}, \&
  {Stellingwerf}}]{Bonoetal2000b}
{Bono}, G., {Marconi}, M., \& {Stellingwerf}, R.~F. 2000{\natexlab{b}}, \aap,
  360, 245

\bibitem[{{Bono} {et~al.}(2013){Bono}, {Matsunaga}, {Inno}, {Lagioia}, \&
  {Genovali}}]{Bonoetal2013}
{Bono}, G., {Matsunaga}, N., {Inno}, L., {Lagioia}, E.~P., \& {Genovali}, K.
  2013, in Astrophysics and Space Science Proceedings, Vol.~34, Cosmic Rays in
  Star-Forming Environments, ed. D.~F. {Torres} \& O.~{Reimer}, 115

\bibitem[{{Castelli} \& {Kurucz}(2004)}]{CastelliKurucz2004}
{Castelli}, F. \& {Kurucz}, R.~L. 2004, ArXiv Astrophysics e-prints
  [\eprint{astro-ph/0405087}]

\bibitem[{{Chiavassa} {et~al.}(2018){Chiavassa}, {Casagrande}, {Collet},
  {Magic}, {Bigot}, {Thevenin}, \& {Asplund}}]{Chiavassaetal2018}
{Chiavassa}, A., {Casagrande}, L., {Collet}, R., {et~al.} 2018, ArXiv e-prints
  [\eprint[arXiv]{1801.01895}]

\bibitem[{{da Silva} {et~al.}(2016){da Silva}, {Lemasle}, {Bono}, {Genovali},
  {McWilliam}, {Cristallo}, {Bergemann}, {Buonanno}, {Fabrizio}, {Ferraro},
  {Fran{\c c}ois}, {Iannicola}, {Inno}, {Laney}, {Kudritzki}, {Matsunaga},
  {Nonino}, {Primas}, {Przybilla}, {Romaniello}, {Th{\'e}venin}, \&
  {Urbaneja}}]{daSilvaetal2016}
{da Silva}, R., {Lemasle}, B., {Bono}, G., {et~al.} 2016, \aap, 586, A125

\bibitem[{{Dekker} {et~al.}(2000){Dekker}, {D'Odorico}, {Kaufer}, {Delabre}, \&
  {Kotzlowski}}]{Dekkeretal2000}
{Dekker}, H., {D'Odorico}, S., {Kaufer}, A., {Delabre}, B., \& {Kotzlowski}, H.
  2000, in \procspie, Vol. 4008, Optical and IR Telescope Instrumentation and
  Detectors, ed. M.~{Iye} \& A.~F. {Moorwood}, 534--545

\bibitem[{{D'Orazi} {et~al.}(2018){D'Orazi}, {Magurno}, {Bono}, {Matsunaga},
  {Braga}, {Elgueta}, {Fukue}, {Hamano}, {Inno}, {Kobayashi}, {Kondo},
  {Monelli}, {Nonino}, {Przybilla}, {Sameshima}, {Saviane}, {Taniguchi},
  {Thevenin}, {Urbaneja-Perez}, {Watase}, {Arai}, {Bergemann}, {Buonanno},
  {Dall'Ora}, {Silva}, {Fabrizio}, {Ferraro}, {Fiorentino}, {Francois},
  {Gilmozzi}, {Iannicola}, {Ikeda}, {Jian}, {Kawakita}, {Kudritzki}, {Lemasle},
  {Marengo}, {Marinoni}, {Martinez-Vasquez}, {Minniti}, {Neeley}, {Otsubo},
  {Prieto}, {Proxauf}, {Romaniello}, {Sanna}, {Sneden}, {Takenaka},
  {Tsujimoto}, {Valenti}, {Yasui}, {Yoshikawa}, \& {Zoccali}}]{DOrazietal2018}
{D'Orazi}, V., {Magurno}, D., {Bono}, G., {et~al.} 2018, ArXiv e-prints
  [\eprint[arXiv]{1802.07314}]

\bibitem[{{Evans}(1992)}]{Evans1992}
{Evans}, N.~R. 1992, \apj, 384, 220

\bibitem[{{Feast} {et~al.}(2008){Feast}, {Laney}, {Kinman}, {van Leeuwen}, \&
  {Whitelock}}]{Feastetal2008}
{Feast}, M.~W., {Laney}, C.~D., {Kinman}, T.~D., {van Leeuwen}, F., \&
  {Whitelock}, P.~A. 2008, \mnras, 386, 2115

\bibitem[{{Fiorentino} {et~al.}(2007){Fiorentino}, {Marconi}, {Musella}, \&
  {Caputo}}]{Fiorentinoetal2007}
{Fiorentino}, G., {Marconi}, M., {Musella}, I., \& {Caputo}, F. 2007, \aap,
  476, 863

\bibitem[{{Follert} {et~al.}(2014){Follert}, {Dorn}, {Oliva}, {Lizon},
  {Hatzes}, {Piskunov}, {Reiners}, {Seemann}, {Stempels}, {Heiter}, {Marquart},
  {Lockhart}, {Anglada-Escude}, {L{\"o}winger}, {Baade}, {Grunhut}, {Bristow},
  {Klein}, {Jung}, {Ives}, {Kerber}, {Pozna}, {Paufique}, {Kaeufl}, {Origlia},
  {Valenti}, {Gojak}, {Hilker}, {Pasquini}, {Smette}, \&
  {Smoker}}]{Follertetal2014}
{Follert}, R., {Dorn}, R.~J., {Oliva}, E., {et~al.} 2014, in \procspie, Vol.
  9147, Ground-based and Airborne Instrumentation for Astronomy V, 914719

\bibitem[{François {et~al.}(2006)François, Schuez, Conn, Monaco, Selman, \&
  Sterzik}]{ferosuserman}
François, P., Schuez, O., Conn, B., {et~al.} 2006, FEROS-II User Manual,
  \url{https://www.eso.org/sci/facilities/lasilla/
  instruments/feros/doc/manual/P78/FEROSII-UserManual-78.0.pdf}

\bibitem[{{Freedman} \& {Madore}(2010)}]{FreedmanMadore2010}
{Freedman}, W.~L. \& {Madore}, B.~F. 2010, \araa, 48, 673

\bibitem[{{Fukue} {et~al.}(2015){Fukue}, {Matsunaga}, {Yamamoto}, {Kondo},
  {Kobayashi}, {Ikeda}, {Hamano}, {Yasui}, {Arasaki}, {Tsujimoto}, {Bono}, \&
  {Inno}}]{Fukueetal2015}
{Fukue}, K., {Matsunaga}, N., {Yamamoto}, R., {et~al.} 2015, \apj, 812, 64

\bibitem[{{Gallenne} {et~al.}(2014){Gallenne}, {Kervella}, {M{\'e}rand},
  {Evans}, {Girard}, {Gieren}, \& {Pietrzy{\'n}ski}}]{Gallenneetal2014}
{Gallenne}, A., {Kervella}, P., {M{\'e}rand}, A., {et~al.} 2014, \aap, 567, A60

\bibitem[{{Genovali} {et~al.}(2014){Genovali}, {Lemasle}, {Bono}, {Romaniello},
  {Fabrizio}, {Ferraro}, {Iannicola}, {Laney}, {Nonino}, {Bergemann},
  {Buonanno}, {Fran{\c c}ois}, {Inno}, {Kudritzki}, {Matsunaga}, {Pedicelli},
  {Primas}, \& {Th{\'e}venin}}]{Genovalietal2014}
{Genovali}, K., {Lemasle}, B., {Bono}, G., {et~al.} 2014, \aap, 566, A37

\bibitem[{{Genovali} {et~al.}(2013){Genovali}, {Lemasle}, {Bono}, {Romaniello},
  {Primas}, {Fabrizio}, {Buonanno}, {Fran{\c c}ois}, {Inno}, {Laney},
  {Matsunaga}, {Pedicelli}, \& {Th{\'e}venin}}]{Genovalietal2013}
{Genovali}, K., {Lemasle}, B., {Bono}, G., {et~al.} 2013, \aap, 554, A132

\bibitem[{{Genovali} {et~al.}(2015){Genovali}, {Lemasle}, {da Silva}, {Bono},
  {Fabrizio}, {Bergemann}, {Buonanno}, {Ferraro}, {Fran{\c c}ois}, {Iannicola},
  {Inno}, {Laney}, {Kudritzki}, {Matsunaga}, {Nonino}, {Primas}, {Romaniello},
  {Urbaneja}, \& {Th{\'e}venin}}]{Genovalietal2015}
{Genovali}, K., {Lemasle}, B., {da Silva}, R., {et~al.} 2015, \aap, 580, A17

\bibitem[{{Gieren} {et~al.}(2013){Gieren}, {G{\'o}rski}, {Pietrzy{\'n}ski},
  {Konorski}, {Suchomska}, {Graczyk}, {Pilecki}, {Bresolin}, {Kudritzki},
  {Storm}, {Karczmarek}, {Gallenne}, {Calder{\'o}n}, \&
  {Geisler}}]{Gierenetal2013}
{Gieren}, W., {G{\'o}rski}, M., {Pietrzy{\'n}ski}, G., {et~al.} 2013, \apj,
  773, 69

\bibitem[{{Gillet} {et~al.}(1999){Gillet}, {Fokin}, {Breitfellner}, {Mazauric},
  \& {Nicolas}}]{Gilletetal1999}
{Gillet}, D., {Fokin}, A.~B., {Breitfellner}, M.~G., {Mazauric}, S., \&
  {Nicolas}, A. 1999, \aap, 344, 935

\bibitem[{{Gilmore} {et~al.}(2012){Gilmore}, {Randich}, {Asplund}, {Binney},
  {Bonifacio}, {Drew}, {Feltzing}, {Ferguson}, {Jeffries}, {Micela}, \&
  et~al.}]{Gilmoreetal2012}
{Gilmore}, G., {Randich}, S., {Asplund}, M., {et~al.} 2012, The Messenger, 147,
  25

\bibitem[{{Gray}(2005)}]{Gray2005}
{Gray}, D.~F. 2005, {The Observation and Analysis of Stellar Photospheres}

\bibitem[{{Grevesse} {et~al.}(2015){Grevesse}, {Scott}, {Asplund}, \&
  {Sauval}}]{Grevesseetal2015}
{Grevesse}, N., {Scott}, P., {Asplund}, M., \& {Sauval}, A.~J. 2015, \aap, 573,
  A27

\bibitem[{{Groenewegen}(2008)}]{Groenewegen2008}
{Groenewegen}, M.~A.~T. 2008, \aap, 488, 25

\bibitem[{{Gustafsson} {et~al.}(2008){Gustafsson}, {Edvardsson}, {Eriksson},
  {J{\o}rgensen}, {Nordlund}, \& {Plez}}]{Gustafssonetal2008}
{Gustafsson}, B., {Edvardsson}, B., {Eriksson}, K., {et~al.} 2008, \aap, 486,
  951

\bibitem[{{Heiter} \& {Eriksson}(2006)}]{HeiterEriksson2006}
{Heiter}, U. \& {Eriksson}, K. 2006, \aap, 452, 1039

\bibitem[{{Hoffmann} {et~al.}(2016){Hoffmann}, {Macri}, {Riess}, {Yuan},
  {Casertano}, {Foley}, {Filippenko}, {Tucker}, {Chornock}, {Silverman},
  {Welch}, {Goobar}, \& {Amanullah}}]{Hoffmannetal2016}
{Hoffmann}, S.~L., {Macri}, L.~M., {Riess}, A.~G., {et~al.} 2016, \apj, 830, 10

\bibitem[{{Ikeda} {et~al.}(2016){Ikeda}, {Kobayashi}, {Kondo}, {Otsubo},
  {Hamano}, {Sameshima}, {Yoshikawa}, {Fukue}, {Nakanishi}, {Kawanishi},
  {Nakaoka}, {Kinoshita}, {Kitano}, {Asano}, {Takenaka}, {Watase}, {Mito},
  {Yasui}, {Minami}, {Izumu}, {Yamamoto}, {Mizumoto}, {Arasaki}, {Arai},
  {Matsunaga}, \& {Kawakita}}]{Ikedaetal2016}
{Ikeda}, Y., {Kobayashi}, N., {Kondo}, S., {et~al.} 2016, in \procspie, Vol.
  9908, Ground-based and Airborne Instrumentation for Astronomy VI, 99085Z

\bibitem[{{Kaufer} {et~al.}(1999){Kaufer}, {Stahl}, {Tubbesing},
  {N{\o}rregaard}, {Avila}, {Francois}, {Pasquini}, \&
  {Pizzella}}]{Kauferetal1999}
{Kaufer}, A., {Stahl}, O., {Tubbesing}, S., {et~al.} 1999, The Messenger, 95, 8

\bibitem[{{Kervella} {et~al.}(2004){Kervella}, {Bersier}, {Mourard},
  {Nardetto}, {Fouqu{\'e}}, \& {Coud{\'e} du Foresto}}]{Kervellaetal2004}
{Kervella}, P., {Bersier}, D., {Mourard}, D., {et~al.} 2004, \aap, 428, 587

\bibitem[{{Kobayashi} {et~al.}(2000){Kobayashi}, {Tokunaga}, {Terada}, {Goto},
  {Weber}, {Potter}, {Onaka}, {Ching}, {Young}, {Fletcher}, {Neil},
  {Robertson}, {Cook}, {Imanishi}, \& {Warren}}]{Kobayashietal2000}
{Kobayashi}, N., {Tokunaga}, A.~T., {Terada}, H., {et~al.} 2000, in \procspie,
  Vol. 4008, Optical and IR Telescope Instrumentation and Detectors, ed.
  M.~{Iye} \& A.~F. {Moorwood}, 1056--1066

\bibitem[{{Kovtyukh} {et~al.}(2016){Kovtyukh}, {Lemasle}, {Chekhonadskikh},
  {Bono}, {Matsunaga}, {Yushchenko}, {Anderson}, {Belik}, {da Silva}, \&
  {Inno}}]{Kovtyukhetal2016}
{Kovtyukh}, V., {Lemasle}, B., {Chekhonadskikh}, F., {et~al.} 2016, \mnras,
  460, 2077

\bibitem[{{Kovtyukh}(2007)}]{Kovtyukh2007}
{Kovtyukh}, V.~V. 2007, \mnras, 378, 617

\bibitem[{{Kovtyukh} \& {Andrievsky}(1999)}]{KovtyukhAndrievsky1999}
{Kovtyukh}, V.~V. \& {Andrievsky}, S.~M. 1999, \aap, 350, L55

\bibitem[{{Kovtyukh} {et~al.}(2005){Kovtyukh}, {Andrievsky}, {Belik}, \&
  {Luck}}]{Kovtyukhetal2005}
{Kovtyukh}, V.~V., {Andrievsky}, S.~M., {Belik}, S.~I., \& {Luck}, R.~E. 2005,
  \aj, 129, 433

\bibitem[{{Kovtyukh} {et~al.}(2003{\natexlab{a}}){Kovtyukh}, {Andrievsky},
  {Luck}, \& {Gorlova}}]{Kovtyukhetal2003a}
{Kovtyukh}, V.~V., {Andrievsky}, S.~M., {Luck}, R.~E., \& {Gorlova}, N.~I.
  2003{\natexlab{a}}, \aap, 401, 661

\bibitem[{{Kovtyukh} \& {Gorlova}(2000)}]{KovtyukhGorlova2000}
{Kovtyukh}, V.~V. \& {Gorlova}, N.~I. 2000, \aap, 358, 587

\bibitem[{{Kovtyukh} {et~al.}(2003{\natexlab{b}}){Kovtyukh}, {Soubiran},
  {Belik}, \& {Gorlova}}]{Kovtyukhetal2003b}
{Kovtyukh}, V.~V., {Soubiran}, C., {Belik}, S.~I., \& {Gorlova}, N.~I.
  2003{\natexlab{b}}, \aap, 411, 559

\bibitem[{{Kraft}(1956)}]{Kraft1956}
{Kraft}, R.~P. 1956, \pasp, 68, 137

\bibitem[{{Kraft}(1957)}]{Kraft1957}
{Kraft}, R.~P. 1957, \apj, 125, 336

\bibitem[{{Krockenberger} {et~al.}(1998){Krockenberger}, {Sasselov}, {Noyes},
  {Korzennik}, {Nisenson}, {Brown}, {Kennelly}, \&
  {Horner}}]{Krockenbergeretal1998}
{Krockenberger}, M., {Sasselov}, D., {Noyes}, R., {et~al.} 1998, in
  Astronomical Society of the Pacific Conference Series, Vol. 154, Cool Stars,
  Stellar Systems, and the Sun, ed. R.~A. {Donahue} \& J.~A. {Bookbinder}, 791

\bibitem[{{Kurucz} {et~al.}(1984){Kurucz}, {Furenlid}, {Brault}, \&
  {Testerman}}]{Kuruczetal1984}
{Kurucz}, R.~L., {Furenlid}, I., {Brault}, J., \& {Testerman}, L. 1984, {Solar
  flux atlas from 296 to 1300 nm}

\bibitem[{{Lemasle} {et~al.}(2007){Lemasle}, {Fran{\c c}ois}, {Bono},
  {Mottini}, {Primas}, \& {Romaniello}}]{Lemasleetal2007}
{Lemasle}, B., {Fran{\c c}ois}, P., {Bono}, G., {et~al.} 2007, \aap, 467, 283

\bibitem[{{Lemasle} {et~al.}(2013){Lemasle}, {Fran{\c c}ois}, {Genovali},
  {Kovtyukh}, {Bono}, {Inno}, {Laney}, {Kaper}, {Bergemann}, {Fabrizio},
  {Matsunaga}, {Pedicelli}, {Primas}, \& {Romaniello}}]{Lemasleetal2013}
{Lemasle}, B., {Fran{\c c}ois}, P., {Genovali}, K., {et~al.} 2013, \aap, 558,
  A31

\bibitem[{{Lemasle} {et~al.}(2008){Lemasle}, {Fran{\c c}ois}, {Piersimoni},
  {Pedicelli}, {Bono}, {Laney}, {Primas}, \& {Romaniello}}]{Lemasleetal2008}
{Lemasle}, B., {Fran{\c c}ois}, P., {Piersimoni}, A., {et~al.} 2008, \aap, 490,
  613

\bibitem[{{Lemasle} {et~al.}(2017){Lemasle}, {Groenewegen}, {Grebel}, {Bono},
  {Fiorentino}, {Fran{\c c}ois}, {Inno}, {Kovtyukh}, {Matsunaga}, {Pedicelli},
  {Primas}, {Pritchard}, {Romaniello}, \& {da Silva}}]{Lemasleetal2017}
{Lemasle}, B., {Groenewegen}, M., {Grebel}, E., {et~al.} 2017, \aap, submitted

\bibitem[{{Li Causi} {et~al.}(2013){Li Causi}, {Antoniucci}, {Bono},
  {Pedicelli}, {Lorenzetti}, {Giannini}, \& {Nisini}}]{Licausietal2013}
{Li Causi}, G., {Antoniucci}, S., {Bono}, G., {et~al.} 2013, \aap, 549, A64

\bibitem[{{Luck} \& {Andrievsky}(2004)}]{LuckAndrievsky2004}
{Luck}, R.~E. \& {Andrievsky}, S.~M. 2004, \aj, 128, 343

\bibitem[{{Luck} {et~al.}(2008){Luck}, {Andrievsky}, {Fokin}, \&
  {Kovtyukh}}]{Lucketal2008}
{Luck}, R.~E., {Andrievsky}, S.~M., {Fokin}, A., \& {Kovtyukh}, V.~V. 2008,
  \aj, 136, 98

\bibitem[{{Luck} {et~al.}(2011){Luck}, {Andrievsky}, {Kovtyukh}, {Gieren}, \&
  {Graczyk}}]{Lucketal2011}
{Luck}, R.~E., {Andrievsky}, S.~M., {Kovtyukh}, V.~V., {Gieren}, W., \&
  {Graczyk}, D. 2011, \aj, 142, 51

\bibitem[{{Luck} \& {Lambert}(2011)}]{LuckLambert2011}
{Luck}, R.~E. \& {Lambert}, D.~L. 2011, \aj, 142, 136

\bibitem[{{Luck} {et~al.}(1998){Luck}, {Moffett}, {Barnes}, \&
  {Gieren}}]{Lucketal1998}
{Luck}, R.~E., {Moffett}, T.~J., {Barnes}, III, T.~G., \& {Gieren}, W.~P. 1998,
  \aj, 115, 605

\bibitem[{{Macri} {et~al.}(2015){Macri}, {Ngeow}, {Kanbur}, {Mahzooni}, \&
  {Smitka}}]{Macrietal2015}
{Macri}, L.~M., {Ngeow}, C.-C., {Kanbur}, S.~M., {Mahzooni}, S., \& {Smitka},
  M.~T. 2015, \aj, 149, 117

\bibitem[{{Marconi} {et~al.}(2005){Marconi}, {Musella}, \&
  {Fiorentino}}]{Marconietal2005}
{Marconi}, M., {Musella}, I., \& {Fiorentino}, G. 2005, \apj, 632, 590

\bibitem[{{Mathias} {et~al.}(2006){Mathias}, {Gillet}, {Fokin}, {Nardetto},
  {Kervella}, \& {Mourard}}]{Mathiasetal2006}
{Mathias}, P., {Gillet}, D., {Fokin}, A.~B., {et~al.} 2006, \aap, 457, 575

\bibitem[{{Mayor} {et~al.}(2003){Mayor}, {Pepe}, {Queloz}, {Bouchy},
  {Rupprecht}, {Lo Curto}, {Avila}, {Benz}, {Bertaux}, {Bonfils}, {Dall},
  {Dekker}, {Delabre}, {Eckert}, {Fleury}, {Gilliotte}, {Gojak}, {Guzman},
  {Kohler}, {Lizon}, {Longinotti}, {Lovis}, {Megevand}, {Pasquini}, {Reyes},
  {Sivan}, {Sosnowska}, {Soto}, {Udry}, {van Kesteren}, {Weber}, \&
  {Weilenmann}}]{Mayoretal2003}
{Mayor}, M., {Pepe}, F., {Queloz}, D., {et~al.} 2003, The Messenger, 114, 20

\bibitem[{{M{\'e}rand} {et~al.}(2015){M{\'e}rand}, {Kervella}, {Breitfelder},
  {Gallenne}, {Coud{\'e} du Foresto}, {ten Brummelaar}, {McAlister}, {Ridgway},
  {Sturmann}, {Sturmann}, \& {Turner}}]{Merandetal2015}
{M{\'e}rand}, A., {Kervella}, P., {Breitfelder}, J., {et~al.} 2015, \aap, 584,
  A80

\bibitem[{{Nardetto} {et~al.}(2009){Nardetto}, {Gieren}, {Kervella},
  {Fouqu{\'e}}, {Storm}, {Pietrzynski}, {Mourard}, \&
  {Queloz}}]{Nardettoetal2009}
{Nardetto}, N., {Gieren}, W., {Kervella}, P., {et~al.} 2009, \aap, 502, 951

\bibitem[{{Origlia} {et~al.}(2016){Origlia}, {Oliva}, {Sanna}, {Mucciarelli},
  {Dalessandro}, {Scuderi}, {Baffa}, {Biliotti}, {Carbonaro}, {Falcini},
  {Giani}, {Iuzzolino}, {Massi}, {Sozzi}, {Tozzi}, {Ghedina}, {Ghinassi},
  {Lodi}, {Harutyunyan}, \& {Pedani}}]{Origliaetal2016}
{Origlia}, L., {Oliva}, E., {Sanna}, N., {et~al.} 2016, \aap, 585, A14

\bibitem[{{Pel}(1978)}]{Pel1978}
{Pel}, J.~W. 1978, \aap, 62, 75

\bibitem[{{Pietrzy{\'n}ski} {et~al.}(2013){Pietrzy{\'n}ski}, {Graczyk},
  {Gieren}, {Thompson}, {Pilecki}, {Udalski}, {Soszy{\'n}ski}, {Koz{\l}owski},
  {Konorski}, {Suchomska}, {Bono}, {Moroni}, {Villanova}, {Nardetto},
  {Bresolin}, {Kudritzki}, {Storm}, {Gallenne}, {Smolec}, {Minniti}, {Kubiak},
  {Szyma{\'n}ski}, {Poleski}, {Wyrzykowski}, {Ulaczyk}, {Pietrukowicz},
  {G{\'o}rski}, \& {Karczmarek}}]{Pietrzynskietal2013}
{Pietrzy{\'n}ski}, G., {Graczyk}, D., {Gieren}, W., {et~al.} 2013, \nat, 495,
  76

\bibitem[{{Preston}(1964)}]{Preston1964}
{Preston}, G.~W. 1964, \araa, 2, 23

\bibitem[{{Preston} {et~al.}(1965){Preston}, {Smak}, \&
  {Paczynski}}]{Prestonetal1965}
{Preston}, G.~W., {Smak}, J., \& {Paczynski}, B. 1965, \apjs, 12, 99

\bibitem[{{Randich} {et~al.}(2013){Randich}, {Gilmore}, \& {Gaia-ESO
  Consortium}}]{Randichetal2013}
{Randich}, S., {Gilmore}, G., \& {Gaia-ESO Consortium}. 2013, The Messenger,
  154, 47

\bibitem[{{Riess} {et~al.}(2016){Riess}, {Macri}, {Hoffmann}, {Scolnic},
  {Casertano}, {Filippenko}, {Tucker}, {Reid}, {Jones}, {Silverman},
  {Chornock}, {Challis}, {Yuan}, {Brown}, \& {Foley}}]{Riessetal2016}
{Riess}, A.~G., {Macri}, L.~M., {Hoffmann}, S.~L., {et~al.} 2016, \apj, 826, 56

\bibitem[{{Romaniello} {et~al.}(2008){Romaniello}, {Primas}, {Mottini},
  {Pedicelli}, {Lemasle}, {Bono}, {Fran{\c c}ois}, {Groenewegen}, \&
  {Laney}}]{Romanielloetal2008}
{Romaniello}, M., {Primas}, F., {Mottini}, M., {et~al.} 2008, \aap, 488, 731

\bibitem[{{Ryabchikova} {et~al.}(2015){Ryabchikova}, {Piskunov}, {Kurucz},
  {Stempels}, {Heiter}, {Pakhomov}, \& {Barklem}}]{Ryabchikovaetal2015}
{Ryabchikova}, T., {Piskunov}, N., {Kurucz}, R.~L., {et~al.} 2015, \physscr,
  90, 054005

\bibitem[{{Sasselov} {et~al.}(1989){Sasselov}, {Fieldus}, \&
  {Lester}}]{Sasselovetal1989}
{Sasselov}, D.~D., {Fieldus}, M.~S., \& {Lester}, J.~B. 1989, \apjl, 337, L29

\bibitem[{{Sasselov} \& {Lester}(1990{\natexlab{a}})}]{SasselovLester1990a}
{Sasselov}, D.~D. \& {Lester}, J.~B. 1990{\natexlab{a}}, \apj, 360, 227

\bibitem[{{Sasselov} \& {Lester}(1990{\natexlab{b}})}]{SasselovLester1990b}
{Sasselov}, D.~D. \& {Lester}, J.~B. 1990{\natexlab{b}}, \apj, 362, 333

\bibitem[{{Scott} {et~al.}(2015{\natexlab{a}}){Scott}, {Asplund}, {Grevesse},
  {Bergemann}, \& {Sauval}}]{Scottetal2015b}
{Scott}, P., {Asplund}, M., {Grevesse}, N., {Bergemann}, M., \& {Sauval}, A.~J.
  2015{\natexlab{a}}, \aap, 573, A26

\bibitem[{{Scott} {et~al.}(2015{\natexlab{b}}){Scott}, {Grevesse}, {Asplund},
  {Sauval}, {Lind}, {Takeda}, {Collet}, {Trampedach}, \&
  {Hayek}}]{Scottetal2015a}
{Scott}, P., {Grevesse}, N., {Asplund}, M., {et~al.} 2015{\natexlab{b}}, \aap,
  573, A25

\bibitem[{Sneden(2002)}]{Sneden2002}
Sneden, C. 2002, The MOOG code, \url{http://www.as.utexas.edu/~chris/moog.html}

\bibitem[{{Soszy{\'n}ski} {et~al.}(2017){Soszy{\'n}ski}, {Udalski},
  {Szyma{\'n}ski}, {Wyrzykowski}, {Ulaczyk}, {Poleski}, {Pietrukowicz},
  {Koz{\l}owski}, {Skowron}, {Skowron}, {Mr{\'o}z}, \&
  {Pawlak}}]{Soszynskietal2017}
{Soszy{\'n}ski}, I., {Udalski}, A., {Szyma{\'n}ski}, M.~K., {et~al.} 2017,
  \actaa, 67, 103

\bibitem[{{Sousa} {et~al.}(2015){Sousa}, {Santos}, {Adibekyan}, {Delgado-Mena},
  \& {Israelian}}]{Sousaetal2015}
{Sousa}, S.~G., {Santos}, N.~C., {Adibekyan}, V., {Delgado-Mena}, E., \&
  {Israelian}, G. 2015, \aap, 577, A67

\bibitem[{{Sousa} {et~al.}(2007){Sousa}, {Santos}, {Israelian}, {Mayor}, \&
  {Monteiro}}]{Sousaetal2007}
{Sousa}, S.~G., {Santos}, N.~C., {Israelian}, G., {Mayor}, M., \& {Monteiro},
  M.~J.~P.~F.~G. 2007, \aap, 469, 783

\bibitem[{{Storm} {et~al.}(2011{\natexlab{a}}){Storm}, {Gieren}, {Fouqu{\'e}},
  {Barnes}, {Pietrzy{\'n}ski}, {Nardetto}, {Weber}, {Granzer}, \&
  {Strassmeier}}]{Stormetal2011a}
{Storm}, J., {Gieren}, W., {Fouqu{\'e}}, P., {et~al.} 2011{\natexlab{a}}, \aap,
  534, A94

\bibitem[{{Storm} {et~al.}(2011{\natexlab{b}}){Storm}, {Gieren}, {Fouqu{\'e}},
  {Barnes}, {Soszy{\'n}ski}, {Pietrzy{\'n}ski}, {Nardetto}, \&
  {Queloz}}]{Stormetal2011b}
{Storm}, J., {Gieren}, W., {Fouqu{\'e}}, P., {et~al.} 2011{\natexlab{b}}, \aap,
  534, A95

\bibitem[{{Strassmeier} {et~al.}(2004){Strassmeier}, {Granzer}, {Weber},
  {Woche}, {Andersen}, {Bartus}, {Bauer}, {Dionies}, {Popow}, {Fechner},
  {Hildebrandt}, {Washuettl}, {Ritter}, {Schwope}, {Staude}, {Paschke},
  {Stolz}, {Serre-Ricart}, {de la Rosa}, \& {Arnay}}]{Strassmeieretal2004}
{Strassmeier}, K.~G., {Granzer}, T., {Weber}, M., {et~al.} 2004, Astronomische
  Nachrichten, 325, 527

\bibitem[{{Strassmeier} {et~al.}(2010){Strassmeier}, {Granzer}, {Weber},
  {Woche}, {Popow}, {J{\"a}rvinen}, {Bartus}, {Bauer}, {Dionies}, {Fechner},
  {Bittner}, \& {Paschke}}]{Strassmeieretal2010}
{Strassmeier}, K.~G., {Granzer}, T., {Weber}, M., {et~al.} 2010, Advances in
  Astronomy, 2010, 970306

\bibitem[{{Struve}(1944)}]{Struve1944}
{Struve}, O. 1944, The Observatory, 65, 257

\bibitem[{{Szabados}(1983)}]{Szabados1983}
{Szabados}, L. 1983, \apss, 96, 185

\bibitem[{{Szabados}(1990)}]{Szabados1990}
{Szabados}, L. 1990, \mnras, 242, 285

\bibitem[{{Szabados}(2003)}]{Szabados2003}
{Szabados}, L. 2003, Information Bulletin on Variable Stars, 5394

\bibitem[{{Taniguchi} {et~al.}(2018){Taniguchi}, {Matsunaga}, {Kobayashi},
  {Fukue}, {Hamano}, {Ikeda}, {Kawakita}, {Kondo}, {Sameshima}, \&
  {Yasui}}]{Taniguchietal2018}
{Taniguchi}, D., {Matsunaga}, N., {Kobayashi}, N., {et~al.} 2018, \mnras, 473,
  4993

\bibitem[{{Vasilyev} {et~al.}(2017{\natexlab{a}}){Vasilyev}, {Ludwig},
  {Freytag}, {Lemasle}, \& {Marconi}}]{Vasilyevetal2017a}
{Vasilyev}, V., {Ludwig}, H.-G., {Freytag}, B., {Lemasle}, B., \& {Marconi}, M.
  2017{\natexlab{a}}, \aap, 606, A140

\bibitem[{{Vasilyev} {et~al.}(2017{\natexlab{b}}){Vasilyev}, {Ludwig},
  {Freytag}, {Lemasle}, \& {Marconi}}]{Vasilyevetal2017b}
{Vasilyev}, V., {Ludwig}, H.-G., {Freytag}, B., {Lemasle}, B., \& {Marconi}, M.
  2017{\natexlab{b}}, ArXiv e-prints [\eprint[arXiv]{1711.00236}]

\bibitem[{{Wallerstein}(1972)}]{Wallerstein1972}
{Wallerstein}, G. 1972, \pasp, 84, 656

\bibitem[{{Wallerstein}(1979)}]{Wallerstein1979}
{Wallerstein}, G. 1979, \pasp, 91, 772

\bibitem[{{Wallerstein} {et~al.}(2015){Wallerstein}, {Albright}, \&
  {Ritchey}}]{Wallersteinetal2015}
{Wallerstein}, G., {Albright}, M.~B., \& {Ritchey}, A.~M. 2015, \pasp, 127, 503

\bibitem[{{Weber} {et~al.}(2012){Weber}, {Granzer}, \&
  {Strassmeier}}]{Weberetal2012}
{Weber}, M., {Granzer}, T., \& {Strassmeier}, K.~G. 2012, in \procspie, Vol.
  8451, Software and Cyberinfrastructure for Astronomy II, 84510K

\end{thebibliography}
\begin{landscape}
\begin{table}
\centering
\caption{New and old Line Depth Ratios (LDRs) adopted for effective temperature estimates.}
\label{teffcaliblist}
{\scriptsize
\begin{tabular}{rcrcccccccccc}
\noalign{\smallskip}\hline\hline\noalign{\smallskip}
$\lambda_1$ [\AA] & Ion & $\lambda_2$ [\AA] & Ion & $\Delta$\teff [K] & a & b & c & d & e & f & Function & Ref. \\
\noalign{\smallskip}\hline\noalign{\smallskip}
5348.30 & \ion{Cr}{i} & 5554.89 & \ion{Fe}{i} & 7200-7700 & 8120         & $-$919.996    & ...            & ...             & ...     & ...        & $a+br$                             & 2 \\
5348.30 & \ion{Cr}{i} & 5565.71 & \ion{Fe}{i} & 7200-7700 & 7940         & $-$646.94     & ...            & ...             & ...     & ...        & $a+br$                             & 2 \\
5373.71 & \ion{Fe}{i} & 5501.46 & \ion{Fe}{i} & 7200-7700 & 6757         & 1603.21       & ...            & ...             & ...     & ...        & $a+br$                             & 2 \\
5410.91 & \ion{Fe}{i} & 5501.46 & \ion{Fe}{i} & 7200-7700 & 6748         & 463.441       & ...            & ...             & ...     & ...        & $a+br$                             & 2 \\
5497.52 & \ion{Fe}{i} & 5554.89 & \ion{Fe}{i} & 7200-7700 & 8065         & $-$295.956    & ...            & ...             & ...     & ...        & $a+br$                             & 2 \\
5501.46 & \ion{Fe}{i} & 5554.89 & \ion{Fe}{i} & 7200-7700 & 8490         & $-$809.128    & ...            & ...             & ...     & ...        & $a+br$                             & 2 \\
5501.46 & \ion{Fe}{i} & 5565.71 & \ion{Fe}{i} & 7200-7700 & 8487         & $-$805.207    & ...            & ...             & ...     & ...        & $a+br$                             & 2 \\
5501.46 & \ion{Fe}{i} & 5633.97 & \ion{Fe}{i} & 7200-7700 & 8473         & $-$557.534    & ...            & ...             & ...     & ...        & $a+br$                             & 2 \\
5506.78 & \ion{Fe}{i} & 5554.89 & \ion{Fe}{i} & 7200-7700 & 8673         & $-$767.096    & ...            & ...             & ...     & ...        & $a+br$                             & 2 \\
5506.78 & \ion{Fe}{i} & 5565.71 & \ion{Fe}{i} & 7200-7700 & 8986         & $-$968.498    & ...            & ...             & ...     & ...        & $a+br$                             & 2 \\
5506.78 & \ion{Fe}{i} & 5633.97 & \ion{Fe}{i} & 7200-7700 & 8639         & $-$521.982    & ...            & ...             & ...     & ...        & $a+br$                             & 2 \\
5578.72 & \ion{Ni}{i} & 5645.62 & \ion{Si}{i} & 5400-6300 & 9486.76      & 0.653644      & 0.283943       & ...             & ...     & ...        & $ab^{1/r}r^{c}$                    & 2 \\
5578.72 & \ion{Ni}{i} & 5805.23 & \ion{Ni}{i} & 5600-6750 & 7343         & $-$975.039    & $-$74.7727     & ...             & ...     & ...        & $a+br+cr^2$                        & 2 \\
5670.86 & \ion{V}{i}  & 5690.43 & \ion{Si}{i} & 3700-6400 & 7113         & $-$7600.38    & 11646.9        & $-$9095.52      & 3251.54 & $-$433.007 & $a+br+cr^{2}+dr^{3}+er^{4}+fr^{5}$ & 2 \\
5754.68 & \ion{Ni}{i} & 5772.15 & \ion{Si}{i} & 7200-7700 & 8101         & $-$941.716    & ...            & ...             & ...     & ...        & $a+br$                             & 2 \\
5754.68 & \ion{Ni}{i} & 5772.15 & \ion{Si}{i} & 5400-7000 & 7553         & $-$906.095    & $-$319.209     & ...             & ...     & ...        & $a+br+cr^{2}$                      & 2 \\
5772.15 & \ion{Si}{i} & 5778.47 & \ion{Fe}{i} & 5000-6500 & 4870.26      & 0.991324      & 0.177759       & ...             & ...     & ...        & $ab^{r}r^{c}$                      & 2 \\
5772.15 & \ion{Si}{i} & 5778.47 & \ion{Fe}{i} & 3600-5000 & 2733         & 1889.61       & ...            & ...             & ...     & ...        & $a+br$                             & 2 \\
5772.15 & \ion{Si}{i} & 5847.00 & \ion{Ni}{i} & 3750-6400 & 5600.3858    & 0.859917      & 7.6436205      & ...             & ...     & ...        & $ab^{1/r}r^{c}$                    & 3 \\
5772.15 & \ion{Si}{i} & 5866.45 & \ion{Ti}{i} & 3750-6400 & 5844.6836    & 0.89753238    & 0.36660117     & ...             & ...     & ...        & $ab^{r}r^{c}$                      & 3 \\
5778.47 & \ion{Fe}{i} & 5793.08 & \ion{Si}{i} & 4700-6550 & 5150.2198    & $-$0.19395103 & $-$0.20263094  & ...             & ...     & ...        & $a(r-b)^{c}$                       & 3 \\
5793.08 & \ion{Si}{i} & 5793.92 & \ion{Fe}{i} & 4600-6900 & 1828         & 4179.03       & $-$1318.34     & 150.31          & ...     & ...        & $a+br+cr^{2}+dr^{3}$               & 2 \\
5793.08 & \ion{Si}{i} & 5847.00 & \ion{Ni}{i} & 3750-6400 & 5292.4677    & 188.04705     & $-$431.74828   & ...             & ...     & ...        & $a+br+c/r^{2}$                     & 3 \\
5793.08 & \ion{Si}{i} & 5866.45 & \ion{Ti}{i} & 3750-6900 & 6385.3741    & 0.85596741    & 0.047715024    & ...             & ...     & ...        & $ab^{1/r}r^{c}$                    & 3 \\
5809.25 & \ion{Fe}{i} & 6046.00 & \ion{S}{i}  & 5300-6900 & 7849         & $-$2708.78    & 1124.64        & $-$187.28       & ...     & ...        & $a+br+cr^{2}+dr^{3}$               & 2 \\
5809.25 & \ion{Fe}{i} & 6052.67 & \ion{S}{i}  & 5450-6900 & 7543         & $-$2173.24    & 763.632        & $-$96.4729      & ...     & ...        & $a+br+cr^{2}+dr^{3}$               & 2 \\
5847.00 & \ion{Ni}{i} & 5862.36 & \ion{Fe}{i} & 4800-6400 & 4732.753     & 0.9950455     & $-$0.14769983  & ...             & ...     & ...        & $ab^{1/r}r^{c}$                    & 3 \\
5847.00 & \ion{Ni}{i} & 5905.67 & \ion{Fe}{i} & 4800-6350 & 6972.3012    & $-$4507.1987  & 3736.2682      & $-$1279.5186    & ...     & ...        & $a+br+cr^{2}+dr^{3}$               & 3 \\
5847.00 & \ion{Ni}{i} & 5987.05 & \ion{Fe}{i} & 4800-6350 & 6952.0126    & $-$5341.52    & 5624.8112      & $-$2521.265     & ...     & ...        & $a+br+cr^{2}+dr^{3}$               & 3 \\
5847.00 & \ion{Ni}{i} & 6003.03 & \ion{Fe}{i} & 4800-6350 & 6988.9083    & $-$6964.1741  & 9499.6138      & $-$5413.4344    & ...     & ...        & $a+br+cr^{2}+dr^{3}$               & 3 \\
5847.00 & \ion{Ni}{i} & 6046.00 & \ion{S}{i}  & 4800-6900 & 5548.4816    & 0.9885989     & $-$0.081945655 & ...             & ...     & ...        & $ab^{r}r^{c}$                      & 3 \\
5862.36 & \ion{Fe}{i} & 5866.45 & \ion{Ti}{i} & 3750-6400 & 6394.1552    & 0.7536621     & 0.033678894    & ...             & ...     & ...        & $ab^{1/r}r^{c}$                    & 3 \\
5866.45 & \ion{Ti}{i} & 5905.67 & \ion{Fe}{i} & 4800-6350 & 6727.9816    & $-$597.326    & $-$1541.8658   & 727.53509       & ...     & ...        & $a+br+cr^{2}+dr^{3}$               & 3 \\
5866.45 & \ion{Ti}{i} & 5983.69 & \ion{Fe}{i} & 4800-6900 & 0.0001341342 & 7.89E$-$05    & $-$1.43E$-$05  & ...             & ...     & ...        & $1/(a+br+cr^{2})$                  & 3 \\
5866.45 & \ion{Ti}{i} & 5984.79 & \ion{Fe}{i} & 4800-6900 & 6164.4205    & 0.79631902    & $-$0.062253009 & ...             & ...     & ...        & $ab^{r}r^{c}$                      & 3 \\
5866.45 & \ion{Ti}{i} & 6003.03 & \ion{Fe}{i} & 4800-6900 & 7219.8944    & $-$3585.8283  & 1210.6378      & ...             & ...     & ...        & $a+br+cr^{2}$                      & 3 \\
5866.45 & \ion{Ti}{i} & 6021.79 & \ion{Mn}{i} & 4800-6700 & 7201.1881    & $-$3234.7364  & 754.27062      & ...             & ...     & ...        & $a+br+cr^{2}$                      & 3 \\
5866.45 & \ion{Ti}{i} & 6046.00 & \ion{S}{i}  & 4700-6900 & 6034.8217    & $-$0.27055898 & $-$0.14115708  & ...             & ...     & ...        & $a(r-b)^{c}$                       & 3 \\
5866.45 & \ion{Ti}{i} & 6055.99 & \ion{Fe}{i} & 4800-6900 & 7099.5674    & $-$2781.3457  & 770.38101      & ...             & ...     & ...        & $a+br+cr^{2}$                      & 3 \\
5934.66 & \ion{Fe}{i} & 6046.00 & \ion{S}{i}  & 7200-7700 & 8390         & $-$978.164    & ...            & ...             & ...     & ...        & $a+br$                             & 2 \\
5934.66 & \ion{Fe}{i} & 6052.67 & \ion{S}{i}  & 7200-7700 & 8221         & $-$918.937    & ...            & ...             & ...     & ...        & $a+br$                             & 2 \\
5956.70 & \ion{Fe}{i} & 5983.69 & \ion{Fe}{i} & 3700-6900 & 7286         & $-$2865.02    & 1353.68        & $-$396.557      & ...     & ...        & $a+br+cr^{2}+dr^{3}$               & 2 \\
5956.70 & \ion{Fe}{i} & 6142.49 & \ion{Si}{i} & 3750-6450 & 7052.7128    & $-$1014.7929  & 97.242351      & $-$3.3812526    & ...     & ...        & $a+br+cr^{2}+dr^{3}$               & 3 \\
5987.05 & \ion{Fe}{i} & 6216.37 & \ion{V}{i}  & 3600-5700 & 1179         & 6441.65       & $-$3564.96     & 709.834         & ...     & ...        & $a+br+cr^{2}+dr^{3}$               & 2 \\
6003.03 & \ion{Fe}{i} & 6052.67 & \ion{S}{i}  & 5550-7000 & 6283.25      & 1.083756      & $-$0.1341008   & ...             & ...     & ...        & $ab^{1/r}r^{c}$                    & 2 \\
6003.03 & \ion{Fe}{i} & 6052.67 & \ion{S}{i}  & 7200-7700 & 8037         & $-$551.182    & ...            & ...             & ...     & ...        & $a+br$                             & 2 \\
6007.96 & \ion{Fe}{i} & 6082.72 & \ion{Fe}{i} & 3750-6350 & 5459.2622    & 305.72716     & $-$787.01683   & ...             & ...     & ...        & $a+br+c/r^{2}$                     & 3 \\
6008.56 & \ion{Fe}{i} & 6046.00 & \ion{S}{i}  & 5600-7000 & 7155.97      & $-$0.241376   & ...            & ...             & ...     & ...        & $ar^{b}$                           & 2 \\
6021.79 & \ion{Mn}{i} & 6046.00 & \ion{S}{i}  & 5700-7000 & 7944         & $-$1437.72    & 260.697        & $-$19.1727      & ...     & ...        & $a+br+cr^{2}+dr^{3}$               & 2 \\
6021.79 & \ion{Mn}{i} & 6052.67 & \ion{S}{i}  & 5700-7000 & 7749         & $-$1312.04    & 221.004        & $-$13.474       & ...     & ...        & $a+br+cr^{2}+dr^{3}$               & 2 \\
6024.07 & \ion{Fe}{i} & 6082.72 & \ion{Fe}{i} & 3750-6550 & 5388.3579    & 150.09207     & $-$1263.9396   & ...             & ...     & ...        & $a+br+c/r^{2}$                     & 3 \\
6039.73 & \ion{V}{i}  & 6046.00 & \ion{S}{i}  & 3700-5400 & 5569         & $-$345.093    & 5.36966        & ...             & ...     & ...        & $a+br+cr^{2}$                      & 2 \\
6039.73 & \ion{V}{i}  & 6078.50 & \ion{Fe}{i} & 3750-5400 & 5590.5654    & $-$1140.171   & $-$197.64575   & 139.85301       & ...     & ...        & $a+br+cr^{2}+dr^{3}$               & 3 \\
6039.73 & \ion{V}{i}  & 6079.02 & \ion{Fe}{i} & 3750-5750 & 5349.3333    & 0.87139818    & $-$0.06060133  & ...             & ...     & ...        & $ab^{r}r^{c}$                      & 3 \\
6039.73 & \ion{V}{i}  & 6091.92 & \ion{Si}{i} & 3800-6050 & 5060.286     & 0.97817533    & $-$0.11128624  & ...             & ...     & ...        & $ab^{r}r^{c}$                      & 3 \\
6039.73 & \ion{V}{i}  & 6145.02 & \ion{Si}{i} & 3750-5650 & 5144.3594    & 0.95420384    & $-$0.073702513 & ...             & ...     & ...        & $ab^{r}r^{c}$                      & 3 \\
6039.73 & \ion{V}{i}  & 6155.14 & \ion{Si}{i} & 3750-5450 & $-$555860.5  & $-$92075392   & $-$16835.861   & $-$3465.0764    & ...     & ...        & $(a+br)/(1+cr+dr^{2})$             & 3 \\
\hline\noalign{\smallskip}
\multicolumn{13}{r}{\it {\footnotesize continued on next page}} \\
\end{tabular}}
\end{table}
\end{landscape}
\addtocounter{table}{-1}
\begin{landscape}
\begin{table}
\centering
\caption[]{continued.}
{\scriptsize 
\begin{tabular}{rcrcccccccccc}
\noalign{\smallskip}\hline\hline\noalign{\smallskip}
$\lambda_1$ [\AA] & Ion & $\lambda_2$ [\AA] & Ion & $\Delta$\teff [K] & a & b & c & d & e & f & Function & Ref. \\
\noalign{\smallskip}\hline\noalign{\smallskip}
6039.73 & \ion{V}{i}  & 6237.33 & \ion{Si}{i} & 3750-5500 & 5188.6032   & 0.89874901    & $-$0.055517417 & ...             & ...     & ...        & $ab^{r}r^{c}$          & 3 \\
6046.00 & \ion{S}{i}  & 6062.89 & \ion{Fe}{i} & 4600-6400 & 5470.5386   & 0.99518466    & 0.10524197     & ...             & ...     & ...        & $ab^{r}r^{c}$          & 3 \\
6046.00 & \ion{S}{i}  & 6081.44 & \ion{V}{i}  & 4450-6400 & 5166.8456   & $-$0.15189352 & 0.10673444     & ...             & ...     & ...        & $a(r-b)^{c}$           & 3 \\
6046.00 & \ion{S}{i}  & 6082.72 & \ion{Fe}{i} & 4450-6550 & 5786.2114   & 0.98639057    & 0.1266466      & ...             & ...     & ...        & $ab^{r}r^{c}$          & 3 \\
6046.00 & \ion{S}{i}  & 6085.27 & \ion{Fe}{i} & 4450-6900 & 5846.0931   & 0.98556293    & 0.13541182     & ...             & ...     & ...        & $ab^{r}r^{c}$          & 3 \\
6046.00 & \ion{S}{i}  & 6086.29 & \ion{Ni}{i} & 4800-6800 & 6156.77     & 0.954343      & 0.218591       & ...             & ...     & ...        & $ab^{r}r^{c}$          & 2 \\
6046.00 & \ion{S}{i}  & 6091.18 & \ion{Ti}{i} & 4450-6350 & 4581.1358   & 619.83166     & $-$82.470124   & 3.8980452       & ...     & ...        & $a+br+cr^{2}+dr^{3}$   & 3 \\
6046.00 & \ion{S}{i}  & 6108.12 & \ion{Ni}{i} & 4800-6700 & 6384.08     & 0.97804       & 0.0987979      & ...             & ...     & ...        & $ab^{1/r}r^{c}$        & 2 \\
6046.00 & \ion{S}{i}  & 6126.22 & \ion{Ti}{i} & 4450-6650 & 4243.1981   & 6981.8997     & 1.0296664      & $-$0.0051885512 & ...     & ...        & $(a+br)/(1+cr+dr^{2})$ & 3 \\
6046.00 & \ion{S}{i}  & 6130.17 & \ion{Ni}{i} & 4600-6350 & 4305.295    & 976.14392     & $-$64.570196   & $-$18.376814    & ...     & ...        & $a+br+cr^{2}+dr^{3}$   & 3 \\
6046.00 & \ion{S}{i}  & 6151.62 & \ion{Fe}{i} & 4450-6900 & 3802.3885   & 16046.227     & 2.3423486      & $-$0.030215793  & ...     & ...        & $(a+br)/(1+cr+dr^{2})$ & 3 \\
6046.00 & \ion{S}{i}  & 6165.37 & \ion{Fe}{i} & 4950-6500 & 5953.195    & 0.9938577     & 0.153777       & ...             & ...     & ...        & $ab^{1/r}r^{c}$        & 2 \\
6046.00 & \ion{S}{i}  & 6176.81 & \ion{Ni}{i} & 7200-7700 & 6075        & 877.871       & ...            & ...             & ...     & ...        & $a+br$                 & 2 \\
6046.00 & \ion{S}{i}  & 6180.22 & \ion{Fe}{i} & 4950-6700 & 6383.75     & 0.96151       & 0.170212       & ...             & ...     & ...        & $ab^{r}r^{c}$          & 2 \\
6046.00 & \ion{S}{i}  & 6215.15 & \ion{Fe}{i} & 7200-7700 & 6896        & 358.445       & ...            & ...             & ...     & ...        & $a+br$                 & 2 \\
6046.00 & \ion{S}{i}  & 6240.66 & \ion{Fe}{i} & 4450-6900 & 4016.0153   & 12659.012     & 1.8004153      & $-$0.011523382  & ...     & ...        & $(a+br)/(1+cr+dr^{2})$ & 3 \\
6046.00 & \ion{S}{i}  & 6243.11 & \ion{V}{i}  & 4450-6350 & 5536.7477   & 0.99676933    & 0.089085045    & ...             & ...     & ...        & $ab^{r}r^{c}$          & 3 \\
6046.00 & \ion{S}{i}  & 6258.10 & \ion{Ti}{i} & 4750-6700 & 6037.88     & 0.98653       & 0.110763       & ...             & ...     & ...        & $ab^{1/r}r^{c}$        & 2 \\
6052.67 & \ion{S}{i}  & 6082.72 & \ion{Fe}{i} & 4600-6550 & 5731.2095   & 0.99327876    & 0.096572399    & ...             & ...     & ...        & $ab^{1/r}r^{c}$        & 3 \\
6052.67 & \ion{S}{i}  & 6108.12 & \ion{Ni}{i} & 5250-6800 & 6293.6      & 0.984529      & 0.129272       & ...             & ...     & ...        & $ab^{r}r^{c}$          & 2 \\
6052.67 & \ion{S}{i}  & 6108.12 & \ion{Ni}{i} & 7200-7700 & 6558        & 246.382       & ...            & ...             & ...     & ...        & $a+br$                 & 2 \\
6052.67 & \ion{S}{i}  & 6122.23 & \ion{Ca}{i} & 7200-7700 & 6905        & 1649.06       & ...            & ...             & ...     & ...        & $a+br$                 & 2 \\
6052.67 & \ion{S}{i}  & 6136.61 & \ion{Fe}{i} & 7200-7700 & 6866        & 1219.4        & ...            & ...             & ...     & ...        & $a+br$                 & 2 \\
6052.67 & \ion{S}{i}  & 6151.62 & \ion{Fe}{i} & 5000-6800 & 6074.89     & 0.984276      & 0.127801       & ...             & ...     & ...        & $ab^{r}r^{c}$          & 2 \\
6052.67 & \ion{S}{i}  & 6162.18 & \ion{Ca}{i} & 7200-7700 & 6821        & 2178.43       & ...            & ...             & ...     & ...        & $a+br$                 & 2 \\
6052.67 & \ion{S}{i}  & 6176.81 & \ion{Ni}{i} & 7200-7700 & 7016        & 265.574       & ...            & ...             & ...     & ...        & $a+br$                 & 2 \\
6052.67 & \ion{S}{i}  & 6180.22 & \ion{Fe}{i} & 5000-6700 & 6224.77     & 0.979038      & 0.142497       & ...             & ...     & ...        & $ab^{r}r^{c}$          & 2 \\
6052.67 & \ion{S}{i}  & 6215.15 & \ion{Fe}{i} & 7200-7700 & 7025        & 239.858       & ...            & ...             & ...     & ...        & $a+br$                 & 2 \\
6052.67 & \ion{S}{i}  & 6219.28 & \ion{Fe}{i} & 7200-7700 & 7140        & 302.498       & ...            & ...             & ...     & ...        & $a+br$                 & 2 \\
6052.67 & \ion{S}{i}  & 6240.66 & \ion{Fe}{i} & 5050-6750 & 6057.77     & 0.986549      & 0.126622       & ...             & ...     & ...        & $ab^{r}r^{c}$          & 2 \\
6052.67 & \ion{S}{i}  & 6258.10 & \ion{Ti}{i} & 5000-6750 & 5924        & 688.557       & ...            & ...             & ...     & ...        & $a+b\log{r}$           & 2 \\
6055.99 & \ion{Fe}{i} & 6062.89 & \ion{Fe}{i} & 3700-4850 & 9851.19     & 0.473424      & $-$0.446733    & ...             & ...     & ...        & $ab^{1/r}r^{c}$        & 2 \\
6055.99 & \ion{Fe}{i} & 6062.89 & \ion{Fe}{i} & 5000-6000 & $-$21.6837  & 418.865       & 4532           & ...             & ...     & ...        & $a+br+cr^{2}$          & 2 \\
6055.99 & \ion{Fe}{i} & 6082.72 & \ion{Fe}{i} & 3750-3550 & 5357.431    & 264.82251     & $-$770.39303   & ...             & ...     & ...        & $a+br+c/r^{2}$         & 3 \\
6055.99 & \ion{Fe}{i} & 6085.27 & \ion{Fe}{i} & 4900-6750 & 5140.04     & 0.96892       & 0.257266       & ...             & ...     & ...        & $ab^{r}r^{c}$          & 2 \\
6055.99 & \ion{Fe}{i} & 6085.27 & \ion{Fe}{i} & 3600-4900 & 2405        & 2480.21       & ...            & ...             & ...     & ...        & $a+br$                 & 2 \\
6055.99 & \ion{Fe}{i} & 6151.62 & \ion{Fe}{i} & 3750-6700 & 5765.9039   & 289.50467     & $-$885.53492   & ...             & ...     & ...        & $a+br+c/r^{2}$         & 3 \\
6055.99 & \ion{Fe}{i} & 6180.22 & \ion{Fe}{i} & 3750-6550 & 6311.9278   & 167.30491     & $-$1248.4987   & ...             & ...     & ...        & $a+br+c/r^{2}$         & 3 \\
6055.99 & \ion{Fe}{i} & 6243.11 & \ion{V}{i}  & 3750-6350 & 5143.6204   & 180.91218     & $-$340.54146   & ...             & ...     & ...        & $a+br+c/r^{2}$         & 3 \\
6062.89 & \ion{Fe}{i} & 6078.50 & \ion{Fe}{i} & 4800-6400 & 7023.5296   & $-$6273.8558  & 6896.536       & $-$2897.7313    & ...     & ...        & $a+br+cr^{2}+dr^{3}$   & 3 \\
6062.89 & \ion{Fe}{i} & 6091.92 & \ion{Si}{i} & 3800-6500 & 5594.0551   & 0.95024411    & $-$0.11399896  & ...             & ...     & ...        & $ab^{r}r^{c}$          & 3 \\
6062.89 & \ion{Fe}{i} & 6145.02 & \ion{Si}{i} & 4750-6500 & 7037.1654   & $-$3469.1523  & 2026.3339      & $-$438.42083    & ...     & ...        & $a+br+cr^{2}+dr^{3}$   & 3 \\
6062.89 & \ion{Fe}{i} & 6237.33 & \ion{Si}{i} & 4750-6350 & 6986.3819   & $-$4788.9303  & 4169.7261      & $-$1434.1114    & ...     & ...        & $a+br+cr^{2}+dr^{3}$   & 3 \\
6078.50 & \ion{Fe}{i} & 6082.72 & \ion{Fe}{i} & 3750-6400 & 5293.1184   & 268.82143     & $-$733.04741   & ...             & ...     & ...        & $a+br+c/r^{2}$         & 3 \\
6078.50 & \ion{Fe}{i} & 6085.27 & \ion{Fe}{i} & 4900-6800 & 5935.8      & 0.824949      & 0.0896904      & ...             & ...     & ...        & $ab^{1/r}r^{c}$        & 2 \\
6078.50 & \ion{Fe}{i} & 6085.27 & \ion{Fe}{i} & 3600-4950 & 2484        & 2374.49       & ...            & ...             & ...     & ...        & $a+br$                 & 2 \\
6078.50 & \ion{Fe}{i} & 6243.11 & \ion{V}{i}  & 3750-6350 & 5152.999    & 173.02512     & $-$349.42793   & ...             & ...     & ...        & $a+br+c/r^{2}$         & 3 \\
6078.50 & \ion{Fe}{i} & 6258.10 & \ion{Ti}{i} & 3750-6400 & 6707.9691   & 0.74607455    & 0.057018238    & ...             & ...     & ...        & $ab^{1/r}r^{c}$        & 3 \\
6078.50 & \ion{Fe}{i} & 6258.71 & \ion{Ti}{i} & 3900-6900 & 8385.9963   & 0.57382889    & ...            & ...             & ...     & ...        & $ar/(b+r)$             & 3 \\
6081.44 & \ion{V}{i}  & 6145.02 & \ion{Si}{i} & 3750-5700 & 547550.54   & 13399692      & 2584.0445      & 211.68131       & ...     & ...        & $(a+br)/(1+cr+dr^{2})$ & 3 \\
6081.44 & \ion{V}{i}  & 6155.14 & \ion{Si}{i} & 3750-5500 & 5751.571    & $-$1528.4472  & 503.44595      & $-$90.6585      & ...     & ...        & $a+br+cr^{2}+dr^{3}$   & 3 \\
6081.44 & \ion{V}{i}  & 6176.81 & \ion{Ni}{i} & 3800-5500 & 5537.8688   & $-$984.85259  & 11.202499      & ...             & ...     & ...        & $a+br+c/r^{2}$         & 3 \\
6081.44 & \ion{V}{i}  & 6237.33 & \ion{Si}{i} & 3750-5900 & 1264810.5   & 55029420      & 10277.115      & 1582.781        & ...     & ...        & $(a+br)/(1+cr+dr^{2})$ & 3 \\
6081.44 & \ion{V}{i}  & 6243.81 & \ion{Si}{i} & 3750-5900 & 1276671.6   & 41437376      & 7844.3312      & 1018.2459       & ...     & ...        & $(a+br)/(1+cr+dr^{2})$ & 3 \\
6082.72 & \ion{Fe}{i} & 6091.92 & \ion{Si}{i} & 3800-6550 & 6255.7979   & 0.91578628    & $-$0.089526206 & ...             & ...     & ...        & $ab^{r}r^{c}$          & 3 \\
6082.72 & \ion{Fe}{i} & 6125.03 & \ion{Si}{i} & 4800-6550 & 7189.7511   & $-$1996.0635  & 425.19285      & ...             & ...     & ...        & $a+br+cr^{2}$          & 3 \\
6082.72 & \ion{Fe}{i} & 6142.49 & \ion{Si}{i} & 3750-6550 & 8026.8322   & $-$1.2495207  & $-$0.4271677   & ...             & ...     & ...        & $a(r-b)^{c}$           & 3 \\
\hline\noalign{\smallskip}
\multicolumn{13}{r}{\it {\footnotesize continued on next page}} \\
\end{tabular}}
\end{table}
\end{landscape}
\addtocounter{table}{-1}
\begin{landscape}
\begin{table}
\centering
\caption[]{continued.}
{\scriptsize 
\begin{tabular}{rcrcccccccccc}
\noalign{\smallskip}\hline\hline\noalign{\smallskip}
$\lambda_1$ [\AA] & Ion & $\lambda_2$ [\AA] & Ion & $\Delta$\teff [K] & a & b & c & d & e & f & Function & Ref. \\
\noalign{\smallskip}\hline\noalign{\smallskip}
6082.72 & \ion{Fe}{i} & 6145.02 & \ion{Si}{i} & 3750-6550 & 6039.2179   & 0.9182164     & $-$0.098249267 & ...            & ... & ... & $ab^{r}r^{c}$          & 3 \\
6082.72 & \ion{Fe}{i} & 6155.14 & \ion{Si}{i} & 4800-6550 & 5145.1625   & $-$0.14258724 & $-$0.20952186  & ...            & ... & ... & $a(r-b)^{c}$           & 3 \\
6082.72 & \ion{Fe}{i} & 6170.49 & \ion{Fe}{i} & 4800-6600 & 5771.2091   & $-$0.55350322 & $-$0.42594585  & ...            & ... & ... & $a(r-b)^{c}$           & 3 \\
6082.72 & \ion{Fe}{i} & 6237.33 & \ion{Si}{i} & 3750-6550 & 7072.4147   & $-$3032.6104  & 1513.9486      & $-$371.04912   & ... & ... & $a+br+cr^{2}+dr^{3}$   & 3 \\
6082.72 & \ion{Fe}{i} & 6243.81 & \ion{Si}{i} & 3750-6550 & 7196.6943   & $-$2751.3448  & 1087.5405      & $-$213.44259   & ... & ... & $a+br+cr^{2}+dr^{3}$   & 3 \\
6085.27 & \ion{Fe}{i} & 6086.29 & \ion{Ni}{i} & 4200-6700 & 7683        & $-$2129.69    & 114.528        & ...            & ... & ... & $a+br+cr^{2}$          & 2 \\
6085.27 & \ion{Fe}{i} & 6091.92 & \ion{Si}{i} & 3800-6650 & 7252.8278   & $-$1703.6908  & 311.72639      & $-$21.895353   & ... & ... & $a+br+cr^{2}+dr^{3}$   & 3 \\
6085.27 & \ion{Fe}{i} & 6142.49 & \ion{Si}{i} & 3750-6650 & 7634.692    & 630.77421     & 0.44872922     & $-$0.018371584 & ... & ... & $(a+br)/(1+cr+dr^{2})$ & 3 \\
6085.27 & \ion{Fe}{i} & 6155.14 & \ion{Si}{i} & 3750-6900 & 5708.4483   & 0.8720183     & $-$0.10554208  & ...            & ... & ... & $ab^{r}r^{c}$          & 3 \\
6085.27 & \ion{Fe}{i} & 6237.33 & \ion{Si}{i} & 3750-6900 & 8222.8194   & $-$1.2533326  & $-$0.56289765  & ...            & ... & ... & $a(r-b)^{c}$           & 3 \\
6090.21 & \ion{V}{i}  & 6091.92 & \ion{Si}{i} & 3700-5800 & 6585        & $-$1215.97    & 185.711        & $-$11.2669     & ... & ... & $a+br+cr^{2}+dr^{3}$   & 2 \\
6090.21 & \ion{V}{i}  & 6155.14 & \ion{Si}{i} & 3700-5750 & 5638.22     & 0.866727      & $-$0.0731712   & ...            & ... & ... & $ab^{r}r^{c}$          & 2 \\
6090.21 & \ion{V}{i}  & 6330.86 & \ion{Fe}{i} & 3700-5550 & 7359        & $-$1844.03    & 213.264        & ...            & ... & ... & $a+br+cr^{2}$          & 2 \\
6091.18 & \ion{Ti}{i} & 6125.03 & \ion{Si}{i} & 4800-6350 & 4277.4717   & 1.139652      & $-$0.22374728  & ...            & ... & ... & $ab^{r}r^{c}$          & 3 \\
6091.18 & \ion{Ti}{i} & 6145.02 & \ion{Si}{i} & 3750-5750 & 4980.2378   & 0.96649545    & $-$0.11349399  & ...            & ... & ... & $ab^{r}r^{c}$          & 3 \\
6091.18 & \ion{Ti}{i} & 6155.14 & \ion{Si}{i} & 3750-5450 & 5153.6733   & 0.8565174     & $-$0.054459935 & ...            & ... & ... & $ab^{r}r^{c}$          & 3 \\
6091.18 & \ion{Ti}{i} & 6237.33 & \ion{Si}{i} & 3750-5750 & 4928.1186   & 0.91700085    & $-$0.1017638   & ...            & ... & ... & $ab^{r}r^{c}$          & 3 \\
6091.92 & \ion{Si}{i} & 6219.28 & \ion{Fe}{i} & 7200-7700 & 6935        & 1346.25       & ...            & ...            & ... & ... & $a+br$                 & 2 \\
6091.92 & \ion{Si}{i} & 6243.11 & \ion{V}{i}  & 3800-6350 & 5625.0556   & 0.96931038    & 0.092095843    & ...            & ... & ... & $ab^{1/r}r^{c}$        & 3 \\
6093.14 & \ion{Co}{i} & 6093.66 & \ion{Fe}{i} & 3700-6200 & 6679        & $-$3868.72    & 3130.89        & $-$1022.74     & ... & ... & $a+br+cr^{2}+dr^{3}$   & 2 \\
6108.12 & \ion{Ni}{i} & 6125.03 & \ion{Si}{i} & 5600-7000 & 7910.42     & 0.821822      & 0.0421413      & ...            & ... & ... & $ab^{r}r^{c}$          & 2 \\
6108.12 & \ion{Ni}{i} & 6145.02 & \ion{Si}{i} & 4000-7000 & 7598        & $-$1309.78    & 90.231         & ...            & ... & ... & $a+br+cr^{2}$          & 2 \\
6108.12 & \ion{Ni}{i} & 6155.14 & \ion{Si}{i} & 4600-6900 & 7023.0844   & $-$614.59734  & $-$1541.5092   & 661.13611      & ... & ... & $a+br+cr^{2}+dr^{3}$   & 3 \\
6108.12 & \ion{Ni}{i} & 6237.33 & \ion{Si}{i} & 5400-7000 & 7549        & $-$1908.86    & 283.413        & ...            & ... & ... & $a+br+cr^{2}$          & 2 \\
6125.03 & \ion{Si}{i} & 6126.22 & \ion{Ti}{i} & 4400-6650 & 5571.4418   & 0.95872488    & 0.2433342      & ...            & ... & ... & $ab^{r}r^{c}$          & 3 \\
6125.03 & \ion{Si}{i} & 6151.62 & \ion{Fe}{i} & 5600-6600 & 6834.79     & 0.892416      & 0.0563776      & ...            & ... & ... & $ab^{1/r}r^{c}$        & 2 \\
6125.03 & \ion{Si}{i} & 6243.11 & \ion{V}{i}  & 4800-6350 & 4414.6542   & 1216.2187     & $-$311.79239   & 31.88682       & ... & ... & $a+br+cr^{2}+dr^{3}$   & 3 \\
6125.03 & \ion{Si}{i} & 6358.69 & \ion{Fe}{i} & 7200-7700 & 8295        & $-$505.937    & ...            & ...            & ... & ... & $a+br$                 & 2 \\
6126.22 & \ion{Ti}{i} & 6145.02 & \ion{Si}{i} & 3750-6900 & 5541.4767   & $-$0.22966992 & $-$0.23199881  & ...            & ... & ... & $a(r-b)^{c}$           & 3 \\
6126.22 & \ion{Ti}{i} & 6155.14 & \ion{Si}{i} & 3750-6550 & 5192.2074   & 0.91920445    & $-$0.12146185  & ...            & ... & ... & $ab^{r}r^{c}$          & 3 \\
6126.22 & \ion{Ti}{i} & 6237.33 & \ion{Si}{i} & 3750-6550 & 5343.9359   & 0.92516671    & $-$0.12840343  & ...            & ... & ... & $ab^{r}r^{c}$          & 3 \\
6128.99 & \ion{Ni}{i} & 6237.33 & \ion{Si}{i} & 3750-6550 & 6934.9822   & $-$3307.7598  & 2018.2402      & $-$577.34869   & ... & ... & $a+br+cr^{2}+dr^{3}$   & 3 \\
6135.36 & \ion{V}{i}  & 6142.49 & \ion{Si}{i} & 3700-5300 & 5575        & $-$656.539    & 35.5005        & ...            & ... & ... & $a+br+cr^{2}$          & 2 \\
6135.36 & \ion{V}{i}  & 6237.33 & \ion{Si}{i} & 3750-5650 & 5193.125    & 0.8986041     & $-$0.044948334 & ...            & ... & ... & $ab^{r}r^{c}$          & 3 \\
6136.61 & \ion{Fe}{i} & 6243.11 & \ion{V}{i}  & 3750-6350 & 5226.9903   & 78.711256     & $-$1353.7489   & ...            & ... & ... & $a+br+c/r^{2}$         & 3 \\
6142.49 & \ion{Si}{i} & 6243.11 & \ion{V}{i}  & 3750-6350 & 5575.6374   & 0.97029759    & 0.097815165    & ...            & ... & ... & $ab^{1/r}r^{c}$        & 3 \\
6145.02 & \ion{Si}{i} & 6151.62 & \ion{Fe}{i} & 3750-6700 & 7414.7921   & 0.24285202    & ...            & ...            & ... & ... & $ar/(b+r)$             & 3 \\
6145.02 & \ion{Si}{i} & 6180.22 & \ion{Fe}{i} & 3700-5000 & 2382        & 5416.72       & ...            & ...            & ... & ... & $a+br$                 & 2 \\
6145.02 & \ion{Si}{i} & 6243.11 & \ion{V}{i}  & 3750-6350 & 5515.3062   & 0.96712503    & 0.09179199     & ...            & ... & ... & $ab^{1/r}r^{c}$        & 3 \\
6145.02 & \ion{Si}{i} & 6258.10 & \ion{Ti}{i} & 3700-6700 & 2870        & 5476.94       & $-$3256.51     & 800.595        & ... & ... & $a+br+cr^{2}+dr^{3}$   & 2 \\
6150.16 & \ion{V}{i}  & 6237.33 & \ion{Si}{i} & 3700-4900 & 5505        & $-$770.29     & ...            & ...            & ... & ... & $a+br$                 & 2 \\
6150.16 & \ion{V}{i}  & 6380.75 & \ion{Fe}{i} & 3700-4800 & 6126        & $-$1611.92    & ...            & ...            & ... & ... & $a+br$                 & 2 \\
6151.62 & \ion{Fe}{i} & 6155.14 & \ion{Si}{i} & 3750-6900 & 6727.799    & 0.7737723     & $-$0.036684239 & ...            & ... & ... & $ab^{r}r^{c}$          & 3 \\
6151.62 & \ion{Fe}{i} & 6237.33 & \ion{Si}{i} & 3750-6900 & 7156.5565   & $-$1767.765   & 146.61902      & ...            & ... & ... & $a+br+cr^{2}$          & 3 \\
6155.14 & \ion{Si}{i} & 6180.22 & \ion{Fe}{i} & 3750-6900 & $-$14407.078& 75426.24      & 10.641479      & $-$0.09935597  & ... & ... & $(a+br)/(1+cr+dr^{2})$ & 3 \\
6155.14 & \ion{Si}{i} & 6240.66 & \ion{Fe}{i} & 3750-6900 & 6784.9875   & 0.76711376    & 0.026378813    & ...            & ... & ... & $ab^{1/r}r^{c}$        & 3 \\
6155.14 & \ion{Si}{i} & 6243.11 & \ion{V}{i}  & 3750-6350 & 5393.2959   & 0.92019335    & 0.073807492    & ...            & ... & ... & $ab^{1/r}r^{c}$        & 3 \\
6155.14 & \ion{Si}{i} & 6358.69 & \ion{Fe}{i} & 3700-6700 & 6642.5      & 0.85674       & 0.086193       & ...            & ... & ... & $ab^{1/r}r^{c}$        & 2 \\
6170.49 & \ion{Fe}{i} & 6180.22 & \ion{Fe}{i} & 5150-6850 & 380         & 7413.41       & $-$3207.31     & 501.652        & ... & ... & $a+br+cr^{2}+dr^{3}$   & 2 \\
6176.81 & \ion{Ni}{i} & 6243.11 & \ion{V}{i}  & 3800-6350 & 5094.4861   & 234.05509     & $-$315.52803   & ...            & ... & ... & $a+br+c/r^{2}$         & 3 \\
6176.81 & \ion{Ni}{i} & 6258.10 & \ion{Ti}{i} & 3700-6200 & 6317.32     & 0.812411      & 0.611525       & ...            & ... & ... & $ab^{r}r^{c}$          & 2 \\
6176.81 & \ion{Ni}{i} & 6261.10 & \ion{Ti}{i} & 4700-6700 & 2011        & 3758          & $-$753.8       & ...            & ... & ... & $a+br+cr^{2}$          & 1 \\
6180.22 & \ion{Fe}{i} & 6237.33 & \ion{Si}{i} & 3750-6900 & 7361.1377   & $-$1749.5269  & 52.616694      & ...            & ... & ... & $a+br+cr^{2}$          & 3 \\
6189.01 & \ion{Co}{i} & 6237.33 & \ion{Si}{i} & 3750-5400 & 5602.0793   & $-$1172.9134  & 370.85881      & $-$109.56462   & ... & ... & $a+br+cr^{2}+dr^{3}$   & 3 \\
6189.01 & \ion{Co}{i} & 6244.48 & \ion{Si}{i} & 3750-5500 & 5421.6284   & $-$616.52128  & 9.4613246      & ...            & ... & ... & $a+br+c/r^{2}$         & 3 \\
6200.32 & \ion{Fe}{i} & 6237.33 & \ion{Si}{i} & 5300-6900 & 7936        & $-$1668.99    & ...            & ...            & ... & ... & $a+br$                 & 2 \\
6219.28 & \ion{Fe}{i} & 6414.99 & \ion{Si}{i} & 7200-7700 & 8068        & $-$364.285    & ...            & ...            & ... & ... & $a+br$                 & 2 \\
\hline\noalign{\smallskip}
\multicolumn{13}{r}{\it {\footnotesize continued on next page}} \\
\end{tabular}}
\end{table}
\end{landscape}
\addtocounter{table}{-1}
\begin{landscape}
\begin{table}
\centering
\caption[]{continued.}
{\scriptsize 
\begin{tabular}{rcrcccccccccc}
\noalign{\smallskip}\hline\hline\noalign{\smallskip}
$\lambda_1$ [\AA] & Ion & $\lambda_2$ [\AA] & Ion & $\Delta$\teff [K] & a & b & c & d & e & f & Function & Ref. \\
\noalign{\smallskip}\hline\noalign{\smallskip}
6232.65 & \ion{Fe}{i} & 6243.11 & \ion{V}{i}  & 3750-6350 & 5105.7991    & 177.4715      & $-$513.57204   & ...           & ...          & ... & $a+br+c/r^{2}$              & 3 \\
6237.33 & \ion{Si}{i} & 6240.66 & \ion{Fe}{i} & 3750-6900 & 32620441     & $-$168828320  & $-$25019.24    & 389.82997     & ...          & ... & $(a+br)/(1+cr+dr^{2})$      & 3 \\
6237.33 & \ion{Si}{i} & 6243.11 & \ion{V}{i}  & 3750-6350 & 5509.5994    & 0.92783258    & 0.074721402    & ...           & ...          & ... & $ab^{1/r}r^{c}$             & 3 \\
6237.33 & \ion{Si}{i} & 6258.10 & \ion{Ti}{i} & 4700-6700 & 6691.94      & 0.808588      & 0.0677634      & ...           & ...          & ... & $ab^{1/r}r^{c}$             & 2 \\
6237.33 & \ion{Si}{i} & 6258.71 & \ion{Ti}{i} & 4800-6700 & 2379         & 5481.69       & $-$2484.04     & 406.352       & ...          & ... & $a+br+cr^{2}+dr^{3}$        & 2 \\
6237.33 & \ion{Si}{i} & 6358.69 & \ion{Fe}{i} & 3750-6900 & 7336.2706    & 0.82039503    & 0.021229756    & ...           & ...          & ... & $ab^{1/r}r^{c}$             & 3 \\
6240.66 & \ion{Fe}{i} & 6243.81 & \ion{Si}{i} & 3750-6900 & 7448.8726    & 0.77089127    & ...            & ...           & ...          & ... & $ab^{r}$                    & 3 \\
6240.66 & \ion{Fe}{i} & 6244.48 & \ion{Si}{i} & 4500-6900 & 7340.2       & 0.793709      & $-$0.0153904   & ...           & ...          & ... & $ab^{r}r^{c}$               & 2 \\
6240.66 & \ion{Fe}{i} & 6414.99 & \ion{Si}{i} & 4000-6900 & 6892.895     & 0.84853114    & $-$0.039046311 & ...           & ...          & ... & $ab^{r}r^{c}$               & 3 \\
6243.11 & \ion{V}{i}  & 6243.81 & \ion{Si}{i} & 4800-5900 & 6561.787     & $-$2261.2405  & 1032.1541      & $-$171.36594  & ...          & ... & $a+br+cr^{2}+dr^{3}$        & 3 \\
6243.11 & \ion{V}{i}  & 6244.48 & \ion{Si}{i} & 4800-5900 & 6564.3353    & $-$2186.4868  & 956.13517      & $-$145.40885  & ...          & ... & $a+br+cr^{2}+dr^{3}$        & 3 \\
6243.11 & \ion{V}{i}  & 6414.99 & \ion{Si}{i} & 4800-5900 & 6478.4211    & $-$1881.5917  & 820.5678       & $-$141.74283  & ...          & ... & $a+br+cr^{2}+dr^{3}$        & 3 \\
6243.11 & \ion{V}{i}  & 6439.08 & \ion{Ca}{i} & 4000-5500 & 6265.953     & $-$4949.8037  & 7534.6692      & $-$4708.8282  & ...          & ... & $a+br+cr^{2}+dr^{3}$        & 3 \\
6243.81 & \ion{Si}{i} & 6261.10 & \ion{Ti}{i} & 3700-5600 & 2952         & 3689.04       & $-$1068.2      & ...           & ...          & ... & $a+br+cr^{2}$               & 2 \\
6243.81 & \ion{Si}{i} & 6358.69 & \ion{Fe}{i} & 3700-7000 & 7757.8       & 0.81203       & 0.00385864     & ...           & ...          & ... & $ab^{1/r}r^{c}$             & 2 \\
6244.48 & \ion{Si}{i} & 6258.10 & \ion{Ti}{i} & 4700-6600 & 2199         & 6096          & 3216           & 634.3         & ...          & ... & $a+br+cr^{2}+dr^{3}$        & 1 \\
6327.60 & \ion{Ni}{i} & 6414.99 & \ion{Si}{i} & 3700-6900 & 6080.41      & 0.906332      & $-$0.0951967   & ...           & ...          & ... & $ab^{r}r^{c}$               & 2 \\
6330.13 & \ion{Cr}{i} & 6330.86 & \ion{Fe}{i} & 4700-6700 & 7190         & $-$2042       & 307.6          & ...           & ...          & ... & $a+br+cr^{2}$               & 1 \\
6330.13 & \ion{Cr}{i} & 6414.99 & \ion{Si}{i} & 3700-6000 & 5526.66      & 0.937233      & $-$0.111614    & ...           & ...          & ... & $ab^{r}r^{c}$               & 2 \\
6355.04 & \ion{Fe}{i} & 6414.99 & \ion{Si}{i} & 3700-7000 & 7780         & $-$1298.65    & ...            & ...           & ...          & ... & $a+br$                      & 2 \\
6355.04 & \ion{Fe}{i} & 6419.98 & \ion{Fe}{i} & 3700-7000 & 7433         & $-$1541.58    & 7.58844        & $-$464.06     & ...          & ... & $a+br+cr^{2}+dr^{3}$        & 2 \\
6358.69 & \ion{Fe}{i} & 6414.99 & \ion{Si}{i} & 4400-7000 & 7527         & $-$1283.15    & 99.1835        & ...           & ...          & ... & $a+br+cr^{2}$               & 2 \\
6358.69 & \ion{Fe}{i} & 6419.98 & \ion{Fe}{i} & 3700-7000 & 7482         & $-$2075.08    & 143.949        & ...           & ...          & ... & $a+br+cr^{2}$               & 2 \\
6392.55 & \ion{Fe}{i} & 6414.99 & \ion{Si}{i} & 3700-6200 & 6889         & $-$4485.89    & 4193.86        & $-$1917.35    & 294.546      & ... & $a+br+cr^{2}+dr^{3}+er^{4}$ & 2 \\
6414.99 & \ion{Si}{i} & 6498.95 & \ion{Fe}{i} & 4200-6550 & 6204.5377    & 0.91929415    & 0.066862673    & ...           & ...          & ... & $ab^{1/r}r^{c}$             & 3 \\
6498.95 & \ion{Fe}{i} & 6597.61 & \ion{Fe}{i} & 4850-6600 & 8132         & $-$4988.98    & 4606.31        & $-$2274.22    & 408.814      & ... & $a+br+cr^{2}+dr^{3}+er^{4}$ & 2 \\
6538.60 & \ion{S}{i}  & 6609.12 & \ion{Fe}{i} & 5600-7000 & 4943         & 4959.69       & $-$4809.9      & 1698.54       & ...          & ... & $a+br+cr^{2}+dr^{3}$        & 2 \\
6597.61 & \ion{Fe}{i} & 6608.03 & \ion{Fe}{i} & 4700-6500 & 3853         & 1195          & $-$225         & 15.9          & ...          & ... & $a+br+cr^{2}+dr^{3}$        & 1 \\
6608.03 & \ion{Fe}{i} & 6721.85 & \ion{Si}{i} & 4700-6550 & 0.0001475568 & 8.19E-05      & $-$2.75E-05    & ...           & ...          & ... & $1/(a+br+cr^{2})$           & 3 \\
6609.12 & \ion{Fe}{i} & 6748.84 & \ion{S}{i}  & 5000-7000 & 6073.03      & 0.994752      & $-$0.118793    & ...           & ...          & ... & $ab^{1/r}r^{c}$             & 2 \\
6609.12 & \ion{Fe}{i} & 6757.17 & \ion{S}{i}  & 4400-6900 & 5845.1056    & 1.0058528     & $-$0.12878833  & ...           & ...          & ... & $ab^{r}r^{c}$               & 3 \\
6680.15 & \ion{Cr}{i} & 6703.57 & \ion{Fe}{i} & 4800-6500 & 3635         & 2633          & $-$967         & 130.6         & ...          & ... & $a+br+cr^{2}+dr^{3}$        & 1 \\
6703.57 & \ion{Fe}{i} & 6721.85 & \ion{Si}{i} & 4700-6550 & 6825.2988    & $-$840.26077  & $-$1038.0854   & 496.98698     & ...          & ... & $a+br+cr^{2}+dr^{3}$        & 3 \\
6710.31 & \ion{Fe}{i} & 6713.76 & \ion{Fe}{i} & 4800-6300 & 6674         & $-$1353       & 233.29         & ...           & ...          & ... & $a+br+cr^{2}$               & 1 \\
6710.31 & \ion{Fe}{i} & 6721.85 & \ion{Si}{i} & 3750-5700 & 6105.508     & $-$1551.6196  & 691.66748      & $-$176.15435  & ...          & ... & $a+br+cr^{2}+dr^{3}$        & 3 \\
6710.31 & \ion{Fe}{i} & 6767.77 & \ion{Ni}{i} & 3700-6000 & 6949         & $-$10279.7    & 20503.9        & $-$15335.9    & ...          & ... & $a+br+cr^{2}+dr^{3}$        & 2 \\
6717.69 & \ion{Ca}{i} & 6757.17 & \ion{S}{i}  & 5300-6800 & 5926.58      & 1.17997       & $-$0.138265    & ...           & ...          & ... & $ab^{1/r}r^{c}$             & 2 \\
6717.69 & \ion{Ca}{i} & 6757.17 & \ion{S}{i}  & 7200-7700 & 8158         & $-$686.136    & ...            & ...           & ...          & ... & $a+br$                      & 2 \\
6721.85 & \ion{Si}{i} & 6771.04 & \ion{Co}{i} & 3750-6250 & 5554.4503    & 0.91377129    & 0.060535721    & ...           & ...          & ... & $ab^{1/r}r^{c}$             & 3 \\
6721.85 & \ion{Si}{i} & 6839.83 & \ion{Fe}{i} & 4800-6400 & 3783.9506    & 1939.4483     & $-$521.88679   & 49.382188     & ...          & ... & $a+br+cr^{2}+dr^{3}$        & 3 \\
6748.84 & \ion{S}{i}  & 6750.15 & \ion{Fe}{i} & 4900-6800 & 6194.06      & 1.004298      & 0.138792       & ...           & ...          & ... & $ab^{1/r}r^{c}$             & 2 \\
6748.84 & \ion{S}{i}  & 6767.77 & \ion{Ni}{i} & 5000-6800 & 6307.54      & 1.005136      & 0.148183       & ...           & ...          & ... & $ab^{1/r}r^{c}$             & 2 \\
6750.15 & \ion{Fe}{i} & 6757.17 & \ion{S}{i}  & 4700-7000 & 6044.21      & 1.00735       & $-$0.125622    & ...           & ...          & ... & $ab^{1/r}r^{c}$             & 2 \\
6757.17 & \ion{S}{i}  & 6767.77 & \ion{Ni}{i} & 4800-6700 & 6169.4       & 1.00416       & 0.145267       & ...           & ...          & ... & $ab^{1/r}r^{c}$             & 2 \\
6806.85 & \ion{Fe}{i} & 6848.57 & \ion{Si}{i} & 4700-6700 & 7116         & $-$790.9      & ...            & ...           & ...          & ... & $a+br$                      & 1 \\
7110.90 & \ion{Ni}{i} & 7022.39 & \ion{Fe}{i} & 5300-6500 & 6492         & $-$791.06341  & ...            & ...           & ...          & ... & $a+b\log(r)$                & 3 \\
7110.90 & \ion{Ni}{i} & 7022.95 & \ion{Fe}{i} & 3500-6600 & 7775         & $-$5150.6755  & 325.81723      & 6084.4549     & $-$4007.2797 & ... & $a+br+cr^{2}+dr^{3}+er^{4}$ & 3 \\
7110.90 & \ion{Ni}{i} & 7034.90 & \ion{Si}{i} & 4500-6700 & 5677.678     & 0.9068567     & $-$0.090936813 & ...           & ...          & ... & $ab^{r}r^{c}$               & 3 \\
7110.90 & \ion{Ni}{i} & 7071.88 & \ion{Fe}{i} & 5050-6500 & 6066         & $-$1076.6458  & ...            & ...           & ...          & ... & $a+b\log(r)$                & 3 \\
7110.90 & \ion{Ni}{i} & 7090.38 & \ion{Fe}{i} & 3900-6500 & 6957         & 796.9239      & $-$16020.09    & 25423.295     & $-$12379.132 & ... & $a+br+cr^{2}+dr^{3}+er^{4}$ & 3 \\
7110.90 & \ion{Ni}{i} & 7130.92 & \ion{Fe}{i} & 4300-7000 & 7740.5691    & $-$8974.467   & 12385.913      & $-$6956.3501  & ...          & ... & $a+br+cr^{2}+dr^{3}$        & 3 \\
7110.90 & \ion{Ni}{i} & 7132.99 & \ion{Fe}{i} & 5500-6050 & 5989.5842    & 748.53279     & $-$1124.968    & ...           & ...          & ... & $a+br+cr^{2}$               & 3 \\
7110.90 & \ion{Ni}{i} & 7181.19 & \ion{Fe}{i} & 4200-7000 & 7926.6838    & $-$8555.0465  & 10974.93       & $-$5542.4826  & ...          & ... & $a+br+cr^{2}+dr^{3}$        & 3 \\
7112.18 & \ion{Fe}{i} & 7022.95 & \ion{Fe}{i} & 4300-6000 & 7924.3779    & $-$8118.8128  & 8946.0425      & $-$4311.6523  & ...          & ... & $a+br+cr^{2}+dr^{3}$        & 3 \\
7112.18 & \ion{Fe}{i} & 7034.90 & \ion{Si}{i} & 3800-6700 & 7348.5613    & $-$5185.4006  & 3691.6343      & $-$967.50273  & ...          & ... & $a+br+cr^{2}+dr^{3}$        & 3 \\
7216.18 & \ion{Ti}{i} & 7132.99 & \ion{Fe}{i} & 3500-5900 & 6365.0976    & $-$2170.5471  & 1242.9243      & $-$491.89608  & ...          & ... & $a+br+cr^{2}+dr^{3}$        & 3 \\
7216.18 & \ion{Ti}{i} & 7181.19 & \ion{Fe}{i} & 3500-5800 & 6463.9587    & $-$4488.1749  & 5325.8248      & $-$2628.69    & ...          & ... & $a+br+cr^{2}+dr^{3}$        & 3 \\
\hline\noalign{\smallskip}
\multicolumn{13}{r}{\it {\footnotesize continued on next page}} \\
\end{tabular}}
\end{table}
\end{landscape}
\addtocounter{table}{-1}
\begin{landscape}
\begin{table}
\centering
\caption[]{continued.}
{\scriptsize 
\begin{tabular}{rcrcccccccccc}
\noalign{\smallskip}\hline\hline\noalign{\smallskip}
$\lambda_1$ [\AA] & Ion & $\lambda_2$ [\AA] & Ion & $\Delta$\teff [K] & a & b & c & d & e & f & Function & Ref. \\
\noalign{\smallskip}\hline\noalign{\smallskip}
7216.18 & \ion{Ti}{i} & 7221.20 & \ion{Fe}{i} & 3500-5800 & 6284.6629 & $-$1527.7419 & ...           & ...          & ...          & ...          & $a+br$                             & 3 \\
7251.71 & \ion{Ti}{i} & 7142.52 & \ion{Fe}{i} & 4200-6800 & 7654.0242 & $-$3858.2697 & 2899.9181     & $-$1135.2843 & ...          & ...          & $a+br+cr^{2}+dr^{3}$               & 3 \\
7251.71 & \ion{Ti}{i} & 7181.19 & \ion{Fe}{i} & 4300-6900 & 5441.5977 & $-$0.4745018 & $-$0.39033324 & ...          & ...          & ...          & $a(r-b)^{c}$                       & 3 \\
7327.65 & \ion{Ni}{i} & 7375.25 & \ion{Si}{i} & 3500-5250 & 5833.1952 & $-$1992.5838 & 1774.5887     & $-$857.75968 & ...          & ...          & $a+br+cr^{2}+dr^{3}$               & 3 \\
7357.73 & \ion{Ti}{i} & 7445.75 & \ion{Fe}{i} & 4400-5400 & 5781.7635 & $-$1570.4442 & ...           & ...          & ...          & ...          & $a+br$                             & 3 \\
7461.52 & \ion{Fe}{i} & 7491.65 & \ion{Fe}{i} & 4400-5800 & 6414.7319 & $-$1800.1392 & ...           & ...          & ...          & ...          & $a+br$                             & 3 \\
7461.52 & \ion{Fe}{i} & 7495.07 & \ion{Fe}{i} & 4400-5800 & 6152.7548 & $-$2318.461  & ...           & ...          & ...          & ...          & $a+br$                             & 3 \\
7461.52 & \ion{Fe}{i} & 7507.27 & \ion{Fe}{i} & 4400-5900 & 7604.3548 & $-$7135.3851 & 8481.292      & $-$4440.1886 & ...          & ...          & $a+br+cr^{2}+dr^{3}$               & 3 \\
7540.43 & \ion{Fe}{i} & 7511.02 & \ion{Fe}{i} & 4700-6000 & 6613.917  & $-$14032.5   & 41178.67      & $-$45064.2   & ...          & ...          & $a+br+cr^{2}+dr^{3}$               & 3 \\
7540.43 & \ion{Fe}{i} & 7563.01 & \ion{Fe}{i} & 5000-6300 & 5198.713  & $-$0.1799965 & ...           & ...          & ...          & ...          & $ar^{b}$                           & 3 \\
7583.79 & \ion{Fe}{i} & 7468.31 & \ion{N}{i}  & 5700-7000 & 7052.4777 & $-$895.819   & 234.25416     & $-$30.425818 & 1.4601492    & ...          & $a+br+cr^{2}+dr^{3}+er^{4}$        & 3 \\
7583.79 & \ion{Fe}{i} & 7531.14 & \ion{Fe}{i} & 3500-6600 & 5973.5196 & 3784.5656    & $-$4235.942   & ...          & ...          & ...          & $a+br+cr^{2}$                      & 3 \\
7583.79 & \ion{Fe}{i} & 7586.02 & \ion{Fe}{i} & 3500-6900 & 11021.68  & $-$38217.372 & 138605.19     & $-$248399.82 & 209495.81    & $-$67865.923 & $a+br+cr^{2}+dr^{3}+er^{4}+fr^{5}$ & 3 \\
7714.27 & \ion{Ni}{i} & 7680.27 & \ion{Si}{i} & 3500-6900 & 7193      & $-$29.77509  & $-$1681.5027  & 731.7729     & $-$95.782326 & ...          & $a+br+cr^{2}+dr^{3}+er^{4}$        & 3 \\
7714.27 & \ion{Ni}{i} & 7780.56 & \ion{Fe}{i} & 5900-7000 & 9319.357  & 0.5872613    & 0.108473      & ...          & ...          & ...          & $ab^{r}r^{c}$                      & 3 \\
7714.27 & \ion{Ni}{i} & 7780.56 & \ion{Fe}{i} & 3600-7000 & 7831.5986 & $-$3202.7715 & 2050.2954     & $-$1218.1437 & ...          & ...          & $a+br+cr^{2}+dr^{3}$               & 3 \\
7723.21 & \ion{Fe}{i} & 7680.27 & \ion{Si}{i} & 3500-6000 & 7282.2329 & $-$4432.9632 & 2956.9517     & $-$776.4689  & ...          & ...          & $a+br+cr^{2}+dr^{3}$               & 3 \\
7748.27 & \ion{Fe}{i} & 7680.27 & \ion{Si}{i} & 6500-7150 & 7979      & $-$1601.28   & ...           & ...          & ...          & ...          & $a+br$                             & 3 \\
7748.27 & \ion{Fe}{i} & 7680.27 & \ion{Si}{i} & 4600-7000 & 13167.773 & $-$21440.335 & 27243.5       & $-$15901.87  & 3287.8501    & ...          & $a+br+cr^{2}+dr^{3}+er^{4}$        & 3 \\
7748.27 & \ion{Fe}{i} & 7832.20 & \ion{Fe}{i} & 3500-6800 & 8443      & $-$5245.24   & 6015.251      & $-$4029.735  & ...          & ...          & $a+br+cr^{2}+dr^{3}$               & 3 \\
7788.95 & \ion{Ni}{i} & 7680.27 & \ion{Si}{i} & 3500-7000 & 7554.168  & $-$2001.119  & 252.7902      & ...          & ...          & ...          & $a+br+cr^{2}$                      & 3 \\
7788.95 & \ion{Ni}{i} & 7780.56 & \ion{Fe}{i} & 3500-7000 & 7889.0498 & $-$4728.7165 & 4151.1333     & $-$2183.3018 & ...          & ...          & $a+br+cr^{2}+dr^{3}$               & 3 \\
7788.95 & \ion{Ni}{i} & 7832.20 & \ion{Fe}{i} & 3500-7000 & 8477.5756 & $-$7634.1137 & 8209.1666     & $-$4105.5848 & ...          & ...          & $a+br+cr^{2}+dr^{3}$               & 3 \\
7788.95 & \ion{Ni}{i} & 7849.97 & \ion{Si}{i} & 5200-7000 & 9151.883  & $-$5400.7313 & 3647.6843     & $-$991.87973 & ...          & ...          & $a+br+cr^{2}+dr^{3}$               & 3 \\
7788.95 & \ion{Ni}{i} & 7932.35 & \ion{Si}{i} & 3500-7000 & 12176.034 & $-$1.8245293 & $-$0.73875711 & ...          & ...          & ...          & $a(r-b)^{c}$                       & 3 \\
7788.95 & \ion{Ni}{i} & 7944.00 & \ion{Si}{i} & 4200-7000 & 11254.539 & $-$30436.921 & 78629.353     & $-$103193.97 & 64744.83     & $-$15594.876 & $a+br+cr^{2}+dr^{3}+er^{4}+fr^{5}$ & 3 \\
7912.87 & \ion{Fe}{i} & 7680.27 & \ion{Si}{i} & 5200-6800 & 7033.8179 & $-$3523.7759 & 2431.6723     & $-$589.24672 & ...          & ...          & $a+br+cr^{2}+dr^{3}$               & 3 \\
7912.87 & \ion{Fe}{i} & 7710.37 & \ion{Fe}{i} & 5300-6800 & 7433.4687 & $-$3131.7646 & 1119.1743     & ...          & ...          & ...          & $a+br+cr^{2}$                      & 3 \\
8426.51 & \ion{Ti}{i} & 7680.27 & \ion{Si}{i} & 5300-6400 & 6963.6582 & $-$2717.0084 & 967.28678     & ...          & ...          & ...          & $a+br+cr^{2}$                      & 3 \\
\hline
\end{tabular}}
\tablefoot{From left to right, the first two columns give the wavelength of
the line pairs adopted for the LDR, while column three gives the range in
effective temperature in which the individual LDRs were calibrated. Columns
four to nine list the coefficients of the analytical relation adopted for
the calibration, while column ten gives the analytical formula. The last
column gives the reference for the calibration of the LDR.}
\tablebib{
\tablefoottext{1}{\citet[KG]{KovtyukhGorlova2000};}
\tablefoottext{2}{\citet[K07]{Kovtyukh2007};}
\tablefoottext{3}{K17: this investigation.}}
\end{table}
\end{landscape}

\begin{table*}
\centering
\caption{Atmospheric parameters, Fe abundances and radial velocities as a
function of the pulsation phase for calibrating CCs.}
\label{paramscontrolsample}
{\scriptsize 
\begin{tabular}{lcc r@{ }l cc r@{ }l c r@{ }l c r@{ }l r@{ }l}
\noalign{\smallskip}\hline\hline\noalign{\smallskip}
Name & Dataset &
\parbox[c]{0.5cm}{\centering MJD [d]} &
\multicolumn{2}{c}{\parbox[c]{0.8cm}{\centering \teff\ $\pm$ $\sigma$ [K]}} &
\logg\ &
\parbox[c]{0.9cm}{\centering \vmic\ [\kms]} &
\multicolumn{2}{c}{\ion{Fe}{i}  $\pm$ $\sigma$} & $N_{\rm{\ion{Fe}{i}}}$ &
\multicolumn{2}{c}{\ion{Fe}{ii} $\pm$ $\sigma$} & $N_{\rm{\ion{Fe}{ii}}}$ &
\multicolumn{2}{c}{[Fe/H] $\pm$ $\sigma$} &
\multicolumn{2}{c}{\parbox[c]{0.9cm}{\centering $RV$ $\pm$ $\sigma$ [\kms]}} \\
\noalign{\smallskip}\hline\noalign{\smallskip}
\object{V340\,Ara}    & FEROS  & 53620.0609687 & 5921 & $\pm$ 154 & 2.0 & 4.9 &    0.29 & $\pm$ 0.22 & 85  &    0.05 & $\pm$ 0.13 & 4  &    0.11 & $\pm$ 0.11 &  $-$76.9 & $\pm$ 5.3  \\
\object{V340\,Ara}    & UVES   & 54708.0671613 & 5181 & $\pm$ 99  & 0.9 & 4.3 &    0.27 & $\pm$ 0.20 & 86  &    0.24 & $\pm$ 0.26 & 7  &    0.26 & $\pm$ 0.16 & $-$95.69 & $\pm$ 0.03 \\
\object{V340\,Ara}    & UVES   & 54709.0803304 & 5095 & $\pm$ 99  & 0.7 & 4.8 &    0.21 & $\pm$ 0.20 & 82  &    0.20 & $\pm$ 0.26 & 6  &    0.21 & $\pm$ 0.16 & $-$91.37 & $\pm$ 0.01 \\
\object{V340\,Ara}    & UVES   & 56137.1372097 & 5172 & $\pm$ 99  & 0.8 & 4.8 &    0.17 & $\pm$ 0.15 & 93  &    0.18 & $\pm$ 0.30 & 6  &    0.17 & $\pm$ 0.14 &  $-$69.0 & $\pm$ 0.2  \\
\object{V340\,Ara}    & UVES   & 56138.0944950 & 5578 & $\pm$ 119 & 0.9 & 4.9 &    0.26 & $\pm$ 0.16 & 118 &    0.25 & $\pm$ 0.06 & 8  &    0.25 & $\pm$ 0.05 &  $-$73.2 & $\pm$ 0.1  \\
\object{V340\,Ara}    & UVES   & 56139.1860724 & 6083 & $\pm$ 133 & 1.6 & 4.9 &    0.36 & $\pm$ 0.16 & 104 &    0.33 & $\pm$ 0.13 & 12 &    0.34 & $\pm$ 0.10 & $-$94.56 & $\pm$ 0.06 \\
\object{V340\,Ara}    & UVES   & 56152.0523667 & 4742 & $\pm$ 108 & 1.1 & 4.8 &    0.11 & $\pm$ 0.18 & 65  & $-$0.18 & $\pm$ 0.35 & 4  &    0.05 & $\pm$ 0.16 &  $-$62.8 & $\pm$ 0.2  \\
\object{eta\,Aql}     & STELLA & 53935.9867868 & 5450 & $\pm$ 97  & 0.9 & 3.4 &    0.20 & $\pm$ 0.18 & 122 &    0.20 & $\pm$ 0.04 & 7  &    0.20 & $\pm$ 0.04 &  $-$15.4 & $\pm$ 0.2  \\
\object{eta\,Aql}     & STELLA & 53936.0890674 & 5469 & $\pm$ 99  & 0.7 & 2.6 &    0.36 & $\pm$ 0.14 & 102 &    0.37 & $\pm$ 0.26 & 14 &    0.36 & $\pm$ 0.12 &  $-$15.3 & $\pm$ 0.2  \\
\object{eta\,Aql}     & STELLA & 53944.0303995 & 5333 & $\pm$ 99  & 1.2 & 4.2 &    0.15 & $\pm$ 0.11 & 84  &    0.15 & $\pm$ 0.14 & 10 &    0.15 & $\pm$ 0.09 &   $-$9.3 & $\pm$ 0.2  \\
\object{eta\,Aql}     & STELLA & 53947.0833325 & 6116 & $\pm$ 192 & 0.7 & 1.9 &    0.37 & $\pm$ 0.20 & 98  &    0.37 & $\pm$ 0.36 & 11 &    0.37 & $\pm$ 0.18 &  $-$30.9 & $\pm$ 0.2  \\
\object{eta\,Aql}     & STELLA & 53949.0641156 & 5810 & $\pm$ 161 & 0.8 & 2.4 &    0.48 & $\pm$ 0.22 & 135 &    0.49 & $\pm$ 0.14 & 9  &    0.49 & $\pm$ 0.12 &  $-$21.6 & $\pm$ 0.2  \\
\object{eta\,Aql}     & STELLA & 53949.9577241 & 5617 & $\pm$ 153 & 1.1 & 3.0 &    0.30 & $\pm$ 0.18 & 121 &    0.31 & $\pm$ 0.15 & 15 &    0.30 & $\pm$ 0.12 &  $-$15.6 & $\pm$ 0.2  \\
\object{eta\,Aql}     & STELLA & 53950.9569429 & 5379 & $\pm$ 100 & 1.2 & 4.4 &    0.16 & $\pm$ 0.20 & 120 &    0.17 & $\pm$ 0.28 & 13 &    0.16 & $\pm$ 0.16 &  $-$12.2 & $\pm$ 0.2  \\
\object{eta\,Aql}     & STELLA & 53953.0029560 & 5659 & $\pm$ 248 & 2.0 & 6.0 & $-$0.03 & $\pm$ 0.26 & 76  & $-$0.03 & $\pm$ 0.17 & 8  & $-$0.03 & $\pm$ 0.14 &      5.4 & $\pm$ 0.2  \\
\object{eta\,Aql}     & STELLA & 53958.9309528 & 5276 & $\pm$ 138 & 0.7 & 2.4 &    0.36 & $\pm$ 0.13 & 82  &    0.37 & $\pm$ 0.11 & 10 &    0.36 & $\pm$ 0.08 &   $-$1.6 & $\pm$ 0.2  \\
\object{eta\,Aql}     & STELLA & 53959.0331372 & 5311 & $\pm$ 145 & 1.8 & 4.9 &    0.10 & $\pm$ 0.21 & 115 &    0.10 & $\pm$ 0.37 & 19 &    0.10 & $\pm$ 0.18 &   $-$0.1 & $\pm$ 0.2  \\
\object{eta\,Aql}     & STELLA & 53962.0178211 & 5985 & $\pm$ 238 & 1.1 & 2.5 &    0.39 & $\pm$ 0.33 & 138 &    0.36 & $\pm$ 0.38 & 15 &    0.37 & $\pm$ 0.25 &  $-$30.0 & $\pm$ 0.2  \\
\object{S\,Cru}       & HARPS  & 53150.1200407 & 6158 & $\pm$ 182 & 1.7 & 2.9 &    0.19 & $\pm$ 0.16 & 129 &    0.18 & $\pm$ 0.19 & 16 &    0.19 & $\pm$ 0.12 &   $-$8.8 & $\pm$ 1.0  \\
\object{S\,Cru}       & HARPS  & 53151.0565227 & 6222 & $\pm$ 61  & 1.7 & 3.0 &    0.12 & $\pm$ 0.12 & 133 &    0.10 & $\pm$ 0.12 & 18 &    0.11 & $\pm$ 0.08 &  $-$18.9 & $\pm$ 0.8  \\
\object{S\,Cru}       & HARPS  & 53152.1277898 & 5732 & $\pm$ 63  & 1.2 & 2.3 &    0.16 & $\pm$ 0.10 & 170 &    0.15 & $\pm$ 0.15 & 18 &    0.15 & $\pm$ 0.08 &   $-$7.7 & $\pm$ 0.7  \\
\object{S\,Cru}       & HARPS  & 53153.0633060 & 5556 & $\pm$ 100 & 1.4 & 3.0 &    0.01 & $\pm$ 0.12 & 153 & $-$0.03 & $\pm$ 0.10 & 15 & $-$0.01 & $\pm$ 0.08 &      3.0 & $\pm$ 0.8  \\
\object{S\,Cru}       & HARPS  & 53154.1283897 & 5612 & $\pm$ 99  & 1.7 & 4.5 & $-$0.03 & $\pm$ 0.17 & 130 & $-$0.04 & $\pm$ 0.16 & 14 & $-$0.03 & $\pm$ 0.12 &     12.4 & $\pm$ 1.2  \\
\object{S\,Cru}       & HARPS  & 53156.1212653 & 6013 & $\pm$ 50  & 1.7 & 2.8 &    0.12 & $\pm$ 0.12 & 152 &    0.14 & $\pm$ 0.11 & 15 &    0.14 & $\pm$ 0.08 &  $-$15.3 & $\pm$ 0.7  \\
\object{S\,Cru}       & HARPS  & 53201.9852976 & 6482 & $\pm$ 98  & 1.9 & 2.8 &    0.12 & $\pm$ 0.12 & 122 &    0.08 & $\pm$ 0.18 & 20 &    0.10 & $\pm$ 0.10 &  $-$19.5 & $\pm$ 1.0  \\
\object{S\,Cru}       & HARPS  & 53202.9803582 & 6058 & $\pm$ 53  & 1.5 & 2.5 &    0.18 & $\pm$ 0.11 & 146 &    0.17 & $\pm$ 0.12 & 18 &    0.17 & $\pm$ 0.08 &  $-$15.8 & $\pm$ 0.7  \\
\object{S\,Cru}       & HARPS  & 53202.9856839 & 6051 & $\pm$ 55  & 1.6 & 2.7 &    0.14 & $\pm$ 0.10 & 145 &    0.17 & $\pm$ 0.10 & 16 &    0.16 & $\pm$ 0.07 &  $-$15.7 & $\pm$ 0.7  \\
\object{S\,Cru}       & HARPS  & 53204.9630967 & 5517 & $\pm$ 97  & 1.4 & 3.6 & $-$0.03 & $\pm$ 0.12 & 140 & $-$0.01 & $\pm$ 0.17 & 20 & $-$0.03 & $\pm$ 0.10 &      5.9 & $\pm$ 0.9  \\
\object{S\,Cru}       & HARPS  & 53205.9747885 & 5784 & $\pm$ 106 & 1.9 & 4.2 &    0.00 & $\pm$ 0.12 & 120 & $-$0.02 & $\pm$ 0.13 & 15 & $-$0.01 & $\pm$ 0.09 &     11.5 & $\pm$ 1.2  \\
\object{S\,Cru}       & HARPS  & 53206.9497352 & 6418 & $\pm$ 98  & 1.7 & 2.7 &    0.09 & $\pm$ 0.11 & 109 &    0.06 & $\pm$ 0.21 & 21 &    0.08 & $\pm$ 0.10 &  $-$21.1 & $\pm$ 0.9  \\
\object{S\,Cru}       & FEROS  & 55284.3152681 & 6472 & $\pm$ 93  & 1.9 & 3.1 & $-$0.05 & $\pm$ 0.19 & 94  & $-$0.10 & $\pm$ 0.28 & 12 & $-$0.07 & $\pm$ 0.16 &  $-$20.5 & $\pm$ 1.0  \\
\object{beta\,Dor}    & HARPS  & 53015.1219391 & 5248 & $\pm$ 93  & 1.5 & 3.9 & $-$0.05 & $\pm$ 0.14 & 165 & $-$0.14 & $\pm$ 0.11 & 13 & $-$0.11 & $\pm$ 0.09 &     14.0 & $\pm$ 0.9  \\
\object{beta\,Dor}    & HARPS  & 53015.1239308 & 5236 & $\pm$ 91  & 1.4 & 3.6 & $-$0.05 & $\pm$ 0.12 & 158 & $-$0.10 & $\pm$ 0.15 & 15 & $-$0.07 & $\pm$ 0.09 &     14.1 & $\pm$ 0.9  \\
\object{beta\,Dor}    & HARPS  & 53015.1259288 & 5249 & $\pm$ 90  & 1.5 & 3.9 & $-$0.05 & $\pm$ 0.13 & 159 & $-$0.12 & $\pm$ 0.14 & 16 & $-$0.08 & $\pm$ 0.09 &     14.1 & $\pm$ 0.9  \\
\object{beta\,Dor}    & HARPS  & 53016.1700025 & 5195 & $\pm$ 98  & 1.4 & 4.9 & $-$0.10 & $\pm$ 0.15 & 138 & $-$0.07 & $\pm$ 0.30 & 18 & $-$0.10 & $\pm$ 0.13 &     22.8 & $\pm$ 1.2  \\
\object{beta\,Dor}    & HARPS  & 53016.1722763 & 5186 & $\pm$ 98  & 1.4 & 4.9 & $-$0.11 & $\pm$ 0.15 & 137 & $-$0.15 & $\pm$ 0.20 & 14 & $-$0.12 & $\pm$ 0.12 &     22.8 & $\pm$ 1.2  \\
\object{beta\,Dor}    & HARPS  & 53016.1745592 & 5203 & $\pm$ 95  & 1.3 & 4.9 & $-$0.13 & $\pm$ 0.15 & 137 & $-$0.08 & $\pm$ 0.27 & 18 & $-$0.12 & $\pm$ 0.13 &     22.8 & $\pm$ 1.2  \\
\object{beta\,Dor}    & HARPS  & 53017.1820488 & 5398 & $\pm$ 160 & 1.1 & 3.7 & $-$0.08 & $\pm$ 0.12 & 108 & $-$0.03 & $\pm$ 0.11 & 12 & $-$0.05 & $\pm$ 0.08 &     26.2 & $\pm$ 1.4  \\
\object{beta\,Dor}    & HARPS  & 53017.1844847 & 5392 & $\pm$ 113 & 1.1 & 4.2 & $-$0.11 & $\pm$ 0.12 & 105 & $-$0.06 & $\pm$ 0.11 & 10 & $-$0.08 & $\pm$ 0.08 &     26.2 & $\pm$ 1.4  \\
\object{beta\,Dor}    & HARPS  & 53017.1869331 & 5388 & $\pm$ 139 & 1.1 & 4.9 & $-$0.11 & $\pm$ 0.13 & 106 & $-$0.12 & $\pm$ 0.21 & 12 & $-$0.11 & $\pm$ 0.11 &     26.2 & $\pm$ 1.4  \\
\object{beta\,Dor}    & HARPS  & 53021.1802715 & 5992 & $\pm$ 77  & 1.1 & 2.8 & $-$0.01 & $\pm$ 0.10 & 149 & $-$0.03 & $\pm$ 0.19 & 17 & $-$0.01 & $\pm$ 0.09 &      1.1 & $\pm$ 0.7  \\
\object{beta\,Dor}    & HARPS  & 53021.1819217 & 5994 & $\pm$ 90  & 1.3 & 3.3 &    0.01 & $\pm$ 0.12 & 152 & $-$0.04 & $\pm$ 0.17 & 19 & $-$0.01 & $\pm$ 0.10 &      1.2 & $\pm$ 0.7  \\
\object{beta\,Dor}    & HARPS  & 53021.1835549 & 6006 & $\pm$ 75  & 1.2 & 3.0 &    0.03 & $\pm$ 0.11 & 151 & $-$0.03 & $\pm$ 0.16 & 15 &    0.01 & $\pm$ 0.09 &      1.2 & $\pm$ 0.7  \\
\object{beta\,Dor}    & HARPS  & 53021.1852025 & 6001 & $\pm$ 85  & 1.2 & 2.8 & $-$0.01 & $\pm$ 0.10 & 143 & $-$0.03 & $\pm$ 0.17 & 16 & $-$0.01 & $\pm$ 0.09 &      1.1 & $\pm$ 0.7  \\
\object{beta\,Dor}    & HARPS  & 53023.1329203 & 5563 & $\pm$ 73  & 1.5 & 3.7 & $-$0.03 & $\pm$ 0.13 & 164 & $-$0.07 & $\pm$ 0.15 & 17 & $-$0.05 & $\pm$ 0.10 &   $-$1.9 & $\pm$ 0.7  \\
\object{beta\,Dor}    & HARPS  & 53023.1345710 & 5556 & $\pm$ 71  & 1.3 & 3.2 & $-$0.00 & $\pm$ 0.12 & 162 &    0.01 & $\pm$ 0.18 & 21 &    0.00 & $\pm$ 0.10 &   $-$1.9 & $\pm$ 0.7  \\
\object{beta\,Dor}    & HARPS  & 53023.1362075 & 5559 & $\pm$ 82  & 1.5 & 3.7 & $-$0.04 & $\pm$ 0.13 & 160 & $-$0.01 & $\pm$ 0.19 & 20 & $-$0.03 & $\pm$ 0.11 &   $-$1.9 & $\pm$ 0.7  \\
\object{beta\,Dor}    & HARPS  & 53023.1378355 & 5560 & $\pm$ 69  & 1.4 & 3.4 & $-$0.03 & $\pm$ 0.12 & 168 & $-$0.01 & $\pm$ 0.17 & 20 & $-$0.02 & $\pm$ 0.10 &   $-$1.9 & $\pm$ 0.7  \\
\object{beta\,Dor}    & HARPS  & 53025.1738720 & 5229 & $\pm$ 93  & 1.3 & 3.6 & $-$0.04 & $\pm$ 0.13 & 158 & $-$0.01 & $\pm$ 0.25 & 20 & $-$0.03 & $\pm$ 0.11 &     15.9 & $\pm$ 1.0  \\
\object{beta\,Dor}    & HARPS  & 53025.1759601 & 5229 & $\pm$ 92  & 1.2 & 3.6 & $-$0.03 & $\pm$ 0.13 & 156 & $-$0.04 & $\pm$ 0.25 & 21 & $-$0.03 & $\pm$ 0.12 &     15.9 & $\pm$ 1.0  \\
\object{beta\,Dor}    & HARPS  & 53025.1780493 & 5227 & $\pm$ 90  & 1.2 & 3.5 & $-$0.03 & $\pm$ 0.14 & 160 & $-$0.02 & $\pm$ 0.26 & 21 & $-$0.03 & $\pm$ 0.12 &     16.0 & $\pm$ 1.0  \\
\object{beta\,Dor}    & HARPS  & 53026.0920026 & 5207 & $\pm$ 97  & 1.5 & 4.9 & $-$0.12 & $\pm$ 0.14 & 135 & $-$0.09 & $\pm$ 0.29 & 19 & $-$0.12 & $\pm$ 0.13 &     23.3 & $\pm$ 1.2  \\
\object{beta\,Dor}    & HARPS  & 53026.0942576 & 5203 & $\pm$ 99  & 1.4 & 4.9 & $-$0.13 & $\pm$ 0.15 & 135 & $-$0.07 & $\pm$ 0.23 & 17 & $-$0.11 & $\pm$ 0.13 &     23.3 & $\pm$ 1.2  \\
\object{beta\,Dor}    & HARPS  & 53028.1694909 & 5692 & $\pm$ 110 & 1.5 & 3.9 & $-$0.05 & $\pm$ 0.10 & 129 & $-$0.03 & $\pm$ 0.13 & 10 & $-$0.04 & $\pm$ 0.08 &     11.0 & $\pm$ 0.9  \\
\object{beta\,Dor}    & HARPS  & 53028.1718352 & 5684 & $\pm$ 129 & 1.5 & 3.8 & $-$0.04 & $\pm$ 0.09 & 126 & $-$0.06 & $\pm$ 0.17 & 13 & $-$0.04 & $\pm$ 0.08 &     10.9 & $\pm$ 0.9  \\
\object{beta\,Dor}    & HARPS  & 53028.1741878 & 5689 & $\pm$ 99  & 1.7 & 4.9 & $-$0.10 & $\pm$ 0.11 & 128 & $-$0.06 & $\pm$ 0.17 & 13 & $-$0.09 & $\pm$ 0.09 &     10.9 & $\pm$ 0.9  \\
\object{beta\,Dor}    & HARPS  & 53029.1306027 & 5814 & $\pm$ 90  & 1.4 & 3.3 & $-$0.01 & $\pm$ 0.10 & 148 &    0.01 & $\pm$ 0.11 & 10 & $-$0.00 & $\pm$ 0.07 &      1.8 & $\pm$ 0.8  \\
\object{beta\,Dor}    & HARPS  & 53029.1325697 & 5809 & $\pm$ 83  & 1.3 & 3.4 & $-$0.02 & $\pm$ 0.11 & 151 &    0.01 & $\pm$ 0.08 & 11 &    0.00 & $\pm$ 0.06 &      1.8 & $\pm$ 0.8  \\
\object{beta\,Dor}    & HARPS  & 53029.1339932 & 5810 & $\pm$ 84  & 1.3 & 3.3 & $-$0.01 & $\pm$ 0.11 & 149 &    0.04 & $\pm$ 0.06 & 10 &    0.03 & $\pm$ 0.05 &      1.7 & $\pm$ 0.8  \\
\object{beta\,Dor}    & HARPS  & 53029.1355791 & 5793 & $\pm$ 86  & 1.4 & 3.5 & $-$0.03 & $\pm$ 0.11 & 154 &    0.00 & $\pm$ 0.11 & 13 & $-$0.02 & $\pm$ 0.08 &      1.7 & $\pm$ 0.8  \\
\object{beta\,Dor}    & HARPS  & 53031.1330521 & 5971 & $\pm$ 89  & 1.3 & 3.0 &    0.01 & $\pm$ 0.11 & 153 &    0.00 & $\pm$ 0.15 & 13 &    0.00 & $\pm$ 0.09 &      1.0 & $\pm$ 0.7  \\
\object{beta\,Dor}    & HARPS  & 53031.1346849 & 5981 & $\pm$ 80  & 1.3 & 3.0 &    0.00 & $\pm$ 0.11 & 152 &    0.04 & $\pm$ 0.07 & 11 &    0.03 & $\pm$ 0.06 &      1.0 & $\pm$ 0.7  \\
\object{beta\,Dor}    & HARPS  & 53031.1363173 & 5991 & $\pm$ 85  & 1.2 & 2.9 &    0.06 & $\pm$ 0.10 & 146 &    0.07 & $\pm$ 0.08 & 10 &    0.07 & $\pm$ 0.06 &      1.0 & $\pm$ 0.7  \\
\object{beta\,Dor}    & HARPS  & 53032.1781164 & 5730 & $\pm$ 73  & 1.6 & 3.7 & $-$0.03 & $\pm$ 0.12 & 150 & $-$0.01 & $\pm$ 0.16 & 18 & $-$0.02 & $\pm$ 0.10 &   $-$5.5 & $\pm$ 0.8  \\
\object{beta\,Dor}    & HARPS  & 53032.1796472 & 5714 & $\pm$ 81  & 1.3 & 3.4 & $-$0.04 & $\pm$ 0.12 & 152 & $-$0.03 & $\pm$ 0.11 & 16 & $-$0.03 & $\pm$ 0.08 &   $-$5.5 & $\pm$ 0.8  \\
\object{beta\,Dor}    & HARPS  & 53032.1811734 & 5732 & $\pm$ 75  & 1.6 & 3.8 & $-$0.03 & $\pm$ 0.12 & 154 & $-$0.02 & $\pm$ 0.16 & 17 & $-$0.02 & $\pm$ 0.10 &   $-$5.5 & $\pm$ 0.8  \\
\object{beta\,Dor}    & HARPS  & 53033.1060160 & 5526 & $\pm$ 74  & 1.4 & 3.2 &    0.01 & $\pm$ 0.11 & 159 &    0.05 & $\pm$ 0.18 & 21 &    0.02 & $\pm$ 0.09 &   $-$1.2 & $\pm$ 0.7  \\
\object{beta\,Dor}    & HARPS  & 53033.1077288 & 5531 & $\pm$ 76  & 1.3 & 3.2 &    0.02 & $\pm$ 0.11 & 165 &    0.01 & $\pm$ 0.18 & 21 &    0.02 & $\pm$ 0.10 &   $-$1.2 & $\pm$ 0.7  \\
\object{beta\,Dor}    & HARPS  & 53033.1094348 & 5525 & $\pm$ 78  & 1.3 & 3.3 & $-$0.03 & $\pm$ 0.12 & 159 &    0.00 & $\pm$ 0.18 & 21 & $-$0.02 & $\pm$ 0.10 &   $-$1.2 & $\pm$ 0.7  \\
\object{beta\,Dor}    & HARPS  & 53034.1394979 & 5340 & $\pm$ 81  & 1.3 & 3.4 & $-$0.01 & $\pm$ 0.13 & 171 & $-$0.10 & $\pm$ 0.14 & 15 & $-$0.05 & $\pm$ 0.09 &      7.8 & $\pm$ 0.9  \\
\object{beta\,Dor}    & HARPS  & 53034.1412812 & 5331 & $\pm$ 74  & 1.3 & 3.3 & $-$0.03 & $\pm$ 0.12 & 164 &    0.01 & $\pm$ 0.22 & 21 & $-$0.02 & $\pm$ 0.10 &      7.8 & $\pm$ 0.9  \\
\object{beta\,Dor}    & HARPS  & 53035.1331423 & 5227 & $\pm$ 92  & 1.5 & 3.8 & $-$0.03 & $\pm$ 0.15 & 167 & $-$0.09 & $\pm$ 0.17 & 15 & $-$0.06 & $\pm$ 0.11 &     17.0 & $\pm$ 1.0  \\
\object{beta\,Dor}    & HARPS  & 53035.1352292 & 5219 & $\pm$ 86  & 1.4 & 4.1 & $-$0.09 & $\pm$ 0.14 & 158 & $-$0.17 & $\pm$ 0.15 & 15 & $-$0.13 & $\pm$ 0.10 &     17.0 & $\pm$ 1.0  \\
\object{beta\,Dor}    & HARPS  & 53036.1369181 & 5210 & $\pm$ 91  & 1.0 & 4.0 & $-$0.12 & $\pm$ 0.14 & 130 & $-$0.08 & $\pm$ 0.21 & 16 & $-$0.11 & $\pm$ 0.12 &     24.5 & $\pm$ 1.2  \\
\object{beta\,Dor}    & HARPS  & 53036.1391838 & 5213 & $\pm$ 96  & 1.3 & 4.9 & $-$0.15 & $\pm$ 0.15 & 132 & $-$0.10 & $\pm$ 0.27 & 15 & $-$0.14 & $\pm$ 0.13 &     24.5 & $\pm$ 1.2  \\
\object{beta\,Dor}    & HARPS  & 53037.1416118 & 5521 & $\pm$ 141 & 1.4 & 4.9 & $-$0.09 & $\pm$ 0.12 & 103 & $-$0.07 & $\pm$ 0.17 & 10 & $-$0.08 & $\pm$ 0.10 &     24.7 & $\pm$ 1.4  \\
\object{beta\,Dor}    & HARPS  & 53037.1440373 & 5494 & $\pm$ 140 & 1.4 & 4.6 & $-$0.09 & $\pm$ 0.12 & 93  & $-$0.08 & $\pm$ 0.15 & 11 & $-$0.09 & $\pm$ 0.10 &     24.7 & $\pm$ 1.4  \\
\object{zeta\,Gem}    & HARPS  & 53015.2071101 & 5271 & $\pm$ 71  & 1.4 & 3.8 &    0.06 & $\pm$ 0.14 & 165 &    0.12 & $\pm$ 0.29 & 20 &    0.07 & $\pm$ 0.13 &     10.4 & $\pm$ 0.9  \\
\object{zeta\,Gem}    & HARPS  & 53015.2093019 & 5253 & $\pm$ 74  & 1.6 & 4.0 & $-$0.00 & $\pm$ 0.13 & 158 & $-$0.10 & $\pm$ 0.12 & 13 & $-$0.05 & $\pm$ 0.09 &     10.4 & $\pm$ 0.9  \\
\object{zeta\,Gem}    & HARPS  & 53015.2115010 & 5254 & $\pm$ 86  & 1.5 & 3.5 &    0.05 & $\pm$ 0.13 & 165 &    0.03 & $\pm$ 0.16 & 14 &    0.04 & $\pm$ 0.10 &     10.4 & $\pm$ 0.9  \\
\hline\noalign{\smallskip}
\multicolumn{17}{r}{\it {\footnotesize continued on next page}} \\
\end{tabular}}
\end{table*}
\addtocounter{table}{-1}
\begin{table*}[p]
\centering
\caption[]{continued.}
{\scriptsize 
\begin{tabular}{lcc r@{ }l cc r@{ }l c r@{ }l c r@{ }l r@{ }l}
\noalign{\smallskip}\hline\hline\noalign{\smallskip}
Name & Dataset &
\parbox[c]{0.5cm}{\centering MJD [d]} &
\multicolumn{2}{c}{\parbox[c]{0.8cm}{\centering \teff\ $\pm$ $\sigma$ [K]}} &
\logg\ &
\parbox[c]{0.9cm}{\centering \vmic\ [\kms]} &
\multicolumn{2}{c}{\ion{Fe}{i}  $\pm$ $\sigma$} & $N_{\rm{\ion{Fe}{i}}}$ &
\multicolumn{2}{c}{\ion{Fe}{ii} $\pm$ $\sigma$} & $N_{\rm{\ion{Fe}{ii}}}$ &
\multicolumn{2}{c}{[Fe/H] $\pm$ $\sigma$} &
\multicolumn{2}{c}{\parbox[c]{0.9cm}{\centering $RV$ $\pm$ $\sigma$ [\kms]}} \\
\noalign{\smallskip}\hline\noalign{\smallskip}
\object{zeta\,Gem}    & HARPS  & 53016.1805731 & 5216 & $\pm$ 100 & 1.4 & 4.9 & $-$0.02 & $\pm$ 0.15 & 144 & $-$0.01 & $\pm$ 0.29 & 15 & $-$0.02 & $\pm$ 0.13 &     17.0 & $\pm$ 1.0  \\
\object{zeta\,Gem}    & HARPS  & 53016.1829469 & 5216 & $\pm$ 97  & 1.2 & 4.4 & $-$0.02 & $\pm$ 0.14 & 135 & $-$0.02 & $\pm$ 0.28 & 17 & $-$0.02 & $\pm$ 0.13 &     17.1 & $\pm$ 1.0  \\
\object{zeta\,Gem}    & HARPS  & 53016.1853382 & 5213 & $\pm$ 71  & 1.6 & 4.9 & $-$0.04 & $\pm$ 0.14 & 143 &    0.06 & $\pm$ 0.30 & 17 & $-$0.02 & $\pm$ 0.13 &     17.1 & $\pm$ 1.0  \\
\object{zeta\,Gem}    & HARPS  & 53017.1932375 & 5322 & $\pm$ 96  & 1.2 & 4.9 & $-$0.04 & $\pm$ 0.14 & 121 & $-$0.04 & $\pm$ 0.24 & 12 & $-$0.04 & $\pm$ 0.12 &     20.2 & $\pm$ 1.2  \\
\object{zeta\,Gem}    & HARPS  & 53017.1957591 & 5305 & $\pm$ 92  & 1.1 & 4.9 & $-$0.04 & $\pm$ 0.13 & 112 & $-$0.05 & $\pm$ 0.21 & 13 & $-$0.05 & $\pm$ 0.11 &     20.2 & $\pm$ 1.2  \\
\object{zeta\,Gem}    & HARPS  & 53017.1982693 & 5337 & $\pm$ 100 & 1.3 & 4.9 &    0.01 & $\pm$ 0.14 & 127 & $-$0.01 & $\pm$ 0.22 & 13 &    0.00 & $\pm$ 0.12 &     20.2 & $\pm$ 1.2  \\
\object{zeta\,Gem}    & HARPS  & 53021.1893175 & 5818 & $\pm$ 64  & 1.1 & 3.0 &    0.10 & $\pm$ 0.11 & 162 &    0.10 & $\pm$ 0.13 & 12 &    0.10 & $\pm$ 0.08 &   $-$0.1 & $\pm$ 0.7  \\
\object{zeta\,Gem}    & HARPS  & 53021.1915396 & 5837 & $\pm$ 60  & 1.2 & 3.1 &    0.15 & $\pm$ 0.11 & 164 &    0.14 & $\pm$ 0.12 & 11 &    0.14 & $\pm$ 0.08 &   $-$0.1 & $\pm$ 0.7  \\
\object{zeta\,Gem}    & HARPS  & 53021.1937450 & 5812 & $\pm$ 80  & 1.1 & 3.0 &    0.13 & $\pm$ 0.12 & 162 &    0.11 & $\pm$ 0.15 & 13 &    0.12 & $\pm$ 0.09 &   $-$0.1 & $\pm$ 0.7  \\
\object{zeta\,Gem}    & HARPS  & 53021.1959571 & 5829 & $\pm$ 61  & 1.3 & 2.8 &    0.14 & $\pm$ 0.11 & 158 &    0.12 & $\pm$ 0.21 & 14 &    0.13 & $\pm$ 0.10 &   $-$0.1 & $\pm$ 0.7  \\
\object{zeta\,Gem}    & HARPS  & 53021.1981818 & 5819 & $\pm$ 59  & 1.2 & 3.0 &    0.14 & $\pm$ 0.12 & 166 &    0.11 & $\pm$ 0.17 & 13 &    0.13 & $\pm$ 0.10 &   $-$0.1 & $\pm$ 0.7  \\
\object{zeta\,Gem}    & HARPS  & 53023.1432110 & 5597 & $\pm$ 86  & 1.4 & 3.3 &    0.11 & $\pm$ 0.12 & 169 &    0.14 & $\pm$ 0.18 & 18 &    0.12 & $\pm$ 0.10 &   $-$4.7 & $\pm$ 0.7  \\
\object{zeta\,Gem}    & HARPS  & 53023.1451185 & 5585 & $\pm$ 93  & 1.5 & 3.7 &    0.05 & $\pm$ 0.13 & 161 &    0.06 & $\pm$ 0.18 & 17 &    0.05 & $\pm$ 0.11 &   $-$4.7 & $\pm$ 0.7  \\
\object{zeta\,Gem}    & HARPS  & 53023.1470277 & 5586 & $\pm$ 74  & 1.4 & 3.5 &    0.06 & $\pm$ 0.13 & 163 &    0.07 & $\pm$ 0.19 & 17 &    0.06 & $\pm$ 0.11 &   $-$4.7 & $\pm$ 0.7  \\
\object{zeta\,Gem}    & HARPS  & 53023.1489287 & 5585 & $\pm$ 63  & 1.5 & 3.6 &    0.05 & $\pm$ 0.13 & 168 &    0.08 & $\pm$ 0.19 & 18 &    0.06 & $\pm$ 0.11 &   $-$4.6 & $\pm$ 0.7  \\
\object{zeta\,Gem}    & HARPS  & 53025.1826223 & 5270 & $\pm$ 74  & 1.1 & 3.2 &    0.07 & $\pm$ 0.12 & 163 & $-$0.01 & $\pm$ 0.18 & 16 &    0.05 & $\pm$ 0.10 &      9.0 & $\pm$ 0.8  \\
\object{zeta\,Gem}    & HARPS  & 53025.1846603 & 5272 & $\pm$ 74  & 1.3 & 3.4 &    0.04 & $\pm$ 0.11 & 153 &    0.02 & $\pm$ 0.20 & 17 &    0.03 & $\pm$ 0.09 &      9.1 & $\pm$ 0.8  \\
\object{zeta\,Gem}    & HARPS  & 53025.1867120 & 5276 & $\pm$ 74  & 1.4 & 3.6 &    0.09 & $\pm$ 0.12 & 156 &    0.00 & $\pm$ 0.11 & 14 &    0.04 & $\pm$ 0.08 &      9.1 & $\pm$ 0.8  \\
\object{zeta\,Gem}    & HARPS  & 53026.0991063 & 5210 & $\pm$ 83  & 1.3 & 3.6 &    0.03 & $\pm$ 0.15 & 149 &    0.02 & $\pm$ 0.27 & 19 &    0.03 & $\pm$ 0.13 &     15.7 & $\pm$ 1.0  \\
\object{zeta\,Gem}    & HARPS  & 53026.1014177 & 5200 & $\pm$ 80  & 1.7 & 4.9 & $-$0.04 & $\pm$ 0.15 & 141 & $-$0.06 & $\pm$ 0.15 & 14 & $-$0.05 & $\pm$ 0.10 &     15.7 & $\pm$ 1.0  \\
\object{zeta\,Gem}    & HARPS  & 53026.1038068 & 5209 & $\pm$ 74  & 1.6 & 4.9 & $-$0.03 & $\pm$ 0.15 & 148 & $-$0.08 & $\pm$ 0.14 & 13 & $-$0.06 & $\pm$ 0.10 &     15.7 & $\pm$ 1.0  \\
\object{zeta\,Gem}    & HARPS  & 53028.1798104 & 5503 & $\pm$ 89  & 1.3 & 4.9 & $-$0.00 & $\pm$ 0.15 & 132 & $-$0.03 & $\pm$ 0.18 & 14 & $-$0.01 & $\pm$ 0.12 &     15.7 & $\pm$ 1.1  \\
\object{zeta\,Gem}    & HARPS  & 53028.1825204 & 5505 & $\pm$ 99  & 1.5 & 4.9 & $-$0.02 & $\pm$ 0.13 & 124 & $-$0.09 & $\pm$ 0.12 & 10 & $-$0.06 & $\pm$ 0.09 &     15.7 & $\pm$ 1.1  \\
\object{zeta\,Gem}    & HARPS  & 53029.1395744 & 5617 & $\pm$ 67  & 1.0 & 3.3 &    0.12 & $\pm$ 0.12 & 143 &    0.10 & $\pm$ 0.12 & 12 &    0.11 & $\pm$ 0.09 &      5.8 & $\pm$ 0.8  \\
\object{zeta\,Gem}    & HARPS  & 53029.1413379 & 5637 & $\pm$ 76  & 1.4 & 4.1 &    0.11 & $\pm$ 0.14 & 155 &    0.09 & $\pm$ 0.18 & 15 &    0.10 & $\pm$ 0.11 &      5.8 & $\pm$ 0.8  \\
\object{zeta\,Gem}    & HARPS  & 53031.1410447 & 5813 & $\pm$ 62  & 1.2 & 3.1 &    0.12 & $\pm$ 0.12 & 168 &    0.10 & $\pm$ 0.15 & 13 &    0.11 & $\pm$ 0.09 &      0.1 & $\pm$ 0.7  \\
\object{zeta\,Gem}    & HARPS  & 53031.1432571 & 5816 & $\pm$ 93  & 1.3 & 3.1 &    0.12 & $\pm$ 0.12 & 164 &    0.16 & $\pm$ 0.13 & 12 &    0.14 & $\pm$ 0.09 &      0.1 & $\pm$ 0.7  \\
\object{zeta\,Gem}    & HARPS  & 53031.1454644 & 5814 & $\pm$ 61  & 1.2 & 3.0 &    0.14 & $\pm$ 0.12 & 165 &    0.15 & $\pm$ 0.12 & 10 &    0.14 & $\pm$ 0.08 &      0.1 & $\pm$ 0.7  \\
\object{zeta\,Gem}    & HARPS  & 53032.2008798 & 5755 & $\pm$ 99  & 1.2 & 3.1 &    0.15 & $\pm$ 0.12 & 157 &    0.15 & $\pm$ 0.07 & 11 &    0.15 & $\pm$ 0.06 &   $-$3.6 & $\pm$ 0.7  \\
\object{zeta\,Gem}    & HARPS  & 53032.2028918 & 5767 & $\pm$ 62  & 1.4 & 3.0 &    0.13 & $\pm$ 0.12 & 163 &    0.15 & $\pm$ 0.15 & 12 &    0.14 & $\pm$ 0.09 &   $-$3.7 & $\pm$ 0.7  \\
\object{zeta\,Gem}    & HARPS  & 53032.2049165 & 5770 & $\pm$ 67  & 1.2 & 3.0 &    0.13 & $\pm$ 0.12 & 166 &    0.13 & $\pm$ 0.12 & 11 &    0.13 & $\pm$ 0.09 &   $-$3.7 & $\pm$ 0.7  \\
\object{zeta\,Gem}    & HARPS  & 53033.1145522 & 5621 & $\pm$ 65  & 1.3 & 3.2 &    0.06 & $\pm$ 0.12 & 158 &    0.07 & $\pm$ 0.18 & 18 &    0.06 & $\pm$ 0.10 &   $-$5.2 & $\pm$ 0.7  \\
\object{zeta\,Gem}    & HARPS  & 53033.1164296 & 5626 & $\pm$ 72  & 1.4 & 3.4 &    0.12 & $\pm$ 0.12 & 158 &    0.10 & $\pm$ 0.19 & 16 &    0.11 & $\pm$ 0.10 &   $-$5.2 & $\pm$ 0.7  \\
\object{zeta\,Gem}    & HARPS  & 53033.1183039 & 5613 & $\pm$ 77  & 1.5 & 3.7 & $-$0.01 & $\pm$ 0.13 & 164 & $-$0.01 & $\pm$ 0.18 & 16 & $-$0.01 & $\pm$ 0.10 &   $-$5.2 & $\pm$ 0.7  \\
\object{zeta\,Gem}    & HARPS  & 53034.1506477 & 5422 & $\pm$ 68  & 1.4 & 3.3 &    0.04 & $\pm$ 0.13 & 178 & $-$0.00 & $\pm$ 0.09 & 14 &    0.01 & $\pm$ 0.08 &      0.2 & $\pm$ 0.7  \\
\object{zeta\,Gem}    & HARPS  & 53034.1525448 & 5429 & $\pm$ 74  & 1.5 & 3.5 &    0.09 & $\pm$ 0.13 & 176 &    0.01 & $\pm$ 0.10 & 13 &    0.04 & $\pm$ 0.08 &      0.3 & $\pm$ 0.7  \\
\object{zeta\,Gem}    & HARPS  & 53034.1544641 & 5428 & $\pm$ 71  & 1.2 & 3.0 &    0.13 & $\pm$ 0.13 & 184 &    0.03 & $\pm$ 0.11 & 14 &    0.07 & $\pm$ 0.09 &      0.3 & $\pm$ 0.7  \\
\object{zeta\,Gem}    & HARPS  & 53035.1428640 & 5288 & $\pm$ 73  & 1.4 & 3.6 &    0.05 & $\pm$ 0.11 & 168 & $-$0.01 & $\pm$ 0.15 & 16 &    0.03 & $\pm$ 0.09 &      7.7 & $\pm$ 0.8  \\
\object{zeta\,Gem}    & HARPS  & 53035.1443945 & 5289 & $\pm$ 74  & 1.3 & 3.2 &    0.09 & $\pm$ 0.12 & 172 &    0.05 & $\pm$ 0.14 & 16 &    0.07 & $\pm$ 0.09 &      7.7 & $\pm$ 0.8  \\
\object{zeta\,Gem}    & HARPS  & 53035.1459297 & 5288 & $\pm$ 73  & 1.2 & 3.1 &    0.10 & $\pm$ 0.12 & 173 &    0.05 & $\pm$ 0.15 & 16 &    0.08 & $\pm$ 0.09 &      7.8 & $\pm$ 0.8  \\
\object{zeta\,Gem}    & HARPS  & 53036.1547434 & 5222 & $\pm$ 95  & 1.4 & 3.6 &    0.01 & $\pm$ 0.14 & 152 & $-$0.04 & $\pm$ 0.14 & 14 & $-$0.02 & $\pm$ 0.10 &     15.1 & $\pm$ 1.0  \\
\object{zeta\,Gem}    & HARPS  & 53036.1569325 & 5216 & $\pm$ 86  & 1.1 & 3.3 &    0.06 & $\pm$ 0.13 & 149 & $-$0.03 & $\pm$ 0.17 & 15 &    0.03 & $\pm$ 0.10 &     15.1 & $\pm$ 1.0  \\
\object{zeta\,Gem}    & HARPS  & 53037.1617356 & 5264 & $\pm$ 95  & 1.3 & 4.9 & $-$0.02 & $\pm$ 0.16 & 128 &    0.01 & $\pm$ 0.28 & 11 & $-$0.01 & $\pm$ 0.14 &     20.0 & $\pm$ 1.2  \\
\object{zeta\,Gem}    & HARPS  & 53037.1642405 & 5247 & $\pm$ 79  & 1.3 & 4.9 & $-$0.03 & $\pm$ 0.15 & 132 &    0.01 & $\pm$ 0.24 & 15 & $-$0.02 & $\pm$ 0.13 &     20.0 & $\pm$ 1.2  \\
\object{zeta\,Gem}    & STELLA & 54348.2456000 & 5560 & $\pm$ 114 & 0.9 & 2.4 &    0.32 & $\pm$ 0.16 & 120 &    0.32 & $\pm$ 0.24 & 13 &    0.32 & $\pm$ 0.13 &      8.7 & $\pm$ 0.2  \\
\object{zeta\,Gem}    & STELLA & 54368.2539931 & 5529 & $\pm$ 98  & 0.9 & 3.0 &    0.20 & $\pm$ 0.17 & 114 &    0.23 & $\pm$ 0.11 & 10 &    0.22 & $\pm$ 0.09 &     11.9 & $\pm$ 0.2  \\
\object{zeta\,Gem}    & STELLA & 54373.2568361 & 5532 & $\pm$ 99  & 1.0 & 2.4 &    0.26 & $\pm$ 0.12 & 128 &    0.36 & $\pm$ 0.15 & 13 &    0.30 & $\pm$ 0.09 &   $-$2.6 & $\pm$ 0.2  \\
\object{zeta\,Gem}    & STELLA & 54417.2497900 & 5245 & $\pm$ 90  & 0.8 & 4.4 &    0.11 & $\pm$ 0.09 & 78  &    0.12 & $\pm$ 0.27 & 10 &    0.11 & $\pm$ 0.08 &     20.7 & $\pm$ 0.2  \\
\object{zeta\,Gem}    & STELLA & 54418.1548977 & 5427 & $\pm$ 100 & 1.9 & 6.4 &    0.03 & $\pm$ 0.11 & 91  &    0.03 & $\pm$ 0.11 & 7  &    0.03 & $\pm$ 0.08 &     20.3 & $\pm$ 0.2  \\
\object{zeta\,Gem}    & STELLA & 54418.1581293 & 5460 & $\pm$ 117 & 1.8 & 4.7 &    0.12 & $\pm$ 0.14 & 96  &    0.12 & $\pm$ 0.06 & 6  &    0.12 & $\pm$ 0.05 &     20.3 & $\pm$ 0.2  \\
\object{zeta\,Gem}    & STELLA & 54418.1730308 & 5404 & $\pm$ 96  & 0.8 & 3.8 &    0.17 & $\pm$ 0.12 & 96  &    0.03 & $\pm$ 0.12 & 7  &    0.10 & $\pm$ 0.08 &     20.2 & $\pm$ 0.2  \\
\object{zeta\,Gem}    & STELLA & 54421.2745932 & 5734 & $\pm$ 140 & 0.8 & 2.5 &    0.29 & $\pm$ 0.15 & 137 &    0.26 & $\pm$ 0.16 & 13 &    0.28 & $\pm$ 0.11 &      0.8 & $\pm$ 0.2  \\
\object{zeta\,Gem}    & STELLA & 54423.2824759 & 5606 & $\pm$ 100 & 1.1 & 2.5 &    0.22 & $\pm$ 0.12 & 129 &    0.23 & $\pm$ 0.18 & 15 &    0.22 & $\pm$ 0.10 &   $-$5.2 & $\pm$ 0.2  \\
\object{zeta\,Gem}    & STELLA & 54438.2882356 & 5339 & $\pm$ 84  & 0.7 & 3.8 &    0.10 & $\pm$ 0.16 & 123 &    0.13 & $\pm$ 0.23 & 11 &    0.11 & $\pm$ 0.13 &     21.5 & $\pm$ 0.2  \\
\object{zeta\,Gem}    & STELLA & 54462.2237082 & 5793 & $\pm$ 92  & 0.8 & 2.2 &    0.32 & $\pm$ 0.10 & 149 &    0.29 & $\pm$ 0.17 & 13 &    0.31 & $\pm$ 0.09 &      0.8 & $\pm$ 0.2  \\
\object{zeta\,Gem}    & STELLA & 54463.0747703 & 5762 & $\pm$ 92  & 0.9 & 2.4 &    0.21 & $\pm$ 0.10 & 144 &    0.22 & $\pm$ 0.15 & 12 &    0.21 & $\pm$ 0.08 &   $-$1.9 & $\pm$ 0.2  \\
\object{zeta\,Gem}    & STELLA & 54465.0558349 & 5455 & $\pm$ 92  & 0.9 & 2.3 &    0.28 & $\pm$ 0.12 & 158 &    0.30 & $\pm$ 0.19 & 20 &    0.28 & $\pm$ 0.10 &      0.5 & $\pm$ 0.2  \\
\object{zeta\,Gem}    & STELLA & 54471.1486908 & 5708 & $\pm$ 94  & 1.0 & 2.4 &    0.23 & $\pm$ 0.11 & 161 &    0.36 & $\pm$ 0.16 & 11 &    0.28 & $\pm$ 0.09 &      1.9 & $\pm$ 0.2  \\
\object{zeta\,Gem}    & STELLA & 54486.1579739 & 5341 & $\pm$ 81  & 0.9 & 2.7 &    0.20 & $\pm$ 0.13 & 161 &    0.16 & $\pm$ 0.26 & 20 &    0.19 & $\pm$ 0.12 &      6.4 & $\pm$ 0.2  \\
\object{zeta\,Gem}    & STELLA & 54501.1521642 & 5693 & $\pm$ 94  & 0.6 & 2.5 &    0.28 & $\pm$ 0.14 & 142 &    0.28 & $\pm$ 0.10 & 10 &    0.28 & $\pm$ 0.08 &      3.6 & $\pm$ 0.2  \\
\object{zeta\,Gem}    & STELLA & 54502.0078883 & 5742 & $\pm$ 100 & 1.0 & 2.5 &    0.22 & $\pm$ 0.11 & 149 &    0.29 & $\pm$ 0.11 & 11 &    0.25 & $\pm$ 0.08 &      1.2 & $\pm$ 0.2  \\
\object{zeta\,Gem}    & STELLA & 54525.0767616 & 5594 & $\pm$ 97  & 0.8 & 2.8 &    0.18 & $\pm$ 0.14 & 151 &    0.19 & $\pm$ 0.04 & 10 &    0.19 & $\pm$ 0.04 &   $-$4.5 & $\pm$ 0.2  \\
\object{zeta\,Gem}    & STELLA & 54557.9765317 & 5256 & $\pm$ 98  & 0.8 & 2.6 &    0.20 & $\pm$ 0.14 & 143 &    0.18 & $\pm$ 0.38 & 21 &    0.20 & $\pm$ 0.13 &     12.7 & $\pm$ 0.2  \\
\object{zeta\,Gem}    & STELLA & 54701.2370873 & 5239 & $\pm$ 91  & 0.9 & 3.5 &    0.13 & $\pm$ 0.14 & 124 &    0.17 & $\pm$ 0.25 & 15 &    0.14 & $\pm$ 0.12 &     20.2 & $\pm$ 0.2  \\
\object{zeta\,Gem}    & STELLA & 54702.2340611 & 5327 & $\pm$ 97  & 0.8 & 3.7 &    0.09 & $\pm$ 0.15 & 111 &    0.14 & $\pm$ 0.17 & 12 &    0.11 & $\pm$ 0.11 &     20.9 & $\pm$ 0.2  \\
\object{zeta\,Gem}    & STELLA & 54702.2373032 & 5380 & $\pm$ 100 & 1.1 & 4.7 &    0.03 & $\pm$ 0.13 & 110 &    0.00 & $\pm$ 0.11 & 11 &    0.01 & $\pm$ 0.08 &     20.9 & $\pm$ 0.2  \\
\object{zeta\,Gem}    & STELLA & 54703.2347503 & 5542 & $\pm$ 107 & 1.0 & 2.9 &    0.15 & $\pm$ 0.13 & 130 &    0.22 & $\pm$ 0.12 & 11 &    0.19 & $\pm$ 0.09 &     11.5 & $\pm$ 0.2  \\
\object{zeta\,Gem}    & STELLA & 54703.2379968 & 5522 & $\pm$ 98  & 1.2 & 3.3 &    0.13 & $\pm$ 0.12 & 123 &    0.15 & $\pm$ 0.12 & 11 &    0.14 & $\pm$ 0.08 &     11.4 & $\pm$ 0.2  \\
\object{zeta\,Gem}    & STELLA & 54704.2296119 & 5670 & $\pm$ 93  & 1.0 & 2.9 &    0.16 & $\pm$ 0.11 & 133 &    0.17 & $\pm$ 0.17 & 12 &    0.16 & $\pm$ 0.09 &      3.1 & $\pm$ 0.2  \\
\object{zeta\,Gem}    & STELLA & 54704.2378082 & 5673 & $\pm$ 89  & 1.2 & 2.7 &    0.21 & $\pm$ 0.14 & 156 &    0.24 & $\pm$ 0.17 & 13 &    0.22 & $\pm$ 0.11 &      3.1 & $\pm$ 0.2  \\
\object{zeta\,Gem}    & STELLA & 54705.2363817 & 5727 & $\pm$ 87  & 1.2 & 2.4 &    0.20 & $\pm$ 0.12 & 152 &    0.23 & $\pm$ 0.18 & 12 &    0.21 & $\pm$ 0.10 &      1.0 & $\pm$ 0.2  \\
\object{zeta\,Gem}    & STELLA & 54705.2396542 & 5734 & $\pm$ 81  & 1.2 & 2.3 &    0.21 & $\pm$ 0.13 & 168 &    0.25 & $\pm$ 0.20 & 12 &    0.22 & $\pm$ 0.11 &      1.0 & $\pm$ 0.2  \\
\object{zeta\,Gem}    & STELLA & 54706.2239765 & 5815 & $\pm$ 100 & 0.8 & 2.4 &    0.19 & $\pm$ 0.12 & 146 &    0.08 & $\pm$ 0.17 & 11 &    0.15 & $\pm$ 0.10 &   $-$0.3 & $\pm$ 0.2  \\
\object{zeta\,Gem}    & STELLA & 54706.2318500 & 5806 & $\pm$ 97  & 0.9 & 2.3 &    0.21 & $\pm$ 0.12 & 148 &    0.11 & $\pm$ 0.19 & 13 &    0.18 & $\pm$ 0.10 &   $-$0.3 & $\pm$ 0.2  \\
\object{zeta\,Gem}    & STELLA & 54706.2350952 & 5840 & $\pm$ 99  & 1.0 & 2.3 &    0.28 & $\pm$ 0.11 & 147 &    0.14 & $\pm$ 0.23 & 13 &    0.25 & $\pm$ 0.10 &   $-$0.3 & $\pm$ 0.2  \\
\object{zeta\,Gem}    & STELLA & 54706.2387765 & 5842 & $\pm$ 98  & 0.7 & 2.1 &    0.33 & $\pm$ 0.13 & 153 &    0.23 & $\pm$ 0.28 & 13 &    0.31 & $\pm$ 0.12 &   $-$0.3 & $\pm$ 0.2  \\
\object{zeta\,Gem}    & STELLA & 54707.2369122 & 5703 & $\pm$ 97  & 1.1 & 2.4 &    0.20 & $\pm$ 0.11 & 143 &    0.18 & $\pm$ 0.18 & 12 &    0.19 & $\pm$ 0.09 &   $-$4.7 & $\pm$ 0.2  \\
\object{zeta\,Gem}    & STELLA & 54708.2184309 & 5550 & $\pm$ 91  & 1.0 & 2.4 &    0.20 & $\pm$ 0.11 & 136 &    0.20 & $\pm$ 0.22 & 17 &    0.20 & $\pm$ 0.10 &   $-$2.5 & $\pm$ 0.2  \\
\object{zeta\,Gem}    & STELLA & 54709.2159272 & 5366 & $\pm$ 85  & 1.2 & 2.8 &    0.14 & $\pm$ 0.10 & 137 &    0.14 & $\pm$ 0.17 & 16 &    0.14 & $\pm$ 0.08 &      4.7 & $\pm$ 0.2  \\
\object{zeta\,Gem}    & STELLA & 54709.2376994 & 5356 & $\pm$ 81  & 1.1 & 2.5 &    0.18 & $\pm$ 0.10 & 145 &    0.22 & $\pm$ 0.20 & 17 &    0.19 & $\pm$ 0.09 &      4.8 & $\pm$ 0.2  \\
\hline\noalign{\smallskip}
\multicolumn{17}{r}{\it {\footnotesize continued on next page}} \\
\end{tabular}}
\end{table*}
\addtocounter{table}{-1}
\begin{table*}[p]
\centering
\caption[]{continued.}
{\scriptsize 
\begin{tabular}{lcc r@{ }l cc r@{ }l c r@{ }l c r@{ }l r@{ }l}
\noalign{\smallskip}\hline\hline\noalign{\smallskip}
Name & Dataset &
\parbox[c]{0.5cm}{\centering MJD [d]} &
\multicolumn{2}{c}{\parbox[c]{0.8cm}{\centering \teff\ $\pm$ $\sigma$ [K]}} &
\logg\ &
\parbox[c]{0.9cm}{\centering \vmic\ [\kms]} &
\multicolumn{2}{c}{\ion{Fe}{i}  $\pm$ $\sigma$} & $N_{\rm{\ion{Fe}{i}}}$ &
\multicolumn{2}{c}{\ion{Fe}{ii} $\pm$ $\sigma$} & $N_{\rm{\ion{Fe}{ii}}}$ &
\multicolumn{2}{c}{[Fe/H] $\pm$ $\sigma$} &
\multicolumn{2}{c}{\parbox[c]{0.9cm}{\centering $RV$ $\pm$ $\sigma$ [\kms]}} \\
\noalign{\smallskip}\hline\noalign{\smallskip}
\object{zeta\,Gem}    & STELLA & 54709.2409483 & 5371 & $\pm$ 78  & 1.0 & 2.3 &    0.21 & $\pm$ 0.12 & 156 &    0.23 & $\pm$ 0.22 & 18 &    0.21 & $\pm$ 0.11 &      4.8 & $\pm$ 0.2  \\
\object{zeta\,Gem}    & STELLA & 54711.2070298 & 5209 & $\pm$ 99  & 0.8 & 3.5 &    0.08 & $\pm$ 0.12 & 125 &    0.15 & $\pm$ 0.24 & 14 &    0.09 & $\pm$ 0.11 &     19.2 & $\pm$ 0.2  \\
\object{zeta\,Gem}    & STELLA & 54711.2102754 & 5239 & $\pm$ 94  & 0.7 & 3.4 &    0.12 & $\pm$ 0.12 & 131 &    0.14 & $\pm$ 0.20 & 13 &    0.12 & $\pm$ 0.10 &     19.2 & $\pm$ 0.2  \\
\object{zeta\,Gem}    & STELLA & 54712.2037665 & 5334 & $\pm$ 87  & 1.2 & 4.4 &    0.03 & $\pm$ 0.13 & 116 &    0.08 & $\pm$ 0.16 & 12 &    0.05 & $\pm$ 0.10 &     21.4 & $\pm$ 0.2  \\
\object{zeta\,Gem}    & STELLA & 54712.2070077 & 5327 & $\pm$ 89  & 1.1 & 5.0 & $-$0.04 & $\pm$ 0.13 & 114 & $-$0.03 & $\pm$ 0.16 & 12 & $-$0.03 & $\pm$ 0.10 &     21.4 & $\pm$ 0.2  \\
\object{zeta\,Gem}    & STELLA & 54712.2427809 & 5331 & $\pm$ 90  & 1.2 & 4.0 &    0.03 & $\pm$ 0.15 & 133 &    0.08 & $\pm$ 0.15 & 12 &    0.05 & $\pm$ 0.11 &     21.4 & $\pm$ 0.2  \\
\object{zeta\,Gem}    & STELLA & 54713.2441852 & 5540 & $\pm$ 93  & 0.9 & 2.9 &    0.18 & $\pm$ 0.10 & 118 &    0.22 & $\pm$ 0.20 & 10 &    0.19 & $\pm$ 0.09 &     13.1 & $\pm$ 0.2  \\
\object{zeta\,Gem}    & STELLA & 54714.1983269 & 5647 & $\pm$ 100 & 1.1 & 2.8 &    0.18 & $\pm$ 0.10 & 118 &    0.23 & $\pm$ 0.14 & 10 &    0.20 & $\pm$ 0.08 &      4.1 & $\pm$ 0.2  \\
\object{zeta\,Gem}    & STELLA & 54714.2015666 & 5639 & $\pm$ 99  & 1.3 & 2.7 &    0.17 & $\pm$ 0.13 & 136 &    0.42 & $\pm$ 0.10 & 8  &    0.34 & $\pm$ 0.08 &      4.1 & $\pm$ 0.2  \\
\object{zeta\,Gem}    & STELLA & 54714.2418747 & 5658 & $\pm$ 100 & 1.1 & 3.0 &    0.16 & $\pm$ 0.13 & 145 &    0.12 & $\pm$ 0.15 & 11 &    0.14 & $\pm$ 0.10 &      3.8 & $\pm$ 0.2  \\
\object{zeta\,Gem}    & STELLA & 54729.1622480 & 5421 & $\pm$ 98  & 1.2 & 2.6 &    0.19 & $\pm$ 0.10 & 150 &    0.21 & $\pm$ 0.17 & 13 &    0.19 & $\pm$ 0.09 &      1.8 & $\pm$ 0.2  \\
\object{zeta\,Gem}    & STELLA & 54750.2108478 & 5304 & $\pm$ 84  & 1.2 & 2.9 &    0.12 & $\pm$ 0.09 & 139 &    0.17 & $\pm$ 0.15 & 15 &    0.13 & $\pm$ 0.08 &      7.8 & $\pm$ 0.2  \\
\object{zeta\,Gem}    & STELLA & 54754.2167227 & 5614 & $\pm$ 99  & 1.2 & 3.1 &    0.15 & $\pm$ 0.12 & 136 &    0.17 & $\pm$ 0.18 & 12 &    0.16 & $\pm$ 0.10 &      8.8 & $\pm$ 0.2  \\
\object{zeta\,Gem}    & STELLA & 54754.2210371 & 5594 & $\pm$ 99  & 1.1 & 2.5 &    0.19 & $\pm$ 0.12 & 131 &    0.25 & $\pm$ 0.26 & 13 &    0.20 & $\pm$ 0.11 &      8.8 & $\pm$ 0.2  \\
\object{zeta\,Gem}    & STELLA & 54760.1902531 & 5331 & $\pm$ 83  & 1.2 & 2.4 &    0.23 & $\pm$ 0.10 & 156 &    0.19 & $\pm$ 0.04 & 9  &    0.20 & $\pm$ 0.04 &      6.5 & $\pm$ 0.2  \\
\object{zeta\,Gem}    & STELLA & 54760.1934999 & 5337 & $\pm$ 91  & 0.7 & 2.3 &    0.26 & $\pm$ 0.12 & 157 &    0.20 & $\pm$ 0.20 & 17 &    0.24 & $\pm$ 0.10 &      6.5 & $\pm$ 0.2  \\
\object{zeta\,Gem}    & STELLA & 54760.1972571 & 5343 & $\pm$ 80  & 1.1 & 2.8 &    0.14 & $\pm$ 0.11 & 158 &    0.14 & $\pm$ 0.12 & 13 &    0.14 & $\pm$ 0.08 &      6.5 & $\pm$ 0.2  \\
\object{zeta\,Gem}    & STELLA & 54760.2005072 & 5331 & $\pm$ 84  & 1.1 & 2.6 &    0.18 & $\pm$ 0.10 & 150 &    0.21 & $\pm$ 0.18 & 18 &    0.19 & $\pm$ 0.09 &      6.6 & $\pm$ 0.2  \\
\object{zeta\,Gem}    & STELLA & 54760.2037406 & 5335 & $\pm$ 96  & 1.1 & 2.6 &    0.20 & $\pm$ 0.12 & 160 &    0.18 & $\pm$ 0.23 & 18 &    0.19 & $\pm$ 0.10 &      6.6 & $\pm$ 0.2  \\
\object{zeta\,Gem}    & STELLA & 54760.2069893 & 5335 & $\pm$ 81  & 1.0 & 2.6 &    0.17 & $\pm$ 0.11 & 155 &    0.13 & $\pm$ 0.17 & 14 &    0.16 & $\pm$ 0.09 &      6.6 & $\pm$ 0.2  \\
\object{zeta\,Gem}    & STELLA & 54760.2260601 & 5334 & $\pm$ 79  & 1.1 & 2.7 &    0.17 & $\pm$ 0.11 & 161 &    0.13 & $\pm$ 0.07 & 11 &    0.15 & $\pm$ 0.06 &      6.7 & $\pm$ 0.2  \\
\object{zeta\,Gem}    & STELLA & 54760.2292960 & 5346 & $\pm$ 76  & 1.1 & 2.5 &    0.21 & $\pm$ 0.11 & 153 &    0.24 & $\pm$ 0.21 & 16 &    0.21 & $\pm$ 0.10 &      6.8 & $\pm$ 0.2  \\
\object{zeta\,Gem}    & STELLA & 54760.2489259 & 5333 & $\pm$ 80  & 1.3 & 2.6 &    0.19 & $\pm$ 0.12 & 165 &    0.24 & $\pm$ 0.12 & 13 &    0.22 & $\pm$ 0.09 &      7.0 & $\pm$ 0.2  \\
\object{zeta\,Gem}    & STELLA & 54760.2521595 & 5328 & $\pm$ 79  & 0.8 & 2.5 &    0.19 & $\pm$ 0.11 & 156 &    0.14 & $\pm$ 0.27 & 17 &    0.18 & $\pm$ 0.10 &      7.0 & $\pm$ 0.2  \\
\object{zeta\,Gem}    & STELLA & 54760.2554083 & 5321 & $\pm$ 77  & 1.0 & 2.8 &    0.16 & $\pm$ 0.11 & 155 &    0.18 & $\pm$ 0.19 & 17 &    0.17 & $\pm$ 0.10 &      7.0 & $\pm$ 0.2  \\
\object{zeta\,Gem}    & STELLA & 54762.2238400 & 5219 & $\pm$ 93  & 0.8 & 3.6 &    0.13 & $\pm$ 0.17 & 130 &    0.14 & $\pm$ 0.22 & 14 &    0.13 & $\pm$ 0.13 &     20.3 & $\pm$ 0.2  \\
\object{zeta\,Gem}    & STELLA & 54777.0470244 & 5800 & $\pm$ 99  & 0.7 & 2.0 &    0.34 & $\pm$ 0.12 & 152 &    0.27 & $\pm$ 0.24 & 13 &    0.32 & $\pm$ 0.11 &      0.1 & $\pm$ 0.2  \\
\object{zeta\,Gem}    & STELLA & 54789.0010630 & 5613 & $\pm$ 81  & 0.8 & 2.5 &    0.21 & $\pm$ 0.11 & 157 &    0.28 & $\pm$ 0.07 & 12 &    0.26 & $\pm$ 0.06 &   $-$4.5 & $\pm$ 0.2  \\
\object{zeta\,Gem}    & STELLA & 54789.0043057 & 5626 & $\pm$ 96  & 0.6 & 2.6 &    0.22 & $\pm$ 0.09 & 146 &    0.19 & $\pm$ 0.01 & 7  &    0.19 & $\pm$ 0.01 &   $-$4.5 & $\pm$ 0.2  \\
\object{zeta\,Gem}    & STELLA & 54789.0128166 & 5609 & $\pm$ 87  & 0.8 & 2.4 &    0.29 & $\pm$ 0.10 & 134 &    0.25 & $\pm$ 0.14 & 16 &    0.28 & $\pm$ 0.08 &   $-$4.4 & $\pm$ 0.2  \\
\object{zeta\,Gem}    & STELLA & 54789.0160527 & 5618 & $\pm$ 94  & 0.8 & 2.5 &    0.22 & $\pm$ 0.09 & 132 &    0.20 & $\pm$ 0.14 & 16 &    0.21 & $\pm$ 0.08 &   $-$4.4 & $\pm$ 0.2  \\
\object{zeta\,Gem}    & STELLA & 54789.0192991 & 5605 & $\pm$ 90  & 0.9 & 2.5 &    0.20 & $\pm$ 0.08 & 130 &    0.23 & $\pm$ 0.11 & 15 &    0.21 & $\pm$ 0.06 &   $-$4.4 & $\pm$ 0.2  \\
\object{zeta\,Gem}    & STELLA & 54792.9898339 & 5241 & $\pm$ 91  & 0.7 & 3.2 &    0.13 & $\pm$ 0.09 & 107 &    0.12 & $\pm$ 0.05 & 9  &    0.12 & $\pm$ 0.04 &     21.3 & $\pm$ 0.2  \\
\object{zeta\,Gem}    & STELLA & 54792.9930840 & 5270 & $\pm$ 85  & 0.9 & 3.9 &    0.07 & $\pm$ 0.14 & 120 &    0.08 & $\pm$ 0.17 & 15 &    0.08 & $\pm$ 0.11 &     21.3 & $\pm$ 0.2  \\
\object{zeta\,Gem}    & STELLA & 54793.0106374 & 5270 & $\pm$ 87  & 0.8 & 3.7 &    0.07 & $\pm$ 0.07 & 87  &    0.08 & $\pm$ 0.22 & 16 &    0.07 & $\pm$ 0.06 &     21.4 & $\pm$ 0.2  \\
\object{zeta\,Gem}    & STELLA & 54793.9876861 & 5470 & $\pm$ 96  & 1.2 & 4.4 &    0.04 & $\pm$ 0.09 & 86  &    0.05 & $\pm$ 0.18 & 12 &    0.04 & $\pm$ 0.08 &     17.9 & $\pm$ 0.2  \\
\object{zeta\,Gem}    & STELLA & 54793.9912187 & 5466 & $\pm$ 99  & 1.3 & 3.9 &    0.09 & $\pm$ 0.12 & 110 &    0.10 & $\pm$ 0.16 & 13 &    0.09 & $\pm$ 0.10 &     17.8 & $\pm$ 0.2  \\
\object{zeta\,Gem}    & STELLA & 54794.9869237 & 5603 & $\pm$ 74  & 1.1 & 3.2 &    0.14 & $\pm$ 0.12 & 132 &    0.15 & $\pm$ 0.13 & 13 &    0.14 & $\pm$ 0.09 &      7.3 & $\pm$ 0.2  \\
\object{zeta\,Gem}    & STELLA & 54797.9939915 & 5797 & $\pm$ 89  & 1.1 & 2.5 &    0.31 & $\pm$ 0.10 & 136 &    0.30 & $\pm$ 0.22 & 15 &    0.31 & $\pm$ 0.09 &   $-$2.2 & $\pm$ 0.2  \\
\object{zeta\,Gem}    & STELLA & 54798.9745787 & 5624 & $\pm$ 93  & 0.8 & 2.5 &    0.31 & $\pm$ 0.11 & 150 &    0.30 & $\pm$ 0.16 & 15 &    0.30 & $\pm$ 0.09 &   $-$4.8 & $\pm$ 0.2  \\
\object{zeta\,Gem}    & STELLA & 54798.9778207 & 5632 & $\pm$ 97  & 0.4 & 2.0 &    0.38 & $\pm$ 0.10 & 120 &    0.39 & $\pm$ 0.13 & 11 &    0.39 & $\pm$ 0.08 &   $-$4.8 & $\pm$ 0.2  \\
\object{zeta\,Gem}    & STELLA & 54799.9715934 & 5466 & $\pm$ 99  & 1.1 & 2.7 &    0.19 & $\pm$ 0.11 & 159 &    0.19 & $\pm$ 0.14 & 18 &    0.19 & $\pm$ 0.08 &      0.3 & $\pm$ 0.2  \\
\object{zeta\,Gem}    & STELLA & 54799.9748388 & 5486 & $\pm$ 90  & 1.0 & 2.5 &    0.23 & $\pm$ 0.07 & 129 &    0.23 & $\pm$ 0.19 & 20 &    0.23 & $\pm$ 0.07 &      0.3 & $\pm$ 0.2  \\
\object{zeta\,Gem}    & STELLA & 54800.9882545 & 5300 & $\pm$ 87  & 0.8 & 2.6 &    0.23 & $\pm$ 0.10 & 145 &    0.24 & $\pm$ 0.24 & 18 &    0.23 & $\pm$ 0.09 &      8.2 & $\pm$ 0.2  \\
\object{zeta\,Gem}    & STELLA & 54822.2375649 & 5241 & $\pm$ 80  & 1.1 & 3.4 &    0.13 & $\pm$ 0.13 & 152 &    0.14 & $\pm$ 0.32 & 18 &    0.13 & $\pm$ 0.12 &     15.3 & $\pm$ 0.2  \\
\object{Y\,Oph}       & HARPS  & 53150.2730135 & 5540 & $\pm$ 96  & 1.1 & 3.1 &    0.00 & $\pm$ 0.14 & 176 &    0.04 & $\pm$ 0.27 & 14 &    0.01 & $\pm$ 0.12 &   $-$4.5 & $\pm$ 0.7  \\
\object{Y\,Oph}       & HARPS  & 53152.2897868 & 5564 & $\pm$ 100 & 1.4 & 3.9 & $-$0.00 & $\pm$ 0.14 & 165 &    0.02 & $\pm$ 0.20 & 11 &    0.00 & $\pm$ 0.12 &   $-$0.1 & $\pm$ 0.8  \\
\object{Y\,Oph}       & HARPS  & 53154.2444076 & 5628 & $\pm$ 94  & 1.3 & 3.6 &    0.01 & $\pm$ 0.15 & 157 &    0.05 & $\pm$ 0.16 & 12 &    0.03 & $\pm$ 0.11 &      1.1 & $\pm$ 0.8  \\
\object{Y\,Oph}       & HARPS  & 53156.2027098 & 5758 & $\pm$ 91  & 1.1 & 3.6 &    0.09 & $\pm$ 0.15 & 165 &    0.15 & $\pm$ 0.12 & 11 &    0.13 & $\pm$ 0.09 &   $-$2.7 & $\pm$ 0.7  \\
\object{Y\,Oph}       & HARPS  & 53201.1235014 & 5562 & $\pm$ 99  & 0.8 & 2.9 &    0.13 & $\pm$ 0.15 & 176 &    0.17 & $\pm$ 0.19 & 14 &    0.14 & $\pm$ 0.12 &   $-$5.8 & $\pm$ 0.7  \\
\object{Y\,Oph}       & HARPS  & 53203.1387889 & 5524 & $\pm$ 75  & 1.1 & 3.2 & $-$0.03 & $\pm$ 0.13 & 166 & $-$0.01 & $\pm$ 0.27 & 14 & $-$0.02 & $\pm$ 0.12 &   $-$1.0 & $\pm$ 0.8  \\
\object{Y\,Oph}       & HARPS  & 53216.2394193 & 5707 & $\pm$ 106 & 0.6 & 2.8 &    0.14 & $\pm$ 0.13 & 185 &    0.09 & $\pm$ 0.20 & 14 &    0.13 & $\pm$ 0.11 &  $-$11.2 & $\pm$ 0.6  \\
\object{Y\,Oph}       & HARPS  & 56213.9822101 & 5663 & $\pm$ 99  & 0.5 & 2.7 &    0.14 & $\pm$ 0.13 & 180 &    0.09 & $\pm$ 0.17 & 12 &    0.13 & $\pm$ 0.11 &   $-$9.6 & $\pm$ 0.6  \\
\object{RS\,Pup}      & HARPS  & 53048.1059728 & 5083 & $\pm$ 103 & 0.4 & 4.9 & $-$0.00 & $\pm$ 0.19 & 72  & $-$0.01 & $\pm$ 0.23 & 7  & $-$0.00 & $\pm$ 0.15 &     44.6 & $\pm$ 2.0  \\
\object{RS\,Pup}      & HARPS  & 53052.1267735 & 6238 & $\pm$ 132 & 2.0 & 4.9 &    0.02 & $\pm$ 0.17 & 82  & $-$0.07 & $\pm$ 0.28 & 15 & $-$0.00 & $\pm$ 0.14 &     17.1 & $\pm$ 2.3  \\
\object{RS\,Pup}      & HARPS  & 53054.1528518 & 6184 & $\pm$ 135 & 1.0 & 3.6 &    0.17 & $\pm$ 0.11 & 104 &    0.24 & $\pm$ 0.29 & 10 &    0.18 & $\pm$ 0.10 &      4.9 & $\pm$ 1.3  \\
\object{RS\,Pup}      & HARPS  & 53056.1793367 & 6037 & $\pm$ 121 & 0.8 & 3.7 &    0.21 & $\pm$ 0.11 & 109 &    0.23 & $\pm$ 0.19 & 8  &    0.22 & $\pm$ 0.10 &      3.1 & $\pm$ 1.1  \\
\object{RS\,Pup}      & HARPS  & 53058.1901193 & 5769 & $\pm$ 133 & 0.8 & 4.9 &    0.14 & $\pm$ 0.14 & 120 &    0.11 & $\pm$ 0.27 & 10 &    0.14 & $\pm$ 0.13 &      4.3 & $\pm$ 1.0  \\
\object{RS\,Pup}      & HARPS  & 53060.1785463 & 5593 & $\pm$ 100 & 0.6 & 4.9 &    0.14 & $\pm$ 0.13 & 129 &    0.17 & $\pm$ 0.04 & 5  &    0.17 & $\pm$ 0.04 &      6.5 & $\pm$ 0.9  \\
\object{RS\,Pup}      & HARPS  & 53062.1660897 & 5438 & $\pm$ 98  & 0.5 & 4.9 &    0.16 & $\pm$ 0.15 & 130 &    0.14 & $\pm$ 0.09 & 9  &    0.14 & $\pm$ 0.07 &      9.3 & $\pm$ 0.9  \\
\object{RS\,Pup}      & HARPS  & 53064.1739999 & 5346 & $\pm$ 99  & 0.6 & 4.9 &    0.14 & $\pm$ 0.14 & 128 &    0.11 & $\pm$ 0.22 & 12 &    0.13 & $\pm$ 0.12 &     12.2 & $\pm$ 0.9  \\
\object{RS\,Pup}      & HARPS  & 53066.1524561 & 5266 & $\pm$ 97  & 0.7 & 4.9 &    0.10 & $\pm$ 0.15 & 142 &    0.12 & $\pm$ 0.27 & 12 &    0.11 & $\pm$ 0.13 &     15.3 & $\pm$ 1.0  \\
\object{RS\,Pup}      & HARPS  & 53149.9663503 & 5206 & $\pm$ 100 & 0.6 & 4.9 &    0.10 & $\pm$ 0.13 & 132 &    0.06 & $\pm$ 0.25 & 10 &    0.09 & $\pm$ 0.12 &     16.1 & $\pm$ 1.0  \\
\object{RS\,Pup}      & HARPS  & 53151.9756287 & 5171 & $\pm$ 82  & 0.8 & 4.9 &    0.08 & $\pm$ 0.15 & 138 &    0.01 & $\pm$ 0.19 & 10 &    0.05 & $\pm$ 0.12 &     19.2 & $\pm$ 1.1  \\
\object{RS\,Pup}      & HARPS  & 53153.9783769 & 5135 & $\pm$ 99  & 1.0 & 4.0 &    0.14 & $\pm$ 0.11 & 126 &    0.10 & $\pm$ 0.15 & 8  &    0.12 & $\pm$ 0.09 &     22.3 & $\pm$ 1.1  \\
\object{RS\,Pup}      & HARPS  & 53155.9674458 & 5089 & $\pm$ 90  & 1.0 & 4.9 &    0.13 & $\pm$ 0.14 & 131 &    0.17 & $\pm$ 0.28 & 13 &    0.14 & $\pm$ 0.13 &     25.4 & $\pm$ 1.2  \\
\object{RS\,Pup}      & HARPS  & 53155.9756820 & 5085 & $\pm$ 95  & 0.9 & 4.9 &    0.11 & $\pm$ 0.13 & 123 &    0.13 & $\pm$ 0.24 & 10 &    0.11 & $\pm$ 0.11 &     25.4 & $\pm$ 1.2  \\
\object{UZ\,Sct}      & FEROS  & 54190.3602653 & 4824 & $\pm$ 121 & 1.3 & 4.9 &    0.14 & $\pm$ 0.28 & 69  &    0.28 & $\pm$ 0.44 & 9  &    0.18 & $\pm$ 0.24 &     52.3 & $\pm$ 1.8  \\
\object{UZ\,Sct}      & FEROS  & 54190.3817051 & 4822 & $\pm$ 99  & 1.3 & 4.9 &    0.02 & $\pm$ 0.26 & 70  &    0.13 & $\pm$ 0.41 & 9  &    0.05 & $\pm$ 0.22 &     52.4 & $\pm$ 1.9  \\
\object{UZ\,Sct}      & UVES   & 54906.4044778 & 5382 & $\pm$ 100 & 1.1 & 4.9 &    0.19 & $\pm$ 0.19 & 102 &    0.14 & $\pm$ 0.21 & 9  &    0.17 & $\pm$ 0.14 &    19.37 & $\pm$ 0.01 \\
\object{UZ\,Sct}      & UVES   & 54923.3689441 & 5088 & $\pm$ 99  & 1.6 & 4.8 &    0.14 & $\pm$ 0.19 & 90  &    0.02 & $\pm$ 0.17 & 7  &    0.07 & $\pm$ 0.13 &    30.95 & $\pm$ 0.01 \\
\object{UZ\,Sct}      & UVES   & 56137.1617631 & 4797 & $\pm$ 98  & 0.9 & 4.8 &    0.11 & $\pm$ 0.17 & 71  & $-$0.01 & $\pm$ 0.19 & 7  &    0.06 & $\pm$ 0.13 &     59.6 & $\pm$ 0.1  \\
\object{UZ\,Sct}      & UVES   & 56152.0662646 & 4769 & $\pm$ 98  & 0.9 & 4.8 & $-$0.01 & $\pm$ 0.22 & 78  & $-$0.19 & $\pm$ 0.33 & 10 & $-$0.07 & $\pm$ 0.18 &     60.5 & $\pm$ 0.1  \\
\object{UZ\,Sct}      & UVES   & 56160.1647217 & 5295 & $\pm$ 99  & 1.3 & 4.9 &    0.16 & $\pm$ 0.15 & 127 &    0.17 & $\pm$ 0.28 & 15 &    0.16 & $\pm$ 0.13 &    19.77 & $\pm$ 0.04 \\
\object{UZ\,Sct}      & UVES   & 56175.0512628 & 5275 & $\pm$ 99  & 1.5 & 4.9 &    0.18 & $\pm$ 0.15 & 120 &    0.20 & $\pm$ 0.29 & 16 &    0.18 & $\pm$ 0.13 &    20.95 & $\pm$ 0.03 \\
\object{AV\,Sgr}      & FEROS  & 53601.1718719 & 6091 & $\pm$ 159 & 1.3 & 4.9 &    0.43 & $\pm$ 0.28 & 81  &    0.34 & $\pm$ 0.16 & 8  &    0.36 & $\pm$ 0.14 &     17.2 & $\pm$ 1.0  \\
\object{AV\,Sgr}      & FEROS  & 53602.1194613 & 6259 & $\pm$ 149 & 1.4 & 5.0 &    0.42 & $\pm$ 0.22 & 75  &    0.37 & $\pm$ 0.03 & 5  &    0.37 & $\pm$ 0.03 &      6.9 & $\pm$ 1.1  \\
\object{AV\,Sgr}      & FEROS  & 54185.3603987 & 5442 & $\pm$ 98  & 0.8 & 4.9 &    0.19 & $\pm$ 0.21 & 86  &    0.18 & $\pm$ 0.07 & 4  &    0.18 & $\pm$ 0.06 &     27.4 & $\pm$ 1.3  \\
\object{AV\,Sgr}      & UVES   & 54923.3508098 & 5062 & $\pm$ 99  & 0.1 & 4.9 &    0.17 & $\pm$ 0.27 & 53  &    0.20 & $\pm$ 0.16 & 4  &    0.20 & $\pm$ 0.14 &    49.19 & $\pm$ 0.03 \\
\object{AV\,Sgr}      & UVES   & 56136.1708156 & 5036 & $\pm$ 98  & 1.6 & 4.8 &    0.25 & $\pm$ 0.17 & 118 &    0.17 & $\pm$ 0.22 & 9  &    0.22 & $\pm$ 0.14 &     22.4 & $\pm$ 0.1  \\
\hline\noalign{\smallskip}
\multicolumn{17}{r}{\it {\footnotesize continued on next page}} \\
\end{tabular}}
\end{table*}
\addtocounter{table}{-1}
\begin{table*}[p]
\centering
\caption[]{continued.}
{\scriptsize 
\begin{tabular}{lcc r@{ }l cc r@{ }l c r@{ }l c r@{ }l r@{ }l}
\noalign{\smallskip}\hline\hline\noalign{\smallskip}
Name & Dataset &
\parbox[c]{0.5cm}{\centering MJD [d]} &
\multicolumn{2}{c}{\parbox[c]{0.8cm}{\centering \teff\ $\pm$ $\sigma$ [K]}} &
\logg\ &
\parbox[c]{0.9cm}{\centering \vmic\ [\kms]} &
\multicolumn{2}{c}{\ion{Fe}{i}  $\pm$ $\sigma$} & $N_{\rm{\ion{Fe}{i}}}$ &
\multicolumn{2}{c}{\ion{Fe}{ii} $\pm$ $\sigma$} & $N_{\rm{\ion{Fe}{ii}}}$ &
\multicolumn{2}{c}{[Fe/H] $\pm$ $\sigma$} &
\multicolumn{2}{c}{\parbox[c]{0.9cm}{\centering $RV$ $\pm$ $\sigma$ [\kms]}} \\
\noalign{\smallskip}\hline\noalign{\smallskip}
\object{AV\,Sgr}      & UVES   & 56136.2000040 & 5051 & $\pm$ 100 & 1.3 & 4.8 &    0.20 & $\pm$ 0.15 & 104 &    0.18 & $\pm$ 0.11 & 7  &    0.19 & $\pm$ 0.09 &     22.5 & $\pm$ 0.1  \\
\object{AV\,Sgr}      & UVES   & 56152.0841941 & 4999 & $\pm$ 97  & 1.3 & 4.8 &    0.20 & $\pm$ 0.18 & 111 &    0.10 & $\pm$ 0.30 & 9  &    0.18 & $\pm$ 0.15 &     25.0 & $\pm$ 0.1  \\
\object{AV\,Sgr}      & UVES   & 56168.0466724 & 4991 & $\pm$ 99  & 1.6 & 4.8 &    0.25 & $\pm$ 0.18 & 106 &    0.16 & $\pm$ 0.34 & 9  &    0.23 & $\pm$ 0.16 &     28.8 & $\pm$ 0.1  \\
\object{VY\,Sgr}      & FEROS  & 53522.2280817 & 6136 & $\pm$ 150 & 1.6 & 5.0 &    0.41 & $\pm$ 0.30 & 100 &    0.33 & $\pm$ 0.05 & 5  &    0.33 & $\pm$ 0.05 &   $-$7.8 & $\pm$ 1.5  \\
\object{VY\,Sgr}      & FEROS  & 53550.1640986 & 5678 & $\pm$ 146 & 1.0 & 4.9 &    0.33 & $\pm$ 0.33 & 73  &    0.27 & $\pm$ 0.14 & 6  &    0.28 & $\pm$ 0.13 &  $-$11.5 & $\pm$ 1.2  \\
\object{VY\,Sgr}      & FEROS  & 53551.2196741 & 5388 & $\pm$ 122 & 0.8 & 4.9 &    0.33 & $\pm$ 0.27 & 75  &    0.21 & $\pm$ 0.16 & 5  &    0.25 & $\pm$ 0.14 &   $-$5.8 & $\pm$ 0.9  \\
\object{VY\,Sgr}      & FEROS  & 53616.0926172 & 6269 & $\pm$ 149 & 1.9 & 5.0 &    0.40 & $\pm$ 0.31 & 77  &    0.43 & $\pm$ 0.22 & 5  &    0.42 & $\pm$ 0.18 &      9.5 & $\pm$ 0.9  \\
\object{VY\,Sgr}      & FEROS  & 54189.2893288 & 5317 & $\pm$ 118 & 0.9 & 4.9 &    0.24 & $\pm$ 0.25 & 118 &    0.30 & $\pm$ 0.15 & 4  &    0.29 & $\pm$ 0.13 &      0.1 & $\pm$ 0.8  \\
\object{VY\,Sgr}      & FEROS  & 54189.3107431 & 5269 & $\pm$ 99  & 0.8 & 3.9 &    0.21 & $\pm$ 0.19 & 114 &    0.26 & $\pm$ 0.29 & 5  &    0.23 & $\pm$ 0.16 &      0.3 & $\pm$ 0.8  \\
\object{VY\,Sgr}      & UVES   & 54923.3591348 & 5107 & $\pm$ 112 & 1.3 & 3.8 &    0.22 & $\pm$ 0.16 & 86  &    0.29 & $\pm$ 0.33 & 10 &    0.23 & $\pm$ 0.14 &    16.44 & $\pm$ 0.02 \\
\object{VY\,Sgr}      & UVES   & 56160.1810996 & 4882 & $\pm$ 86  & 0.3 & 4.8 &    0.07 & $\pm$ 0.18 & 79  &    0.11 & $\pm$ 0.23 & 5  &    0.09 & $\pm$ 0.14 &     38.7 & $\pm$ 0.1  \\
\object{VY\,Sgr}      & UVES   & 56162.1638617 & 5202 & $\pm$ 97  & 0.3 & 4.9 &    0.10 & $\pm$ 0.17 & 69  &    0.12 & $\pm$ 0.08 & 6  &    0.12 & $\pm$ 0.07 &     45.6 & $\pm$ 0.1  \\
\object{VY\,Sgr}      & UVES   & 56168.0638766 & 5419 & $\pm$ 99  & 0.9 & 3.8 &    0.21 & $\pm$ 0.15 & 127 &    0.28 & $\pm$ 0.17 & 11 &    0.25 & $\pm$ 0.11 &  $-$4.59 & $\pm$ 0.05 \\
\object{XX\,Sgr}      & UVES   & 54599.4044771 & 6313 & $\pm$ 167 & 1.8 & 3.5 &    0.01 & $\pm$ 0.14 & 82  & $-$0.07 & $\pm$ 0.15 & 15 & $-$0.03 & $\pm$ 0.10 &     6.01 & $\pm$ 0.04 \\
\object{XX\,Sgr}      & UVES   & 56054.2339884 & 5867 & $\pm$ 91  & 1.3 & 2.6 &    0.15 & $\pm$ 0.12 & 157 &    0.08 & $\pm$ 0.14 & 17 &    0.12 & $\pm$ 0.09 &    11.48 & $\pm$ 0.02 \\
\object{XX\,Sgr}      & UVES   & 56136.2215481 & 6259 & $\pm$ 76  & 1.1 & 2.3 &    0.10 & $\pm$ 0.11 & 118 &    0.07 & $\pm$ 0.15 & 20 &    0.09 & $\pm$ 0.09 &     0.91 & $\pm$ 0.05 \\
\object{XX\,Sgr}      & UVES   & 56152.0466585 & 5511 & $\pm$ 86  & 1.2 & 3.0 &    0.13 & $\pm$ 0.14 & 139 &    0.07 & $\pm$ 0.11 & 15 &    0.10 & $\pm$ 0.08 &    21.53 & $\pm$ 0.01 \\
\object{XX\,Sgr}      & UVES   & 56159.1275487 & 5465 & $\pm$ 90  & 1.1 & 3.5 &    0.02 & $\pm$ 0.16 & 131 & $-$0.07 & $\pm$ 0.15 & 20 & $-$0.03 & $\pm$ 0.11 &    29.60 & $\pm$ 0.01 \\
\object{Y\,Sgr}       & HARPS  & 53149.2916913 & 5405 & $\pm$ 91  & 1.2 & 4.1 & $-$0.04 & $\pm$ 0.16 & 102 & $-$0.05 & $\pm$ 0.21 & 13 & $-$0.05 & $\pm$ 0.13 &      8.1 & $\pm$ 1.3  \\
\object{Y\,Sgr}       & HARPS  & 53149.2960838 & 5447 & $\pm$ 99  & 1.3 & 4.9 & $-$0.04 & $\pm$ 0.17 & 100 & $-$0.03 & $\pm$ 0.20 & 10 & $-$0.04 & $\pm$ 0.13 &      8.1 & $\pm$ 1.3  \\
\object{Y\,Sgr}       & HARPS  & 53150.2840780 & 5672 & $\pm$ 120 & 2.0 & 4.9 & $-$0.00 & $\pm$ 0.14 & 70  &    0.02 & $\pm$ 0.18 & 10 &    0.01 & $\pm$ 0.11 &     17.8 & $\pm$ 1.9  \\
\object{Y\,Sgr}       & HARPS  & 53150.2888491 & 5602 & $\pm$ 193 & 1.8 & 4.9 & $-$0.04 & $\pm$ 0.14 & 78  & $-$0.06 & $\pm$ 0.02 & 5  & $-$0.06 & $\pm$ 0.02 &     17.8 & $\pm$ 1.9  \\
\object{Y\,Sgr}       & HARPS  & 53151.2398841 & 6239 & $\pm$ 145 & 2.0 & 3.4 &    0.11 & $\pm$ 0.13 & 100 &    0.11 & $\pm$ 0.07 & 13 &    0.11 & $\pm$ 0.06 &  $-$15.3 & $\pm$ 1.3  \\
\object{Y\,Sgr}       & HARPS  & 53151.2439506 & 6339 & $\pm$ 100 & 2.0 & 3.1 &    0.18 & $\pm$ 0.10 & 86  &    0.14 & $\pm$ 0.08 & 13 &    0.16 & $\pm$ 0.06 &  $-$15.4 & $\pm$ 1.3  \\
\object{Y\,Sgr}       & HARPS  & 53152.2965307 & 6049 & $\pm$ 131 & 2.0 & 4.4 &    0.04 & $\pm$ 0.17 & 122 &    0.02 & $\pm$ 0.14 & 14 &    0.03 & $\pm$ 0.11 &  $-$15.6 & $\pm$ 1.0  \\
\object{Y\,Sgr}       & HARPS  & 53152.2992080 & 6036 & $\pm$ 104 & 2.0 & 4.2 &    0.01 & $\pm$ 0.14 & 119 & $-$0.01 & $\pm$ 0.14 & 15 & $-$0.00 & $\pm$ 0.10 &  $-$15.5 & $\pm$ 1.0  \\
\object{Y\,Sgr}       & HARPS  & 53156.3224439 & 5891 & $\pm$ 208 & 1.9 & 4.9 &    0.06 & $\pm$ 0.14 & 75  &    0.07 & $\pm$ 0.23 & 9  &    0.06 & $\pm$ 0.12 &     14.2 & $\pm$ 1.9  \\
\object{Y\,Sgr}       & HARPS  & 53156.3273170 & 5906 & $\pm$ 201 & 2.0 & 3.7 &    0.12 & $\pm$ 0.11 & 74  &    0.10 & $\pm$ 0.11 & 9  &    0.11 & $\pm$ 0.08 &     14.0 & $\pm$ 1.9  \\
\object{Y\,Sgr}       & HARPS  & 53202.1441349 & 5616 & $\pm$ 164 & 1.8 & 4.9 & $-$0.02 & $\pm$ 0.13 & 77  & $-$0.03 & $\pm$ 0.15 & 11 & $-$0.03 & $\pm$ 0.10 &     17.5 & $\pm$ 1.8  \\
\object{Y\,Sgr}       & HARPS  & 53202.1489274 & 5613 & $\pm$ 125 & 1.6 & 4.4 & $-$0.03 & $\pm$ 0.11 & 74  & $-$0.05 & $\pm$ 0.11 & 8  & $-$0.04 & $\pm$ 0.08 &     17.5 & $\pm$ 1.8  \\
\object{Y\,Sgr}       & HARPS  & 53203.1454884 & 6305 & $\pm$ 99  & 2.0 & 3.3 &    0.16 & $\pm$ 0.11 & 96  &    0.18 & $\pm$ 0.11 & 13 &    0.17 & $\pm$ 0.08 &  $-$13.9 & $\pm$ 1.3  \\
\object{Y\,Sgr}       & HARPS  & 53203.1495670 & 6249 & $\pm$ 106 & 2.0 & 3.7 &    0.11 & $\pm$ 0.12 & 98  &    0.11 & $\pm$ 0.09 & 12 &    0.11 & $\pm$ 0.07 &  $-$14.0 & $\pm$ 1.3  \\
\object{Y\,Sgr}       & HARPS  & 53204.1180208 & 6132 & $\pm$ 140 & 2.0 & 4.9 & $-$0.03 & $\pm$ 0.17 & 102 & $-$0.10 & $\pm$ 0.17 & 15 & $-$0.06 & $\pm$ 0.12 &  $-$16.6 & $\pm$ 1.1  \\
\object{Y\,Sgr}       & HARPS  & 53204.1206929 & 6112 & $\pm$ 99  & 1.9 & 4.1 &    0.15 & $\pm$ 0.16 & 110 &    0.08 & $\pm$ 0.15 & 16 &    0.11 & $\pm$ 0.11 &  $-$16.6 & $\pm$ 1.1  \\
\object{Y\,Sgr}       & HARPS  & 53205.1591202 & 5728 & $\pm$ 87  & 2.0 & 4.9 &    0.02 & $\pm$ 0.16 & 133 & $-$0.07 & $\pm$ 0.23 & 16 & $-$0.01 & $\pm$ 0.13 &   $-$8.2 & $\pm$ 1.0  \\
\object{Y\,Sgr}       & STELLA & 54249.1753617 & 5826 & $\pm$ 274 & 0.6 & 1.9 &    0.34 & $\pm$ 0.26 & 114 &    0.34 & $\pm$ 0.20 & 12 &    0.34 & $\pm$ 0.16 &  $-$15.8 & $\pm$ 0.2  \\
\object{Y\,Sgr}       & STELLA & 54266.1156498 & 5962 & $\pm$ 287 & 0.7 & 1.8 &    0.30 & $\pm$ 0.20 & 82  &    0.31 & $\pm$ 0.34 & 15 &    0.30 & $\pm$ 0.17 &  $-$18.0 & $\pm$ 0.2  \\
\object{Y\,Sgr}       & STELLA & 54271.1069764 & 6182 & $\pm$ 161 & 1.9 & 4.1 &    0.17 & $\pm$ 0.20 & 82  &    0.15 & $\pm$ 0.23 & 14 &    0.16 & $\pm$ 0.15 &  $-$11.5 & $\pm$ 0.2  \\
\object{Y\,Sgr}       & HARPS  & 56212.9902382 & 5734 & $\pm$ 98  & 1.9 & 4.9 & $-$0.05 & $\pm$ 0.17 & 126 & $-$0.05 & $\pm$ 0.25 & 17 & $-$0.05 & $\pm$ 0.14 &   $-$9.5 & $\pm$ 0.9  \\
\object{Y\,Sgr}       & HARPS  & 56240.0036931 & 6249 & $\pm$ 97  & 2.0 & 5.0 &    0.07 & $\pm$ 0.17 & 108 &    0.04 & $\pm$ 0.16 & 14 &    0.06 & $\pm$ 0.11 &  $-$15.3 & $\pm$ 1.3  \\
\object{R\,TrA}       & HARPS  & 53150.1353086 & 5934 & $\pm$ 143 & 2.0 & 4.9 &    0.04 & $\pm$ 0.13 & 93  & $-$0.18 & $\pm$ 0.11 & 10 & $-$0.09 & $\pm$ 0.08 &      2.6 & $\pm$ 1.4  \\
\object{R\,TrA}       & HARPS  & 53150.1453265 & 5910 & $\pm$ 143 & 2.0 & 4.9 & $-$0.01 & $\pm$ 0.13 & 89  & $-$0.18 & $\pm$ 0.11 & 13 & $-$0.11 & $\pm$ 0.08 &      2.4 & $\pm$ 1.4  \\
\object{R\,TrA}       & HARPS  & 53152.1394611 & 5814 & $\pm$ 94  & 1.9 & 3.4 & $-$0.01 & $\pm$ 0.13 & 136 & $-$0.03 & $\pm$ 0.08 & 15 & $-$0.02 & $\pm$ 0.07 &  $-$11.6 & $\pm$ 0.9  \\
\object{R\,TrA}       & HARPS  & 53152.1469512 & 5824 & $\pm$ 95  & 1.8 & 3.2 &    0.04 & $\pm$ 0.12 & 139 &    0.04 & $\pm$ 0.09 & 17 &    0.04 & $\pm$ 0.07 &  $-$11.4 & $\pm$ 0.9  \\
\object{R\,TrA}       & HARPS  & 53154.1400956 & 6467 & $\pm$ 84  & 2.0 & 3.7 &    0.02 & $\pm$ 0.13 & 101 & $-$0.09 & $\pm$ 0.10 & 17 & $-$0.05 & $\pm$ 0.08 &  $-$22.7 & $\pm$ 1.3  \\
\object{R\,TrA}       & HARPS  & 53154.1496573 & 6452 & $\pm$ 96  & 2.0 & 2.9 &    0.07 & $\pm$ 0.10 & 96  &    0.02 & $\pm$ 0.10 & 14 &    0.04 & $\pm$ 0.07 &  $-$23.0 & $\pm$ 1.3  \\
\object{R\,TrA}       & HARPS  & 53156.1326474 & 5698 & $\pm$ 102 & 2.0 & 4.9 & $-$0.08 & $\pm$ 0.12 & 104 & $-$0.07 & $\pm$ 0.14 & 15 & $-$0.08 & $\pm$ 0.09 &   $-$1.6 & $\pm$ 1.1  \\
\object{R\,TrA}       & HARPS  & 53156.1414131 & 5745 & $\pm$ 96  & 2.0 & 4.8 & $-$0.02 & $\pm$ 0.13 & 106 & $-$0.03 & $\pm$ 0.19 & 17 & $-$0.03 & $\pm$ 0.11 &   $-$1.5 & $\pm$ 1.1  \\
\object{R\,TrA}       & HARPS  & 53201.0286305 & 5966 & $\pm$ 109 & 2.0 & 4.3 &    0.03 & $\pm$ 0.11 & 89  & $-$0.03 & $\pm$ 0.09 & 11 & $-$0.01 & $\pm$ 0.07 &      1.7 & $\pm$ 1.3  \\
\object{R\,TrA}       & HARPS  & 53202.0246343 & 6277 & $\pm$ 80  & 2.0 & 3.5 &    0.03 & $\pm$ 0.11 & 96  & $-$0.05 & $\pm$ 0.14 & 18 & $-$0.00 & $\pm$ 0.08 &  $-$25.6 & $\pm$ 1.2  \\
\object{R\,TrA}       & HARPS  & 53203.0415339 & 5799 & $\pm$ 100 & 2.0 & 3.5 &    0.01 & $\pm$ 0.13 & 133 &    0.00 & $\pm$ 0.07 & 14 &    0.00 & $\pm$ 0.06 &  $-$10.5 & $\pm$ 0.9  \\
\object{R\,TrA}       & HARPS  & 53204.0101104 & 5836 & $\pm$ 97  & 2.0 & 4.9 &    0.04 & $\pm$ 0.11 & 96  & $-$0.10 & $\pm$ 0.18 & 14 & $-$0.00 & $\pm$ 0.09 &      3.0 & $\pm$ 1.3  \\
\object{R\,TrA}       & HARPS  & 53205.0362291 & 6450 & $\pm$ 98  & 2.0 & 3.5 &    0.01 & $\pm$ 0.10 & 98  & $-$0.06 & $\pm$ 0.16 & 20 & $-$0.01 & $\pm$ 0.09 &  $-$24.1 & $\pm$ 1.3  \\
\object{R\,TrA}       & HARPS  & 53206.0233396 & 5941 & $\pm$ 79  & 1.8 & 2.8 &    0.12 & $\pm$ 0.11 & 129 &    0.10 & $\pm$ 0.14 & 15 &    0.12 & $\pm$ 0.09 &  $-$17.5 & $\pm$ 0.9  \\
\object{R\,TrA}       & FEROS  & 55283.3432676 & 6096 & $\pm$ 90  & 2.0 & 4.9 & $-$0.04 & $\pm$ 0.21 & 86  & $-$0.22 & $\pm$ 0.20 & 11 & $-$0.13 & $\pm$ 0.14 &  $-$22.0 & $\pm$ 1.1  \\
\object{RZ\,Vel}      & HARPS  & 53149.9869296 & 5321 & $\pm$ 97  & 0.7 & 3.3 &    0.13 & $\pm$ 0.10 & 150 &    0.04 & $\pm$ 0.14 & 9  &    0.10 & $\pm$ 0.08 &     16.7 & $\pm$ 0.9  \\
\object{RZ\,Vel}      & HARPS  & 53151.9991635 & 5163 & $\pm$ 97  & 1.2 & 4.1 &    0.07 & $\pm$ 0.12 & 152 &    0.05 & $\pm$ 0.07 & 7  &    0.05 & $\pm$ 0.06 &     25.1 & $\pm$ 1.1  \\
\object{RZ\,Vel}      & HARPS  & 53153.9963816 & 5041 & $\pm$ 78  & 1.3 & 4.9 &    0.06 & $\pm$ 0.14 & 125 &    0.02 & $\pm$ 0.19 & 10 &    0.04 & $\pm$ 0.11 &     33.5 & $\pm$ 1.3  \\
\object{RZ\,Vel}      & HARPS  & 53155.9860314 & 4981 & $\pm$ 85  & 1.6 & 4.9 &    0.02 & $\pm$ 0.20 & 103 & $-$0.13 & $\pm$ 0.25 & 7  & $-$0.04 & $\pm$ 0.16 &     41.9 & $\pm$ 1.8  \\
\object{RZ\,Vel}      & HARPS  & 53200.9387705 & 5409 & $\pm$ 106 & 1.0 & 4.9 &    0.12 & $\pm$ 0.14 & 77  &    0.05 & $\pm$ 0.23 & 10 &    0.10 & $\pm$ 0.12 &     41.3 & $\pm$ 1.8  \\
\object{RZ\,Vel}      & HARPS  & 53201.9423125 & 5422 & $\pm$ 100 & 0.9 & 4.9 &    0.08 & $\pm$ 0.16 & 121 &    0.07 & $\pm$ 0.13 & 9  &    0.08 & $\pm$ 0.10 &     36.7 & $\pm$ 1.4  \\
\object{RZ\,Vel}      & HARPS  & 53202.9406629 & 6245 & $\pm$ 98  & 1.2 & 3.8 &    0.11 & $\pm$ 0.10 & 102 &    0.10 & $\pm$ 0.11 & 12 &    0.10 & $\pm$ 0.07 &     29.6 & $\pm$ 1.4  \\
\object{RZ\,Vel}      & HARPS  & 53203.9388780 & 6314 & $\pm$ 192 & 1.2 & 5.0 &    0.06 & $\pm$ 0.20 & 99  &    0.07 & $\pm$ 0.15 & 9  &    0.06 & $\pm$ 0.12 &     13.9 & $\pm$ 1.2  \\
\object{RZ\,Vel}      & HARPS  & 53204.9383376 & 6474 & $\pm$ 120 & 1.4 & 4.9 &    0.02 & $\pm$ 0.15 & 101 &    0.04 & $\pm$ 0.10 & 10 &    0.03 & $\pm$ 0.08 &      0.6 & $\pm$ 1.2  \\
\object{RZ\,Vel}      & HARPS  & 53205.9402008 & 6225 & $\pm$ 100 & 1.5 & 4.9 &    0.14 & $\pm$ 0.14 & 93  &    0.16 & $\pm$ 0.08 & 8  &    0.16 & $\pm$ 0.07 &   $-$2.6 & $\pm$ 1.1  \\
\object{RZ\,Vel}      & FEROS  & 55280.0861358 & 5199 & $\pm$ 99  & 1.2 & 4.9 & $-$0.05 & $\pm$ 0.28 & 93  &    0.08 & $\pm$ 0.17 & 6  &    0.04 & $\pm$ 0.15 &     49.1 & $\pm$ 2.3  \\
\object{RZ\,Vel}      & HARPS  & 56606.1850289 & 5308 & $\pm$ 110 & 1.6 & 4.9 &    0.18 & $\pm$ 0.20 & 54  &    0.16 & $\pm$ 0.27 & 5  &    0.18 & $\pm$ 0.16 &     47.9 & $\pm$ 2.6  \\
\hline
\end{tabular}}
\tablefoot{The first three columns give the name of the target, the
spectroscopic dataset, and the Modified Julian Date at which the spectrum was
collected. Columns 4, 5, and 6 give the effective temperature and its
standard deviation, the surface gravity, and the microturbulent velocity.
The columns 7--8 and 9--10 list both \ion{Fe}{i} and \ion{Fe}{ii}
abundances and their standard deviations together with the number of lines
adopted for the measurements. Column eleven gives the weighted mean
of \ion{Fe}{i} and \ion{Fe}{ii} abundances (weighted by 1/$\sigma^2$)
with its intrinsic error, while the last column gives the
radial-velocity measurement and the respective uncertainty.}
\end{table*}

\begin{table*}
\centering
\caption{High resolution spectra only adopted for radial-velocity
measurements.}
\label{noparamscontrolsample}
{\scriptsize 
\begin{tabular}{lcc  r@{ }l}
\noalign{\smallskip}\hline\hline\noalign{\smallskip}
Name & Dataset &
\parbox[c]{0.5cm}{\centering MJD [d]} &
\multicolumn{2}{c}{\parbox[c]{0.9cm}{\centering $RV$ $\pm$ $\sigma$ [\kms]}} \\
\noalign{\smallskip}\hline\noalign{\smallskip}
\object{V340\,Ara}    & FEROS  & 53520.2213553 &   $-$99.7 & $\pm$ 7.5   \\
\object{V340\,Ara}    & FEROS  & 53521.2126174 &   $-$96.7 & $\pm$ 5.8   \\
\object{V340\,Ara}    & FEROS  & 53522.2102258 &   $-$93.9 & $\pm$ 0.8   \\
\object{V340\,Ara}    & FEROS  & 53523.2653732 &   $-$89.8 & $\pm$ 0.7   \\
\object{V340\,Ara}    & FEROS  & 53524.1628015 &   $-$85.8 & $\pm$ 0.8   \\
\object{V340\,Ara}    & FEROS  & 53549.3036618 &   $-$63.6 & $\pm$ 14.4  \\
\object{V340\,Ara}    & FEROS  & 53550.1532155 &   $-$64.6 & $\pm$ 0.6   \\
\object{V340\,Ara}    & FEROS  & 53551.1938192 &   $-$60.0 & $\pm$ 0.6   \\
\object{V340\,Ara}    & FEROS  & 53551.2732169 &   $-$59.1 & $\pm$ 0.5   \\
\object{V340\,Ara}    & FEROS  & 53552.3038235 &   $-$55.0 & $\pm$ 0.8   \\
\object{V340\,Ara}    & FEROS  & 53556.3161638 &   $-$65.7 & $\pm$ 1.2   \\
\object{V340\,Ara}    & FEROS  & 53556.3192516 &   $-$65.9 & $\pm$ 1.4   \\
\object{V340\,Ara}    & FEROS  & 53597.0374439 &   $-$63.3 & $\pm$ 1.5   \\
\object{V340\,Ara}    & FEROS  & 53599.0975102 &   $-$73.3 & $\pm$ 5.2   \\
\object{V340\,Ara}    & FEROS  & 53601.1158191 &   $-$93.3 & $\pm$ 17.8  \\
\object{V340\,Ara}    & FEROS  & 53602.1093876 &   $-$98.8 & $\pm$ 13.7  \\
\object{V340\,Ara}    & FEROS  & 53603.1099409 &  $-$100.2 & $\pm$ 7.6   \\
\object{V340\,Ara}    & FEROS  & 53609.0378955 &   $-$80.0 & $\pm$ 0.5   \\
\object{V340\,Ara}    & FEROS  & 53609.0424397 &   $-$80.0 & $\pm$ 0.5   \\
\object{V340\,Ara}    & FEROS  & 53615.0266904 &   $-$54.1 & $\pm$ 0.8   \\
\object{V340\,Ara}    & FEROS  & 53616.0536654 &   $-$51.5 & $\pm$ 1.0   \\
\object{V340\,Ara}    & FEROS  & 53617.0696186 &   $-$55.9 & $\pm$ 1.3   \\
\object{V340\,Ara}    & FEROS  & 53618.0534958 &   $-$64.5 & $\pm$ 1.3   \\
\object{V340\,Ara}    & FEROS  & 53619.0473470 &   $-$67.9 & $\pm$ 1.0   \\
\object{V340\,Ara}    & FEROS  & 53621.1257226 &   $-$81.0 & $\pm$ 17.6  \\
\object{RS\,Pup}      & HARPS  & 53051.1389691 &      33.0 & $\pm$ 4.4   \\
\object{UZ\,Sct}      & FEROS  & 53520.2424019 &      16.3 & $\pm$ 1.0   \\
\object{UZ\,Sct}      & FEROS  & 53521.2331590 &      21.1 & $\pm$ 0.9   \\
\object{UZ\,Sct}      & FEROS  & 53522.2357550 &      26.9 & $\pm$ 1.0   \\
\object{UZ\,Sct}      & FEROS  & 53523.2847746 &      33.3 & $\pm$ 1.1   \\
\object{UZ\,Sct}      & FEROS  & 53524.1926044 &      39.3 & $\pm$ 1.0   \\
\object{UZ\,Sct}      & FEROS  & 53549.3253449 &      16.3 & $\pm$ 1.1   \\
\object{UZ\,Sct}      & FEROS  & 53550.1730666 &      18.3 & $\pm$ 1.0   \\
\object{UZ\,Sct}      & FEROS  & 53551.2382008 &      24.2 & $\pm$ 1.0   \\
\object{UZ\,Sct}      & FEROS  & 53551.2431731 &      24.1 & $\pm$ 1.0   \\
\object{UZ\,Sct}      & FEROS  & 53551.2475029 &      24.4 & $\pm$ 1.0   \\
\object{UZ\,Sct}      & FEROS  & 53551.2518292 &      24.4 & $\pm$ 0.9   \\
\object{UZ\,Sct}      & FEROS  & 53552.3179119 &      30.9 & $\pm$ 1.0   \\
\object{UZ\,Sct}      & FEROS  & 53556.3564325 &      53.8 & $\pm$ 1.6   \\
\object{UZ\,Sct}      & FEROS  & 53556.3615687 &      54.1 & $\pm$ 1.3   \\
\object{UZ\,Sct}      & FEROS  & 53577.2848080 &      26.4 & $\pm$ 1.0   \\
\object{UZ\,Sct}      & FEROS  & 53602.1340975 &      59.5 & $\pm$ 2.2   \\
\object{UZ\,Sct}      & FEROS  & 53603.1347598 &      60.2 & $\pm$ 2.1   \\
\object{UZ\,Sct}      & FEROS  & 53609.0764160 &      18.2 & $\pm$ 0.8   \\
\object{UZ\,Sct}      & FEROS  & 53609.0815589 &      17.7 & $\pm$ 0.9   \\
\object{UZ\,Sct}      & FEROS  & 53615.0471044 &      51.5 & $\pm$ 1.6   \\
\object{UZ\,Sct}      & FEROS  & 53616.1000454 &      56.4 & $\pm$ 2.0   \\
\object{UZ\,Sct}      & FEROS  & 53617.1424676 &      60.6 & $\pm$ 2.2   \\
\object{UZ\,Sct}      & FEROS  & 53618.1711558 &      58.5 & $\pm$ 2.1   \\
\object{UZ\,Sct}      & FEROS  & 53619.1658348 &      48.0 & $\pm$ 1.3   \\
\object{UZ\,Sct}      & FEROS  & 53620.1685455 &      41.8 & $\pm$ 1.3   \\
\object{UZ\,Sct}      & FEROS  & 53621.0360640 &      34.0 & $\pm$ 0.9   \\
\object{AV\,Sgr}      & FEROS  & 53520.2272967 &      42.6 & $\pm$ 2.0   \\
\object{AV\,Sgr}      & FEROS  & 53521.2192697 &      45.1 & $\pm$ 2.4   \\
\object{AV\,Sgr}      & FEROS  & 53522.2164015 &      34.9 & $\pm$ 1.4   \\
\object{AV\,Sgr}      & FEROS  & 53523.2705083 &      21.9 & $\pm$ 1.2   \\
\object{AV\,Sgr}      & FEROS  & 53524.1671280 &      16.9 & $\pm$ 1.2   \\
\object{AV\,Sgr}      & FEROS  & 53550.1575431 &      39.0 & $\pm$ 1.6   \\
\object{AV\,Sgr}      & FEROS  & 53551.1663433 &      43.5 & $\pm$ 1.4   \\
\object{AV\,Sgr}      & FEROS  & 53551.1768545 &      43.2 & $\pm$ 2.0   \\
\object{AV\,Sgr}      & FEROS  & 53551.2794534 &      43.0 & $\pm$ 2.3   \\
\object{AV\,Sgr}      & FEROS  & 53552.3294632 &      44.3 & $\pm$ 2.8   \\
\object{AV\,Sgr}      & FEROS  & 53556.3258803 &    $-$2.0 & $\pm$ 1.2   \\
\object{AV\,Sgr}      & FEROS  & 53556.3363054 &    $-$1.3 & $\pm$ 1.2   \\
\object{AV\,Sgr}      & FEROS  & 53577.2606543 &      17.5 & $\pm$ 1.0   \\
\object{AV\,Sgr}      & FEROS  & 53597.0421718 &      42.0 & $\pm$ 1.9   \\
\object{AV\,Sgr}      & FEROS  & 53599.1027847 &      36.9 & $\pm$ 1.6   \\
\object{AV\,Sgr}      & FEROS  & 53603.1196862 &    $-$7.9 & $\pm$ 1.5   \\
\object{AV\,Sgr}      & FEROS  & 53609.0474987 &      24.0 & $\pm$ 1.1   \\
\object{AV\,Sgr}      & FEROS  & 53609.0537392 &      23.0 & $\pm$ 1.3   \\
\object{AV\,Sgr}      & FEROS  & 53615.0318342 &      28.6 & $\pm$ 1.4   \\
\object{AV\,Sgr}      & FEROS  & 53616.0863475 &      22.1 & $\pm$ 1.1   \\
\object{AV\,Sgr}      & FEROS  & 53617.1247625 &      11.1 & $\pm$ 1.3   \\
\object{AV\,Sgr}      & FEROS  & 53618.1537678 &       3.5 & $\pm$ 2.2   \\
\object{AV\,Sgr}      & FEROS  & 53619.1482114 &    $-$7.9 & $\pm$ 1.3   \\
\object{AV\,Sgr}      & FEROS  & 53620.1526565 &    $-$3.7 & $\pm$ 1.1   \\
\object{VY\,Sgr}      & FEROS  & 53520.2344426 &      17.0 & $\pm$ 0.9   \\
\object{VY\,Sgr}      & FEROS  & 53521.2256094 &       9.0 & $\pm$ 1.1   \\
\object{VY\,Sgr}      & FEROS  & 53523.2768347 &   $-$10.6 & $\pm$ 1.1   \\
\object{VY\,Sgr}      & FEROS  & 53524.1851391 &    $-$4.8 & $\pm$ 0.9   \\
\hline\noalign{\smallskip}
\multicolumn{5}{r}{\it {\footnotesize continued on next page}} \\
\end{tabular}}
\end{table*}
\addtocounter{table}{-1}
\begin{table*}[p]
\centering
\caption[]{continued.}
{\scriptsize 
\begin{tabular}{lcc  r@{ }l}
\noalign{\smallskip}\hline\hline\noalign{\smallskip}
Name & Dataset &
\parbox[c]{0.5cm}{\centering MJD [d]} &
\multicolumn{2}{c}{\parbox[c]{0.9cm}{\centering $RV$ $\pm$ $\sigma$ [\kms]}} \\
\noalign{\smallskip}\hline\noalign{\smallskip}
\object{VY\,Sgr}      & FEROS  & 53549.3172633 &    $-$7.4 & $\pm$ 1.7   \\
\object{VY\,Sgr}      & FEROS  & 53551.2057981 &    $-$5.7 & $\pm$ 0.9   \\
\object{VY\,Sgr}      & FEROS  & 53551.2141933 &    $-$5.7 & $\pm$ 0.9   \\
\object{VY\,Sgr}      & FEROS  & 53551.2252183 &    $-$5.9 & $\pm$ 1.0   \\
\object{VY\,Sgr}      & FEROS  & 53552.3099053 &       2.6 & $\pm$ 0.7   \\
\object{VY\,Sgr}      & FEROS  & 53556.3425215 &      32.3 & $\pm$ 1.1   \\
\object{VY\,Sgr}      & FEROS  & 53556.3485575 &      32.2 & $\pm$ 1.5   \\
\object{VY\,Sgr}      & FEROS  & 53577.2677235 &   $-$10.9 & $\pm$ 1.4   \\
\object{VY\,Sgr}      & FEROS  & 53597.0481329 &      30.8 & $\pm$ 1.6   \\
\object{VY\,Sgr}      & FEROS  & 53599.1090806 &      39.3 & $\pm$ 2.5   \\
\object{VY\,Sgr}      & FEROS  & 53601.1784165 &      18.9 & $\pm$ 1.0   \\
\object{VY\,Sgr}      & FEROS  & 53602.1267713 &      14.0 & $\pm$ 0.9   \\
\object{VY\,Sgr}      & FEROS  & 53603.1265701 &       3.5 & $\pm$ 1.0   \\
\object{VY\,Sgr}      & FEROS  & 53609.0614284 &      21.0 & $\pm$ 1.0   \\
\object{VY\,Sgr}      & FEROS  & 53615.0381043 &      16.7 & $\pm$ 1.0   \\
\object{VY\,Sgr}      & FEROS  & 53617.1312705 &    $-$7.3 & $\pm$ 1.7   \\
\object{VY\,Sgr}      & FEROS  & 53618.1641875 &   $-$10.4 & $\pm$ 1.3   \\
\object{VY\,Sgr}      & FEROS  & 53619.1559350 &    $-$4.5 & $\pm$ 0.9   \\
\object{VY\,Sgr}      & FEROS  & 53620.1598186 &       3.1 & $\pm$ 0.9   \\
\object{Y\,Sgr}       & HARPS  & 56553.1423824 &   $-$13.4 & $\pm$ 1.0   \\
\hline\noalign{\smallskip}
\end{tabular}}
\tablefoot{From left to right the columns give the name of the target, the
spectroscopic dataset, the Modified Julian Date, and the
radial-velocity measurements together with their standard deviations.}
\end{table*}


\end{document}